\documentstyle[12pt,psfig]{article}
\oddsidemargin--1mm
\newcommand{\Rs}{\relax{\rm I\kern-.18em R}}
\newcommand{\<}{\langle}
\def\>{\rangle}
  
\newcommand{\Z}{Z\!\!\! Z}
\newcommand{\pl}{\partial}
\newcommand{\tr}{{\rm tr}\,}
\newcommand{\be}{\begin{equation}}
\newcommand{\ee}{\end{equation}}
\newcommand{\ba}{\begin{eqnarray}}
\newcommand{\ea}{\end{eqnarray}}
\def\1{\relax{\rm 1\kern-.27em I}}
\newcommand{\ph}{{\rm PS}_{{\rm phys}}}
\newcommand{\mbf}[1]{\mbox{\boldmath$ #1$}}

\begin{document}
\begin{center}{\Large\bf
GEOMERTY OF THE PHYSICAL PHASE SPACE

\vskip 0.5cm
 IN QUANTUM GAUGE SYSTEMS}

\vskip 2cm

{\bf Sergei V. SHABANOV}


\vskip 0.5cm
{\em Department of Mathematics, University of Florida,
Gainesville, FL 32611-2085, USA}
\footnote{on leave from Laboratory of Theoretical Physics,
Joint Institute for Nuclear Research, Dubna, Russia; 
email: shabanov@phys.ufl.edu}

\end{center}

\vskip 3cm
\begin{abstract}
The physical phase space in gauge systems is studied.
Effects caused by a non-Euclidean geometry of the physical
phase space in quantum gauge models are described in the
operator and path integral formalisms. The projection on
the Dirac gauge invariant states is used to derive a necessary 
modification of the Hamiltonian path integral in gauge 
theories of the Yang-Mills type with fermions  that takes 
into account the non-Euclidean
geometry of the physical phase space. 
The new path integral is
applied to resolve the Gribov obstruction. Applications to the 
Kogut-Susskind lattice gauge theory are given. The basic ideas
are illustrated with examples accessible for non-specialists. 

\end{abstract}

\newpage

{\small\tableofcontents}

\vskip 2cm
\section{Introduction}

Yang-Mills theory and gauge theories in general play the most
profound role in our present understanding of the universe. Nature
is quantum in its origin so any classical gauge model should 
be promoted to its quantum version in order to be used as
a model of the physical reality. We usually do this by applying
one or another quantization recipe which we believe to lead
to a consistent quantum theory. In general, quantization is by no means
unique and should be regarded as a theoretical way to guess the
true theory. We certainly expect any quantization procedure to
comply with some physical principles, like the correspondence
principle, gauge invariance, etc. And finally, the resulting quantum 
theory should not have any internal contradiction. All these conditions
are rather loose to give us a unique quantization recipe.

The simplest way to quantize a theory is to use canonical
quantization based on the Hamiltonian formalism of the classical
theory. Given a set of canonical coordinates and momenta,
one promotes them into a set of self-adjoint operators satisfying
the Heisenberg commutation relations. Any classical observable,
as a function on the phase space, becomes
a function of the canonical operators. Due to the noncommutativity
of the canonical operators, 
there is no unique correspondence between classical
and quantum observables. One can modify a quantum observable
by adding some operators proportional to commutators of the 
canonical operators. This will not make any difference in the formal
limit when the Planck constant, which ``measures'' 
the noncommutativity of the canonical variables, vanishes.
In classical mechanics, the Hamiltonian equations of motion
are covariant under general canonical transformations. So there is
no preference of choosing  a particular set of canonical variables
to span the phase space of the system. It was, however, found
in practice that canonical quantization would be successful only when
applied with the phase space coordinates referring to a Cartesian
system of axes and not to more general curvilinear coordinates
\cite{dirac}. On the other hand, a global Cartesian coordinate
system can be found only 
if the phase space of the system is {\em Euclidean}.
This comprises a fundamental restriction on the canonical quantization
recipe.  

Another quantization method is due to Feynman \cite{feyn}
which, at first
sight, seems to avoid the use of noncommutative phase space
variables. Given a classical action for a system in the Lagrangian
form, which is usually assumed to be quadratic in velocities, 
the quantum mechanical transition
amplitude between two fixed points of the configuration space is
determined by a sum over all continuous paths connecting these points
with weight being the phase exponential of the classical action
divided by the Planck constant. Such
a sum is called the Lagrangian path integral.
If the action is taken in the 
Hamiltonian form, the sum 
is extended over all phase-space trajectories connecting the initial
and final states of the system and, in addition, this sum also involves
integration over the momenta of the final and initial states. Recall
that a phase-space point specifies uniquely a state of a Hamiltonian
system in classical theory. Such a sum is
called the Hamiltonian path integral. One should however keep in mind
that such a definition of the Hamiltonian path integral (as a sum
over paths in a phase space) is formal. One usually defines it
by a specific finite dimensional integral on the time lattice
rather than a sum over paths in a phase space. 
The correspondence principle
follows from the stationary phase approximation to the sum 
over paths when the classical action is much greater than
the Planck constant. 
The stationary point, if any, of the action is a classical trajectory.
So the main contribution to the sum over paths comes from
paths fluctuating around the classical trajectory.
But again, one could add some terms
of higher orders in the Planck constant to the classical action
without changing the classical limit.
  
Despite this ambiguity, Feynman's sum over paths looks like
a miracle because no noncommutative phase-space variables are
involved in the quantum mechanical description. It just seems
like the knowledge of a classical theory is sufficient to obtain
the corresponding quantum theory.
Moreover, the phase-space
path integral with the local Liouville measure seems to enjoy 
another wonderful property of being invariant
under general canonical transformations. Recall that the Liouville
measure is defined as a volume element on the phase space
which is invariant under canonical transformations.  
One may tend to the conclusion that the phase-space path integral
provides a resolution of the aforementioned problem of the canonical
quantization. This is, however, a trap hidden by
the {\em formal} definition of the path integral measure
as a product of the Liouville measures at each moment of time. 
For systems with 
one degree of freedom one can easily find a canonical transformation
that turns a generic Hamiltonian into one for a free particle or
harmonic oscillator. It is obvious that the quantum mechanics
of a generic one-dimensional system is not that of the harmonic
oscillator. From this point of view the Feynman integral should
also be referred to the Cartesian  coordinates on the phase space,
unless the formal measure is properly modified \cite{kld,kld2,kl88}.

So, we conclude that the existence of the Cartesian coordinates 
that span the phase space is indeed important for both the canonical
and path integral quantization. When quantizing a system by 
one of the above methods, one often makes an {\em implicit}
assumption that the phase space of the physical degrees of
freedom is Euclidean, i.e., it admits a global Cartesian system of 
coordinates. We will show that, in general, this assumption is
not justified for physical degrees of freedom in 
systems with gauge symmetry. Hence,
all the aforementioned subtleties of the path integral
formalism play a major role in the path integral quantization
of gauge systems. The true geometry of the physical phase space 
must be taken into account
in quantum theory, which significantly affects 
the corresponding  path integral formalism.

Gauge theories have a characteristic property that the 
Euler-Lagrange equations of motion 
are covariant under symmetry transformations whose
parameters are general functions of time. Therefore
the equations of motion
do not determine completely
the time evolution of all degrees of freedom. A solution under
specified initial conditions on the positions and velocities
would contain a set of general functions of time, which is
usually called gauge arbitrariness \cite{diraclec}. 
Yet, some of the equations
of motion have no second time derivatives, so they are {\em constraints}
on the initial positions and velocities. In the Hamiltonian formalism,
one has accordingly constraints on the canonical variables \cite{diraclec}.
The constraints in gauge theories enjoy an additional property.
Their Poisson bracket with the canonical Hamiltonian as well
as among themselves 
 vanishes on the surface of the constraints in the 
phase space (first-class constraints according to the Dirac
terminology \cite{diraclec}). 
Because of this property,
the Hamiltonian can be modified by adding to it
a linear combination of the constraints with general coefficients,
called the Lagrange multipliers of the constraints or 
just gauge functions or variables.
This, in turn, implies that the Hamiltonian equations of motion 
would also contain a gauge arbitrariness associated with each
independent constraint.  By changing the gauge functions 
one changes the state of the system if the latter is defined as
a point in the phase space. These are the gauge transformations
in the phase space. On the other hand,
the physical state of the system cannot
depend on the gauge arbitrariness. If one wants to associate a
{\em single} point of the phase space with each physical state of the
system, one is necessarily led to the conclusion that the physical
phase space is a subspace of the constraint surface in the total
phase space of the system.  Making it more precise, the physical
phase space should be the quotient of the constraint surface 
by the gauge transformations generated by {\em all}
independent constraints. Clearly, the quotient space will
generally not be a Euclidean space. One can naturally expect 
some new phenomena in quantum gauge theories associated with
a non-Euclidean geometry of the phase space of the physical
degrees of freedom because 
quantum theories determined by the same Hamiltonian
as a function of canonical variables may be different if they
have different phase spaces, e.g., the plane and spherical phase spaces. 
This peculiarity of the Hamiltonian dynamics of gauge systems
looks interesting and quite unusual for dynamical models 
used in fundamental physics, and certainly deserves a better
understanding.
 
In this review we study the geometrical structure of the 
physical phase space in gauge theories and its role in the
corresponding quantum dynamics. Since the path integral
formalism is the main tool in modern fundamental physics,
special attention is paid to 
the path integral formalism for gauge models whose physical phase space
is not Euclidean. This would lead us to a modification of the 
conventional Hamiltonian path integral used in gauge theories,
which takes into account the geometrical structure of the 
physical phase space. We also propose a general method to
derive such a path integral that is in a full correspondence
with the Dirac operator formalism for gauge theories.
Our analysis is mainly focused on soluble  gauge models
where the results obtained by different methods, say, by the operator  
or path integral formalisms, are easy to compare, and thereby,
one has a mathematical control of the formalism being developed.
In realistic gauge theories, a major problem is to make the quantum 
theory well-defined nonperturbatively. Since the perturbation
theory is not sensitive to the global geometrical properties of the 
physical phase space -- which is just a fact for the theory in hand --
we do not go into speculations about the realistic case, because
there is  an unsolved problem of the nonperturbative definition of the path 
integral in a strongly interacting field theory, and limit the discussion
to reviewing existing approaches to this hard problem. However, 
we consider a Hamiltonian lattice gauge theory due to Kogut
and Susskind and extend the concepts developed for low-dimensional
gauge models to it. In this case we have a rigorous
definition of the path integral measure because the 
system has a finite number of degrees of freedom. The continuum
limit still remains as  a problem to reach the goal of constructing
a nonperturbative path integral in gauge field theory that takes
into account the non-Euclidean geometry of the physical
phase space. Nevertheless from the analysis of simple gauge models,
as well as from the general method we propose to derive the path
integral, one might anticipate some new properties of the modified
path integral that would essentially be due to the non-Euclidean
geometry of the physical phase space.

The review is organized as follows. In section 2 a definition
of the physical phase space is given. Section 3 is devoted
to mechanical models with one physical degree of freedom. 
In this example, the physical phase space is shown to be a cone unfoldable
into a half-plane. Effects of the conic phase space on classical
and quantum dynamics are studied. In section 4 we discuss
the physical phase space structure of gauge systems
with several physical degrees of freedom. Special attention
is paid to a new dynamical phenomenon which we call a kinematic
coupling. The point being is that, though physical degrees of
freedom  are not coupled in the Hamiltonian, i.e., they
are dynamically decoupled, nonetheless their dynamics
is not independent due to a non-Euclidean structure of their
phase (a kinematic coupling). This phenomenon is analyzed
as in classical mechanics as in quantum theory. It is shown
that the kinematic coupling has a significant effect
on the spectrum of the physical quantum Hamiltonian.   
In section 5 the physical phase space of Yang-Mills theory
in a cylindrical spacetime is studied. A physical configuration
space, known as the gauge orbit space, is also analyzed in detail.
Section 6 is devoted to artifacts which one may encounter upon
a dynamical description that uses a gauge fixing
(e.g., the Gribov problem). We emphasize the importance of establishing
the geometrical structure of the physical phase space {\em prior} to
fixing a gauge to remove nonphysical degrees of freedom.
With simple examples, we illustrate dynamical artifacts
that might occur through a bad, though formally admissible,
choice of the gauge. A relation between the Gribov problem,
topology of the gauge orbits and coordinate singularities of  
the symplectic structure on the physical phase space is discussed
in detail.

In section 7 the Dirac quantization method is applied to all 
the models. Here we also compare
the so called reduced phase space quantization (quantization
after eliminating all nonphysical degrees of freedom)
and the Dirac approach. Pitfalls of the reduced phase space
quantization are listed and illustrated with examples.
Section 8 is devoted to the path integral formalism in gauge
theories. The main goal is a general method which 
allows one to develop a path integral formalism {\em equivalent}
to the Dirac operator method. The new path integral formalism
is shown to resolve the Gribov obstruction to the conventional
Faddeev-Popov path integral quantization of gauge theories
of the Yang-Mills type (meaning that the gauge transformations
are {\em linear} in the total phase space). For soluble
gauge models, the spectra and partition functions are calculated
by means of the Dirac operator method and the new path integral
formalism. The results are compared and shown to be the same.
The path integral formalism developed is applied to instantons
and minisuperspace cosmology. In section 9 fermions are 
included into the path integral formalism. We
observe that the kinematic coupling induced by
a non-Euclidean structure of the physical phase space
occurs for {\em both} fermionic and bosonic physical degrees of
freedom, which has an important effects on quantum dynamics
of fermions. In particular, the modification of fermionic Green's functions
in quantum theory is studied in detail.   
Section 10 contains a review of geometrical properties of
the gauge orbit space in realistic classical Yang-Mills theories.
Various approaches to describe the effects of the non-Euclidean
geometry of the orbit space in quantum theory are discussed.
The path integral formalism of section 8 is applied to the 
Kogut-Susskind lattice Yang-Mills theory. Conclusions are given
in section 11.

The material of the review is presented in a pedagogical fashion
and is believed to be easily accessible for nonspecialists.
However a basic knowledge of quantum mechanics and group theory
might be useful, although the necessary facts from the group
theory are provided and explained as needed. For readers who
are not keen to look into technical details and would only be
interested to glean the basic physical and mathematical ideas
discussed in the review, it might be convenient to look through
sections 2, 3, 6.1, 7.1, 7.2, section 8 (without 8.5 and 8.6),
9.1, 9.3 and sections 10, 11. 

One of the widely used quantization techniques, the BRST quantization
(see, e.g., \cite{marc}) is not discussed in the review. Partially, this
is because it is believed that on the operator level the 
BRST formalism is equivalent to the Dirac method and, hence,
the physical phenomena associated with a non-Euclidean
geometry of the physical phase space can be studied by either
of these techniques. The Dirac method is technically simpler,
while the BRST formalism is more involved as it requires
an extension of the original phase space rather than its reduction.
The BRST formalism has been proved to be useful when an
explicit relativistic invariance of the perturbative path integral
has to be maintained. Since the discovery of the BRST symmetry \cite{215,216}
of the Faddeev-Popov effective action and its successful
application to perturbation theory \cite{219}, there existed a believe
that the path integral for theories with local symmetries can
be defined as a path integral for an effective theory
with the global BRST symmetry. It was pointed out \cite{neub,fuji83} 
that this equivalence breaks down beyond the perturbation
theory. The conventional BRST action may give rise to a zero
partition function as well as to
vanishing expectation values of physical operators.
The reason for such a failure boils down to the nontrivial
topology of the gauge orbit space. Therefore
a study of the role of the gauge orbit
space in the BRST formalism is certainly important. In this
regard one should point out the following. There is a mathematical
problem within the BRST formalism of constructing a proper inner
product for physical states \cite{brst1}. This problem appears to be
relevant for the BRST quantization scheme when the Gribov
problem is present \cite{brst2}. An interesting approach 
to the inner product BRST quantization has been proposed in
\cite{brst3,brst4} (cf. also \cite{marc}, Chapter 14) 
where the norm of physical states is
regularized. However if the gauge orbits possess a nontrivial
topology, it can be shown that there may exist a topological
obstruction to define the inner product \cite{brst5}.
There are many proposals 
to improve a formal BRST path integral
\cite{brst6}. They will not be discussed here.
The BRST path integral measure is usually ill-defined, or
defined as a perturbation expansion around the Gaussian measure,
while the effects in question are nonperturbative.
Therefore the validity of any modification of the BRST
path integral should be tested by comparing it with (or deriving it from)
the corresponding operator formalism. It is important 
that the gauge invariance is preserved in any modification
of the conventional BRST scheme. 
 As has been already mentioned, the BRST operator formalism
needs a proper inner product, and a construction of such
an inner product can be tightly related to the gauge orbit space
geometry. It seems that more studies are still needed
to come to a definite conclusion about the role of the 
orbit space geometry in the BRST quantization.

\section{The physical phase space}
\setcounter{equation}0

As has been emphasized in the preceding remarks, solutions
to the equations of motion of gauge systems are not fully
determined by the initial conditions and depend on arbitrary
functions of time. 
Upon varying these functions the solutions undergo
gauge transformations. Therefore at any
moment of time, the state of the system can only be determined modulo
gauge transformations. Bearing in mind that the gauge system
never leaves the constraint surface in the phase space, we are
led to the following definition of the physical phase space.
The physical phase space is a quotient space of the constraint
surface relative to the action of the gauge group generated by
all independent constraints. Denoting the gauge group by
${\cal G}$, and the set of constraints by $\sigma_a$,
the definition can be written in the compact form
\begin{equation}
{\rm PS}_{\rm phys} = \left. {\rm PS}\right|_{\sigma_a=0}/{\cal G}\ ,
\label{1}
\end{equation}
where PS is the total phase space of the gauge system, usually 
assumed to be a Euclidean space.
If the gauge transformations do not mix  generalized coordinates
and momenta, one can also define the physical configuration space
\begin{equation}
{\rm CS}_{\rm phys} =  {\rm CS}/{\cal G}\ .
\label{2}
\end{equation}
As they stand, the definitions (\ref{1}) and (\ref{2}) do not
depend on any para\-met\-ri\-za\-tion (or local coordinates) of the
configuration or phase space. In practical applications, one
always uses some particular sets of local coordinates to span
the gauge invariant spaces (\ref{1}) and (\ref{2}). The choice
can be motivated by a physical interpretation of the preferable
set of physical variables or, e.g., by simplicity of calculations, etc.
So our first task is to learn how 
the geometry of the physical 
phase space is manifested in a coordinate description.
Let us turn to some examples of gauge systems to illustrate formulas
(\ref{1}) and (\ref{2}) and to gain some experience in classical
gauge dynamics on the physical phase space.

\section{A system with one physical degree of freedom}
\setcounter{equation}0

Consider the Lagrangian
\begin{equation}
L = \frac{1}{2}\left(\dot{\bf x} - y^a
T_a {\bf x}\right)^2 -
V({\bf x}^2)\ .
\label{so.1}
\end{equation}
Here ${\bf x}$ is an N-dimensional real vector, $T_a$ real
N$\times$N antisymmetric matrices, generators of SO(N) and 
$(T_a{\bf x})^i=(T_a)^{i}{}_{j}x^j$. Introducing the
notation $y = y^a T_a$ for an antisymmetric real matrix
(an element of the Lie algebra of SO(N)), 
the gauge transformations under which the
Lagrangian (\ref{so.1}) remains invariant can be written in the
form
\be
{\bf x}\rightarrow  \Omega {\bf x}\ ,\ \ \ \ 
y\rightarrow \Omega y\Omega^{T} - \Omega\dot{\Omega}^{T}\ ,
\label{so.2}
\ee
where $\Omega= \Omega(t)$ is an element of the gauge group
SO(N), $\Omega^{T}\Omega=\Omega\Omega^{T}=1$, and $\Omega^T$
is the transposed matrix. In fact, the Lagrangian (\ref{so.1})
is invariant under a larger group O(N). As we learn shortly
(cf. a discussion after (\ref{so.9})),
only a connected component of the group O(N), i.e. SO(N),
can be identified as the gauge group. Recall that a connected
component of a group is obtained by the exponential map of the
corresponding Lie algebra. We shall also return to this point
in section 7.1 when discussing the gauge invariance of physical
states in quantum theory.

The model
has been studied in various aspects \cite{christ,jackiw,prokhorov,ufn}.
For our analysis, the work \cite{prokhorov} of Prokhorov will be the
most significant one.
The system under consideration 
can be thought as the (0+1)-dimensional Yang-Mills theory with
the gauge group SO(N) coupled to a scalar field in the fundamental
representation. The real antisymmetric
matrix $y(t)$ plays
the role of the time-component $A_0(t)$ of the Yang-Mills potential
(in fact, the only component available in (0+1)-spacetime), while the
variable ${\bf x}(t)$ is the scalar field in (0+1)-spacetime. The analogy
becomes more transparent if one introduces the covariant derivative
$D_t {\bf x} \equiv \dot{\bf x} -y{\bf x}$ so that the Lagrangian
(\ref{so.1}) assumes the form familiar in gauge field theory
\begin{equation}
L =\frac{1}{2}\left(D_t {\bf x}\right)^2 - V({\bf x}^2)\ .
\label{so.4}
\end{equation}

\subsection{Lagrangian formalism}

The Euler-Lagrange equations are
\begin{eqnarray}
\frac{d}{dt}\frac{\pl L}{\pl \dot{\bf x}} -\frac{\pl L}{\pl {\bf x}}&=&
D^2_t {\bf x}+2{\bf x} V'({\bf x}^2)=0\ ;
\label{so.5}\\
\frac{d}{dt}\frac{\pl L}{\pl \dot{y }^a} -\frac{\pl L}{\pl y^a}&=&
(D_t{\bf x},T_a{\bf x})=0\ .
\label{so.6}
\end{eqnarray}
The second equation in this system is nothing but a constraint associated
with the gauge symmetry. In contrast to Eq. (\ref{so.5}) it
does not contain a second derivative in time and, hence, 
serves as a {\em restriction} (or {\em constraint}) 
on the admissible initial values
of the velocity $\dot{\bf x}(0)$ and position ${\bf x}(0)$
with which the dynamical equation (\ref{so.5}) is to be solved.
The variables $y^a$ are the Lagrange multipliers for the constraints
(\ref{so.6}). 

Any solution to the equations of
motion is determined up to the gauge transformations (\ref{so.2}). 
The variables $y^a=y^a(t)$ remain unspecified by the equation 
of motion. Solutions associated with various choices of $y^a(t)$
are related to one another by gauge transformations.
The dependence of the solution on the functions $y^a(t)$ can be
singled out by means of the following change of variables 
\begin{equation}
x^i(t)=\left[
{\rm T}\exp \int_{0}^{t}y(\tau)d\tau \right]^i_j\, z^j(t)\ ,
\label{so.7}
\end{equation}
where ${\rm T}\exp$ stands for the time-ordered exponential.
Indeed, in the new variables 
the system (\ref{so.5}), (\ref{so.6}) becomes independent of 
the gauge functions $y^a(t)$
\begin{eqnarray}
&\ &\ddot{\bf z}=  -2V'({\bf z}^2){\bf z}\ ;
\label{so.8} \\
&\ &(\dot{\bf z},T_a{\bf z})=  0\ .
\label{so.9}
\end{eqnarray}
The matrix given by the time-ordered exponential in
(\ref{so.7}) is orthogonal and, therefore, ${\bf x}^2 = {\bf z}^2$.
When transforming the equations of motion, we have used 
some properties of the time-ordered exponential which are described
below.
Consider a solution to the equation
\be
\left[\frac{d}{dt} - y(t)\right]_i^j\varphi^i =0\ .
\label{texp1}
\ee
The vectors $\varphi^i(t_1)$ and $\varphi^i(t_2)$ are related 
as
\be
\varphi^i(t_2)=\Omega^i{}_j(t_2,t_1)\varphi^j(t_1)\ ,
\label{texp2}
\ee
where
\be
\Omega(t_2,t_1) = {\rm T}\,\exp\int_{t_1}^{t_2}y(\tau)d\tau\ .
\label{texp3}
\ee
Relations (\ref{texp1}) and (\ref{texp2}) can be regarded as the
definition of the time-ordered exponential (\ref{texp3}). The matrix
$\Omega$ can also be represented as a power series
\be 
\Omega^j{}_i(t_2,t_1)=\sum_{n=0}^{\infty}\int d\tau_1\cdots d\tau_n
\left[y(\tau_1)\cdots y(\tau_n)\right]^j{}_i\ ,
\label{texp4}
\ee
where the integration is carried out over the domain 
$t_2\geq\tau_1\geq\cdots\geq\tau_n\geq t_1$.
If $y$ is an antisymmetric matrix, then from (\ref{texp4}) 
it follows that the time-ordered exponential in (\ref{so.7}) is  
an element of SO(N), that is, the gauge arbitrariness
is exhausted by the SO(N) transformations of ${\bf x}(t)$ 
rather than by those from the larger group O(N).

Since the matrices $T_a$ are antisymmetric, 
the constraint equation (\ref{so.9}) 
is fulfilled for the states in which the velocity vector is
proportional to the position vector
\begin{equation}
\dot{\bf z}(t)=\lambda (t){\bf z}(t)\ ,
\label{so.10}
\end{equation}
and $\lambda (t)$ is to be determined from the dynamical 
equation (\ref{so.8}). A
derivation  of the relation (\ref{so.10})
relies on a simple observation
that equation (\ref{so.9}) means the vanishing of  
all components of the angular momentum of a point-like particle
whose positions are labeled by the N-dimensional radius-vector ${\bf z}$.
Thus, the physical motion is the radial motion for which Eq. (\ref{so.10})
holds and vice versa. Substituting (\ref{so.10}) into (\ref{so.8})
and multiplying the latter by ${\bf z}$, we infer
\begin{equation}
\dot{\lambda}+\lambda ^2=-2V'({\bf z}^2)\ .
\label{so.11}
\end{equation}
Equations (\ref{so.10}) and (\ref{so.11}) form a system of first-order
differential equations to be solved under the initial conditions
$\lambda(0)=\lambda _0$ and ${\bf z}(0)={\bf x}(0) ={\bf x}_0$. 
According to (\ref{so.10}) the relation
$\dot{\bf z}(0)= \dot{\bf x}(0) 
=\lambda _0{\bf x}_0$ specifies initial values of the velocity
allowed by the constraints.

In the case of a harmonic oscillator $V=\frac{\omega
^2}{2}{\bf x}^2=\frac{\omega ^2}{2}{\bf z}^2$, 
Eq. (\ref{so.11}) is easily solved
\begin{equation}
\lambda (t)=-\omega \tan (\omega t+\varphi _0)\ ,\ \ \ \
\varphi_0\in (-\pi/2,\pi/2)\ ,
\label{so.12}
\end{equation}
thus leading to
\begin{equation}
{\bf z}(t)={\bf x}_0\cos (\omega t+\varphi _0)/\cos\varphi_0\ ,
\label{so.13}
\end{equation}
where the initial condition is taken into account. 
A general solution ${\bf x}(t)$ is obtained from (\ref{so.13})
by means of the gauge transformation (\ref{so.7}) where
components of the matrix $y(t)$ play the role of the gauge
transformation parameters. In particular, 
one can always choose $y(t)$ to direct the vector ${\bf x}$
along, say, the first axis $x^i(t)=x(t)\delta^{i1}$
for all moments of time. That is, the first coordinate axis can
always be chosen to {\em label} physical states and to describe the
physical motion of the gauge system. This is, in fact, a general
feature of gauge theories: By specifying the Lagrange multipliers
one fixes a supplementary (gauge) condition to be fulfilled
by the solutions of the Euler-Lagrange equations. 
The gauge fixing surface in the configuration
(or phase) space is used to label physical states of the gauge theory.
In the model under consideration, we have chosen the gauge $x^i =0$,
for all $i\neq 1$.   
Furthermore, for those moments of time when
$x(t)<0$ one can find $y(t)$ such that
\begin{equation}
x(t)\rightarrow -x(t)\ ,
\label{so.14}
\end{equation}
being the SO(N) rotations of the vector ${\bf x}$ through the 
angle $\pi$.
The physical motion is described by a non-negative variable
$r(t)=|x(t)|\geq 0$ because there is no further gauge equivalent 
configurations among those satisfying the chosen gauge condition.
The physical configuration space is isomorphic to a
half-line
\begin{equation}
{\rm CS}_{\rm phys}=\Rs ^N/{\rm SO(N)}\sim \Rs_+\ .
\label{so.15}
\end{equation}
It should be remarked that the residual gauge transformations
(\ref{so.14}) cannot decrease the number of physical degrees
of freedom, but they do reduce the ``volume'' of the physical 
configuration space. 

The physical configuration space can be regarded as the 
gauge orbit space whose elements are gauge orbits.
In our model the gauge orbit space is the space of
concentric spheres.
By having specified the gauge
we have chosen the Cartesian coordinate $x^1$ to parameterize
the gauge orbit space. 
It appears however that our gauge is incomplete.
Among configurations belonging to the gauge fixing surface, 
there are configurations related to one another by gauge
transformations, thus describing the same physical state.
Clearly, the $x^1$ axis intersects each sphere (gauge orbit)
twice so that the points $x^1$ and $-x^1$ belong to the 
same gauge orbit. Thus, the gauge orbit space can be 
parameterized by non-negative $x^1$. In general,
given a gauge condition and a configuration satisfying it,
one may find other configurations that satisfy the gauge
condition and belong to the gauge orbit passing through
the chosen configuration. 
Such configurations are called Gribov copies. This phenomenon
was first observed by Gribov in Yang-Mills theory in the Coulomb
gauge \cite{gribov}. At this point we shall only remark that the Gribov
copying depends on the gauge, although it is unavoidable
and  always present in any gauge in Yang-Mills theory \cite{singer}.
The existence of the Gribov copying is directly related to a 
non-Euclidean geometry of the
gauge orbit space \cite{singer,babelon}.
For the latter reason, this phenomenon is important
in gauge systems and deserves further study. 

As the Gribov copying is gauge-dependent, one
can use gauge-invariant variables to avoid it.
This, however, does not always provide us with a description
of the physical motion free of ambiguities.
For example, for our model problem
let the physical motion be  described by 
the gauge invariant variable $r(t)=|x(t)|= |{\bf x}(t)|$.
If the trajectory
goes through the origin at some moment of time $t_0$, 
i.e., $r(t_0)=0$, the velocity
$\dot{r}(t)$ suffers a jump as if the particle hits a wall at
$r=0$.
Indeed, $\dot{r}(t)=\varepsilon (x(t))\dot{x}(t)$
where $\varepsilon (x)$ is the
sign function, $\varepsilon (x)=+1$ if $x>0$ and
$\varepsilon(x)=-1$ for $x<0$.
Setting $v_0=\dot{x}(t_0)$, we find
$\dot{r}(t_0-\epsilon)-\dot{r}(t_0+\epsilon )\rightarrow 2v_0$ as
$\epsilon\rightarrow 0$. On the other hand, the potential $V(r^2)$ 
is smooth and regular
at the origin and, therefore, cannot cause 
any infinite force acting on the
particle passing through the origin. So, despite  using the 
{\em gauge-invariant} variables to 
describe the physical motion, we may encounter 
non-physical singularities which are not at all anticipated
for smooth potentials. Our next
step is therefore to establish a description 
where the ambiguities are absent. This can be achieved in 
the framework of the Hamiltonian dynamics to which we now turn.   

\subsection{Hamiltonian dynamics and the physical phase space}

The canonical momenta for the model (\ref{so.1}) read
\begin{eqnarray}
{\bf p}&= & \frac{\partial L}{\partial \dot{\bf x}}=D_t{\bf x}\ ,
\label{so.16} \\
\pi_a&= & \frac{\partial L}{\partial \dot{y}^a}=0\ .
\label{so.17}
\end{eqnarray}
Relations (\ref{so.17}) are primary constraints \cite{diraclec}. 
A canonical Hamiltonian is
\begin{equation}
H=\frac{1}{2}{\bf p}^2 +V({\bf x}^2)-y^a\sigma _a\ ,
\label{so.18}
\end{equation}
where
\begin{equation}
\sigma _a=\{\pi_a,H\} =-({\bf p},T_a{\bf x})=0
\label{so.19}
\end{equation}
are secondary constraints. Here $\{\,,\,\}$ denotes the Poisson
bracket. By definitions (\ref{so.16}) and (\ref{so.17}) we
set $\{x^i,p_j\}=\delta_j^i$ and $\{y^a,\pi_b\}=\delta_b^a$,
while the other Poisson bracket of the canonical variables vanish.
The constraints (\ref{so.19}) ensure that the primary constraints
hold as time proceeds, $\dot{\pi}_a =\{\pi_a ,H\}=0 $.  
All the constraints are in
involution
\begin{equation}
\{\pi_a,\pi_b\}=0\ ,\ \ \ \{\pi_a,\sigma_a\} =0\ ,
\ \ \ \{\sigma _a,\sigma _b\}=
f_{ab}{}^{c}\sigma_c\ ,
\label{so.20}
\end{equation}
where $f_{ab}{}^{c}$ are the structure constraints of SO(N),
$[T_a,T_b]= f_{ab}{}^{c} T_c$.
There is no further restriction on the canonical variables because
$\dot{\sigma}_a$ {\em weakly} vanishes, $\dot{\sigma}_a = \{\sigma_a, H\}
\sim \sigma_a \approx 0$, i.e., it vanishes on the surface 
of constraints \cite{diraclec}.

Since $\pi_a=0$, one can consider a generalized Dirac dynamics \cite{diraclec}
which is obtained by replacing the {\em canonical} Hamiltonian
(\ref{so.18}) by a generalized Hamiltonian $H_T = H +\xi^a\pi_a$
where $\xi^a$ are the Lagrange multipliers for the primary 
constraints. The Hamiltonian equations of motion 
$\dot{F} = \{F,H_T\}$ will contain
two sets of gauge functions, $y^a$ and $\xi^a$ (for primary and
secondary constraints). However, the primary constraints
$\pi_a=0$ generate only shifts of $y^a:\, \delta y^a =
\delta\xi^b\{\pi_b,y^a\}=-\delta\xi^a$ with $\delta\xi^a$ being 
infinitesimal parameters of the gauge transformation.
In particular, $\dot{y}^a =\{y^a,H_T\}= -\xi^a$. 
The  degrees of freedom $y^a$ turn out to be
purely nonphysical (their dynamics is fully determined by
arbitrary functions $\xi^a$). For this reason, 
we will not introduce generalized
Dirac dynamics \cite{diraclec}, rather we discard the variables $y^a$ as  
independent canonical variables and consider them as 
the Lagrange multipliers for the secondary constraints $\sigma_a$.
That is, 
in the Hamiltonian equations of motion
$\dot{\bf p} = \{{\bf p},H\}$ and $\dot{\bf x} = \{{\bf x},H\}$,
which we can write in the form covariant under the gauge transformations,
\begin{equation}
D_t{\bf p}=-2{\bf x}V'({\bf x}^2)\ ,\ \ \ \ D_t{\bf x}={\bf p}\ ,
\label{so.21}
\end{equation}
the variables
$y^a$ will be regarded as arbitrary functions of time and 
canonical variables ${\bf p}$ and ${\bf x}$.
The latter is consistent with the Hamiltonian form 
of the equations of motion because for any $F=F({\bf p},{\bf x})$
we get $\{F,y^a\sigma_a\}=\{F,y^a\}\sigma_a + y^a\{F,\sigma_a\}
\approx  y^a\{F,\sigma_a\}$. Thus, even though 
the Lagrange multipliers are allowed
to be general functions not only of time, but also of the canonical
variables, the Hamiltonian equations of motion are equivalent
to (\ref{so.21}) on the surface of constraints.
The constraints $\sigma_a$ generate simultaneous rotations
of the vectors ${\bf p}$ and ${\bf x}$ because
\be
 \{ {\bf p},\sigma_a\} =
T_a{\bf p}\ ,\ \ \ \ \{{\bf x},\sigma_a\}=T_a{\bf x}\ .
\label{so21a}
\ee
Thus, the last term in the Hamiltonian (\ref{so.18}) generates
rotations of the classical trajectory  at each moment of
time. A finite gauge transformation is built by
successive infinitesimal rotations, that is, the gauge group
generated by the constraints is SO(N), not O(N).

The time evolution
of a quantity $F$
does not depend on arbitrary functions $y$, provided
$\{F,\sigma_a\}\approx 0$, i.e., $F$ is gauge invariant
on the surface of constraints. The quantity $F$ is gauge invariant
in the total phase space if $\{F,\sigma_a\}=0$.
The constraints (\ref{so.19}) mean that all components of the angular
momentum are zero. The physical motion is the radial motion for 
which the following relation holds
\begin{equation}
{\bf p}(t)=\lambda (t){\bf x}(t)\ .
\label{so.22}
\end{equation}
As before, the scalar function $\lambda (t)$ is determined by the dynamical
equations (\ref{so.21}). Applying the covariant derivative to
(\ref{so.7}), we find
\begin{equation}
{\bf p}(t)=
\left[{\rm T}\exp \int_{0}^{t}y(\tau)d\tau\right]\dot{\bf z}(t)\ ,
\label{so.23}
\end{equation}
where ${\bf z}(t)$ and $\lambda (t)$ are solution to the system (\ref{so.8}),
(\ref{so.11}). Now we can analyze the motion in the phase space spanned by
variables ${\bf p}$ and ${\bf x}$. 
The trajectories lie on the surface
of constraints (\ref{so.22}).  Although the constraints are fulfilled
by the actual motion, trajectories still have gauge arbitrariness 
which corresponds to various choices of $y^a(t)$. 
\begin{figure}
\centerline{\psfig{figure=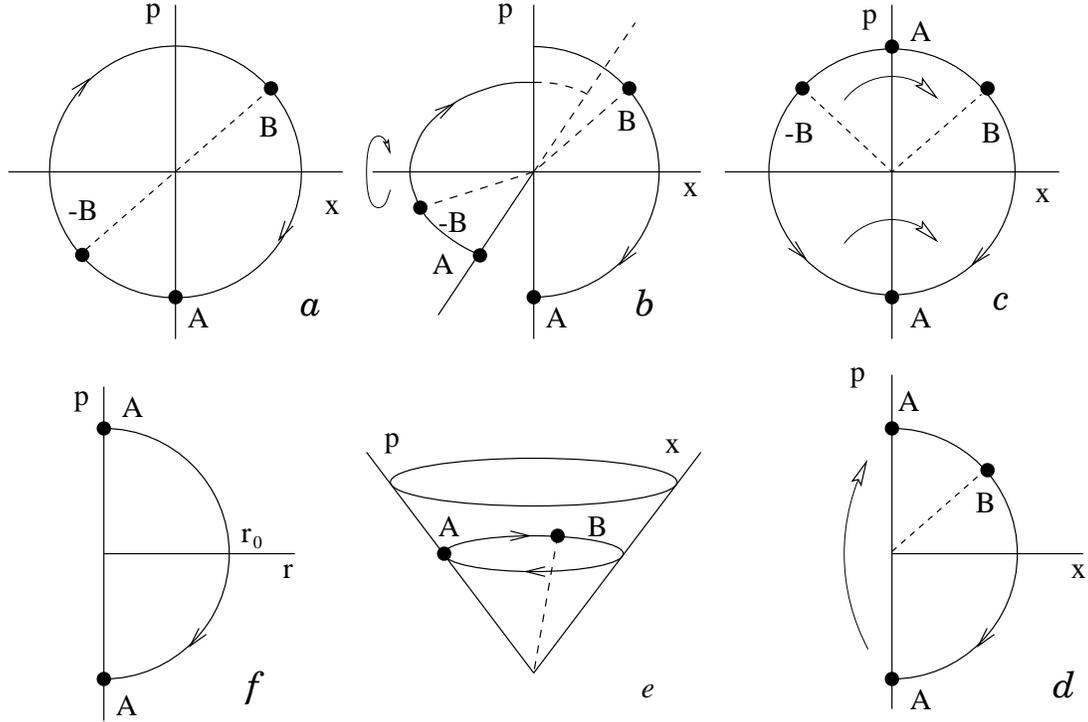}}
\caption{\small {\em a}. The phase-space plane $(p,x)$ and the
oscillator trajectory on it. The states $B=(p,x)$ and $-B =(-p,-x)$
are gauge equivalent and to be identified; \newline
{\em b}. The phase-space plane is cut along the $p$-axis. The
half-plane $x<0$ is rotated relative to the $x$-axis 
through the angle $\pi$.\newline 
{\em c}. The resulting plane is folded
along the $p$-axis so that the states $B$ and $-B$ get identified;\newline
{\em d}. Two copies of each state on the $p$-axis, which occur
upon the cut (e.g., the state $A$), are glued back to remove this
doubling;\newline
{\em e}. The resulting conic phase space. Each point of it corresponds
to one physical state of the gauge system. The oscillator trajectory
does not have any discontinuity;\newline
{\em f}. The physical motion of the harmonic oscillator in
the local gauge invariant variables $(p_r,r)$. The trajectory
has a discontinuity at the state A. The discontinuity occurs
through the cut of the cone along the momentum axis. The cut
is associated with the $(p_r,r)$ parameterization of the cone. }
\end{figure}
Variations
of $y^a$ generate {\em simultaneous} SO(N)-rotations of 
the vectors ${\bf x}(t)$ and
${\bf p}(t)$ 
as follows from the representations (\ref{so.7}) and (\ref{so.22}).
Therefore, 
with 
an appropriate choice of the arbitrary functions $y^a(t)$,
the physical motion can be described 
in two-dimensional phase space
\begin{equation}
x^i(t)=x(t)\delta^{i1}\ ,\ \ p_i(t)=\lambda (t) x(t) \delta _{i1}\equiv
p(t)\delta _{i1}\ .
\label{so.24}
\end{equation} 
An important
observation is the following \cite{prokhorov}. 
Whenever the variable $x(t)$ changes 
sign under the gauge transformation (\ref{so.14}), so does 
the canonical momentum
$p(t)$  because of the constraint
(\ref{so.22}) or (\ref{so.24}). In other words, for any motion in the 
phase-space plane  two
states $(p,x)$ and $(-p,-x)$ are physically indistinguishable.
Identifying these points on the plane,
we obtain the physical phase space of the system which is
a cone unfoldable into a half-plane \cite{prokhorov,ufn}
\begin{equation}
{\rm PS}_{\rm phys}={\rm PS}\vert _{\sigma _a=0}/{\rm SO(N)} 
\sim \Rs ^2/\Z_2\sim {\rm cone} (\pi)\ .
\label{so.25}
\end{equation}
Figure 1 illustrates how the phase-space plane turns
into the cone upon the identification of the points $(p,x)$
and $(-p,-x)$.

Now we can address the above issue about nonphysical singularities of the
gauge invariant velocity $\dot{r}$. 
To simplify the discussion and to make it transparent, let us
first take a harmonic oscillator as an example. 
To describe the physical motion, we
choose gauge-invariant canonical coordinates $r(t)=|{\bf x}(t)|$ and
$p_r(t)=({\bf x},{\bf p})/r$. The gauge invariance means that
\begin{equation}
\{r,\sigma _a\}=\{p_r,\sigma _a\}=0\ ,
\label{so.26}
\end{equation}
i.e., the evolution of the canonical pair $p_r,r$ does not depend on
arbitrary functions $y^a(t)$. Making use of (\ref{so.12}) and
(\ref{so.13}) we find
\begin{eqnarray}
r(t)& =&r_0|\cos \omega t| \ ;\label{so.28}\\
p_r(t)&= & \lambda (t)r(t)=\dot{r}(t)=-\omega r_0 \sin \omega t\, 
\,\varepsilon(\cos \omega t)\ .
\label{so.27} 
\end{eqnarray}
Here the constant $\varphi _0$ has 
been set to zero, and $r_0=|{\bf{x}}_0|$.
The trajectory  starts at the phase-space point $(0,r_0)$ and goes down
into the area of negative momenta as shown in Fig. 1f. At the time
$t_A=\pi /2\omega $, the trajectory reaches the half-axis $p_r<0,r=0$ (the
state $A$ in Fig. 1f). The physical momentum $p_r(t)$ has the sign flip
as if the particle hits a wall. At that instant the 
acceleration is infinite because $\Delta
p_r(t_A)=p_r(t_A+\epsilon )-p_r(t_A-\epsilon )\rightarrow 2r_0\omega
\ ,\epsilon \rightarrow 0$, which is not possible as the
oscillator potential vanishes at the origin.  
Now we recall that the physical phase space
of the model is a cone unfoldable into a half-plane. 
To parameterize the cone by the local gauge-invariant
phase-space coordinates (\ref{so.27}), (\ref{so.28}), one has to
make a cut of the cone along the momentum axis, which
is readily seen from the comparison of figures 1d and 1f where
the same motion is represented. 
The states
$(r_0\omega ,0)$ and $(-r_0\omega ,0)$ are two images of one state that
lies on the cut made on the cone.  
Thus, in the conic phase space, 
the trajectory is smooth and does not contains any
discontinuities. The  nonphysical ``wall'' force is absent (see Fig.1e). 

In our discussion, a particular form of the 
potential $V$ has been assumed. This
restriction can easily be dropped. Consider a trajectory
$x^i(t)=x(t)\delta ^{i1}$ passing through the origin at $t=t_0,\
x(t_0)=0$. 
In the physical variables the trajectory is $r(t)=|x(t)|$ and
$p_r(t)=\dot{r}(t)=p(t)\varepsilon (x(t))$ where $p(t)=\dot{x}(t)$. Since
the points $(p,x)$ and $(-p,-x)$ correspond to the same physical state, we
find that the phase-space points
$(p_r(t_0-\epsilon ),x(t_0-\epsilon))$ and 
$(p_r(t_0+\epsilon ),x(t_0+\epsilon))$
approach the same physical state as $\epsilon$ 
goes to zero. So, for any trajectory and
any regular potential the discontinuity $|p_r(t_0-\epsilon
)-p_r(t_0+\epsilon)|\rightarrow 2|p(t_0)|$, as $\epsilon \rightarrow
0$, is removed by going over to the conic phase space.

The observed singularities of the phase-space trajectories 
are essentially artifacts of 
the {\em coordinate description} 
and, hence, depend on the parameterization of
the physical phase space. 
For instance, the cone can be parameterized
by another set of {\em canonical} gauge-invariant variables
\begin{equation}
p_r=|{\bf p}|\geq 0\ ,\ \ r =\frac{({\bf p},{\bf x})}{p_r}\ ,\ \ 
\{r,p_r\}=1\ .
\label{so.29}
\end{equation}
It is easy to convince oneself 
that $r(t)$ would have discontinuities, rather than the momentum
$p_r$. This set of local coordinates on the physical phase space is
associated with the cut on the cone along the coordinate axis. In general,
local canonical coordinates on the physical phase space 
are determined up to canonical transformations
\begin{equation}
(p_r,r)\rightarrow  (P_R,R)=(P_R(r,p_r),R(p_r,r))\ ,
\ \ \ \
\{R,P_R\}= 1\ .
\label{so.31}
\end{equation}
The coordinate singularities associated with arbitrary
local canonical coordinates on the physical phase space 
may be tricky to analyze.
However, the motion considered on the true physical phase space
is free of these ambiguities. That is why it is important to
establish the geometry of the physical phase space before studying
Hamiltonian dynamics in some local formally gauge invariant 
canonical coordinates.

It is also of interest to find out whether there exist a set
of {\em canonical} variables in which  the discontinuities
of the classical phase-space  trajectories  do not
occur. Let us return to the local coordinates where the momentum
 $p_r$ changes sign as the trajectory passes through
the origin $r=0$. The sought-for new canonical variables must be
even functions of $p_r$ when $r=0$ and be {\em regular}
on the half-plane $r\geq0$. Then the trajectory in the 
new coordinates will not suffer the discontinuity. 
In the vicinity of the origin, we set
\begin{equation}
 R = a_0(p_r^2)+\sum_{n=1}^\infty a_n(p_r)r^n\ ,\ \ \ \ 
P_R = b_0(p_r^2)+\sum_{n=1}^\infty b_n(p_r)r^n \ .
\label{add.1}
\end{equation} 
Comparing the coefficients of powers of $r$ in the 
Poisson bracket (\ref{so.31}) we find, in particular,
\begin{equation}
2p_r\left[a_1(p_r)b_0^\prime(p_r^2) - 
a_0^\prime(p_r^2) b_1(p_r)\right] = 1\ .
\label{add.2}
\end{equation}
Equation (\ref{add.2}) has no solution for regular functions $a_{0,1}$
and $b_{0,1}$. 
By assumption the functions $a_n$ and $b_n$ are regular and so
should be $a_1b_0'-a_0'b_1 = 1/(2p_r)$, 
but the latter is not true at $p_r=0$
as follows from (\ref{add.2}).
A solution exists only for functions singular at
$p_r=0$. For instance, one can take $R=r/p_r$ and $P_R=p_r^2/2$,
$\{R,P_R\}=1$ which is obviously singular at $p_r=0$. In these
variables the evolution of the canonical momentum does not have
abrupt jumps, however, the new canonical coordinate does have 
jumps as the system goes through the states with $p_r=0$.

In general, the existence of singularities are due to the condition
that $a_0$ and $b_0$ must be even functions of $p_r$. This latter
condition leads to the factor $2p_r$ in the left-hand side of
Eq.(\ref{add.2}), thus making  it impossible for $b_1$ and $a_1$ to
be regular everywhere.  
We conclude that, although in the conic phase space the trajectories 
are regular, the motion always exhibits singularities when 
described in any local canonical coordinates on the phase space.

Our analysis of the simple gauge model reveals an important and
rather general feature of gauge theories. The physical phase 
space in gauge theories may have a non-Euclidean geometry.
The phase-space trajectories are smooth in the 
physical phase space. However, when described in local
canonical coordinates, the motion may exhibit nonphysical
singularities.  In Section 6 we show that the impossibility of
constructing canonical (Darboux) coordinates on the physical
phase space, which would provide a classical description without
singularities, is essentially due to the nontrivial topology
of the gauge orbits (the concentric spheres in this model). 
The singularities
fully depend on the choice of local canonical coordinates,
even though this choice is made in a
gauge-invariant way. 
What remains coordinate- and gauge-independent 
is the geometrical structure of the physical
phase space which, however, may
reveal itself through the coordinate singularities
occurring in any particular parameterization of the 
physical phase space by local canonical variables.
One cannot assign any
direct physical meaning to the singularities, but 
their presence indicates that 
the phase space of the physical
degrees of freedom is not Euclidean.
At this stage of our discussion it becomes
evident that it is of great importance  to find a quantum formalism
for gauge theories which does not depend on local
parameterization of the physical phase space and 
takes into account its genuine geometrical structure.    

\subsection{Symplectic structure on the physical phase space}

The absence of local canonical coordinates in which the
dynamical description does not have singularities may 
seem to look rather disturbing. This is partially because of
our custom to often identify canonical variables with 
physical quantities which can be directly measured,
like, for instance, positions and momenta of particles
in classical mechanics. In gauge theories canonical
variables, that are defined through the Legendre transformation
of the Lagrangian, cannot always be measured and, in fact,
may not even be physical quantities. For example, canonical variables
in electrodynamics are components of the electrical field
and vector potential. The vector potential is subject to
the gradient gauge transformations. So it is a nonphysical 
quantity. 

The simplest gauge invariant
quantity that can be built of the vector potential is the magnetic field.
It can be measured. Although the electric and magnetic 
fields are not canonically conjugated variables, we may
calculate the Poisson bracket of them and determine the evolution
of all gauge invariant quantities (being functions of the electric
and  magnetic fields) via the Hamiltonian equation motion
with the new Poisson bracket. 
Extending this analogy further we may try to find a new set 
of physical variables in the SO(N) model
that are not necessarily canonically
conjugated but have a smooth time evolution. A simple choice
is
\begin{equation}
Q = {\bf x}^2\ ,\ \ \ P = ({\bf p},{\bf x})\ .
\label{333}
\end{equation}
The variables (\ref{333}) 
are gauge invariant and in a one-to-one correspondence
with the canonical variables $r,p_r$ parameterizing the 
physical (conic) phase space: $Q=r^2, P = p_rr, r\geq 0$.
Due to analyticity in the original phase space variables,
they also have a smooth time evolution $Q(t), P(t)$.
However, we find
\begin{equation}
\{Q,P\} = 2Q\ ,
\label{PQ}
\end{equation} 
that is, the symplectic structure is no longer canonical.
The new symplectic structure is also acceptable to formulate
Hamiltonian dynamics of physical degrees of freedom. The Hamiltonian
assumes the form
\begin{equation}
H = \frac{1}{2Q}\,P^2 + V(Q)\ . 
\end{equation} 
Therefore 
\begin{equation}
\dot{Q} = \{Q,H\} = 2P\ ,\ \ \ \dot{P}=\{P,H\}=
\frac{P^2}{Q} -2QV'(Q)\ .
\end{equation}
The solutions $Q(t)$ and $P(t)$ are regular for a sufficiently
regular $V$, and there is no need to ``remember'' where
the cut on the cone has been made. 

The Poisson bracket ({\ref{PQ}) can be regarded as a skew-symmetric
product (commutator) of two basis elements of the Lie algebra
of the dilatation group. This observation allows one to quantize 
the symplectic structure. The representation of the corresponding
quantum commutation relations is realized by the
so called affine coherent states.
Moreover the coherent-state representation of the path integral
can also be developed \cite{kldil}, which is not a canonical path
integral when compared with the standard lattice treatment.

\subsection{The phase space in curvilinear coordinates}

Except the simplest case when the gauge transformations are translations in
the configuration space, physical variables are non-linear functions of
the original variables of the system. The separation of local coordinates
into the physical and pure gauge ones can be done by means of going
over to curvilinear coordinates such that
some of them span gauge orbits,
while the others change along the directions transverse to the gauge
orbits and, therefore, label physical states. 
In the example considered above, the gauge orbits are spheres
centered at the origin. An appropriate coordinate system to
separate physical and nonphysical variables is the spherical coordinate
system. It is clear that dynamics of angular variables is fully arbitrary
and determined by the choice of functions $y^a(t)$.  In contrast the
temporal evolution of the radial variable does not depend on $y^a(t)$.
The phase space of the only physical degree of freedom turns out
to be  a cone unfoldable into a half-plane. 

Let us forget about the gauge symmetry in the model for a moment. 
Upon a canonical
transformation induced by going over to the spherical coordinates, the
radial degree of freedom seems to have a phase space being a half-plane
because $r=|{\bf x}|\geq 0$, and the corresponding
canonical momentum would have an abrupt sign flip when the system passes
through the origin.
It is then natural to put forward the question
whether the conic structure of the physical phase 
space is really due to the
gauge symmetry, and may not emerge upon a certain canonical
transformation.
We shall argue that without the gauge symmetry, the {\em full} phase-space
plane $(p_r, r)$ is required to uniquely 
describe the motion of the system \cite{ufn}. 
As a general remark, we point out that the phase-space structure 
cannot be changed by any canonical transformation. The curvature 
of the conic phase space, 
which is concentrated on the tip of the cone,
cannot be introduced or even eliminated by any coordinate transformation. 

For the sake of simplicity, the discussion is restricted to the simplest case
of the $SO(2)$ group \cite{ufn}. 
The phase space is a four-dimensional Euclidean space spanned by the
canonical coordinates ${\bf p}\in \Rs ^2$ and ${\bf x}\in \Rs^2$.
For the polar coordinates $r$ and $\theta$ introduced by
\begin{equation}
x^1=r \cos \theta \ ,\ \ \ \ x^2=r \sin \theta\ ,
\label{pc.1}
\end{equation}
the canonical momenta are
\begin{equation}
p_r=\frac{({\bf x},{\bf p})}{r}\ ,\ \ \ \ p_\theta =({\bf p},T{\bf x})
\label{pc.2}
\end{equation}
with $T_{ij}=-T_{ji},\ 
T_{12}=1$, being the only generator of SO(2). The
one-to-one correspondence between the Cartesian and polar coordinates is
achieved if the latter are restricted to non-negative values for $r$ and to
the segment $[0,2\pi )$ for $\theta$. 

To show that the full plane $(p_r,r)$  is  necessary for a unique
description of  the motion, we compare the motion
of a particle through the origin in Cartesian and polar coordinates,
assuming the potential to be regular at the origin. Let
the particle move along the $x^1$ axis. As long as the particle moves
along the positive semiaxis the equality $x^1=r$ is satisfied and  no
paradoxes arise. As the particle moves through the origin, $x^1$ changes
sign, $r$ does not change sign, and $\theta$ and $p_r$ change abruptly:
$\theta \rightarrow\theta +\pi ,
\ p_r=|p|\cos \theta \rightarrow -p_r$. Although these jumps
are not related with the action of any forces, they are consistent with
the equations of motion. The kinematics of the system admits an
interpretation in which the discontinuities are avoided. As follows from
the transformation formulas (\ref{pc.1}), the Cartesian coordinates
$x^{1,2}$ remains unchanged under the transformations
\begin{eqnarray}
\theta& \rightarrow & \theta +\pi\ ,\ \ \ \ \ \ r\rightarrow -r\ ;
\label{pc.3} \\
\theta&\rightarrow & \theta +2\pi \ ,\ \ \ \ \  r\rightarrow r\ .
\label{pc.4}
\end{eqnarray}
This means that the motion with values of the polar coordinates $\theta
+\pi$ and $r>0$ is indistinguishable from the motion with values of the
polar coordinates $\theta$ and $r<0$. Consequently, the phase-space points
$(p_r,r; p_\theta ,\theta )$ and $(-p_r,-r;\theta +\pi ,p_\theta )$
correspond to the same state of the system. 
Therefore, the state $(-p_r,r;p_\theta,\theta +\pi)$ 
the particle attains after passing through the origin is
equivalent to $(p_r,-r;p_\theta ,\theta )$. As expected, the phase-space
trajectory will be identical in both the $(p_r,r)-${\em plane} and the
$(p_1,x^1)-$plane.
\begin{figure}
\centerline{\psfig{figure=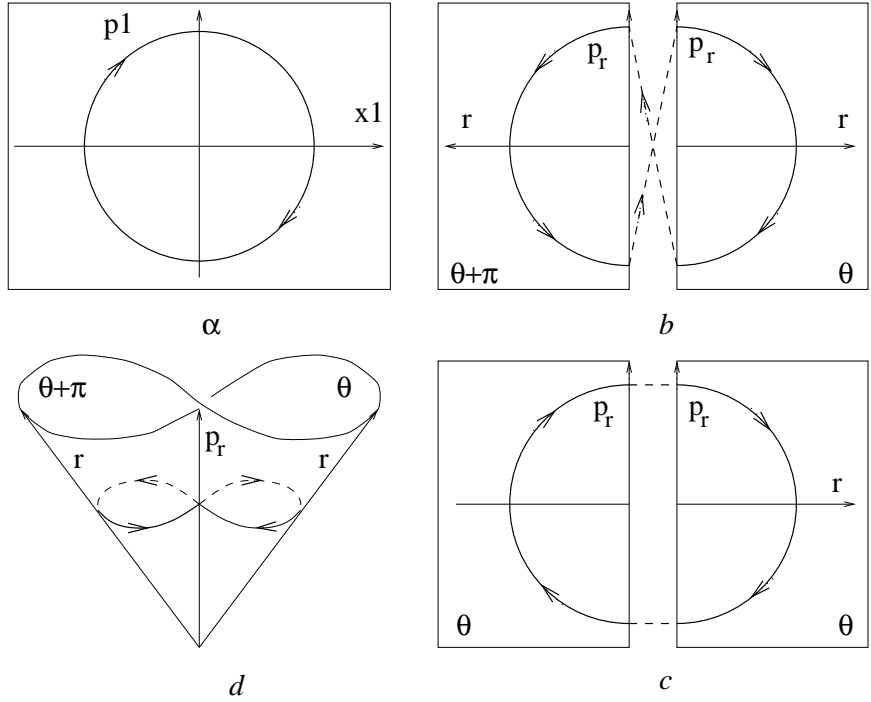}}
\caption{\small {\em a}. A phase-space trajectory of a harmonic
oscillator. The initial condition are such that
$x^2=p_2=0$ for all moments of time. 
The system moves through the origin $x^1=0$;
\newline
{\em b}. The same motion is represented in the canonical 
variables associated with the polar coordinates. 
When passing the origin $r=0$, the trajectory suffers a discontinuity
caused by the jump of the canonical momenta. 
The discontinuity can be removed in two ways:\newline
{\em c}. One can convert the motion with values of
the canonical coordinates $(-p_r,r;$ $p_\theta,\theta +\pi)$
into the equivalent motion $(p_r,-r;p_\theta,\theta)$, thus
making  a full phase-space plane out of two half-planes.\newline
{\em d}. Another possibility is to glue directly the points 
connected by the dashed lines. The resulting surface is the 
Riemann surface with two conic leaves. It
has no curvature at the origin because
the phase-space radius vector $(p_r,r)$ sweeps 
the total angle $2\pi$ around the two
conic leaves before returning to the initial state.}
\end{figure}

In Fig.2 it is shown how the continuity of the phase-space 
trajectories can be maintained in the canonical variables 
$p_r$ and $r$. The original trajectory in the Cartesian
variables is mapped into two copies of the half-plane $r\geq 0$.
Each half-plane corresponds to the states of the system with
values of $\theta$ differing by $\pi$ (Fig. 2b). Using the equivalence
between the states $(-p_r, r;p_\theta, \theta+\pi)$ and
$(p_r, -r;p_\theta, \theta)$, the half-plane corresponding to the 
value of the angular value $\theta +\pi$ can be viewed 
as the half-plane with negative values of $r$ so that the
trajectory is continuous on the $(p_r,r)$-plane and the
angular variables does not change when the system passes
through the origin (Fig. 2c).

Another possibility to keep the trajectories continuous
under the canonical transformation, while maintaining the
positivity of $r$, is to glue the edges of the half-planes
connected by the dashed lines in Fig. 2b. The resulting
surface resembles the Riemann surface with two conic leaves
(Fig. 2d). The curvature at the origin of this surface 
is zero because for any periodic motion the trajectory goes around
both conic leaves before it returns to the initial state, i.e.,
the phase-space radius-vector $(r,p_r)$ sweeps the total
angle $2\pi$.
Thus, the motion is indistinguishable from the motion 
in the phase-space plane. 

When the gauge symmetry is switched on, the angular variable $\theta$
becomes nonphysical, the constraint is determined by $p_\theta=0$.
The states which differ only by values of $\theta$ must be identified.
Therefore two conic leaves of the $(p_r,r)$-Riemann surface become two
images of the physical phase space. By identifying them, the Riemann
surface turns into a cone unfoldable into a half-plane.
In the representation given in Fig. 2c, the cone emerges upon
the familiar identification of the points $(-p_r,-r)$ with $(p_r,r)$.
This follows from the equivalence of the states
$(-p_r,-r;p_\theta =0,\theta)\sim (p_r,r;p_\theta = 0,\theta +\pi)\sim
(p_r,r;p_\theta =0,\theta)$, where the first one is due to the symmetry 
of the change of variables, while the second one is due to
the gauge symmetry: States differing by values of $\theta$
are physically the same.

\subsection{Quantum mechanics on a conic phase space}

It is clear from the correspondence principle that 
quantum theory should, in general, 
depend on the geometry of the phase space. 
It is most naturally exposed 
in the phase-space path integral  representation of
quantum mechanics. Before
we proceed with establishing  the path integral
formalism for gauge theories whose physical phase space differs
from a Euclidean space, let us first use simpler tools, like 
Bohr-Sommerfeld semiclassical quantization, to get 
an idea of how the phase space geometry in gauge theory may affect quantum
theory \cite{prokhorov}, \cite{ufn}.

Let the potential $V$ of the system  be such that there exist periodic
solutions of the classical equations of motion. According to the
Bohr-Sommerfeld quantization rule, the energy levels can be
determined by solving the equation
\begin{equation}
W(E)=\oint_{}^{}pdq=\int_{0}^{T}p\dot{q}dt=
2\pi \hbar\left(n+\frac{1}{2}\right)\ ,\ \ n=0,1,\ldots\ ,
\label{bs.1}
\end{equation}
where the integral is taken over a periodic phase-space trajectory with the
period $T$ which may depend on the energy $E$ of the system. 
The quantization rule
(\ref{bs.1}) does not depend on the parameterization of the phase space
because the functional 
$W(E)$ is invariant under canonical transformations:$\oint pdq=
\oint PdQ$ and, therefore, {\em coordinate-free}. 
For this reason we adopt it to analyze quantum mechanics on
the conic phase space. For a harmonic oscillator of frequency $\omega$
and having a Euclidean phase space, 
the Bohr-Sommerfeld rule gives exact
energy levels. Indeed, classical trajectories are
\begin{equation}
q(t)=\frac{\sqrt{2E}}{\omega}\sin \omega t\ ,\ \ p(t)=
\sqrt{2E} \cos \omega t\ ,
\label{bs.2}
\end{equation}
thus leading to
\begin{equation}
E_n=\hbar \omega \left(n+\frac{1}{2}\right)\ ,\ \ n=0,1,\ldots\ .
\label{bs.3}
\end{equation}
In general, the Bohr-Sommerfeld quantization determines the spectrum in
the semiclassical approximation
(up to higher orders of $\hbar$) \cite{maslov}. 
So our consideration is not yet
a full quantum theory. Nonetheless it
will be sufficient to qualitatively distinguish between  
the influence of the non-Euclidean
geometry of the physical phase space 
and the effects of potential forces
on quantum gauge dynamics.

Will the spectrum (\ref{bs.3}) be modified if 
the phase space of the system is changed to a cone unfoldable into a
half-plane? The answer is affirmative \cite{prokhorov,ufn,book}. 
The cone is obtained by identifying points on the
plane related by reflection with respect to the origin, ${\rm
cone}(\pi)\sim \Rs ^2/\Z_2$. Under the residual gauge
transformations $(p,q) \rightarrow (-p,-q)$, the oscillator trajectory
maps into itself. Thus on the conic phase space it remains a periodic
trajectory. However the period is twice less than that of the oscillator
with a flat phase space because the states the oscillator passes at
$t\in [0, \pi /\omega )$ are physically indistinguishable from those at
$t\in[\pi/\omega,2\pi/\omega)$. Therefore the oscillator 
with the conic phase space returns to the
initial state in two times faster than the ordinary oscillator:
\begin{equation}
T_c=\frac{1}{2}T=\frac{\pi}{\omega}\ .
\label{bs.4}
\end{equation}
The Bohr-Sommerfeld quantization rule leads to
the spectrum
\begin{equation}
E^c_n=2E_n=2\hbar \omega \left(n+\frac{1}{2}\right)\ ,\ \ n=0,1,\ldots\ .
\label{bs.5}
\end{equation}
The distance between energy levels is doubled as though the physical
frequency of the oscillator were $\omega _{\rm phys}=2\omega $. 
Observe that the frequency as the {\em parameter} of the Hamiltonian
is {\em not} changed. The entire effect is therefore due to the conic
structure of the physical phase space.

Since the
Bohr-Sommerfeld rule does not depend on the parameterization of the phase
space, one can also apply it directly to the conic phase space. We
introduce the polar coordinates on the phase space \cite{prokhorov}
\begin{equation}
q=\sqrt{\frac{2P}{\omega}} \cos Q\ ,\ \ p=\sqrt{2\omega P}\sin Q\ .
\label{bs.6}
\end{equation}
Here $\{Q,P\}=1$. If the variable $Q$ ranges from $0$ to $2\pi$, then
$(p,q)$ span the entire plane $\Rs ^2$. The local variables $(p,q)$
would span a cone unfoldable into a half-plane if one restricts $Q$ to the
interval $[0,\pi )$ and identify the phase-space points $(p,q)$
of the rays $Q=0$ and $Q=\pi$. From (\ref{bs.2}) it follows that the new
canonical momentum $P$ is proportional to the total energy of the
oscillator
\begin{equation}
E=\omega P\ .
\label{bs.7}
\end{equation}
For the oscillator trajectory on the conic phase space, we have
\begin{equation}
W_c(E)=\oint\limits_{}^{}pdq=\oint\limits_{}^{}PdQ=\frac{E}{\omega}
\int_{0}^{\pi}dQ =\frac{\pi E}{\omega}= 2\pi\hbar
\left(n+\frac 12\right)\ ,
\label{bs.8}
\end{equation}
which leads to the energy spectrum (\ref{bs.5}).

The curvature of the conic phase space is localized at the origin.
One may expect that the conic singularity of the phase space does
not affect motion localized in phase-space regions which do not
contain the origin. Such motion would be indistinguishable from the motion in
the flat phase space. The simplest example of this kind is the
harmonic oscillator whose equilibrium is not located at the origin
\cite{prokhorov}. In the original gauge model, we take the potential
\begin{equation}
V = \frac{\omega^2}{2}\left( |{\bf x}| - r_0\right)^2\ .
\label{bs.9}
\end{equation}
The motion is easy to analyze in the local gauge invariant
variables $(p_r, r)$, when the cone  is cut along the momentum
axis.
\begin{figure}
\centerline{\psfig{figure=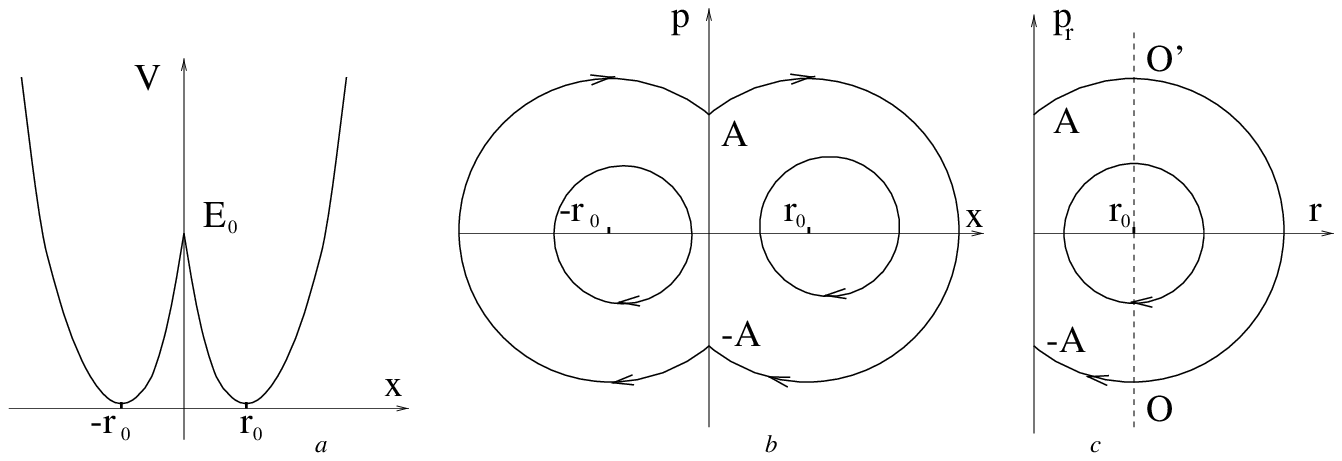}}
\caption{\small {\em a}. Oscillator double-well potential;
\newline
{\em b}. Phase-space trajectories in the flat phase space.
For $E <E_0$ there are two periodic trajectories associated 
with two minima of the double-well potential.  
\newline
{\em c}. The same motion in the conic phase space. It is obtained
from the corresponding motion in the flat phase space by identifying
the points $(p,x)$ with $(-p,-x)$. The local coordinates $p_r$ and
$r$ are related to the parameterization of the cone when the cut
is made along the momentum axis (the states A and -A are the same). 
}
\end{figure}

As long as the energy does not exceed a critical value $E_0=
\omega^2r^2_0/2$, i.e., the oscillator cannot reach the
origin $r =0$, the period of classical trajectory remains
$2\pi /\omega$. The Bohr-Sommerfeld quantization yields the
spectrum of the ordinary harmonic oscillator (\ref{bs.3}).
However the gauge system differs from the corresponding
system with the phase space being a full plane.
As shown in Fig. 3b, the latter system has two periodic trajectories
with the energy $E<E_0$ associated with two minima of the oscillator
double-well potential. Therefore in quantum theory the low energy
levels must be doubly degenerate. Due to the tunneling effect the
degeneracy is removed. Instead of one degenerate level with $E<E_0$
there must be two close levels (we assume $(E_0\, -\, E)/E_0$ $ <\!\!< 1$ to
justify the word ``close''). In contrast, there is no doubling of
classical trajectories in the conic phase space (see Fig. 3c),
and no splitting of the energy levels should be expected.
These qualitative arguments can also be given a rigorous derivation
in the framework of the instanton calculus. We shall return to
this issue after establishing the path integral formalism
for the conic phase space (see section 8.9).

When the energy is greater than $E_0$, the particle can go over
the potential barrier. In the flat phase space there would be only
one trajectory with fixed energy $E$ exceeding $E_0$. From the
symmetry arguments it is also clear that this trajectory is mapped
onto itself upon the reflection $(p,x)\rightarrow (-p,-x)$.
Identifying these points of the flat phase space, we observe that
the trajectory on the conic phase space with $E>E_0$ is continuous
and periodic. In Fig. 3c the semiaxes $p_r <0$ and $p_r>0$ on
the line $r=0$ are identified in accordance with the
chosen parameterization of the cone.

Assume the initial state of the gauge system to be at the
phase space point $O$ in Fig. 3c, i.e. $r(0) =r_0$. Let $t_A$ be
the time when the system approaches the state $-A$.
In the next moment of time the system leaves the state $A$.
The states $A$ and $-A$ lie on the cut of the cone and, hence,
correspond to the same state of the system. There is no jump
of the physical momentum at $t=t_A$. 
From symmetry arguments it follows that
the system returns to the initial state
in the time
\begin{equation}
T_c = \frac{\pi}{\omega} + 2t_A\ .
\label{bs.9a}
\end{equation}
It takes $t=2t_A$ to go from the state $O$ to $-A$ and then
from $A$ to $O'$. From the state $O'$ the system reaches the
initial state $O$ in half of the period of the harmonic oscillator, 
$\pi/\omega$. The time $t_A$ depends on the energy of the system
and is given by
\begin{equation}
t_A = \frac{1}{\omega} \sin^{-1}\sqrt{\frac{E_0}{E}} \leq
\frac{\pi}{2\omega}\ , \ \ \ \ \ E\geq E_0\ .
\label{bs.10}
\end{equation}
The quasiclassical quantization rule yields the equation for
energy levels
\begin{eqnarray}
W_c(E) &=& W(E) - 2E\int_{t_A}^{\frac{\pi}{\omega}-t_A}
\cos^2 \omega t\, dt  
\nonumber\\
&=& W(E)\left(\frac 12 + \frac{\omega t_A}{\pi} +\frac{1}{2\pi}\,
\sin 2\omega t_A\right)= 2\pi\hbar\left(n +\frac 12\right)\ . 
\label{bs.11}
\end{eqnarray}
Here $W(E)=2\pi E/\omega$ is the Bohr-Sommerfeld functional for 
the harmonic oscillator of frequency $\omega$.
The function $W_c(E)$ for the conic phase space is 
obtained by subtracting a contribution of the portion 
of the ordinary oscillator trajectory between
the states $-A$ and $A$ for negative values of the canonical
coordinate, i.e., for $t\in [t_A, \pi/\omega -t_A]$.
When the energy is sufficiently large, $E>\!\!>E_0$, the time
$2t_A$ is much smaller than the half-period $\pi/\omega$, and
$W_c(E) \sim \frac 12 W(E)$, leading to the doubling of
the distance between the energy levels. In this case typical fluctuations
have the amplitude much larger than the 
distance from the classical vacuum to
the singular point of the phase space. The system ``feels'' the
curvature of the phase space localized at the origin. 
For small energies as compared with $E_0$, typical quantum fluctuations
do not reach the singular point of the phase space. The dynamics
is mostly governed by the potential force, i.e., 
the deviation of 
the phase space geometry from the Euclidean one 
does not affect much the low energy dynamics (cf. (\ref{bs.11}) for
$t_A\approx \pi/(2\omega)$).
 As soon as the energy attains the critical value
$E_0$ the distance between energy levels starts growing, tending
to its asymptotic value $\Delta E = 2\hbar\omega$.

The quantum system may penetrate into classically forbidden 
domains. The wave functions of the states with $E < E_0$
do not vanish under the potential barrier. So even for
$E<E_0$ there are fluctuations that can reach the conic
singularity of the phase space. 
As a result a small shift of the oscillator energy 
levels for $E <\!\!<E_0$ occurs.
The shift can be calculated by means of the instanton technique.
It is easy to see that there should exist 
an instanton solution that starts at the
classical vacuum $r=r_0$, goes to the origin and then returns
back to the initial state. We postpone the instanton calculation 
for later. Here we only draw the attention to the fact that,
though in some regimes the classical dynamics may not be
sensitive to the phase space structure, in the quantum theory
the influence of the phase space geometry on dynamics 
may be well exposed.  

The lesson we could learn from this simple qualitative consideration
is that both the potential force and the phase space geometry
affect the behavior of the gauge system. In some regimes the
dynamics is strongly affected by the non-Euclidean geometry
of the phase space. But there might also be regimes where the
potential force mostly determines the evolution of the gauge system,
and only a little of the phase-space structure influence can be seen.
Even so, the quantum dynamics may be more sensitive to the non-Euclidean
structure of the physical phase space than the classical one.

\section{Systems with many physical degrees of freedom}
\setcounter{equation}0

So far only gauge systems with a single physical degree of
freedom have been considered. A non-Euclidean geometry 
of the physical configuration or phase spaces may cause 
a specific {\em kinematic} coupling between physical degrees of
freedom \cite{ijmp91}. The coupling does not depend on details of
dynamics governed by some local Hamiltonian. One could say that
the non-Euclidean geometry of the physical configuration or phase
space reveals itself through observable effects caused by this
kinematic coupling.  We now turn to studying this new feature
of gauge theories.

\subsection{Yang-Mills theory with adjoint scalar matter in
(0+1) spacetime}

Consider Yang-Mills potentials $A_\mu({\bf x},t)$.
They are elements of a Lie algebra $X$ of a semisimple
compact Lie group $G$. In the (0+1) spacetime, the vector
potential has one component, $A_0$, which can depend only on
time $t$. This only component is denoted by $y(t)$. Introducing
a scalar field in (0+1) spacetime in the adjoint representation
of $G$, $x=x(t) \in X$, we can construct a gauge invariant Lagrangian
using a simple dimensional reduction of  the Lagrangian for
Yang-Mill fields coupled to a scalar 
field in the adjoint representation \cite{tmp89,plb89}
\begin{eqnarray}
L&=& \frac 12 \left(D_t x,D_t x\right) - V(x)\ , \label{ym1}\\
D_t x &=& \dot{x} +i [y,x]\ .\label{ym2}
\end{eqnarray} 
Here $(\,,\,)$ stands for an invariant scalar product for the adjoint
representation of the group.
Let $\lambda_a$ be a matrix representation of an orthonormal basis
in $X$ so that $\tr\lambda_a\lambda_b = \delta_{ab}$. Then we can 
make decompositions $y=y^a\lambda_a$ and $x=x^a\lambda_a$ with
$y^a$ and $x^a$ being real. The  invariant scalar product 
can be normalized on the trace $(x,y)=\tr xy$. 
The commutator in (\ref{ym2}) is specified
by the commutation relation of the basis elements 
\begin{equation}
[\lambda_a,\lambda_b]= if_{ab}{}^c\lambda_c\ ,
\label{ym3}
\end{equation}
where $f_{ab}{}^c$ are the structure constants of the Lie algebra.

The Lagrangian (\ref{ym1}) is invariant under the gauge transformations
\begin{equation}
x\rightarrow x^\Omega = \Omega x\Omega^{-1}\ ,\ \ \ \ 
y\rightarrow y^\Omega= \Omega y\Omega^{-1} + i\dot{\Omega}\Omega^{-1}\ ,
\label{ym4}
\end{equation}
where $ \Omega= \Omega(t)$ is an element of the group $G$. Here the 
potential $V$ is also assumed to be invariant under the adjoint action of
the group on its argument, $V(x^\Omega)=V(x)$. The Lagrangian does not
depend on the velocities $\dot{y}$. Therefore the 
corresponding Euler-Lagrange equations yield a constraint
\begin{equation}
-\frac{\pl L}{\pl y} = i [x,D_tx]=0\ .
\label{ym5}
\end{equation}
This is the Gauss law for the model (cf. with the Gauss
law in the electrodynamics or Yang-Mills theory). Note that it involves 
no second order time derivatives of the dynamical variable $x$
and, hence, only implies restrictions on admissible initial 
values of the velocities and positions with which the dynamical
equation
\begin{equation}
D_t^2x = - V^\prime_x
\label{ym6}
\end{equation}
is to be solved.
The Yang-Mills degree of freedom $y$ appears to be purely
nonphysical; its evolution is not determined by the equations
of motion. It can be removed from them 
and the constraint (\ref{ym5}) by
the substitution
\begin{equation} 
x(t) = U(t) h(t) U^{-1}(t)\ ,\ \ \ \ U(t) = 
{\rm T}\exp\left\{ -i\int_0^t
d\tau y(\tau)\right\}\ .
\label{ym7}
\end{equation}
In doing so, we get
\begin{equation}
[h,\dot{h}]=0\ ,\ \ \ \ 
\ddot{h} = - V_h'\ .
\label{ym9}
\end{equation}
The freedom in choosing the function $y(t)$ can be used 
to remove some components of $x(t)$ (say, to set them to zero
for all moments of time). This would imply the removal of
nonphysical degrees of freedom of the scalar field by means
of {\em gauge fixing}, just as we did for the SO(N) model above.
Let us take $G=SU(2)$. The orthonormal basis reads $\lambda_a=
\tau_a/\sqrt{2}$, where $\tau_a, a=1,2,3,$ are the Pauli matrices,
$\tau_a\tau_b = \delta_{ab} +i\varepsilon_{abc}\tau_c$; 
$\varepsilon_{abc}
$ is the totally antisymmetric structure
constant tensor of SU(2), $\varepsilon_{123}=1$.
The variable $x$ is a hermitian traceless $2\times2$ matrix
which can be diagonalized by means of the adjoint transformation 
(\ref{ym7}). Therefore one may always set $h=h^3\lambda_3$.
All the continuous gauge arbitrariness is exhausted, and the real
variable $h^3$ describes the only physical degree of freedom.
However, whenever this variable attains, say, negative values
as time proceeds, the gauge transformation
$h\rightarrow -h$ can still be made. For example, taking $U=e^{i\pi\tau_2/2}$
one find $U\tau_3U^{-1} = \tau_2\tau_3\tau_2=-\tau_3$.
Thus, the physical values of $h^3$ lie on the positive half-axis.
We conclude that
\begin{equation}
{\rm CS}_{\rm phys} = su(2)/{\rm ad} SU(2) \sim \Rs_+\ ,\ \ \ \ \ 
{\rm CS}=X=su(2) \sim \Rs^3\ .
\label{ym10}
\end{equation}

It might look surprising that the system has physical degrees of
freedom at all because the number of gauge variables $y^a$ exactly equals
the number of degrees of freedom of the scalar field $x^a$. The point is
that the variable $h$ has a stationary group formed by
the group elements $e^{i\varphi}, [\varphi,h]=0$ and, hence, so does
a generic element of the Lie algebra $x$. The stationary
group is a subgroup of the gauge group. So the elements $U$ in (\ref{ym7}) 
are specified modulo right multiplication on elements from the
stationary group of $h$, $U\rightarrow Ue^{i\varphi}$. In the SU(2)
example, the stationary group of $\tau_3$ is isomorphic to U(1),
therefore the group element $U(t)$ in 
(\ref{ym7}) belongs to SU(2)/U(1) and has only
two independent parameters, i.e., the scalar field $x$ carries 
one physical and two nonphysical degrees of freedom.    
From the point of view of the general constrained dynamics,
the constraints (\ref{ym5}) are not all independent.
For instance, $\tr(\varphi [x,D_tx]) = 0$ for all $\varphi$
commuting with $x$. 
Such constraints are called {\em reducible} (see
\cite{marc,jan} for a general discussion of constrained systems).
Returning to the SU(2) example, one can see
that among the three constraints only two are independent,
which indicates that there are 
only two nonphysical degrees of freedom contained
in $x$.

To generalize our consideration to an arbitrary group $G$, we would need
some mathematical facts from group theory. The reader familiar
with group theory may skip the following section.

\subsection{The Cartan-Weyl basis in Lie algebras}

Any simple Lie algebra $X$ is characterized by a set of linearly
independent $r$-di\-men\-sio\-nal vectors $\vec{\omega}_j,\ j= 1,2,...
, r= {\rm rank}\ X$, called simple roots. The simple roots form a basis
in the root system of the Lie algebra. Any root $\vec{\alpha}$ is a linear
combination of $\vec{\omega}_j$ with either non-negative integer
coefficients ($\vec{\alpha}$ is said to be a positive root) or non-positive
integer coefficients ($\vec{\alpha}$ is said to be a negative root). 
Obviously,
all simple roots are positive. If $\vec{\alpha}$ is a root then
$-\vec{\alpha}$ is also a root. The root system is completely
determined  by the
Cartan matrix $c_{ij}=-2(\vec{\omega}_i,\vec{\omega}_j)/
(\vec{\omega}_j,\vec{\omega}_j)$ (here $(\vec{\omega}_i,
\vec{\omega}_j)$ is a usual Euclidean scalar product of two $r$-vectors)
 which has a graphic representation known as the Dynkin diagrams 
 \cite{hel,zhel}.
Elements of the Cartan matrix are integers.
For any two roots $\vec{\alpha}$ and $\vec{\beta }$, the cosine of the
angle between them can take only the following values $(\vec{\alpha},
\vec{\beta})[(\vec{\alpha},\vec{\alpha}) (\vec{\beta},\vec{\beta}
)]^{-1/2} = 0,\pm 1/2,\pm 1/\sqrt{2},
\pm \sqrt{3}/2$. By means of this fact 
the whole root system can be restored from the
Cartan matrix \cite{hel}, p.460.

For any two elements $x,\ y$ of $X$, the Killing form is defined as
$(x,y) = \tr({\rm ad}\,{x}{\rm ad}\,{y}) = (y,x)$ 
where the operator ${\rm ad}\,{x}$
acts on any element $y\in X$ as ${\rm ad}\,{x}(y) =[x,y]$ where
$[x,y]$ is a skew-symmetric Lie algebra product that satisfies
the Jacobi identity $[[x,y],z]+[[y,z],x]+[[z,x],y]=0$ for
any three elements of the Lie algebra.
A maximal Abelian subalgebra $H$
in $X$ is called the Cartan subalgebra, $\dim H= {\rm rank}\ X =r$. There
are $r$ linearly independent elements $\omega_j$ in $H$ such that
$(\omega_i,\omega_j) = (\vec{\omega}_i,\vec{\omega}_j)$. We shall
also call the algebra elements $\omega_i$ simple roots. It will
not lead to any confusing in what follows because the root space 
$\Rs^r$ and the Cartan
subalgebra are isomorphic, but we shall keep arrows over elements of
$\Rs^r$. The corresponding elements of $H$ have no over-arrow.

A Lie algebra $X$ is decomposed into the direct sum $X=
H\oplus\sum_{\alpha >0} (X_\alpha\oplus X_{-\alpha}),\ \alpha$ ranges
over the positive roots, $\dim X_{\pm\alpha} = 1$. Simple roots form a
basis (non-orthogonal) in $H$. Basis elements $e_{\pm\alpha}$ of
$X_{\pm\alpha}$ can be chosen such that \cite{hel}, p.176,
\begin{eqnarray}
\left[e_\alpha ,e_{-\alpha}\right]& =& \alpha\ ,\label{A.1}\\
\left[h,e_\alpha \right] &=& (\alpha, h)e_\alpha\ ,\label{A.2}\\
\left[e_\alpha,e_\beta\right] &=& N_{\alpha,\beta}e_{\alpha+\beta}\ ,
\label{A.3}
\end{eqnarray}
for all $\alpha, \beta$ belonging to the root system and for any $h
\in H$, where the constants $N_{\alpha,\beta}$ satisfy
$N_{\alpha,\beta}= -N_{-\alpha,-\beta}$. For any such choice
$N_{\alpha,\beta}^2 = 1/2q(1-p)(\alpha,\alpha)$ where $\beta +n\alpha\
(p\leq n \leq q)$ is the $\alpha$-series of roots containing $\beta$;
$N_{\alpha,\beta}=0$ if $\alpha+\beta$ is not a root.
Any element $x\in X$ can be decomposed over the Cartan-Weyl basis
(\ref{A.1})--(\ref{A.3}),
\begin{equation}
x = x_H + \sum_{\alpha >0}(x^\alpha e_\alpha + x^{-\alpha}e_{-\alpha})
\label{A.4}
\end{equation}
with $x_H$ being the Cartan subalgebra component of $x$.

The commutation relations (\ref{A.1})--(\ref{A.3})
imply a definite choice of the norms
of the elements $e_{\pm\alpha}$, namely, $(e_{\pm\alpha},
e_{\pm\alpha})= 0$ and $(e_\alpha,e_{-\alpha})= 1$ \cite{hel}, p.167.
Norms of simple roots are also fixed in (\ref{A.1})--(\ref{A.3}). 
Consider, for
instance, the su(2) algebra. There is just one positive root
$\omega$. Let its norm be $\gamma = (\omega,\omega)$. The Cartan-Weyl
basis reads $[e_\omega,e_{-\omega}] = \omega$ and
$[\omega,e_{\pm\omega}] = \pm \gamma e_{\pm\omega}$. Let us calculate
$\gamma$ in this basis.  By definition $\gamma =
\tr({\rm ad}\,{\omega})^2$. 
The operator ${\rm ad}\,{\omega}$ is a $3\times3$ 
diagonal matrix with $0,\pm\gamma$ being its diagonal elements as
follows from the basis commutation relations and the definition of the
operator ${\rm ad}\,{\omega}$. Thus, $\tr({\rm ad}\,
{\omega})^2 = 2\gamma^2 =\gamma$, i.e. $\gamma =1/2$.

The su(3) algebra has two equal-norm simple roots $\vec{\omega}_1$
and $\vec{\omega}_2$ with the angle between them equal to $2\pi/3$.
For the corresponding Cartan subalgebra elements we have $(\omega_1,
\omega_1)= (\omega_2,\omega_2) =\gamma$ and $(\omega_1,\omega_2) =
-\gamma/2$. The whole root system is given by six elements
$\pm\omega_1,
\pm\omega_2$ and $\pm(\omega_1+\omega_2) \equiv \pm \omega_{12}$. It is
readily seen that $(\omega_{12},\omega_{12})= \gamma$ and $(\omega_1,
\omega_{12})= (\omega_2,\omega_{12}) = \gamma/2$. All the roots have
the same norm and the angle between two neighbor roots is equal to
$\pi/3$. Having obtained the root pattern, we can evaluate the number
$\gamma$. The (non-orthogonal) basis consists of eight elements
$\omega_{1,2}, \ e_{\pm 1},\ e_{\pm 2}$ and $ e_{\pm 12}$ where we
have introduced simplified notations $e_{\pm\omega_1} \equiv e_{\pm
1}$, etc. The operators ${\rm ad}{\omega}_{1,2}$ are $8\times 8$ diagonal
matrices as follows from (\ref{A.2}) and $[\omega_1,\omega_2]= 0$. Using
(\ref{A.2}) we find $\tr({\rm ad}\,{\omega}_{1,2})^2 = 
3\gamma^2 = \gamma$ and,
therefore, $\gamma= 1/3$. As soon as root norms are established, one
can obtain the structure constants $N_{\alpha,\beta}$. For $X=su(3)$
we have $N_{1,2}^2 = N_{12,-1}^2 =N_{12,-2}^2 = 1/6$ and all others
vanish (notice that $N_{\alpha,\beta} =-N_{-\alpha,-\beta}$ and
$N_{\alpha,\beta}= -N_{\beta,\alpha}$). The latter determines the
structure constants up to a sign. The transformation
$e_\alpha\rightarrow -e_\alpha,\ N_{\alpha,\beta} \rightarrow
-N_{\alpha,\beta}$ leaves the Cartan-Weyl 
commutation relations unchanged. Therefore,
only relative signs of the structure constants must be fixed.
Fulfilling the Jacobi identity for elements $e_{-1},\ e_1,\ e_2$ and
$e_{-2},\ e_1,\ e_2$ results in $N_{1,2}= -N_{12,-1}$ and $N_{1,2} =
N_{12,-2}$, respectively. Now one can set $N_{1,2}=N_{12, -2} =
-N_{12,-1} = 1/\sqrt{6}$, which completes determining the structure
constants for $su(3)$.

One can construct a basis orthonormal with respect to the Killing
form. With this purpose we introduce the elements \cite{hel}, p.181,
\begin{equation}
s_\alpha = i(e_\alpha - e_{-\alpha})/\sqrt{2},\ \ \ \ \ \ c_\alpha =
(e_\alpha + e_{-\alpha})/\sqrt{2}\ 
\label{A.5}
\end{equation}
so that
\be
[h,s_\alpha]=i(h,\alpha)c_\alpha\ ,\ \ \ [h,c_\alpha]=-i(h,\alpha)s_\alpha\ ,
\ \ \ h\in H\ .
\label{adda6}
\ee
Then $(s_\alpha,s_\beta) = (c_\alpha,c_\beta) = \delta_{\alpha\beta}$
and $(c_\alpha,s_\beta) = 0$. Also,
\begin{equation}
(x,x)=\sum_{\alpha >0} \left[(x^\alpha_s)^2 + (x_c^\alpha)^2\right] +
(x_H,x_H)\ ,
\label{A.6}
\end{equation}
where $x_{s,c}^\alpha$ are {\em real} 
decomposition coefficients of $x$ in the
orthonormal basis (\ref{A.5}). Supplementing (\ref{A.5}) by an 
orthonormal basis
$\lambda_j,\ (\lambda_j,\lambda_i)=
\delta_{ij}$, of the Cartan subalgebra (it might be
obtained by orthogonalizing the simple root basis of $H$), we get an
orthonormal basis in $X$; 
we shall denote it $\lambda_a $, that is, for $a=j$,  $\lambda_a $
ranges over the orthonormal basis in the Cartan subalgebra,
and for $a=\alpha$ over the set $s_\alpha, c_\alpha$.

Suppose we have a matrix representation of $X$. Then $(x,y)= c_r\,\tr\
(xy)$ where $xy$ means a matrix multiplication. The number $c_r$ depends
on $X$.  For classical Lie algebras, the numbers $c_r$ are listed in
\cite{hel}, pp.187-190. For example, $c_r = 2(r+1)$ for $X= su(r+1)$.
Using this, one can establish a relation of the orthonormal basis
constructed above for su(2) and su(3) with the Pauli matrices
and the Gell-Mann matrices \cite{hua}, p.17,
respectively. For the Pauli matrices we have
$[\tau_a,\tau_b]=2i\varepsilon_{abc}\tau_c$,
hence, $(\tau_a,\tau_b) = -4\varepsilon_{ab'c'}\varepsilon_{bc'b'} = 8
\delta_{ab} = 4 \tr\ \tau_a\tau_b$ in full accordance with $c_r= 2(r+1),\
r=1$. One can set $\omega = \tau_3/4,\ s_\omega= \varphi\tau_1$ and
$c_\omega = \varphi\tau_2$ where $1/\varphi = 2\sqrt{2}$. A similar
analysis of the structure constants for the Gell-Mann matrices
$\lambda_a$
\cite{hua}, p.18, yields $\omega_1 = \lambda_3/6,\ s_1 =\varphi\lambda_1,\
c_1= \varphi\lambda_2,\ \omega_2= (\sqrt{3}\lambda_8 - \lambda_3)/12,\
s_2= \varphi\lambda_6,\ c_2= \varphi\lambda_7\ , \omega_{12} =
(\sqrt{3}\lambda_8 + \lambda_3)/12,\ s_{12}=\varphi\lambda_5$ and
$c_{12}= -\varphi\lambda_{4}$ where $1/\varphi = 2\sqrt{3}$. This
choice is not unique. Actually, the identification of non-diagonal
generators $\lambda_a,\ a\neq 3,8$ with (\ref{A.5}) depends on a
representation of the simple roots $\omega_{1,2}$ by the diagonal
matrices $\lambda_{3,8}$.  One could choose $\omega_1=\lambda_3/6$ and
$\omega_2=-(\sqrt{3}\lambda_8 + \lambda_3)/12$, which would lead to
another matrix realization of the elements (\ref{A.5}). 

Consider the adjoint action of the group $G$ on its Lie algebra $X$:
$x\rightarrow {\rm ad}\, U(x)$. Taking $U=e^z,\, z\in X$, the 
adjoint action can be written in the form ${\rm ad}\, 
U =\exp(i{\rm ad}\, z)$.
In a matrix representation it has a more familiar form,
$x\rightarrow UxU^{-1}$. The Killing form is invariant under the
adjoint action of the group
\be
\left({\rm ad}\,U(x),{\rm ad}\,U(y)\right) =(x,y)\ .
\label{kf1} 
\ee
In a matrix representation this is a simple statement:
$\tr(UxU^{-1}UyU^{-1})=\tr(xy)$. The Cartan-Weyl basis
allows us to make computations {\em without} referring to
any particular representation of a Lie algebra. This
great advantage will often be exploited in what follows. 

\subsection{Elimination of nonphysical degrees of freedom. 
An arbitrary gauge group case.}

The key fact for the subsequent analysis will be the following formula for
a representation of a generic element of a Lie algebra \cite{zhel} 
\begin{equation}
x = {\rm ad}\, U(h)\ ,\ \ \ U= U(z)=e^{iz}\ ,\ \ \ ({\rm or}\ \ 
x = Uh U^{-1})\ ,
\label{huz}
\end{equation}
in which $h = h^i\lambda_i$ is an element of the Cartan subalgebra $H$
with an orthonormal basis $\lambda_i,\, i=1,2,...,r={\rm rank}\, G$
and the group element $U(z)$ is obtained by the exponential
map of $z = z^\alpha \lambda_\alpha \in X\ominus H$ to the 
group $G$. Here $\alpha = r+1, r+2, ..., N={\rm dim}\, G$ and
$z^\alpha$ are real. The $r$ variables $h^i$ are analogous to
$h^3$ from the SU(2) example, while the variables $z^\alpha$
are nonphysical and
can be removed by a suitable choice of the gauge variables $y^a$
for any actual motion
as follows from a comparison of (\ref{huz}) and (\ref{ym7}).
Thus the rank of the Lie algebra specifies the number of physical
degrees of freedom. The function $h(t)\in H$ describes the time
evolution of the physical degrees of freedom. Note that the constraint
in (\ref{ym9}) is fulfilled identically, $[h,\dot{h}]\equiv 0$,
because both the velocity and position are elements of the maximal
Abelian (Cartan) subalgebra. We can also conclude that the original 
constraint (\ref{ym5}) contains only $N -r$ independent equations.

There is still a gauge arbitrariness left. Just like in the SU(2)
model, we cannot reduce the number of physical degrees of freedom,
but a further reduction of the configuration space of the variable
$h$ is possible. It is known \cite{zhel} that a Lie group contains
a discrete finite subgroup $W$, called the Weyl group, whose
elements are compositions of reflections in hyperplanes 
orthogonal to simple roots of the Cartan subalgebra.
The group $W$ is isomorphic to the group of permutations
of the roots, i.e., to a group that preserves the root system.
The gauge $x=h$ is called an {\em incomplete global gauge with
the residual symmetry group} $W\subset G$ \footnote{The incomplete 
global gauge does not exist for the vector potential (connection) in four
dimensional Yang-Mills theory \cite{soloviev}.See also section 
10.4 in this regard.}.
The residual gauge symmetry can be used for a further 
reduction of the configuration  space. The residual gauge
group of the SU(2) model is $\Z_2$ (the Weyl group for SU(2)) 
which identifies 
the mirror points $h^3$ and $-h^3$ on the real axis. One can
also say that this group ``restores'' the real axis (isomorphic
to the Cartan subalgebra of SU(2)) from the modular domain
$h^3>0$. Similarly, the Weyl group $W$ restores the Cartan
subalgebra from the modular domain called the Weyl {\em chamber},
$K^+\subset H$ \cite{zhel} (up to the boundaries of the 
Weyl chamber being a zero-measure set in $H$). 

The generators of the Weyl group are easy to construct in the
Cartan-Weyl basis.
The reflection of a simple root $\omega$ is given 
by the adjoint transformation: $\hat{R}_\omega \omega \equiv
e^{i\varphi s_\omega}\omega e^{-i\varphi s_\omega}= -\omega$
where $\varphi = \pi/\sqrt{(\omega,\omega)}$. Any element of
$W$ is obtained by a composition of $\hat{R}_\omega$ with
$\omega$ ranging over the set of simple roots.
The action of the generating elements of the  Weyl group  
on an arbitrary element of the Cartan subalgebra reads
\be
\hat{R}_\omega h = \Omega_\omega h\Omega_\omega^{-1}
=h - \frac{2(h,\omega)}{(\omega,\omega)}\,\omega
\ ,\ \ \ \ \Omega_\omega\in G\ .
\label{hhg}
\ee
The geometrical meaning of (\ref{hhg}) is transparent.
It describes a reflection of the vector $h$ in the 
hyperplane orthogonal to the simple root $\omega$. 
In what follows
we assume the Weyl chamber to be an intersection of all
positive half-spaces bounded by hyperplanes orthogonal to
simple roots (the positivity is determined relative to
the root vector direction). The Weyl chamber is said to be
an open convex cone \cite{loosbook}. For any element $h\in K^+$,
we have $(h,\omega) >0$ where $\omega$ ranges over all simple
roots. Thus we conclude that
\begin{equation}
{\rm CS}_{\rm phys} = X/{\rm ad}\, G \sim H/W \sim K^+\ .
\label{hwk}
\end{equation}    
The metric on the physical configuration space can be 
constructed as the induced metric on the surface $U(z)=e$ where
$e$ is the group unity.  
First, the Euclidean metric 
$ds^2 = (dx,dx)\equiv dx^2$ 
is written in the new curvilinear variables (\ref{huz}).
Then one takes its inverse. The induced physical metric is identified 
with the inverse of the $hh$-block of the inverse of the total metric
tensor. In doing so, we find
\begin{eqnarray}
dx&=& {\rm ad}\, U(dh + [h,U^{-1}dU])\ ,\nonumber\\
ds^2 
&=& dh^2+ [h,U^{-1}dU]^2 = 
\delta_{ij}dh^idh^j + \tilde{g}_{\alpha\beta}(h,z) dz^\alpha dz^\beta\ ,
\label{zdz}
\end{eqnarray}
where we have used (\ref{kf1}) and the fact that $[h,U^{-1}dU]
\in X\ominus H$ (cf. (\ref{A.2})) and hence $(dh,[h,U^{-1}dU])=0$.
The metric has a block-diagonal structure and so has its inverse. 
Therefore, the physical metric (the induced metric
on the surface $z=0$) is the
Euclidean one. The physical configuration space is
a Euclidean space with boundaries (cf. (\ref{hwk})). 
It has the  structure of an orbifold \cite{orbi}.

The above procedure of determining the physical
metric is general for first-class constrained systems whose
constraints are linear in momenta. The latter condition
insures that the gauge transformations do not mix up the
configuration and momentum space variables in the total phase space.
There is an equivalent method of calculating the metric
on the orbit space \cite{babelon} which uses only a gauge condition.
One takes the (Euclidean) metric on the original configuration space
and obtains the physical metric by projecting tangent vectors
(velocities) onto the subspace defined by the constraints.    
Since in what follows this procedure will also be used, we give here a
brief description. Suppose we have independent
first-class constraints $\sigma_a = F_a^i(q)p_i$. Consider the 
kinetic energy $H_0 =
g^{ij}p_ip_j/2 \equiv g_{ij}v^iv^j/2$, where $g_{ij}$
is the metric on the total configuration space,
$v^i=g^{ij}p_j$ tangent vectors, and $g^{ij}$
the inverse of the metric. We split the set of the canonical coordinates
$q^i$ into two subsets $h^\nu$ and $\bar{q}{}^a$ such that
the matrix $\{\bar{q}{}^a,\sigma_b\}= F_b^a(q)$ is not
degenerate  on the surface $\bar{q}^a=0$ except, maybe,
on a set of zero measure. Then the physical phase space
can be parameterized by canonical coordinates $p_\nu$ and $h^\nu$.
Denoting $\bar{F}_a^b (h) = F_a^b\vert_{\bar{q}=0}$,
similarly $\bar{F}_a^\nu$ and $\bar{g}{}^{ij}$, we solve the
constraints for nonphysical momenta $\bar{p}_a 
= -(\bar{F}^{-1})_a^b \bar{F}_b^\nu p_\nu
\equiv \gamma_b^\nu p_\nu$ and substitute
the result into the kinetic energy:
\ba
H_0 &=& \frac 12 g_{ph}^{\mu\nu}p_\nu p_\mu =
\frac 12 g^{ph}_{\mu\nu} v^\mu v^\nu\ ,\\
g_{ph}^{\mu\nu} &=& \bar{g}{}^{\mu\nu}- \gamma_a^\mu \bar{g}{}^{a\nu}
-\bar{g}{}^{\mu a}\gamma_a^\nu +\gamma_a^\mu \bar{g}{}^{ab}\gamma_b^\nu\ ,
\label{genmet}
\ea
where $g^{ph}_{\mu\nu}$ is the inverse of $g^{\mu\nu}_{ph}$;
it is the metric on the orbit space which determines the
norm of the corresponding tangent vectors $v^\mu$ (physical velocities).
Instead of conditions $\bar{q}=0$, one can use  general
conditions $\chi^a(q)=0$, which means that locally $\bar{q}=\bar{q}(h)$,
where $h$ is a set of parameters to span the surface $\chi^a(q)=0$,
instead of $\bar{q}=0$ in the above formulas.    
In the model under consideration, we set $x = h +z$, $h\in H$, and impose
the condition $z=0$. Then setting $z$ equal zero in the  constraint 
we obtain $[h,p_z]=0$ ($p_z\equiv \bar{p}$), which leads
to $p_z=0$ as one can see from the commutation relation (\ref{A.2}).
Therefore $g_{ph}=1$ because $g=\bar{g}=1$.

It is also of interest to calculate the induced volume
element $\mu(h)d^rh$ in ${\rm CS}_{\rm phys}$. 
In the curvilinear coordinates (\ref{huz}), the variables $z$ 
parameterize a gauge orbit through 
a point $x=h$. For $h\in K^+$, the gauge orbit 
is a compact manifold of dimension $N -r,\ 
\dim X =N$,
and isomorphic to $G/G_H$ where $G_H$ is the maximal
Abelian subgroup of $G$, the Cartan subgroup.
The variables $h$ span the space locally transverse to
the gauge orbits. So, the induced volume element
can be obtained from the decomposition 
\begin{equation}
d^Nx = 
\sqrt{\det g}\, d^{N-r}z d^rh = \mu(h)d^rh\tilde{\mu}(z)d^{N-r}z\ .
\label{nrz}
\end{equation}
Here $g$ is the metric tensor in (\ref{zdz}).
Making use of the orthogonal basis constructed in the previous
subsection, the algebra element $U^{-1}dU$ can be represented in the
form $-i\lambda_a F^a_\alpha(z) dz^\alpha$ with $F^a_\alpha$ being
some functions of $z$. Their explicit form will not be relevant to us.
Since the commutator $[\lambda_i,\lambda_\alpha]$ always belongs
to $X\ominus H$ and the $\lambda_i$'s are commutative, we find
$[h,U^{-1}dU] = \lambda_\gamma h^i f_{i\alpha}{}^\gamma F^\alpha_\beta
dz^\beta$. Hence,
\begin{equation}
\tilde{g}_{\alpha\beta} = F_\alpha^\gamma(z) G_{\gamma\delta}(h)
 F_\beta^\delta(z)\ ,\ \ \ \ 
G_{\alpha\beta} = \omega_{\alpha\gamma} \omega_{\beta}{}^\gamma\ ,\ \ \ \
\omega_{\alpha\beta} = h^i f_{i\alpha\beta}\ ,
\label{ifi}
\end{equation}
and the Cartesian metric $\delta_{ab}= (\lambda_a,\lambda_b)$ 
is used to lower and rise the indices of the structure constants.
Substituting these relations
into the volume element (\ref{nrz}) we obtain $\mu(h) = 
\det \omega(h)=\det (i{\rm ad}\, h)$.
The latter determinant is quite easy to calculate in the orthogonalized
Cartan-Weyl basis.  Indeed, from (\ref{A.2}) it 
follows that $[h,\lambda_\alpha]
= i\omega_{\alpha}{}^{\beta}(h) \lambda_\beta$ and $\lambda_\alpha$ 
is the set (\ref{A.5}). Let us order the basis elements $\lambda_a$
so that the first $r$ elements form the basis in the Cartan subalgebra,
while $\lambda_a = s_\alpha$ and $\lambda_{a+1} = c_\alpha$ for
$a=r+1, r+3, ..., N-1$.
An explicit form of the matrix 
$i\omega_{\alpha}{}^{\beta}$ is obtained from the 
commutation relations (\ref{adda6}). It is block-diagonal,
and each block is associated with the corresponding positive
root $\alpha$ and equals $i(h,\alpha)\tau_2$ ($\tau_2$ being
the Pauli matrix). Thus,
\begin{equation}
\mu(h) =\det(i{\rm ad}\, h) = {\kappa}^2(h) \ ,\ \ \ 
\kappa(h) = \prod_{\alpha>0}(\alpha,h)\ .
\label{hhh}
\end{equation}   
The density $\mu$ is invariant under
permutations and reflections of the roots, i.e., with
respect to the Weyl group: $\mu(\hat{R}_\omega h)=\mu(h)$
for any simple root $\omega$. It also {\em vanishes} at
the boundary of the Weyl chamber, $(\omega,h)=0$.

One should draw attention to the fact that the determinant
of the induced metric on the physical configuration space  does
{\em not} yield the density. This is a generic
situation in gauge theories \cite{babelon}: In addition to the 
square root of the determinant of the physical metric, the 
density also contains a factor being the volume of the gauge orbit
associated with each point of the gauge orbit space. In the model
under consideration the physical configuration space has a Euclidean
metric, and $\mu(h)$ determines the volume of the gauge orbit through the 
point $x=h$ up to a factor ($\int_{G/G_H} dz\tilde{\mu}(z)$) which is
independent of $h$. For example, the adjoint action of SU(2)
in its Lie algebra can be viewed as rotations in three dimensional
Euclidean space. The gauge orbits are concentric two-spheres.
In the spherical coordinates we have $d^3x =\sin\theta d\theta d\phi
r^2dr$. The volume of a gauge orbit through $x^i=\delta^{i1}r$ is 
$4\pi r^2$. In (\ref{nrz}) $z^\alpha$ are the angular variables
$\theta$ and $\phi$, while $h$ is $r$, and $\tilde{\mu}=\sin\theta$,
$\mu = r^2$.  

\subsection{Hamiltonian formalism}

Now we develop the Hamiltonian formalism for the model
and describe the structure of the physical phase space.
The system has $N$ primary constraints $\pi_a= \pl L/\pl\dot{y}^a=0$.
Its canonical Hamiltonian reads
\begin{equation}
H = \frac 12 p^2 + V(x) + y^a \sigma_a\ ,
\label{vxy}
\end{equation}
where $p^2=(p,p)$, $p =\pl L/\pl\dot{x} = D_t x$ is the momentum
conjugate to $x$ and 
\begin{equation}
\sigma_{a} = i (\lambda_a,[x,p])=0\ ,\ \ \ \{\sigma_{a},\sigma_{b}\}
=if_{ab}{}^c \sigma_{c}
\label{pxf}
\end{equation} 
are the secondary constraints. They generate the gauge transformations
on phase space given by the adjoint action of the group $G$ on its
Lie algebra
\begin{equation}
p\rightarrow p^\Omega = \Omega p\Omega^{-1}\ ,\ \ \ \
x\rightarrow x^\Omega = \Omega x\Omega^{-1}\ ,
\label{xxx}
\end{equation}
because $\{p,\sigma_a\}=i[\lambda_a,p]$ and  $\{x,\sigma_a\}=i[\lambda_a,x]$.
The Hamiltonian equations of motion do not specify the
time evolution of the gauge variable $y$. So the phase
space trajectory described by the pair $p(t),\, x(t)$
depends on the choice of $y(t)$. Trajectories associated
with different functions $y(t)$ are related to one
another by gauge transformations.
Just like in the Lagrangian formalism, this gauge arbitrariness
can be used to suppress dynamics of some degrees of freedom of the
scalar field $x(t)$. 

We choose the $y(t)$ so that $x(t)=h(t)\in H$.   
The constraint (\ref{pxf}) means that the momentum and
position should commute as Lie algebra elements, $[p,x]=0$.
Therefore on the constraint  surface, the canonical momentum
$p_h$ conjugate to $h$ must commute with $h$, $[p_h,h]=0$.
This is a simple consequence of the gauge transformation law
(\ref{xxx}): If the variable $x(t)$ is brought to the Cartan subalgebra
by a gauge transformation, then the same gauge 
transformation simultaneously
applies to $p(t)$ turning it into $p_h(t)$. Since the constraint
is covariant under gauge transformations, the new canonical variables
$h$ and $p_h$ should also fulfill the constraint. Thus, we are led
to the conclusion that $p_h$ is an element of the Cartan subalgebra
because it commutes with a generic element $h\in H$. There is
no more continuous gauge arbitrariness left, but a further reduction of
the phase space is still possible. The variable $h$ has gauge equivalent
configurations related to one another by the Weyl transformations.
In the phase space spanned by the Cartan algebra elements $p_h$ and
$h$, the Weyl group acts simultaneously on the momentum and position
variables in accordance with the gauge transformation law (\ref{xxx}).
Thus,
\begin{equation}
{\rm PS}_{\rm phys} \sim H\oplus H/W \sim \Rs^{2r}/W\ . 
\label{rrw}
\end{equation}
By identifying the points $(\hat{R}p_h, \hat{R}h), \, 
\hat{R}\in W$, the Euclidean
space $\Rs^{2r}$ turns into a $2r$-dimensional hypercone 
which, after an appropriate cut, is unfoldable into $\Rs^r\oplus K^+$.   

For generic configurations 
$h\in K^+$ the physical phase space has no singularities
and is locally flat.
When $h$ approaches a generic point on the boundary $(h,\omega)=0$ of 
the Weyl chamber, the physical phase space exhibits a conic singularity.
Indeed, we may always make a linear canonical transformation such that
one of the canonical coordinates, say $h^\perp$, 
varies along the line perpendicular to
the boundary, while the others span hyperplanes 
parallel to the hyperplane  
$(h,\omega)=0$ being a part of the Weyl chamber boundary. 
In the new variables, the Weyl transformation
that flips  sign of the root $\omega$ will change signs of
$h^\perp$ and its canonical momentum, while leaving the other canonical
variables unchanged. Thus, at a generic point of the Weyl chamber boundary,
the physical phase space has a {\em local} structure $\Rs^{2(r-1)}\oplus
cone(\pi)$. 

The Weyl chamber boundary is not a smooth manifold and 
contains  intersections
of two hyperplanes $(\omega_1,h)=0$ and $(\omega_2,h)=0$. At these
points, the two local conic singularities of the physical phase space
associated with simple roots $\omega_{1,2}$ would merge, forming
locally a 4-dimensional hyperconic singularity. This singularity 
cannot be simply described as a direct product of two    
cones $cone(\pi)$. It would only be the case if the roots $\omega_{1,2}$
are orthogonal. In general, the tip of the hypercone would be ``sharper''
than that of $cone(\pi)\oplus cone(\pi)$, meaning that the hypercone
can always be put inside of  $cone(\pi)\oplus cone(\pi)$ when their tips
are at the same point. This can be understood again in the local canonical
variables where the coordinates $h$ are split into a pair $h^\perp$ that
spans a plane perpendicular to the intersection of two hyperplanes 
$(\omega_{1,2},h)=0$ and the others orthogonal to $h^\perp$. The root
pattern in any plane containing at least two roots (e.g., a plane
through the origin and parallel to 
the $h^\perp$- plane) is isomorphic to one of the root patterns
of the groups of rank two, i.e., SU(3), Sp(4)$\sim$SO(5), G${}_2$ or
just SU(2)$\times$SU(2). 
In the latter case the simple roots $\omega_{1,2}$ 
are orthogonal. A modular domain of $h^\perp$ coincides with
the Weyl chamber of one of these groups and is contained in the positive 
quadrant being the Weyl chamber for SU(2)$\times$ SU(2).
That is, a solid region bounded by 
the hypercone spanned by $p^\perp$ and $h^\perp$ and 
isomorphic to the quotient space $\Rs^4/W$
is contained in the solid region
bounded by $cone(\pi)\oplus cone(\pi)$. The procedure is straightforward to
generalize it to the boundary points belonging to intersections
of three hyperplanes $(\omega_{1,2,3},h)=0$, etc. At the origin,
the physical phase space has the most singular point being the tip
of $2r$-dimensional hypercone which is ``sharper'' than 
$[\oplus cone(\pi)]^r$. 

We shall see that the impossibility to split globally the physical
degrees of freedom into ``conic'' and ``flat'' ones, which is 
 due to the non-Euclidean (hyperconic) structure 
of the physical phase space, will have 
significant dynamical consequences. For example, the physical
frequencies of an {\em isotropic} oscillator turn out to be
proportional to orders of the independent invariant (Casimir) 
polynomials of the corresponding Lie algebra, rather than being
equal as one might naively expect after fixing the gauge $x=h$.
In the coordinate representation of quantum theory, 
the existence of the boundaries in the configuration space
of the physical variable $h$ will also have important
consequences.

\subsection{Classical dynamics for groups of rank 2.}

To find out what kind
of dynamical effects are caused by the hyperconic structure
of the phase space, we analyze an isotropic
harmonic oscillator for groups of rank 2, i.e., for SU(3),
SO(5)$\sim$Sp(4) and G${}_2$. Eliminating the nonphysical degrees
of freedom by choosing $y(t)$ so that $x(t)=h(t)\in H$, the Hamiltonian
for physical degrees of freedom assumes the form
\begin{equation}
H= \frac 12  \left(p_h^2 + h^2\right)\ .
\label{phh}
\end{equation}
In this parameterization of the physical phase space the canonical
coordinates are restricted to the Weyl chamber.   
For the sake of simplicity we set the oscillator frequency,
as the {\em parameter} of the Hamiltonian (\ref{phh}),  to {\em one}.
The physical configuration space, i.e., the Weyl chamber, is a sector
with angle $\pi/\nu$ on the plane, 
where $\nu = 3,4,6$ for SU(3), SO(5)$\sim$Sp(4)
and G${}_2$, respectively. A trajectory of the oscillator for
the group SU(3) is shown in Fig. 4. The initial conditions are chosen
so that the solutions of equations of motion have the form
\begin{equation}
h_2(t) = A_2\cos t\ ,\ \ \ \ h_1(t) = A_1\sin t\ ,
\label{tat}
\end{equation}
and $A_2 > A_1$. The Weyl chamber is a sector with 
angle $\pi/3$ which is shown as a grey area in the figure.
The ray $OO'$ is its symmetry axis. 

In the initial moment of time
$t=0$ the oscillator is located at the point $A$. Then it follows
the elliptic trajectory extended along the axis $h_2$ and at $t=\pi/6$
reaches the point $B$, i.e., the boundary of the Weyl chamber.
The further motion along the ellipse in the sector bounded by the rays
$O\gamma$ and $O\gamma_1$ is gauge equivalent to the motion
from $B$ to $C$ in the Weyl chamber $K^+$. It looks like the oscillator
hits the boundary, reflects from it and  arrives
to the point $C$ at time $t=\pi/3$. 
Though at the point $B$ the oscillator momentum
abruptly changes its direction,  it is important to realize 
that there is {\em no} force causing this change
because the oscillator potential is smooth on the entire plane.
The momenta right before and after hitting
the boundary wall are {\em gauge equivalent}. They are related by
the Weyl transformation being
the reflection $\hat{R}_{\omega_2}$ relative to the line $\gamma'\gamma$ 
perpendicular to the root $\omega_2$. So there is no dramatic 
change of the physical state of the system at the moment of reaching the
boundary. Just like in the SO(N)
model of section 3, the trajectory is {\em smooth}
on the hyperconic physical phase space (\ref{rrw}). The momentum jump is a
{\em coordinate artifact} occurring through a cut made on the physical
phase space to parameterize it by a particular set of
local canonical coordinates $p_h\in H$
and $h\in K^+$.  
\begin{figure}
\centerline{\psfig{figure=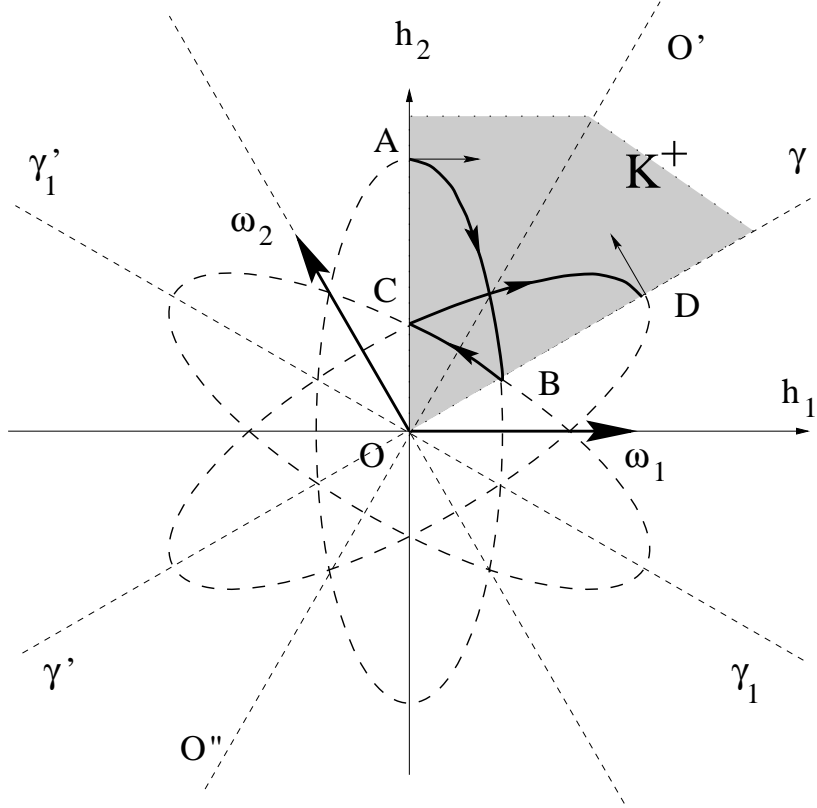}}
\caption{\small Classical dynamics in the Weyl chamber of SU(3)}
\end{figure}
Within the path integral formalism for the model, we shall see
that the phase of the wave function does {\em not} change
under such a reflection, in full contrast with the realistic
reflection from an infinite potential wall where the phase 
would be shifted by $\pi$.

At the point $C$ the oscillator hits 
the boundary of the Weyl chamber one more time and follows the elliptic
segment $C\rightarrow D$. Again, at the very moment of
the collision, no abrupt change of the physical state occurs.  
Finally, at $t=\pi$ the oscillator
reaches the point $D$, reflects from the boundary  and
goes the same way back to the point $A$, returning there at
$t=2\pi$.

What are independent frequencies of this two-dimensional 
isotropic oscillator? It is quite surprising that they
do not equal just one (the frequency that enters into the Hamiltonian),
but rather 2 and $\nu =3$.
By definition the angular frequency is $2\pi/T$ where $T$ is the
time in which the system returns into the initial state upon 
periodic motion. The system state is specified by values of
the momentum and position of the physical degree of freedom 
in question. Let us decompose the motion of the system into
oscillations along the axis $O'O''$ and the angular motion
about the origin $O$. After passing the segments
$A\rightarrow B\rightarrow C$ by the oscillator, the angular
variable attains its initial value since the angles $O'OC$ and
$O'OA$ coincide, while the equality of the corresponding 
(angular) canonical momentum at the points $A$ and $C$ follows from its
conservation law. The angular degree of freedom  
returns to the initial state two more times as the oscillator
follows the path $C\rightarrow D\rightarrow C$, and then returns
to the initial state after passing the segments
$C\rightarrow B\rightarrow A$. Thus, the period
of the angular variable is three times less than that of
the angular variable of an ordinary two-dimensional isotropic
oscillator, i.e., the physical frequency is tripled.
From Fig. 4 one can easily see that the states of the radial
degree of freedom at points $A$ and $D$ are the same, so the 
physical frequency of the radial degree of freedom is doubled.

A similar analysis can also be done for the groups SO(5)
$\sim$ Sp(4) and G${}_2$. For them the independent frequencies
appear to be $2$ and $\nu$. Note that the Weyl chamber is
a sector with angle $\pi/\nu$.
The numbers 2 and $\nu$ are, in fact, fine
characteristics of the groups, namely, they are degrees
of two independent invariant (Casimir) polynomials,
which are $\tr x^2$ and $\tr x^\nu$ in a matrix representation. 
Any  regular  function $f(x)$ invariant under the adjoint action of
the group on its argument is a function of these two independent
polynomials. This fact holds for an arbitrary semisimple
compact gauge group G: The independent frequencies of the 
isotropic harmonic oscillator are determined by degrees
of independent Casimir polynomials. The number of the
independent Casimir polynomials equals the rank of the group G, i.e.,
the number of physical degrees of freedom.
The list of degrees of the independent Casimir polynomials  
for each group can be found in
\cite{zhel}. In the next subsection we shall develop a Hamiltonian
formalism in explicitly gauge invariant variables and 
see the relation between the physical frequencies and orders of the 
independent Casimir polynomials  once again.

So, in the classical theory the hyperconic structure of the 
physical phase space reveals itself through the effect
of reflections of physical trajectories from the boundaries
of the physical configuration space when the latter is parameterized
by elements of the Cartan subalgebra. One should stress again that
the effect of changing the physical frequencies of the oscillator
does not depend on the choice of local canonical variables and
is essentially due to the hyperconic structure of the physical
phase space. To calculate the effect, we used the above 
parameterization of the physical phase space. The choice
of the parameterization is, in fact,
a matter of convenience. Had we taken another set of local 
canonical coordinates, say, by making cuts of the hyperconic
phase space such that the momentum variable $p_h$ is restricted
to the Weyl chamber, we would have arrived to the very same 
conclusion about the oscillator frequencies. The message
is therefore: Whatever
local canonical coordinates are assumed, the coordinate singularities
associated with them should be carefully taken into account when solving
the dynamical problem because they may contain information about
the geometry of the physical phase space.

Another important observation is that the geometry of the physical
phase space does not permit excitations
of the Cartesian degrees of freedom $h^i$ independently, even though the 
Hamiltonian does not contain any interaction between them.
This effect can be anticipated from the fact that the residual
Weyl transformations mix the $h^i$.  
Such a {\em kinematic} coupling between the physical degrees of freedom 
appears to be crucial for constructing a 
correct path integral formalism for gauge systems \cite{plb89,tmp89}.
If the Hamiltonian does not contain any coupling between the
degrees of freedom, then the transition amplitude in quantum mechanics
is factorized over the degrees of freedom. This, however, is not
the case if the phase space is not Euclidean. It
can already be seen from the correspondence principle. Let us, for example,
set $V=0$ in the above model. In addition to the straight trajectory
connecting the initial point $h_{1}\in K^+$ and the final
point $h_{2}\in K^+$, there are $2\nu -1$ trajectories which
involve several reflections from the boundary of the Weyl chamber.
Since no change of the physical state occurs at the very moment
of the reflection, these trajectories would also be acceptable
classical trajectories contributing to the semiclassical 
transition amplitude at the same footing as the straight one. 
The reflected trajectories
can be viewed as {\em straight} lines connecting the points 
$\hat{R}h_{1},\ \hat{R}\in W$, 
and $h_{2}$.  Hence, they satisfy the classical equations of motion. 
Using the Weyl symmetry they can be mapped
into piecewise straight continuous trajectories inside of the Weyl
chamber. The contribution of these trajectories makes it
impossible to factorize the semiclassical transition amplitude
because the reflected trajectories cannot be associated with
an excitation of any particular Cartesian degrees of freedom $h^i$
(see Fig. 9 in section 8.5).

\subsection{Gauge invariant canonical variables for groups of rank 2.}

The analysis of classical dynamics of the isotropic harmonic
oscillator shows that independent excitations of the Cartesian
degrees of freedom $h^1$ and $h^2$ are impossible due to
the non-Euclidean structure of their physical phase space.
If, say, $p_{h2}=\dot{h}^2=0$ in the initial moment of time, then
after hitting the boundary of the Weyl chamber, the momentum
$p_{h1}=\dot{h}^1$ will be re-distributed between both physical
degrees of freedom, thus exciting the $h^2$-degree of freedom. 
This occurs not due to an action of any local
potential force (it can even be zero), but rather due to the non-Euclidean
structure of the physical phase space.   
This specific kinematic coupling implies that the independent
physical excitations must
be {\em collective} excitations of the original degrees of freedom.
Here we show that the collective excitations are described by 
composite gauge invariant variables. The goal is therefore to demonstrate
that the kinematic coupling is important
to maintain the gauge invariance of the Hamiltonian dynamics
of physical canonical variables.   

In the Hamiltonian (\ref{vxy}) for groups of rank $r=2$
we introduce new gauge invariant variables \cite{tmp89}
\begin{equation}
\Phi_1 = (\tr x^2)^{1/2}\ ,\ \ \ \ \Phi_2 = \Phi_1^{-\nu}\tr x^\nu\ ,
\label{phi12}
\end{equation}
where $\nu$ is the degree of the second independent Casimir polynomial.
The use of a matrix representation is just a matter of 
technical convenience.
The invariant {\em independent} (Casimir) polynomials can also be
written via symmetric invariant {\em irreducible} tensors,
$\tr x^2 \sim \delta_{ab}x^ax^b$ and $\tr x^\nu\sim
d^{(\nu)}_{a_1\cdots a_\nu}x^{a_1}\cdots x^{a_\nu}$.
Every symmetric invariant tensor in a Lie algebra can be
decomposed over the basis formed by irreducible symmetric
invariant tensors \cite{zhel}. Ranks of the irreducible
tensors are orders of independent Casimir polynomials in the 
Lie algebra.
The canonical momenta conjugate to the new variables read
\begin{equation}
\pi_i = \frac{\tr( pe_i)}{ \tr e_i^2}\ ,\ \ \ i=1,2\ ;\ \ \ 
e_i = \frac{\pl \Phi_i}{\pl x^a}\, \lambda_a\equiv 
\frac{\pl \Phi_i}{\pl x}\ .
\label{pi12}
\end{equation}
By straightforward computation one can convince oneself that
the elements $e_i$ possess the following properties:
\begin{equation}
\tr(e_1e_2) = 0 \ ,\ \ \ \ [e_1,e_2] = 0\ .
\label{e1e2}
\end{equation}
Therefore they can serve as the local basis in the Cartan subalgebra
$H$. One can also show that 
\begin{equation} 
\tr e_1^2 = 1\ ,\ \  \tr e_2^2 = \frac{\nu^2}{\Phi_1^{2}}\,\left(
c_2 +c_1\Phi_2 - \Phi_2^2\right) \equiv 
\frac{\nu^2}{\Phi_1^{2}}\,\left(a - \left[\Phi_2 -b\right]^2 
\right)\, ,\ \
\label{tre}
\end{equation}
where $b= c_1/2,\, a = c_2 + c_1^2/4$, and the constants
$c_{1,2}$ depend on the structure constants and specify the 
decomposition of the gauge invariant polynomial 
$[\tr(\lambda_ax^{\nu-1})]^2
= (c_1\Phi_2 +c_2)\Phi_1^{2(\nu-1)}$ over the basis
polynomials $\tr x^2$ and $\tr x^\nu$. For example, for SU(3) we have
$\nu=3$ and
$c_1=0$, $c_2 =1/6$. This can be verified by a straightforward
computation in the matrix representation. 
 
Let us decompose the canonical momentum $p$ over the basis $e_i$,
$p= \pi_ie_i +\tilde{p}$ where $\tr e_i\tilde{p}=0$. A solution 
to the constraint equation $[p,x]=[\tilde{p},x]=0$ is $\tilde{p}=0$.
That is, all the components of $p$ orthogonal to the Cartan basis elements
$e_i$ must vanish since the commutator of $\tilde{p}$
and $x\sim e_1$ does not belong to the Cartan subalgebra.
The physical Hamiltonian of an isotropic harmonic oscillator,
$V=\tr x^2/2=\Phi_1^2/2$, assumes the form
\begin{equation} 
H_{ph} = \frac 12 \pi_1^2 + \frac{\nu^2\pi_2^2}{2\Phi_1^2}\,
\left(a - \left[b- \Phi_2\right]^2
\right) +\frac 12 \Phi_1^2\ .
\label{hab}
\end{equation}
From positivity of the norm $\tr e^2_2\geq 0$ we infer the condition
\begin{equation}
-1\leq (\Phi_2 - b)/\sqrt{a}\leq 1\ .
\label{phi2}
\end{equation}
The Hamiltonian equations of motion are
\begin{eqnarray}
\dot{\pi}_1 &=&\{\pi_1,H_{ph}\} = -\Phi_1 + 
\frac{\nu^2\pi_2^2}{\Phi_1^3}\,\left(a - \left[b- \Phi_2\right]^2
\right)\, ,\nonumber\\
\dot{\Phi}_1 &=&\{\Phi_1,H_{ph}\} = \pi_1\, ;\label{eq1}\\
\dot{\pi}_2 &=&\{\pi_2,H_{ph}\} =
\frac{\nu^2\pi_2^2}{\Phi_1^2}\, \left[b- \Phi_2\right]\, ,
\nonumber\\
\dot{\Phi}_2 &=&\{\Phi_2,H_{ph}\} = 
\frac{\nu^2\pi_2}{\Phi_1^2}\,\left(a - \left[b- \Phi_2\right]^2
\right)\, .
\label{eq2}
\end{eqnarray}
They admit the following oscillating solutions {\em independently} for 
each degree of freedom
\begin{eqnarray}
\Phi_2(t)&=& \pi_2(t)=0\, ;\label{sol11}\\
\Phi_1(t)&=& \sqrt{E}|\cos t|\, ,\ \ \ \pi_1(t)=- \sqrt{E}\sin t \,
\varepsilon(\cos t)\ ,
\label{sol12}
\end{eqnarray}
where $E$ is the energy and $\varepsilon$ denotes the sign function;
and
\begin{eqnarray}
\Phi_1(t)&=& \sqrt{E} \ ,\ \ \ \ \ \ \ \  \pi_1(t)=0\ ;\label{sol21}\\
\Phi_2(t)&=& b + \sqrt{a}\cos\nu t\ ,\ \ \ 
 \pi_2(t)=-\frac{E}{\nu\sqrt{a}\sin{\nu t}}\ .
\label{sol22}
\end{eqnarray}
Absolute value bars in (\ref{sol12}) are necessary because
$\Phi_1$ is positive. One can easily see that the independent
frequencies are degrees of the independent Casimir polynomials,
2 and $\nu$. Clearly, the variable $\cos^{-1}[(\Phi_2 -b)/\sqrt{a}]$
can be associated with the angular variable introduced in the
previous section and $\Phi_1$ with the radial variable.

Thanks to the gauge invariance of the new variables $\Phi_{1,2}$
we may always set $x=h$ in (\ref{phi12}) and $p=p_h$ in
(\ref{pi12}), thus establishing the canonical transformation 
between the two sets of canonical variables. The kinematic
coupling between $\Phi_{1,2}$ is absent. However, to excite
either of $\Phi_{1,2}$ {\em independently}, excitations
of {\em both} Cartesian degrees of freedom $h_{1,2}$ are
needed. Thus, the removal of the kinematic coupling is
{\em equivalent} to restoration of the explicit gauge invariance.
In sections 7.3 and 7.4 we show that this remarkable feature
has an elegant group theoretical explanation based on the 
theorem of Chevalley. The mathematical fact is that, if one attempts to
construct all polynomials of $h$ invariant relative to
the Weyl group, which specify wave functions of the
physical excitations of the harmonic oscillator, then one
would find that all such polynomials are polynomials
of the elementary ones $\tr h^2$ and $\tr h^\nu$ \cite{zhel}.
Since orders of the polynomials determines the energy levels
of the harmonic oscillator, we anticipate that the spectrum
must be of the form $2n_1+\nu n_2$, where $n_{1,2}$ are nonnegative
integers. 

{\em Remark}. The canonical variables (\ref{phi12}) and (\ref{pi12}),
though being explicitly gauge invariant
and describing independent physical excitations of the harmonic oscillator, 
can be regarded as just another possible
set of the local canonical coordinates on the 
non-Euclidean physical phase space.
As one can see from (\ref{sol12}) and (\ref{sol22}), there are singularities
in the phase space trajectories in these variables too. One can actually
find arguments similar to those given at the end section 3.2 to show that 
there are no {\em canonical} coordinates on the hyperconic phase space
in which the phase space trajectories are free of singularities.
The singularities can be removed 
by introducing a {\em noncanonical}
symplectic structure on the physical phase space (cf. section 3.3
and see section 6.4 for a generalization). 

\subsection{Semiclassical quantization}

Having chosen the set $h,p_h$ of local canonical variables
to describe elementary excitations of physical degrees
of freedom, we have found a specific kinematic coupling
as a consequence of the non-Euclidean structure 
of the physical phase space. If now we proceed to quantize
the system in these variables, it is natural to expect
some effects caused by the kinematic coupling. Let us
take a closer look on them.  

The Bohr-Sommerfeld quantization rule is 
coordinate-free, i.e., invariant
under canonical transformations. We take advantage
of this property and go over to the new canonical
variables $\Phi_i,\pi_i$ from $p_h,h$. 
Note that due to the gauge invariance of the new 
variables we can always replace $x$ and $p$ in (\ref{phi12})
and (\ref{pi12}) by $h$ and $p_h$, respectively.
We have  
\begin{equation}
W =\oint(p_h,dh) = \oint \left(\pi_1d\Phi_1 + \pi_2d\Phi_2\right)
=2\pi\hbar n\ ,
\label{wkb}
\end{equation}
where $n$ is a non-negative integer. Here we have also
omitted the vacuum energy \cite{maslov}.
For an ordinary isotropic oscillator of unit frequency,
one can find that $E= n\hbar =(n_1 + n_2)\hbar$, where
$n_{1,2}$ are non-negative integers, just by applying the
rule (\ref{wkb}) for an independent periodic motion of
each degree of freedom. That is, the functional $W$ is
calculated for the motion of one degree of freedom of the
energy $E$, while the motion of the other degree of freedom
is suppressed by an appropriate choice of the initial conditions.
Then the same procedure applies to the other degree of freedom.
So the total energy $E$ of the system is attained through
exciting only one degree of freedom in the above procedure.

Although the independent excitations of the components of $h$
are impossible, the new canonical variables can be excited
{\em independently}. Denoting $\oint \pi_id\Phi_i =W_i(E)$ (no
summation over $i$), we take the phase-space trajectory  
(\ref{sol12}) and find
\be
W(E)=W_1(E)=\pi E = 2\pi \hbar n_1\ .
\label{n1}
\ee
For the other degree of freedom we have the trajectory
(\ref{sol22}), which leads to
\be
W(E)=W_2(E) = 2\pi \nu^{-1}E = 2\pi \hbar n_2\ .
\label{n2} 
\ee
Therefore we conclude that
\be
E=\hbar(2n_1 + \nu n_2)\ .
\label{nn12}
\ee
Up to the ground state energy the spectrum coincides with
the spectrum of two harmonic oscillators with frequencies
$2$ and $\nu$, being degrees of the independent Casimir polynomials
for groups of rank 2. We will see that the same conclusion
follows from the Dirac quantization method for gauge systems
without an explicit parameterization of the physical
phase space.

\subsection{Gauge matrix models. Curvature of the orbit space
and the kinematic coupling}

So far we have considered gauge models whose physical configuration
space is flat. Here we give a few simple examples of gauge models
with a curved gauge orbit space. Another purpose of considering
these models is to elucidate the role of a non-Euclidean 
metric on the physical configuration space
in the kinematic coupling between the physical degrees of
freedom. 

To begin with let us take a system of two particles
in the plane  with the Lagrangian being
the sum of the Lagrangian (\ref{so.1}), where $N=2$,  
\cite{ijmp91,book}
\be 
L = \frac 12 \left(D_t{\bf x}_1\right)^2 + 
\frac 12 \left(D_t{\bf x}_2\right)^2
+ V_1({\bf x}_1^2) +  V_2({\bf x}_2^2)\ ,
\label{mat2}
\ee
which is invariant under the gauge transformations
\be
{\bf x}_{q}\rightarrow e^{T\omega}{\bf x}_{q}\ ,\ \ \ \ 
y\rightarrow y + \dot{\omega}\ ,\ \ \ q=1,2\ .
\ee
The gauge transformations are simultaneous rotations of the
vectors ${\bf x}_{1,2}$.
By going over to the Hamiltonian formalism one easily
finds that the system has two first-class constraints
\be
\pi = \frac{\pl L}{\pl \dot{y}}=0\ ,\ \ \ 
\sigma = ({\bf p}_1,T{\bf x}_1) + ({\bf p}_2,T{\bf x}_2) =0 \ .
\label{mat2b}
\ee
The second constraint means that the physical motion
has zero total angular momentum. The Hamiltonian of the
system reads
\be
H= \frac 12\left({\bf p}_1^2 +{\bf p}_2^2\right) +
V_1({\bf x}_1^2) +  V_2({\bf x}_2^2) + y\sigma\equiv H_1 + H_2\ ,
\label{mat2a}
\ee
where each $H_i$ coincides with (\ref{so.18}). The coupling
between the degrees of freedom occurs {\em only} through the constraint.

The physical phase space of the system is the quotient
$\Rs^2\oplus\Rs^2\vert_{\sigma=0}/SO(2)$ where the gauge transformations
are simultaneous SO(2) rotations of all four vectors
${\bf x}_q$ and ${\bf p}_q$. To introduce a local 
parameterization of the physical phase by canonical coordinates,
we observe that by a suitable gauge transformation the vector
${\bf x}_1$ can be directed along the first coordinate axis, i.e.,
$x_1^{(2)}=0$. Here we label the components of the vector ${\bf x}_q$
as $x_q^{(i)}$. So, the phase space of physical degrees of freedom
can be determined by two conditions
\be
x_1^{(2)}=0\ ,\ \ \ 
p_1^{(2)} = -\frac{1}{x_1^{(1)}}({\bf p}_2,T{\bf x}_2)\ .
\label{mat3}
\ee
The second equation follows from the constraint $\sigma=0$.
The gauge condition still allows discrete gauge transformations
generated by the rotations through the angles $n\pi$ ($n$ is
an integer).
It is important to understand that the residual gauge transformations
on the hypersurface (\ref{mat3}) do not act only on 
$p_{1}^{(1)}$ and $x_1^{(1)}$ changing their sign, but rather
they apply to all degrees of freedom simultaneously:
${\bf x}_q \rightarrow \pm {\bf x}_q$ and ${\bf p}_q
\rightarrow \pm {\bf p}_q$. The physical phase space cannot
be split into a cone and two planes. It is isomorphic to
the quotient
\be
{\rm PS}_{\rm phys}\sim \Rs^3\oplus\Rs^3/\Z_2\ .
\ee 
The residual gauge symmetry 
forbids independent excitations of the chosen canonical variables. 
Only pairwise excitations, like $x_1^{(1)}x_2^{(1)}$,
are invariant under the residual gauge transformations.
Thus, we have the familiar kinematic coupling of the physical degrees of
freedom. Accordingly, if one takes the potentials
$V_{1,2}$ as those of the harmonic oscillators, 
only pairwise collective excitations of the oscillators
are allowed by the gauge symmetry, which is most easily seen in the Fock
representation of the quantum theory (see section 7.1).
 
In addition to the kinematic coupling induced by the 
non-Euclidean structure of the physical phase space,
there is another source for the kinematic
coupling which often occurs
in gauge theories. Making use of (\ref{genmet})
we calculate the metric on the orbit space in the parameterization
(\ref{mat3}). Let us introduce
a three-vector ${\bf q}$ whose components $q^a$, $a=1,2,3$, are,
respectively, $x_1^{(1)}$,  $x_2^{(1)}$, $x_2^{(2)}$. Then \cite{ijmp91}
\be
g^{ph}_{ab} = \frac{1}{{\bf q^2}}
\left(
\begin{array}{ccc}
{\bf q}^2 & 0 & 0\\
0& {\bf q}^2 -(q^3)^2 & q^3q^2\\
0& q^3q^2 & {\bf q}^2 - (q^2)^2
\end{array}
\right)\ .
\label{mat4}
\ee
The metric (\ref{mat4}) is not flat. The scalar curvature
is $R=6/{\bf q}^2$. 
Since the metric is {\em not} diagonal,
the reduction of the kinetic energy onto the physical
phase space spanned by the chosen canonical variables
will induce the coupling between physical degrees of
freedom: ${\bf p}^2_1 +{\bf p}^2_2= g_{ph}^{ab}p_ap_b$,
where $p_a$ are canonical momenta for $q^a$. 
Thus, the physical Hamiltonian is no longer a sum
of the Hamiltonians of each degree of freedom. The degrees
of freedom described by ${\bf q}$ {\em cannot} be excited
independently. It is possible to find new parameterization
of the orbit space where the kinematic coupling caused
by both the non-Euclidean structure of the phase space
and the metric of the orbit space is absent \cite{ijmp91}
(the metric is diagonal in the new variables).
The new variables are related to the ${\bf q}$'s by a non-linear
transformation and naturally associated with the independent
Casimir polynomials in the model (see section 7.1). 
In this regard, the 
model under consideration and the one discussed in section 4.5
are similar. So we will not go into technical details. 

Let us calculate the induced volume element on the orbit
space, $\mu({\bf q})d{\bf q}$. 
As before, the density $\mu({\bf q})$ 
does not coincide with $\sqrt{\det (g_{ab}^{ph})}
=q^1/\sqrt{{\bf q}^2}$ because the volume of the gauge orbit
through a configuration space point depends on that point \cite{babelon}.
Consider a matrix $x$ with the components $x_{ij} =x_j^{(i)}$, i.e.,
the columns of $x$ are vectors ${\bf x}_j$. Then the gauge transformation
law is written in a simple form $x\rightarrow \exp(\omega T)x$.
For this reason we will also refer to the model (\ref{mat2})
as a gauge matrix model.
For a generic point $x$ of the configuration space one can find
a gauge transformation such that the transformed configuration
satisfies the condition $x_{21}=0$. Therefore
\be
x=e^{\theta T}\left(\begin{array}{cc} 
q^1 & q^2\\
0& q^3
\end{array}\right)\equiv e^{\theta T}\rho\ ,
\label{mat5}
\ee
where the coordinates $q^a$ span the gauge orbit space.
The volume element $\mu({\bf q})d{\bf q}d\theta$ can found
by taking the square root of the determinant of the 
Euclidean metric $\tr (dx^Tdx)= \tr\{(d\rho + T\rho d\theta)^T
(d\rho + T\rho d\theta)\}$, where $x^T$ is the transposed
matrix $x$, in the new curvilinear coordinates. After a modest
computation, similar to (\ref{zdz}), we get 
the Jacobian $\mu({\bf q})=q^1$. Note that the variable
${\bf q}$ is gauge invariant in this approach, while $\theta$ spans the
gauge orbits.
We have ${\bf q}^2=\tr(x^Tx)$ and therefore
the scalar curvature can also be written in the gauge
invariant way $R = 6/\tr(x^Tx)$. Clearly, the curvature
must be gauge invariant because it is a parameterization
independent characteristic of the gauge orbit space. 

The Jacobian $\mu$ vanishes at $q^1=0$. Its zeros form a plane
in the space of $q^a$. On this plane the change of
variables (\ref{mat5}) is degenerate, which also indicates that
the gauge $x_{21}=0$ is not complete on the plane $x_{11}=0$. Indeed, 
at the singular points $x_{11}=x_1^{(1)}=0$ the 
constraint cannot be solved for the nonphysical
momentum $p_{1}^{(2)}$ and is reduced to $\sigma
=({\bf x}_2,T{\bf p}_2)$ which generates the SO(2)
{\em continuous} rotations on the plane ${\bf x}_1=0$.
Such gauge transformations are known as the residual
gauge transformations {\em within} the Gribov horizon 
\cite{soloviev87,baal92,zw5}. Given a set of constraints
$\sigma_a$ and the gauge conditions $\chi_a=0$ (such
that $\{\chi_a,\chi_b\}=0$ \cite{fd68}), the Faddeev-Popov
determinant is $\det\{\chi_a,\sigma_b\}\equiv \Delta_{FP}$.
Zeros of $\Delta_{FP}$ on the gauge fixing surface $\chi_a=0$
form the Gribov horizon (or horizons, if the set of zeros
is disconnected). It has a codimension one (or higher)
on the surface $\chi_a=0$. Within the Gribov horizon
the gauge is not complete, and continuous gauge transformations
may still be allowed \cite{soloviev87}. 
Consequently, there are identifications within the Gribov
horizon, which may, in general, lead 
to a nontrivial topology of the gauge
orbit space \cite{baal92} (see an example in section
10.3). In our case, $\chi = x_1^{(2)}$
and, hence, $\Delta_{FP}=x_1^{(1)}$, i.e., it coincides
with the Jacobian $\mu$. This is a generic
feature of gauge theories: The Faddeev-Popov determinant
specifies the volume element on the gauge orbit space \cite{babelon}.

In our parameterization, the orbit space is isomorphic
to the half-space $x_1^{(1)}>0$ modulo boundary identifications.
To make the latter we can, e.g., make an additional
gauge fixing on the plane $x_1^{(1)}=0$, say, by requiring
$x_2^{(2)}=0$ \cite{heinzl}. We are left with {\em discrete} gauge
transformations $x_2^{(1)}\rightarrow  -x_2^{(1)}$.
Therefore every half-plane formed by positive values
of  $x_1^{(1)}$ and values of $x_2^{(1)}$ would have
the gauge equivalent half-axes 
$x_2^{(1)}>0$ and $x_2^{(1)}<0$ on its edge $x_1^{(1)}=0$. Identifying
them we get the cone unfoldable into a half-plane.
So the orbit space has no boundaries, and there is
one singular point (the origin) where the curvature is
infinite.  The topology of the gauge orbit space is trivial.

On the Gribov horizon, the physical phase space
of the model also exhibits the conic structure. 
On the horizon ${\bf x}_1=0$ the constraint is reduced to $\sigma=
({\bf x}_2,T{\bf p}_2)$, so we get the familiar
situation discussed in section 3.2: One particle
on the plane with the gauge group SO(2).
The corresponding physical phase space is a cone
unfoldable into a half-plane, $\Rs^4\vert_{\sigma=0}/SO(2)\sim cone(\pi)$. 

Another interesting matrix gauge model can be obtained
from the Yang-Mills theory under the condition
that all vector potentials depend only on time \cite{sav1,sav2}.
The orbit space in this model has been studied by Soloviev
\cite{soloviev87}. The analysis of the physical phase
space structure and its effects on quantum
theory can be found in \cite{jinrlec,ufn,book}. The orbit
space of several gauge matrix models is discussed in the work
of Pause and Heinzl \cite{heinzl}. It is also noteworthy
that gauge matrix models appear in the theory
of eleven-dimensional supermembranes \cite{11dm,11dma},
in the dynamics on D-particles \cite{dpat} and in
the matrix theory \cite{mtheory} describing
some important properties of the superstring theory. The
geometrical structure of the physical configuration and phase
space of these models does not exhibit essentially new features.
The details are easy to obtain by the method discussed above.

\section{Yang-Mills theory in a cylindrical spacetime}
\setcounter{equation}0

The definition of the physical phase space as the quotient space
of the constraint surface relative to the gauge group 
holds for gauge
field theories, i.e., for systems with an infinite number of degrees
of freedom. 
The phase space in a field theory is a functional space, and this
gives rise to considerable technical difficulties when calculating
the quotient space. One has to specify a functional class to which
elements of the phase space, being a pair of functions of the spatial
variables, belong. In classical theory it can be a space of smooth
functions \cite{singer} (e.g., to make
the energy functional finite). However, in quantum field theory
the corresponding quotient space appears to be of no use,
say, in the path integral formalism because the support of the
path integral measure typically lies in a Sobolev functional 
class \cite{sob1,sob2,soloviev}, 
i.e., in the space of distributions, where smooth
classical configuration form a zero-measure subset.
To circumvent this apparent difficulty, one can, for instance,
discretize the space or compactify it into a torus (and
truncate the number of Fourier modes), thus
making the number of degrees of freedom finite. 
This would make a gauge field model
looking more like mechanical models considered above where
the quotient space can be calculated.

The simplest example of this type is
the Yang-Mills theory on
a cylindrical spacetime (space is compactified to a circle ${\bf S}^1$)
\cite{mig,raj,het,mic,lan,lan2,plb93,het2,2d}. 
Note that in two dimensional spacetime Yang-Mills theory 
does not have physical degrees of freedom, unless the 
spacetime has a nontrivial topology \cite{t1,t2,t3,t4,t5,t6,t7}. 
In the Hamiltonian approach, only space is compactified, thus
leading to a cylindrical spacetime. We shall
establish the ${\rm PS}_{{\rm phys}}$ structure of this theory 
in the case of an arbitrary compact semisimple gauge group \cite{plb93,2d}.
The Lagrangian reads
\be
L=-\frac14\int_0^{2\pi l}dx (F_{\mu\nu},F^{\mu\nu})
\equiv - \frac14\langle F_{\mu\nu},F^{\mu\nu}\rangle\ ,
\label{fff}
\ee
where $F_{\mu\nu}= \pl_{\mu} A_{\nu}-\pl_{\nu} A_{\mu} - ig[A_{\mu},
A_{\nu}],\ g$ a coupling constant, $\mu,\nu = 0,1$; the Yang-Mills
potentials $A_\mu$, being elements of a Lie algebra $X$, are periodic
functions of a spatial coordinate, $A_\mu(t,x+2\pi l)=A_{\mu}(t,x)$,
i.e. $l$ is the space radius; the parenthesis $(,)$ 
in the integrand (\ref{fff}) stand for the
invariant inner product in $X$. We assume it to be the Killing
form introduced in section 3.2.
In a matrix representation, one can always normalize it to be
a trace. 
Since the vector potential is a periodic function in space,
it can be decomposed into a Fourier series. The Fourier components
of $A_\mu$ are regarded as independent 
(Cartesian) degrees of freedom in the theory.

To go over to the Hamiltonian formalism, we determine the canonical momenta
$E_\mu = \delta L/\delta\dot{A}^\mu = F_{0\mu}$; the overdot denotes
the time derivative. The momentum conjugated to $A_0$ vanishes, $E_0=0$,
forming the primary constraints. The canonical Hamiltonian has the
form 
\be
H=\langle E_\mu,A^\mu\rangle - L= \langle E_1,E_1\rangle/2 -
\langle A_0,\sigma\rangle\ ,
\label{eea}
\ee
where $ \sigma= \nabla(A_1)E_1$ with
$\nabla(A_1)= \pl_1- ig[A_1,\ ]$ being the covariant derivative
in the adjoint representation. The primary constraints must be satisfied
during the time evolution. This yields the secondary
constraints
\be
\dot{E}_0 = \{E_0,H\} =  \pl_1E_1 - ig[A_1,E_1] = \sigma = 0\ ,
\label{gae}
\ee
where the standard symplectic structure 
\be
\{A^{a\mu}(x), E_\nu^b(y)\} =
\delta^{ab}\delta^\mu_\nu\delta(x-y)\ ,\ \ \ \ \ x,y\in {\bf S}^1\ ,
\label{xys}
\ee
has been
introduced, and the suffices $a,b$ refer to the adjoint representation of
the  Lie algebra. The constraints are in involution
\be
\{\sigma_a(x),\sigma_b(y)\}= if_{ab}{}^{c}\delta(x-y)\sigma_c(x)\ ,
\ \ \ \ \{\sigma_a , H\} =
-f_{ab}{}^{c}A_0^b\sigma_c\ ,
\label{hfa}
\ee
with $f_{ab}{}^{c}$ being the structure constants of $X$. 
There are no more constraints in the theory, and all
constraints are of the first class.

The primary and secondary (first-class) constraints are 
independent generators of gauge transformations. As in the 
mechanical models,
the primary constraints $E_0^a=0$ generate shifts of the Lagrange
multipliers
$A_0^a$, $\delta A_0^a(x)= \{A_0^a,\langle\omega_0,E_0\rangle\} =
\omega_0^a(x)$, and leave the phase space variables $E_\mu^a$ and
$A_1^a$ unchanged. Therefore the hyperplane $E_0^a=0$ 
($E_1$ and $A_1$ are fixed) spanned by $A^a_0$
in the total phase space is the gauge orbit. 
We can discard $A^a_0$ and $E^a_0$ as
pure nonphysical degrees of freedom and concentrate
our attention on the remaining variables.

To simplify the notation, from now on we omit the Lorentz suffix
``1'' of the field variables, i.e., instead of $E_1$ and
$A_1$ we write just $E$ and $A$. 
The constraints (\ref{gae}) generate the following gauge transformations
\be
E\rightarrow \Omega E\Omega^{-1}=E^\Omega\ ,\ \ \ \
A\rightarrow \Omega A\Omega^{-1} + \frac ig\Omega\pl\Omega^{-1}
=A^\Omega\ .
\ee 
Here and below $\pl_1\equiv \pl$, while the overdot is used
to denote the time derivative $\pl_0$;
$\Omega = \Omega(x)$ takes its values in a semisimple compact
group $G$ ($X$ is its Lie algebra). The gauge transformed variables
$E^\Omega$ and $A^\Omega$ must also be periodic functions of $x$.
This results in the periodicity of $\Omega$ modulo the center $Z_G$
of $G$
\be
\Omega(x+2\pi l)=z\Omega(x)\ ,\ \ \ \ \  z\in Z_G\ .
\label{center}
\ee
Indeed, by definition an element $z$ from the center
 commutes with any element of $X$ and, therefore,
$E^\Omega$ and $A^\Omega$ are invariant under the shift $x
\rightarrow x + 2\pi l$.
The relation (\ref{center}) is called a 
twisted boundary condition \cite{thooft}.
The twisted gauge transformations
(i.e., satisfying (\ref{center}) 
with $z\neq e$, $e$ a group unit) form distinct
homotopy classes. Therefore they cannot be continuously deformed towards
the identity. On the other hand, gauge transformations generated by the
constraints (\ref{gae}) are homotopically trivial because they are built
up by {\em iterating} the infinitesimal transformations \cite{jackiw}:
$\delta E= \{E,\langle\omega,\sigma\rangle\} = ig [E,\omega]$
and $\delta A= \{A, \langle\omega,\sigma\rangle\} =
-\nabla(A)\omega$ with $\omega$ being an $X$-valued periodic function
of $x$. Thus, we are led to the following
conclusion. When determining ${\rm PS}_{{\rm phys}}$ 
as the quotient space, one should
restrict oneself by {\em periodic} (i.e. homotopically trivial)
gauge transformations. Such transformations determine
a mapping ${\bf S}^1\rightarrow G$. A collection of all such 
transformations is called a gauge group and will be denoted
${\cal G}$, while an abstract group $G$ is usually called
a structure group of the gauge theory.
Yet we shall see that quantum states annihilated by the operators
of the constraints -- these are the Dirac physical states -- are not
invariant under the twisted gauge transformations.

Consider a periodic function $f(x)$ taking its values in $X$.
It is expanded into a Fourier series
\be
f(x)=f_0 +\sum\limits_{n=1}^\infty\left(f_{s,n}\sin\frac{nx}{l}
+f_{c,n}\cos\frac{nx}{l}\right)\ .
\label{fxx}
\ee
We denote a space of functions (\ref{fxx}) 
${\cal F}$ and its finite dimensional
subspace formed by constant functions ${\cal F}_0$ so that
$A= A_{0} + \tilde{A}$, where $A_{0}\in {\cal F}_0$ and
$\tilde{A}\in {\cal F}\ominus {\cal F}_0$.
For a generic connection $A(x)$, there
exists a {\em periodic} gauge element $\Omega(x)$
such that the gauge transformed connection $A^\Omega$ is
homogeneous in space,
\be
\pl A^\Omega =0\ .
\label{coulomb}
\ee
This means that the Coulomb gauge fixing surface
$\pl A=0$ intersects each gauge orbit at least once.
To find $\Omega(x)$, we set
\be 
\omega = -\frac{i}{g} \Omega^{-1}\pl \Omega \in X
\label{omega}
\ee
and, hence,
\be
\Omega(x)= {\rm P}\exp ig\int_x^0\omega(x') dx'\ .
\label{group}
\ee
The path-ordered exponential (\ref{group}) is defined
similarly to the time-ordered exponential in section 3.1.
They differ only by the integration variables.
After simple algebraic transformations, Eq. (\ref{coulomb})
can be written in the form
\be
\nabla(A)\omega = \pl\omega - ig[A,\omega] = -\pl {A}\ ,
\label{coulomb2}
\ee
which has to be solved for the Lie algebra element
$\omega(x)$. It is a linear nonhomogeneous differential
equation of first order. So its general solution is a sum
of a general solution of the corresponding homogeneous
equation and a particular solution of the nonhomogeneous
equation. Introducing the group element
\be
U_A(x) = {\rm P}\exp ig\int_0^xdx' A(x')\ ,
\ee
that has simple properties $\pl U_A = igAU_A$ and
$\pl U^{-1}_A = -ig U^{-1}_A A$, 
the general solution can be written as
\be
\omega(x)=U_A(x)\omega_0 U_A^{-1}(x) - A(x)\ .
\label{solution}
\ee
The first term containing an arbitrary {\em constant} Lie algebra
element $\omega_0$ represents a solution of the homogeneous
equation, while the second term is obviously a particular solution
of the nonhomogeneous equation. 
The constant $\omega_0$ should be
chosen so that the group element (\ref{group}) would satisfy the
periodicity condition, which yields
\be
\Omega(2\pi l) = {\rm P}\exp ig\oint dx\,\omega = e\ ,
\label{pcond}
\ee
where $e$ is the group unit.
This specifies completely the function $\omega(x)$, and, hence,
$\Omega(x)$ for any
generic $A(x)$.
So, any configuration $A\in {\cal F}$ can be reduced
towards a spatially homogeneous configuration by means of
a gauge transformation.

Now we shall prove that the gauge reduction of $A$ to a homogeneous
connection $A_{0}\in {\cal F}_0$
leads to a simultaneous gauge reduction of the momentum $E$
to $E_{0}\in {\cal F}_0$ on the constraint surface. 
To this end, we substitute the gauge transformed canonical
pair $A^\Omega = A_0\in  {\cal F}_0$, $E^\Omega$ into the
constraint equation $\nabla(A)E=0$ and obtain
\be
\nabla(A^\Omega)E^\Omega =
\nabla(A_0)E^\Omega=0\ .
\label{a0e}
\ee
The momentum variable is then divided into a homogeneous
part $E_0$ and a nonhomogeneous one 
$\tilde{E}^\Omega = E^\Omega -E_0$. 
For these two components one obtains
two independent equations from
Eq. (\ref{a0e}):
\ba
\sigma_0\equiv [A_{0},E_{0}]&=& 0\ ,\label{homc}\\
\pl \tilde{E}^\Omega - ig[A_{0},\tilde{E}^\Omega] &=&
\nabla (A_{0})\tilde{E}^{\Omega}=0\ .
\label{nonhomc}
\ea
The first equation stems from the ${\cal F}_0$-component
of the constraint equation (\ref{a0e}), while the second one
is the constraint in the subspace ${\cal F}\ominus{\cal F}_0$.
A general solution of Eq. (\ref{nonhomc}) can written in the
form $\tilde{E}^\Omega(x) = U_0(x)\tilde{E}^\Omega_0 U_0^{-1}(x)$
where $U_0(x)=\exp[igA_0x]$ and $\pl\tilde{E}^\Omega_0=0$.
For a generic $A_0$, the solution is {\em not} periodic
in $x$ for all constants $\tilde{E}^\Omega_0\neq 0$. 
Since  $\tilde{E}^\Omega(x)$ must be a periodic function,
the constant $\tilde{E}^\Omega_0$ should necessarily vanish.
Thus, Eq. (\ref{nonhomc}) has only a 
trivial solution $\tilde{E}^\Omega = 0$,
and $E^\Omega =E_0\in {\cal F}_0$.

A useful observation following from the above analysis is that
the operator $\nabla(A_0)$ has no zero modes in the subspace 
${\cal F}\ominus{\cal F}_0$ and, hence, is invertible. The determinant
of the operator $\nabla(A_0)$ restricted 
on ${\cal F}\ominus{\cal F}_0$ does
not vanish. We shall calculate it later when studying the metric
on the physical configuration space.

We are led to a redundant system with $N=\dim X$ degrees of freedom
and the constraint (\ref{homc}) which generates {\em homogeneous} gauge
transformations of the phase-space variables $A_{0}$ and $E_{0}$
($\pl\Omega\equiv 0$)
\footnote{In section 10.3 we discuss the special role
of constant gauge transformations in detail in relation with a 
general analysis due to Singer \cite{singer}. 
Here we proceed to calculate 
the physical phase space as the quotient space (\ref{1})
with respect to the full gauge group of the Lagrangian (\ref{fff}).
In fact, in the path integral formalism we develop in sections 8 and 9,
there is no need to pay a special attention to the constant gauge
transformations and neither to the so-called reducible connections 
\cite{fiber} which have a nontrivial
stabilizer in the gauge group and, therefore, 
play a special role in Singer's analysis
of the orbit space.}.
 This mechanical system has been
studied in Section 3. The system is shown to have $r={\rm rank}\ X$
physical degrees of freedom which can be described by Cartan
subalgebra components of $A_{0}$ and $E_{0}$.
Since any element of $X$ can be represented in the form 
$A_{0}=\Omega_Aa\Omega_A^{-1}$, $a$ an element of the Cartan
subalgebra $H$, $\Omega_A\in G$,
configurations $A_{0}$ and $a$ belong to the same gauge
orbit. Moreover, a spatially homogeneous
gauge transformation with $\Omega =
\Omega_A^{-1}$ brings the momentum $E_{0}$ on the
constraint surface (\ref{homc}) to the Cartan subalgebra. Indeed,
from (\ref{homc}) we derive $[a,\Omega_A^{-1}E_{0}\Omega_A] =0$
and conclude that $p_a=\Omega_A^{-1}E_{0}\Omega_A \in H$
by the definition of $H$. The element $a$ has a stationary
group being the Cartan subgroup of $G$. This means that
not all of the constraints (\ref{homc}) are independent. 
There are just $N-r,\ r =\dim H$, independent constraints
among (\ref{homc}). The continuous gauge arbitrariness is
exhausted in the theory.

\subsection{The moduli space}

We expect the existence of the residual gauge freedom
which cannot decrease the number of physical degrees of
freedom, but might change the geometry of their configuration
and phase spaces. If two homogeneous connections from the
Cartan subalgebra, $a$ and $a_s$, belong to the same gauge
orbit, then there should exists a gauge group element
$\Omega_s(x)$ such that 
\be
a_s = \Omega_s a \Omega_s^{-1} 
-\frac{i}{g}\Omega_s\pl\Omega_s^{-1}\ ,\ \ \ 
\pl a_s=\pl a=0\ ,\ \ \ a_s,a\in H\ .
\label{copies}
\ee
There are two types of solutions to this equation for 
$\Omega_s$. First, we can take {\em homogeneous} gauge group
elements, $\pl \Omega_s =0$. This problem has already been 
solved in Section 3. The homogeneous residual
gauge transformations form the Weyl group.
Thus, we conclude that the phase-space points $\hat{R}p_a,\ \hat{R}a$,
where $ \hat{R}$ ranges over the Weyl group, are gauge equivalent 
and should be
identified when calculating the quotient space ${\rm PS}_{{\rm phys}}$. 
To specify the corresponding 
modular domain in the configuration space,  we recall
that the Weyl group acts simply
transitively on the set of Weyl chambers \cite{hel},
p.458. Any element of $H$ can be obtained from an element
of the positive Weyl chamber $K^+$ ($a\in K^+$ if $(a,\omega)
> 0$, for all simple roots $\omega$) by a certain transformation
from $W$. In other words, the Weyl chamber $K^+$ is isomorphic to 
the quotient $H/W$.

In contrast with the mechanical model of Section 3,
the Weyl group does not cover the whole admissible discrete
gauge arbitrariness in the 2D Yang-Mills theory. 
To find {\em nonhomogeneous} solutions to Eq. (\ref{copies}),
we take the derivative of it, thus arriving at the equation
\be 
\pl\Omega_s a \Omega_s^{-1}+\Omega_s a\pl \Omega_s^{-1}
-\frac{i}{g}\pl\left(\Omega_s \pl \Omega_s^{-1}\right) =0\ .
\label{omegas}
\ee
To solve this equation, we introduce an auxiliary Lie algebra
element $\omega_s = -\frac{i}{g}\Omega_s^{-1} \pl \Omega_s$.
From (\ref{omegas}) we infer that it satisfies the equation
\be 
\nabla(a)\omega_s =0\ .
\label{nabla}
\ee
For a generic $a$ from the Cartan subalgebra
this equation has only a homogeneous
solution which we write in the form
\be 
\omega_s = a_0\eta\ ,\ \ \ a_0 = (gl)^{-1}\ ,\ \ \ \eta \in H\ .
\label{slh}
\ee
Note that Eq.(\ref{nabla}) can always be transformed into 
two independent equations by setting $\omega_s = a_0\eta
+ \tilde{\omega}_s$, where $\eta \in {\cal F}_0$
and $\tilde{\omega}_s\in{\cal F}\ominus {\cal F}_0 $.
As has been shown above, the operator $\nabla(A_0)$
has no zero modes in the space ${\cal F}\ominus {\cal F}_0 $,
and, hence, so does $\nabla(a)= 
\Omega_A^{-1}\nabla(A_0)\Omega_A$ which means
that $\det\nabla(a)=\det\nabla(A_0)\neq 0$.
So  $\tilde{\omega}_s=0$, whereas the homogeneous
component satisfies the equation $\nabla(a)\eta
=-ig[a,\eta]=0$, that is, $\eta$ must
be from the Cartan subalgebra because it commutes
with a generic $a$. From the relation 
$\pl\Omega_s = ig\Omega_s\omega_s$ we find that
\be
\Omega_s(x) = \exp\left(iga_0\eta x\right)\ .
\label{gxl}
\ee
This is still not the whole story because the group
element we have found must to obey the periodicity condition
otherwise it does not belong to the gauge group.
The periodicity condition yields the restriction
on the admissible values of $\eta$:
\be
\Omega_s(2\pi l) = \exp(2\pi i\eta)= e\ ,
\label{unit}
\ee
where $e$ stands for the group unit.
The set of elements $\eta$ obeying this condition is 
called the unit lattice
in the Cartan subalgebra \cite{hel}, p.305.
The nonhomogeneous residual gauge transformations
do not change the canonical momentum $p_a$,
since $[p_a,\eta]=0$, and shift the canonical
coordinate $a\rightarrow a + a_0\eta$, 
along the unit lattice in the Cartan subalgebra.

Consider a diagram $D(X)$ being a union of a finite number of families
of equispaced hyperplanes in $H$ determined by $(\alpha,a)\in a_0\Z,\
\alpha$ ranges over the root system and $\Z$ stands for the set of
all integers. Consider then a group $T_e$ of
translations in $H$, $a\rightarrow a+ a_0\eta,$ where $ \eta$ belongs
to the unit lattice. The group $T_e$ leaves the diagram $D(X)$ invariant
\cite{hel}, p.305. The diagram $D(X)$ is also invariant with respect
to  Weyl group transformations. Since $W$ is generated by the
reflections (\ref{hhg}) in the hyperplanes orthogonal to simple
roots,
it is sufficient to prove the invariance of $D(X)$
under them. We have $(\alpha,\hat{R}_\omega a) = a_0n_\omega$ where
$n_\omega = n - 2k_\omega(\omega,\alpha)/(\omega,\omega)$ is an integer
as $(a,\omega)= k_\omega a_0,\ k_\omega\in \Z$ because $a\in D(X)$. 
We recall that any root $\alpha$ can be decomposed over
the basis formed by simple roots. The coefficients of this
decomposition are all either nonnegative or nonpositive integers.
Therefore the number $- 2(\omega,\alpha)/(\omega,\omega)$
is a sum of integers since the elements of the Cartan
matrix $- 2(\omega,\omega')/(\omega,\omega)$ are integers.
So, $\hat{R}_\omega D(X)= D(X)$.
Now we take the complement $H\ominus D(X)$. It consists of equal
polyhedrons whose walls form the diagram $D(X)$. Each polyhedron
is called a cell. A cell inside of the positive Weyl chamber
$K^+$ such that its closure contains the origin is called
the Weyl cell $K_W^+$. 

The Weyl cell will play an important role in the subsequent
analysis, so we turn to examples before studying the problem
in general.
The diagram $D(su(2))$ consists of
points $na_0\omega/(\omega,\omega),\ n\in \Z$ with $\omega$ being
the only positive root of $su(2)$, $(\omega,\omega)= 1/2$ ($\omega
= \tau_3/4$ in the matrix representation). A cell of $H_{su(2)}\ominus
D(su(2))$ is an open
interval between two neighbor points of $D(su(2))$. 
Assuming the orthonormal basis in the Cartan subalgebra,
we can write $a=\sqrt{2}
a_3\omega,\ (a,a)= a_3^2$.
Since
the Weyl chamber 
$K^+$ is isomorphic to the positive half-line $ \Rs^+$, 
we conclude that $a$ belongs to the Weyl cell $ K^+_W$ if
$a_3$ lies in the open interval $(0,\sqrt{2}a_0)$. 
The translations $a \rightarrow
a+ 2na_0\omega/(\omega,\omega),\ n\in \Z$, form the group
$T_e$, and $W=\Z_2,\ \hat{R}_\omega a = -a$. Thus, $D(su(2))$
is invariant under translations from $T_e$ and the reflection
from the Weyl group $W$.

For $X=su(3)$ we have three positive roots, $\omega_1,\ \omega_2$
and $\omega_{12}= \omega_1 + \omega_2$ which have the same norms.
\begin{figure}
\centerline{\psfig{figure=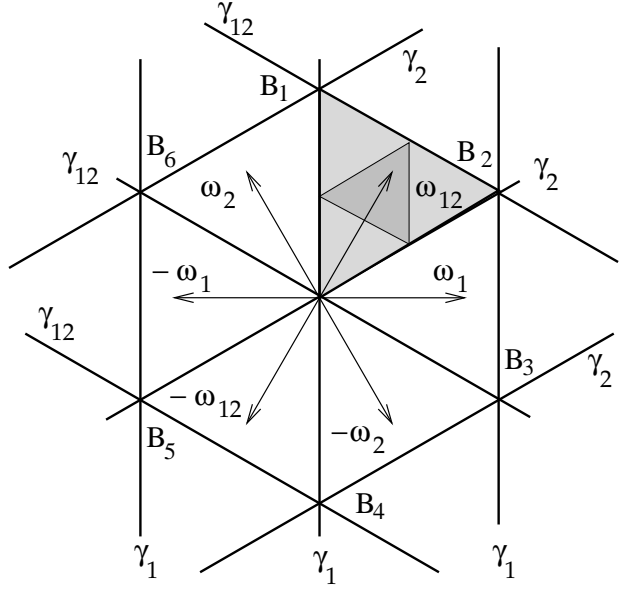}}
\caption{\small The root pattern of SU(3). The diagram 
$D(su(3))$ is formed by three families 
of straight lines perpendicular to
the simple roots $\omega_1, \omega_2$ and the root
$\omega_{12}= \omega_1+ \omega_2$. These families
are denoted as $\gamma_1, \gamma_2$ and $\gamma_{12}$,
respectively. The grayed equilateral triangle is the 
Weyl cell of su(3) which is the moduli space
of the su(3) connections with respect to homotopically
trivial gauge transformations. Had we included the 
homotopically nontrivial transformations into the gauge
group, the moduli space would have been four times less
and  isomorphic to the equilateral triangle whose vertices
are the mid-points of the Weyl cell boundaries
(see section 7.6 for details).}
\end{figure}
The angle between any two neighbor roots  equals $\pi/3$.
The root pattern of SU(3) is plotted in Fig. 5.
The diagram $D(su(3))$ consists of three families of equispaced
straight lines $(\omega_{1,2,12},a) = a_0n_{1,2,12},\ n\in \Z$, on
the plane $H_{su(3)}\sim \Rs^2$. The lines are perpendicular
to the roots $\omega_{1,2,12}$, respectively. The complement
$H_{su(3)}\ominus D(su(3))$ is a set of equilateral triangles
covering the plane $H_{su(3)}$. The Weyl cell $K^+_W$ is
the triangle bounded by lines $(\omega_{1,2},a)= 0$ (being
the boundary of $ K^+$) and
$(\omega_{12},a)= a_0$. The group $T_e$ is generated by integral
translations through the vectors $2a_0\alpha/(\alpha,\alpha),\
\alpha$ ranges over $\omega_{1,2,12}$, and $(\alpha,\alpha)=1/3$ (see
section 3.2 for details of the matrix representation
of the roots).

Let $W_A$ denote the group of linear transformations of $H$
generated by the reflections in all the hyperplanes in the
diagram $D(X)$. This group is called the affine Weyl group
\cite{hel}, p.314. $W_A$ preserves $D(X)$ and, hence,
\be
K^+_W\sim H/W_A\ ,
\ee
i.e. the Weyl cell is isomorphic to 
a quotient of the Cartan subalgebra by
the affine Weyl group. Consider a group $T_r$ of translations
$$
a\rightarrow a + 2a_0\sum_{\alpha >0}n_\alpha\alpha/(\alpha,
\alpha)\equiv a+ a_0\sum_{\alpha >0}
n_\alpha\eta_\alpha,\ \ \ \ \  n_\alpha\in \Z\ .
$$ 
Then $W_A$ is a semidirect product
of $T_r$ and $W$ \cite{hel}, p.315. For the element $\eta_\alpha$
we have the following equality \cite{hel}, p.317,
\be
\exp(2\pi i\eta_\alpha)= \exp\frac{4\pi i\alpha}{(\alpha,\alpha)} = e\ ,
\label{eta}
\ee
Comparing it with (\ref{unit}) we conclude that
the residual discrete gauge transformations form the
affine Weyl group. 

The space of all periodic connections $A(x)$ is 
${\cal F}$. Now we can calculate the {\em moduli} space 
of connections relative to the gauge group, i.e., obtain the
physical configuration space, or the gauge orbit space
\be 
{\rm CS}_{{\rm phys}} = {\cal F}/{\cal G}
\sim H/W_A \sim  K^+_W\ .
\label{csphys}
\ee
Similarly, the original phase space is  
$ {\cal F}\oplus {\cal F}$ because it is formed by pairs
of Lie-algebra-valued periodic functions $A(x)$ and $E(x)$.
The quotient with respect to the gauge group reads
\be
{\rm PS}_{{\rm phys}} =
{\cal F}\oplus {\cal F}/{\cal G}\sim
 \Rs^{2r}/W_A\ ,
\label{psphys}
\ee
where the action of $W_A$ on $H\oplus H\sim \Rs^{2r}$ is
determined by all possible compositions of the following transformations
\ba
\hat{R}_{\alpha,n}p_a&=&\hat{R}_\alpha p_a =
p_a -\frac{2(\alpha,p_a)}{(\alpha,\alpha)}\,\alpha\ ,
\label{wap}\\
\hat{R}_{\alpha,n}a &=&\hat{R}_\alpha a + \frac{2n_\alpha a_0}
{(\alpha,\alpha)}\,\alpha\ ,
\label{waa}
\ea
where the element $\hat{R}_{\alpha,n}\in W_A$ acts on $a$ as a
reflection in the hyperplane $(\alpha,a)= n_\alpha a_0,\ n_\alpha
\in \Z$, and $\alpha$ is any root.

To illustrate the formula (\ref{psphys}), let us construct 
${\rm PS}_{{\rm phys}} $ for
the simplest case $X=su(2)$. We have $r=1, \ W=\Z_2,\ (\omega,\omega)
=1/2$. 
\begin{figure}
\centerline{\psfig{figure=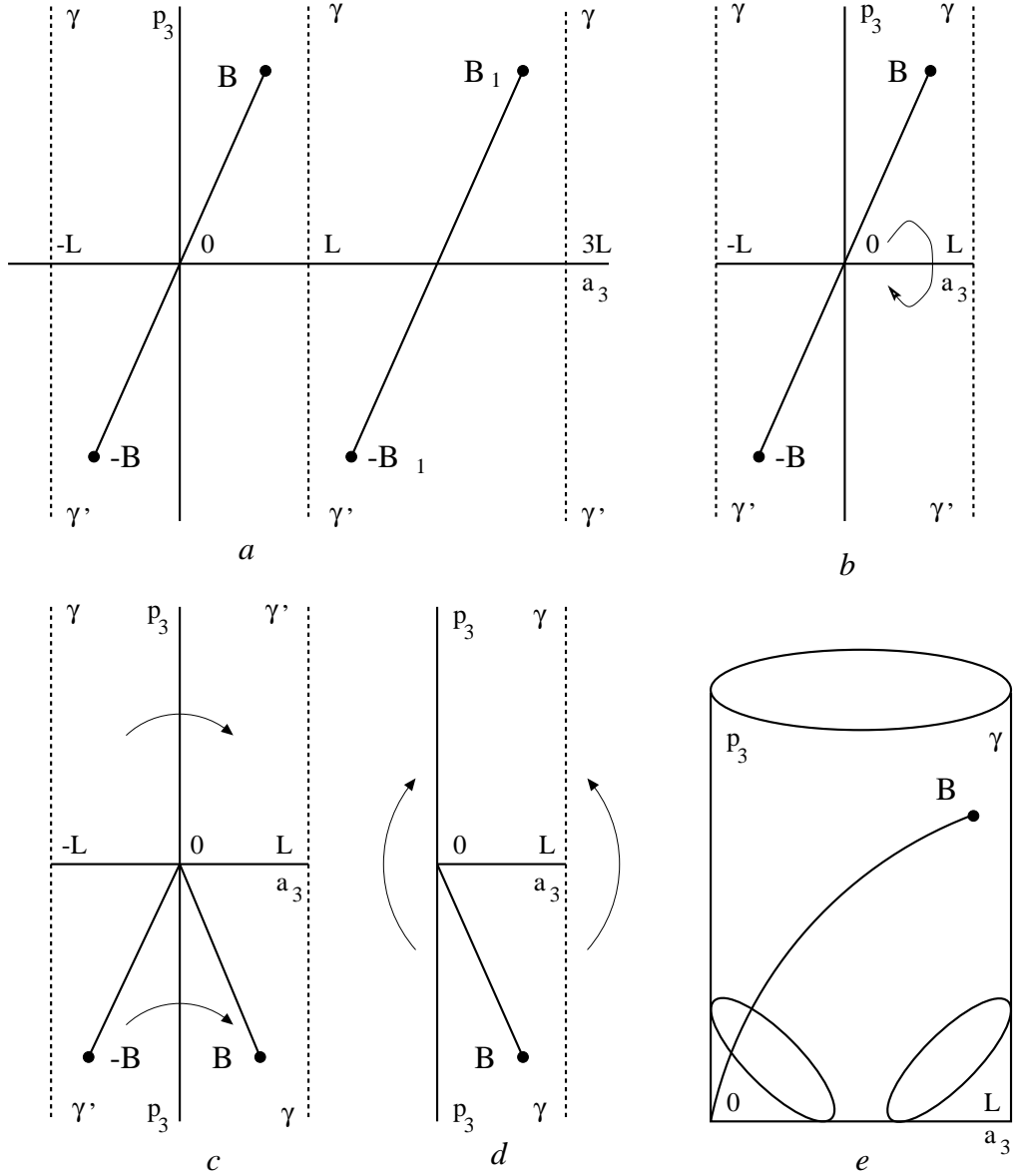}}
\caption{\small The physical phase space of the SU(2) Yang-Mills
theory on a cylindrical spacetime. It is a half-cylinder
with two conic horns attached to it. It is flat 
everywhere except the conic singularities where the curvature
is infinite. Here $L=\sqrt{2}a_0$.}
\end{figure}
The group $T_r=T_e$ acts on the phase plane $\Rs^2$ spanned
by the coordinates $p_3,a_3$ (we have introduced the orthonormal
basis in $H_{su(2)}$; see the discussion of $D(su(2))$ above) as
$p_3,a_3\rightarrow p_3, a_3+ 2\sqrt{2}na_0$. 
In Figure 6 we set $L=\sqrt{2}\,a_0$. The points $B$ and $B_1$
are related by the gauge transformation from $T_e$. The strips
bounded by the vertical lines $(\gamma\gamma')$ are gauge 
equivalent through the translations from $T_e$. 
The boundary lines $(\gamma\gamma')$ are
gauge equivalent to one another, too.  
So, $\Rs^2/T_e$ is
a cylinder. After an appropriate cut, this cylinder
can be unfolded into the strip $p_3\in \Rs,\ a_3\in (-L,
L)$ as shown in Fig. 6b. The
boundary lines $a_3=\pm L$ are edges of the cut. They 
contain the same physical
states and later will be identified.
On the strip one should stick together the points $p_3,a_3$ and
$-p_3,-a_3$ connected by the reflection from the Weyl group
(the points $B$ and $-B$ in Fig. 6b).
This converts the cylinder into a half-cylinder ended by two
conic horns at the points $p_3=0, a_3=0,L$. 
Indeed, we can cut the strip along the $p_3$-line and rotate
the right half (the strip $0<a_3<L$) relative to the coordinate
axis $a_3$ through the angle $\pi$ (cf. a similar procedure
in Fig. 1b). The result is shown in Fig. 6c. It is important to
observe that the half-axis $(L\gamma)$ is gauge equivalent
to $(-L\gamma)$ and  $(-L\gamma')$ to $(L\gamma')$, while
the positive and negative momentum half-axes in Fig. 6c
are edges of the cut and therefore to be identified too
(cf. Fig. 1c). Next, we fold the strip in Fig. 6c along
the momentum axis to identify the points $B$ and $-B$.
Finally, we glue together the half-lines $(\gamma L)$
with $(L\gamma)$ and $(p_3O)$ with $(Op_3)$ in Fig. 6d,
thus obtaining the physical phase space (Fig. 6e). In
neighborhoods of the singular conic points, $\ph$ looks locally like
$cone(\pi)$ studied in Section 2 because 
$W_A$ acts as the $\Z_2$-reflections
(\ref{wap}) and (\ref{waa}) with $\alpha = \omega$ 
and $n=0,1$ near $a_3= 0, \sqrt{2}a_0$, respectively. 

For groups of rank 2, all conic (singular) points of $\ph$ are
concentrated on a triangle being the boundary $\pl K_W^+$ of the
Weyl cell (if $X=su(3)$, $\pl K_W^+$ is an equilateral triangle
with side length $\sqrt{3}a_0$ in the orthonormal basis
defined in section 3.2). Let us  introduce local symplectic
coordinates $p_a^\bot,\ a^\bot$ and $p_a^\| ,\ a^\|$ in a
neighborhood of a point of $\pl K_W^+$ (except the triangle
vertices) such that $a^\bot $ and $a^\|$ vary along lines
perpendicular and parallel
to $\pl K_W^+$, respectively. The $W_A$-reflection in
the boundary of $\pl K^+_W$ going through this
neighborhood leaves $p_a^\|,\ a^\|$ invariant, while it
changes the sign of the other symplectic pair, $p_a^\bot,
a^\bot\rightarrow -p_a^\bot,-a^\bot$. Therefore ${\rm PS}_{{\rm phys}}$
locally coincides with $\Rs^2\oplus cone(\pi)$. At the
triangle vertices, two conic singularities going along
two triangle edges merge. If those edges are
perpendicular, ${\rm PS}_{{\rm phys}}$ is locally $cone(\pi)\oplus
cone(\pi)$. If not, ${\rm PS}_{{\rm phys}}$ is a $4D-hypercone$. The tip
of the $4D-hypercone$ is ``sharper'' than the tip of
$cone(\pi)\oplus cone(\pi)$, meaning that the $4D-hypercone$
can be always put inside of $cone(\pi)\oplus cone(\pi)$
when the tips of both the hypercones are placed at the same
point. Obviously, a lesser angle between the triangle edges
corresponds to a ``sharper'' hypercone (cf. section 4.4).

A generalization of this pattern of singular points 
in ${\rm PS}_{{\rm phys}}$ to
gauge groups of an arbitrary rank is trivial. The Weyl cell
is an $rD$-polyhedron. ${\rm PS}_{{\rm phys}}$ has
the most singular local $2rD-hypercone$ structure 
at the polyhedron vertices. On the
polyhedron edges it is locally viewed as an $\Rs^2\oplus
2(r-1)D-hypercone$. Then on the polyhedron faces, being
polygons, the local ${\rm PS}_{{\rm phys}}$ structure is 
an $\Rs^4\oplus 2(r-2)D-hypercone$, etc.

{\em Remark}. As in the mechanical models studied 
earlier one can choose various
ways to parameterize the physical phase space. When calculating
a quotient space, one can, for instance, restrict the values of the
canonical momentum $E(x)$. This is equivalent to imposing a gauge
condition on the field variables rather than on the connection
\cite{fd79,tmap}. By a gauge rotation $E$ can be brought to
the Cartan subalgebra at each point $x$. So we set $E(x)= E_H(x)
\in H$. Decomposing the connection $A$ into the Cartan component
$A_H(x)$ and $\bar{A}(x)\in X\ominus H$, we find that the constraint
$\nabla(A)E_H=0$ is equivalent to two independent constraints:
$\pl E_H =0$ and $[\bar{A}, E_H]=0$. These are two components
of the original constraint in $H$ and $X\ominus H$, respectively.
From the Cartan-Weyl commutation relations follows that $\bar{A}(x)=0$.
The residual constraints $\pl E_H =0$ generate the gradient shifts
of the corresponding canonical variables: $A_H\rightarrow
A_H + \pl \omega$, where $\pl\omega$ is a {\em periodic} function
of $x$. We have obtained the so called 
Abelian projection of the theory \cite{tmap}. 
Therefore the physical degrees of freedom can again be described
by the pair $E_H(x)= p_a$ and $A_H= a$. Now $p_a$ can be taken 
into the Weyl chamber by an appropriate Weyl transformation,
while $a$ is determined modulo shifts on the periods of the 
group torus (the shifts along the group unit lattice).
Note that we can take $\omega =\eta x$ since $\pl\omega=\eta$
is periodic as is any constant. The necessary restrictions on $\eta$
follow from the periodicity condition on the corresponding
gauge {\em group} element. Thus, we have another parameterization
of the {\em same} physical phase space 
such that $p_a\in K^+$ and $a\in H/T_e$, which, obviously,
corresponds to another cut of the $rD$-hypercone. The 
quantum theory of some topological field models 
in the momentum representation
has been studied in \cite{mom1,mom2,mom3}. 

\subsection{Geometry of the gauge orbit space}

Let us find the metric and the induced volume element
of the physical configuration space. They will be used 
in quantum mechanics of the Yang-Mills theory under consideration. 
It is useful to introduce the following decomposition
of the functional space (\ref{fxx})
\be
{\cal F}=\sum\limits_{n=0}^\infty\oplus{\cal F}_n =
\sum\limits_{n=0}^\infty\oplus({\cal F}_n^H\oplus\bar{\cal F}_n)\ ,
\label{geo1}
\ee
where ${\cal F}_0$ is a space of constant Lie algebra-valued functions
(the first term in the series (\ref{fxx})), 
${\cal F}_n, n\neq 0$, is a space
of functions with the fixed $n$ in the sum (\ref{fxx}).
Each subspace ${\cal F}_n$ is finite-dimensional,
$\dim {\cal F}_0=\dim X,\ \dim {\cal F}_n=2\dim X, n\neq 0$ (we recall
that Lie algebra-valued functions are considered). Functions belonging
to ${\cal F}_n^H$ take their values in the Cartan subalgebra $H$, while
functions from $\bar{\cal F}_n $ take 
their values in $X\ominus H$. All subspaces introduced are orthogonal with
respect to the scalar product $\langle,\rangle=\int_0^{2\pi l}dx(,)$.

From the above analysis of the moduli space of Yang-Mills connections 
follows a local parameterization of a generic connection
\be
A = \Omega a \Omega^{-1} + ig^{-1}\Omega \pl \Omega^{-1}\ ,\ \ 
\ \pl a=0\ ,\ \ a\in H\ ,
\label{aah}
\ee
where $\Omega \in {\cal G}/G_H$, and $G_H$ is the Cartan subgroup
(the maximal Abelian subgroup of $G$) which is isomorphic to
the stationary group of the homogeneous connection $a$.
By definition the connection remains invariant under gauge transformations
from its stationary group. 
Eq. (\ref{aah}) can be regarded as a change of variables in the 
functional space ${\cal F}$. In the new variables
the functional differential $\delta A\in {\cal F}$ 
can be represented in the form
\be
\delta A  = \Omega\left(da - ig^{-1} \nabla(a)
\delta w\right)\Omega^{-1}\ ,
\label{dif}
\ee
where by the definition of the parameterization (\ref{aah}) 
$\delta a=da\in {\cal F}_0^H$ and
$\delta w(x)= i\Omega^{-1}\delta \Omega\in {\cal F}\ominus{\cal F}_0^H$.
Therefore the metric tensor reads
\ba
\langle\delta A,\delta A\rangle&=& 2\pi l(da,da) - g^{-2}
\langle\delta w,\nabla^2(a)\delta w\rangle
\label{metric}\\
&\equiv& (da,g_{aa}da) + \langle\delta w,g_{ww}\delta w\rangle\ .
\nonumber
\ea
Equality (\ref{metric}) results from
(\ref{dif}) and the relation that $\langle da,
\nabla(a)\delta w\rangle = $ 
$-\langle\nabla(a)da, \delta w\rangle = 0$
which is due to $\pl da=0$ and $[da,a]=0$.
The operator $\nabla(a)$ acts in the subspace ${\cal F}\ominus
{\cal F}_0^H$. It has no zero mode in this subspace if $a\in K_W^+$ and,
hence, is {\em invertible}. Its determinant is computed below.
The metric tensor has the block-diagonal form. The physical
block is proportional to the $r\times r$ unit matrix $g_{aa}=
2\pi l$. For the nonphysical sector we have an infinite dimensional 
block represented by the kernel of the differential
operator: $g_{ww} = -g^{-2}\nabla^2(a)$, and $g_{aw}=g_{wa}=0$.
Taking the inverse of the $aa$-block of the inverse total metric,
we find that the physical metric $g_{aa}^{ph}$ coincides with $g_{aa}$.
That is, the physical configuration
space is a flat manifold with (singular) boundaries. It has the
structure of an orbifold \cite{orbi}. 

To obtain the induced volume element, one has to calculate the Jacobian
of the change of variables (\ref{aah})
\ba
\int_{{\cal F}}\prod\limits_{x\in {S}^1}dA(x)\Phi &=&
\int_{{\cal G}/G_H}\prod_x dw(x)\int_{K^+_W}da J(a)\Phi
\rightarrow \int_{K_W^+}da\kappa^2(a)\Phi\ ,
\label{vym}\\
J^2(a) &=& \det g_{aa}\det g_{ww} = (2\pi l)^r
\det\left[-g^{-2}\nabla^2(a)\right]\ . 
\ea
Here $\Phi=\Phi(A)=\Phi(a)$ is a gauge invariant functional of $A$.
The induced volume element does {\em not} coincide with the 
square root of the determinant of the induced metric on the 
orbit space. It contains an additional factor, 
$(\det g_{ww})^{1/2}$, being the volume
of the gauge orbit through a generic configuration $A(x)=a$, in the
full accordance with the general analysis given in \cite{babelon}
for Yang-Mills theories (see also section 10.1). 
Consider the orthogonal decomposition
\be
\bar{\cal F}_n = \sum\limits_{\alpha>0}\oplus{\cal F}_n^\alpha\ ,
\label{geo2}
\ee
where ${\cal F}_n^\alpha$ contains only functions
taking their values in the two-di\-men\-sion\-al subspace $X_\alpha\oplus
X_{-\alpha}$ of the Lie algebra $X$.
The subspaces ${\cal F}^H_n,\ {\cal F}^\alpha _n$ are invariant subspaces of
the operator $\nabla (a)$, that is, $\nabla (a){\cal F}^H_n$ is a subspace
of ${\cal F}^H_n$, and $\nabla (a){\cal F}^\alpha_n$ is a subspace
of ${\cal F}^\alpha_n$. We conclude that the operator  $\nabla (a)$
has a block-diagonal form in the
decomposition (\ref{geo1}) and (\ref{geo2}).
Indeed, we have $\nabla (a)
=\pl -ig{\rm ad}\,{a}$, where ${\rm ad}\,{a} =[a,\ ]$ 
is the adjoint operator acting in $X$.
The operator $\pl $ is diagonal in the algebra space,
and its action does not change periods of functions, 
i.e. ${\cal F}^{H,\alpha}_n$ are its invariant spaces.
Obviously, ${\rm ad}\,{a} {\cal F}^H_n=0$ and ${\rm ad}\,{a}
{\cal F}^\alpha _n={\cal F}^\alpha _n$ if $(\alpha ,a)\neq 0$ in accordance
with the Cartan-Weyl commutation relation (\ref{A.2}). 
Therefore an action of the operator
$\nabla (a)$ on ${\cal F}\ominus {\cal F}^H_0$
is given by an infinite-dimensional,
block-diagonal matrix. In the real basis $\lambda_i$ introduced
after Eq. (\ref{A.6}), its blocks have the form
\ba
\nabla ^H_n(a)&\equiv &\nabla (a)\vert _{{}_{{\cal F}^H_n}}=\pl \vert _{{}_
{{\cal F}^H_n}}=
\left (\otimes \frac{n}{l}\varepsilon\right )^r,\ \ n\neq 0,\ \ 
r={\rm rank}\, X, \label{geo311}\\
\nabla ^\alpha _0(a)&\equiv & \nabla (a)\vert _{{}_{{\cal F}^\alpha _0}}=
-ig{\rm ad}\,{a}\vert _{{}_{{\cal F}^\alpha _0}}=
g(a,\alpha )\varepsilon\ ,\\
\nabla ^\alpha _n(a)&\equiv& \nabla (a)\vert _{{}_{{\cal F}^\alpha _n}}=\1
\otimes
\frac{n}{l}\varepsilon + g(a,\alpha )\varepsilon\otimes \1\ ,
\label{geo3}
\ea
where $\varepsilon$ is a 2$\times$2 totally antisymmetric
matrix, $\varepsilon_{ij}=- \varepsilon_{ji}, 
\varepsilon_{12}=1$; and $\1$ is the $2\times 2$
unit matrix. In (\ref{geo3}) the first components in the tensor products
 correspond to the algebra indices, while the second ones determine the
 action of $\nabla (a)$ on the functional basis $\sin (xn/l),\ \ 
\cos (nx/l)$. The vertical bars at the operators in Eqs. 
(\ref{geo311})--(\ref{geo3}) mean a restriction of the corresponding
operator onto a specified finite dimensional subspace of ${\cal F}$.
An explicit matrix form of the restricted operator is easily obtained
by applying $\pl$ to the Fourier basis, and the action of ${\rm ad}\, a$,
$a\in H$, is computed by means of (\ref{adda6}). Since $\varepsilon^2=-1$,
we have for the Jacobian
\ba
J^2(a)&=& (2\pi l)^r\prod \limits _{\alpha > 0}
\det (ig^{-1}\nabla ^\alpha _0)^2\prod \limits _{n=1}^{\infty}
\left[\det (ig^{-1}\nabla ^H_n)^2\prod \limits _{\alpha > 0}
 \det (ig^{-1}\nabla ^\alpha _n)^2\right]=\nonumber \\
&=& (2\pi l)^r\prod \limits _{\alpha > 0}(a,\alpha )^4\prod
\limits _{n=1}^{\infty} \left[ \left( \frac{n}{gl}\right)^{4r}
\prod \limits _{\alpha > 0}\left(
\frac{n^2}{g^2l^2}-(a,\alpha )^2\right)^4\right]\ .
\label{geo4}
\ea
Set $J(a)=C(l)\kappa ^2(a)$.
Including all divergences of the product 
(\ref{geo4}) into $C(l)$ we get 
\ba
\kappa (a)&=& \prod \limits _{\alpha > 0}\left[
\frac{\pi (a,\alpha)}{a_0}
\prod \limits _{n=1}^{\infty}\left(1-
\frac{(a,\alpha )^2}{a^2_0n^2}\right)\right]=
\prod \limits _{\alpha > 0}\sin \frac{\pi (a,\alpha )}{a_0}\ ,
\label{geo5}\\
C(l)&=& (2\pi l)^{r/2}\left (\frac{a_0}{\pi }\right ) ^{N_+}\prod
\limits _{n=1}^{\infty}(n^2a_0^2)^{r+2}\ ,
\label{geo6}
\ea
where $a_0=(gl)^{-1}$, the integer $N_+=(N-r)/2$ is 
the number of positive roots in $X$;
 the last equality in (\ref{geo5}) results from a product formula given 
in \cite{ryz}, p.37. The induced volume element is $da\kappa^2(a)$.
It vanishes at the boundaries of the Weyl cell (at the 
boundaries of the physical configuration space in the
parameterization considered) since 
$(a,\alpha)/a_0 \in \Z$ for all $a\in \pl K_W^+$. 
Zeros of the function $\kappa(a)$ extended to the whole 
Cartan subalgebra form the diagram $D(X)$. This fact will
be important for quantization of the model in section 8.6.

\subsection{Properties of the measure on the gauge orbit space}

We will need a few mathematical facts about the function $\kappa$
which are later proved to be useful when solving quantum Yang-Mills theory
on a cylindrical spacetime in the operator and path integral approaches.

The first remarkable fact is that
the function (\ref{geo5}) is proportional to the Weyl determinant 
\cite{burb}, p.185 
\ba
(2i)^{N_+}\kappa (a)&=&\prod\limits_{\alpha >0}^{} 
\left( e^{i\pi (a,\alpha )/a_0}-e^{-i\pi (a,\alpha )/a_0}\right)
\nonumber \\
&=&\sum\limits_{\hat{R}\in W}^{} \det \hat{R}\exp \left[
\frac{2\pi i}{a_0}(\hat{R}\rho ,a)\right]\ .
\label{c.2}
\ea 
Here we have introduced the parity $\det \hat{R}$ of the elements
of the Weyl group. It is 1 if $\hat{R}$ contains even number of
the generating elements $\hat{R}_\omega$ and $-1$ if this number 
is odd. Recall that in the root space $\Rs^r$ the reflection $\hat{R}_\omega$ 
in the hyperplane orthogonal to a simple
root $\vec{\omega}$ can be thought as an $r\times r$-matrix from
the orthogonal group O($r$) such that $\det \hat{R}_\omega =-1$. 
The element $\rho$ is a half-sum of all positive roots:
\be
\rho =  \frac{1}{2}\sum\limits_{\alpha >0} \alpha \ .
 \label{c.1}
 \ee
 The relation between  $\kappa$ and the Weyl determinant allows
 us to establish the transformation properties of $\kappa$ relative
to the action of the affine Weyl group on its argument. 
 From  (\ref{waa}) and (\ref{c.2}) we infer 
 \ba
&\,& (2i)^{N_+}\kappa (\hat{R}_{\beta,n}a) \nonumber\\ 
 &=&
 \sum\limits_{\hat{R}\in W}\det \hat{R}\exp \left[
\frac{2\pi i}{a_0}(\hat{R}\rho ,\hat{R}_\beta a)\right]
 \exp \left[\frac{4\pi in_\beta}{(\beta,\beta)}\, 
(\hat{R}\rho ,\beta)\right]\\
 &=&
 \det\hat{R_\beta}\sum\limits_{\hat{R}\in W} \det \hat{R}\exp \left[
\frac{2\pi i}{a_0}(\hat{R}\rho ,  a)\right]
 \exp \left[-\, \frac{4\pi in_\beta}{(\beta,\beta)}\, 
(\hat{R}\rho ,\beta)\right]\ ,
 \label{c.3}
\ea
where we have rearranged the sum over the Weyl group by the change
$\hat{R}\rightarrow \hat{R}_\beta\hat{R}$ and made use of the 
properties that $\hat{R}_\beta^2 =1$ and $\hat{R}_\beta \beta = 
-\beta$. Next we show that the second exponential in (\ref{c.3})
is 1 for any $\beta$ and $\hat{R}$. 

To this end, we observe that
$(\hat{R}\rho,\beta)=(\rho,\beta')$ where $\beta'=\hat{R}^T\beta$
is also a root that has the same norm as $\beta$ because the
Weyl group preserves the root pattern. Therefore we have to prove
that 
\be
n_\rho(\beta) = \frac{2(\rho,\beta)}{(\beta,\beta)}
\ee
is an integer.
The half-sum of the positive roots has the following properties
\cite{hel}, p.461, 
\begin{eqnarray} 
&\ &\frac{2(\omega ,\rho )}{(\omega ,\omega )} = 1\ ,
\label{c.6}\\ 
&\ &\hat{R}_\omega \rho =\rho -\omega \ ,
\label{c.7}
\end{eqnarray} 
for any simple root $\omega$. 
Since the Weyl group $W$ preserves the root 
system and the reflection $\hat{R}_\beta $ in the hyperplane $(\beta ,a)=0$ 
is a composition of reflections $\hat{R}_\omega $, 
there exists an element $\hat{R} \in W$ and a simple root $\omega 
_\beta $ such that $\hat{R}\omega _\beta =\beta $. 
The statement that
$n_\rho (\beta)$ is an integer follows from the relation
\begin{equation} 
n_\rho(\beta)=\frac{2(\beta ,\rho )}{(\beta ,\beta )}=\frac{2(\omega 
_\beta , \hat{R}^T\rho )}{(\omega _\beta ,\omega _\beta )} \in \Z\ . 
\label{c.8}
\end{equation} 
Indeed, representing $\hat{R}^T $ as a product of the generating 
elements $\hat{R}_\omega $ and applying (\ref{c.6}) and (\ref{c.7}) we 
obtain (\ref{c.8}) from the fact that 
$2(\omega_\beta,\alpha)/(\omega_\beta,\omega_\beta) $
is an integer for any root $\alpha$.  Recall that a root $\alpha$
can be decomposed into a sum over simple roots with integer
valued coefficients, and the Cartan matrix 
$2(\omega,\omega')/(\omega,\omega)$ is also integer valued.

Thus, we arrive at the simple property 
\be 
\kappa(\hat{R}_{\beta,n}a)= \det{\hat{R}_\beta}\,\kappa(a)=-\kappa(a)\ 
\label{c.9}
\ee
 for any root $\beta$.
 Since any elements of the affine Weyl group $W_A$ is a composition of the 
reflections (\ref{waa}), we conclude that
\begin{equation} 
\kappa (\hat{R}a)=\det{\hat{R}}\,\kappa(a) = \pm \kappa (a) ,\ \ \hat{R}\in 
W_A\ , 
\end{equation} 
where by definition $\det \hat{R} =-1$ if $\hat{R}$ 
contains an odd number of the 
reflections (\ref{waa}) and $\det \hat{R}=1$ for an even number. 
The Jacobian $\mu =\kappa^2$ is invariant under
the affine Weyl group transformations.
 
 The second remarkable property of the function $\kappa(a)$ is that
 it is an eigenfunction of the $r$--dimensional  Laplace operator 
\begin{equation} 
(\pl _a,\pl _a)\kappa (a)\equiv
\Delta_{(r)}\kappa(a)
=-\frac{ 4\pi ^2(\rho ,\rho )}{a_0^2}\,\kappa (a)
=-\frac{\pi^2 N}{6a_0^2}\,\kappa(a)\  ,
\label{c.11}
\end{equation}
where the relation $(\rho,\rho)=N/24$ \cite{zhel} between the 
norm of $\rho$ and the dimension $N$ of the Lie algebra has been used.  
A straightforward calculation of the action of the Laplace operator on 
$\kappa (a)$ leads to the equality 
\ba
\Delta_{(r)}\kappa (a)=&-& \frac{4\pi ^2}{a_0^{2}}
(\rho ,\rho )\,\kappa (a)
\nonumber\\ 
&+&\frac{\pi ^2}{a_0^2} \sum\limits_{\alpha \ne \beta >0}^{}(\alpha ,\beta 
)\left [\cot \frac{\pi (a,\alpha )}{a_0}\cot \frac{\pi (a,\alpha 
)}{a_0}+1\right]\kappa (a)\ . 
\label{c.12}
\ea 
The sum over positive roots in (\ref{c.12}) can be transformed into
a sum over the roots $\alpha\neq\beta$ in a plane $P_{\alpha\beta}$
and the sum over all planes $P_{\alpha\beta}$. Each plane
contains at least two positive roots. 
Relation (\ref{c.11}) follows from 
\begin{equation} 
{\sum\limits_{\alpha \neq \beta >0\in P_{\alpha\beta}}}
(\alpha ,\beta ) [\cot (b,\alpha ) \cot ( 
b,\beta ) +1]=0\ , 
\label{c.13}
\end{equation} 
for any $b\in H$.
To prove the latter relation, we remark that 
the root pattern in each plane 
coincides with one of the root patterns for algebras of rank 2,
su(3),
sp(4)$\sim$ so(5) and g${}_2$, because the absolute value of cosine of an 
angle between any two roots $\alpha $ and $\beta $ may take only 
four values $|\cos \theta _{\alpha \beta }|=\ 
0,\ 1/\sqrt{2},\ 1/2,\ \sqrt{3}/2$. For 
the algebras of rank 2, equality (\ref{c.13}) can be verified by an explicit 
calculation. For example, in the case of the su(3) algebra, 
the sum (\ref{c.13}) is proportional to 
\begin{eqnarray*} 
-\cot  b_1\cot  b_2+\cot  b_1 \cot 
 (b_1+b_2)+\cot  b_2\cot  (b_1+b_2)+1&= &0\ , 
\end{eqnarray*} 
where 
$b_{1,2}=(b,\omega _{1,2})$, and $\omega _1, \omega _2$ and $\omega _1 
+\omega _2$ constitute all positive roots of SU(3).
 
\section{Artifacts of gauge fixing in classical theory}
\setcounter{equation}0

The definition of $\ph$ is independent of the choice of local symplectic 
coordinates and explicitly gauge-invariant. However, upon a dynamical 
description (quantum or classical) of constrained systems, we often need 
to introduce coordinates on $\ph$, which means fixing a gauge or 
choosing a $\ph$ parameterization. The choice of the parameterization is 
usually motivated 
by physical reasons. If one deals with gauge fields, one may describe 
physical degrees of freedom by transverse components ${\bf A}^\bot$ 
of the vector potential and their canonically conjugated momenta 
${\bf E}^\bot$, i.e. the Coulomb gauge 
$\pl_iA_i=0$ is imposed 
to remove nonphysical degrees of freedom. This choice comes
naturally from 
our experience in QED where two independent polarizations of 
a photon are described by the transverse vector-potential.
The Coulomb gauge is a complete global gauge condition in QED. 
Apparently, the phase space of each physical degree of freedom
in the theory is a Euclidean space. 

In the high-energy limit of non-Abelian gauge theories
like QCD the physical picture of self-interacting transverse
gluons works extremely well. However, in the infrared domain 
where the coupling constant becomes big and dynamics favors
large fluctuations of the fields,
transverse gauge fields do not serve any longer as good 
variables parameterizing $\ph$. It appears that 
there are gauge-equivalent configurations in the
functional hyperplane $\pl_iA_i=0$, 
known as Gribov's copies \cite{gribov}. 
Moreover, this gauge fixing ambiguity always occurs and has an 
intrinsic geometric origin \cite{singer} related to the topology of 
the gauge orbit space and cannot 
be avoided if gauge potentials are assumed 
to vanish at spatial infinity. This makes a substantial
difficulty for developing a consistent nonperturbative path
integral formalism for gauge theories (see section 10 for
details).
 
To illustrate the Gribov copying phenomenon in the Coulomb
gauge, one can take  
the 2D Yang-Mills theory considered above. The spatially homogeneous 
Cartan subalgebra components of  the vector potential $A=a$ and field 
strength $E=p_a$ can be regarded as symplectic coordinates on $\ph$. 
In fact, this implies the Coulomb gauge condition $\pl A=0$.
This condition is not complete in the two-dimensional case because 
there are some nonphysical degrees of freedom 
left \footnote{The Coulomb gauge would have been complete, had we removed
the constant gauge transformations from the gauge group, which,
however, would have been rather artificial since the Lagrangian
of the theory has the gauge invariance relative to spatially homogeneous
gauge transformations (see also section 10.3 in this regard).}. 
They are removed by imposing an additional gauge condition 
$(e_{\pm\alpha}, A)=0$, i.e. $A\in H$. Gribov copies of a 
configuration $A=a \in H \sim \Rs^r$ are obtained by applying 
elements of the affine Weyl group $W_A$ to $a$. The modular domain 
coincides with the Weyl cell. By definition, the modular
domain of the gauge fixing surface consists of configurations
whose Gribov copies, if any, lie outside it.  
We will see that the residual transformations
from the affine Weyl group 
are important for constructing the Hamiltonian path
integral in the Coulomb gauge for the model in question.
In fact, if we ignore them and calculate the path integral 
as if there were no Gribov copies, the answer would appear
in conflict with the explicitly gauge invariant approach 
due to Dirac. 

From the geometrical point of view \cite{singer}, the absence
of a ``good'' gauge condition $\chi({\bf A})=0$ is due to
nontriviality of the fiber bundle with the base being space
(compactified into a sphere by imposing zero boundary conditions
on the connection ${\bf A}$ at the spatial infinity) and the 
fibers being the group G (see also a comprehensive work \cite{mitter}).
For this reason, the Gribov problem is often identified with the
absence of the global cross-section on the non-trivial fiber bundle.
However, one could look at this problem differently. Gribov found
the obstruction to the nonperturbative extension of the Faddeev-Popov
path integral \cite{fp}. To give an operator interpretation to 
the Lagrangian gauge-fixed (formal) 
path integral, a more general Hamiltonian
path integral has been developed by Faddeev \cite{fd68}. The construction
is based on an explicit parameterization of the physical phase space,
which is introduced by imposing supplementary (gauge) conditions 
on the {\em canonical} variables. We have discussed such 
parameterizations of the physical phase space in the SO(N)
model (the gauge $x_i=x\delta_{1i}$), in the Yang-Mills
mechanics (the gauge $x=h\in H$), or in the 2D Yang-Mills
theory (the gauge $A(x)=a\in H$). 
The singularities discovered by Gribov are associated with
the particular choice of the supplementary conditions imposed on
the canonical {\em coordinates} (connections ${\bf A}$). 
In Yang-Mills theory this
particular class of gauge conditions
is indeed subject to the 
mathematical ``no-go'' theorem due to Singer. As has later been proposed
by Faddeev with collaborators, this mathematical problem of constructing
a cross-section on the non-trivial fiber bundle can be circumvented if the 
supplementary condition is imposed on the {\em momentum} variables
\cite{fd79}. One could even construct a set of local gauge invariant
canonical variables to span the physical phase space \cite{gj,mat}.  
The gauge fixing in the space of the canonical momenta ${\bf E}$ 
is an algebraic
(local) problem similar to the one discussed in section 3 because
under the gauge transformations, ${\bf E}\rightarrow \Omega{\bf E}
\Omega^{-1}$ \cite{jinrlec}. 

The physical phase space structure does not depend whether
one uses canonical coordinates or momenta 
to remove the gauge arbitrariness.  
The problem of constructing the 
correct path integral measure on the physical phase space parameterized
in either way would still remain because there would be singularities
in the canonical momentum space or in the configuration space 
as the consequence of the non-Euclidean
structure of the physical phase space. 't Hooft  considered
gauge fixing for the field variables rather than for the vector
potentials \cite{tmap}. He identified the singularities occurring 
in such a gauge with topological defects in gauge fields that carry 
quantum numbers of magnetic monopoles with respect to
the residual Abelian gauge group. The existence of singularities
in the momentum space were also stressed in \cite{fd79}.

In view of these arguments, we consider the Gribov obstruction as
a part of a much more {\em general and fundamental} quantization problem: 
Quantization on non-Euclidean phase spaces.
The phase space of physical degrees of freedom may not be
Euclidean even if one can find a global
cross section in the fiber bundle associated with a gauge model.
In fact, it is the geometry of the phase space that lies at the heart
of the canonical or path integral quantization because the Heisenberg
commutation relations and their representation strongly depend
on it. The quantization problem of non-Euclidean phase spaces 
is known since the birth of quantum mechanics. Yang-Mills theory
has given us a first example of the fundamental theory where  
such an unusual feature of the Hamiltonian dynamics may have
significant physical consequences.

An explicit parameterization of 
the physical phase space by local canonical coordinates is
often used in gauge theories, e.g., in the path integral
formalism. Although a particular set of canonical variables
may look preferable from the physical point of view, it
may not always appear reasonable from the mathematical point 
of view as a natural and convenient set of local canonical
coordinates on a non-Euclidean phase space because it may
create artificial (coordinate dependent) singularities in a 
dynamical description. On the other hand, it could also
happen that the physical phase space is hard to compute
and find mathematically most convenient coordinates to describe
dynamics. Therefore it seems natural  to take a closer look 
at possible ``kinematic'' effects caused by the coordinate
singularities in a {\em generic}
parameterization of the physical phase space.
Here we investigate classical Hamiltonian dynamics. A quantum
mechanical description will be developed in next section. 
  
\subsection{Gribov problem and the topology of gauge orbits}

Gribov copies themselves do not have much physical meaning because 
they strongly depend on a concrete choice of a gauge fixing condition 
that is rather arbitrary. An ``inappropriate'' choice of the gauge 
condition can complicate a dynamical description. 
To illustrate what we mean by this statement, let us take a simple
gauge model with three degrees of freedom whose dynamics is 
governed by the Lagrangian
\be
L= \frac 12\dot{x}^2_1 + \frac 12 (\dot{x}_2-y)^2 - V(x_1)\ .
\label{tra}
\ee
The Lagrangian is invariant under the gauge transformations
$x_2 \rightarrow x_2+\xi$, $y \rightarrow y +\dot{\xi}$, while
the variable $x_1$ remains invariant. The variable $y$ is 
the Lagrange multiplier since the Lagrangian
does not depend on the velocity $\dot{y}$. We can exclude it
from consideration at the very beginning. On the plane spanned
by the other two variables $x_{1,2}$, the gauge orbits are
straight lines parallel to the $x_2$ axis. 
Therefore any straight line that is not parallel to the $x_2$ axis
can serve as a unique gauge fixing condition because it intersects
each orbit precisely once. 

However, one is free to choose any gauge fixing condition,
$\chi (x_1,x_2)=0$, to 
remove the gauge arbitrariness. A necessary condition on the
gauge fixing curve is that it should intersect each gauge orbit
at least once. 
In the dynamical description, this amounts to a specific 
choice of the function $y(t)$ in the Euler-Lagrange equations
of motion. Recall that the equations of motion do not impose
any restrictions on the Lagrange multipliers in gauge models
because of their covariance under gauge transformations. Therefore the
solutions depend on generic functions of time, the Lagrange
multipliers, which can be specified so that the solutions
would fulfill a supplementary (or gauge) condition. 
Now let us take the parametric equations of the gauge fixing
curve $x_{1,2} = f_{1,2}(u)$ and let $u$ range over the real line.
One can, for instance, set $f_{1,2}(0)=0$ and let $u$ equal
the arc length of the curve counted in one direction from the origin
and  negative of the arc length, 
when the latter counted in the other direction traced out
by the curve from the origin. The parameter $u$ describes
the only physical degree of freedom in the model. 
\begin{figure}
\centerline{\psfig{figure=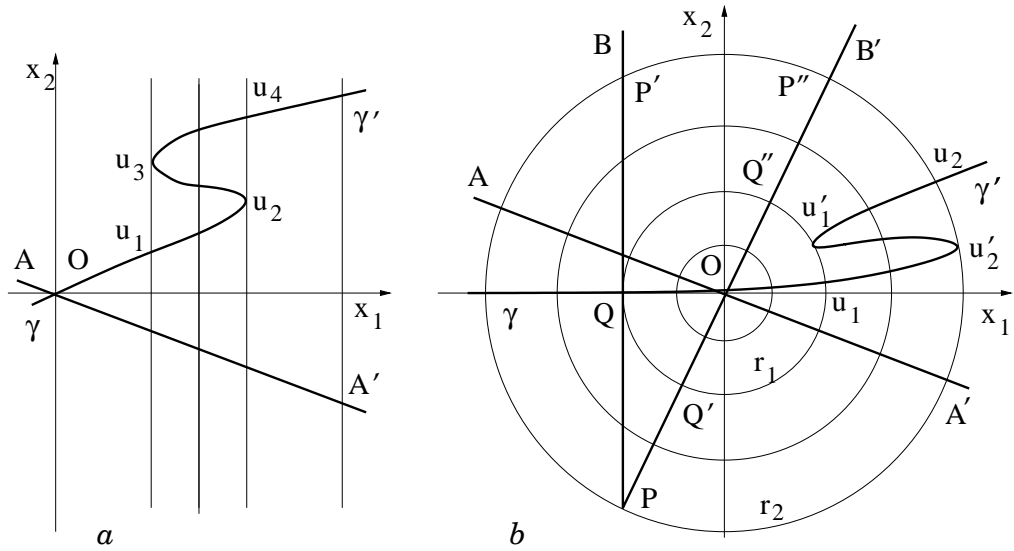}}
\caption{\small {\em a}). An illustration to 
an artificially created Gribov copying.
The gauge fixing curve $AA'$ intersects each gauge
orbit, being vertical parallel straight lines, 
precisely once. There is a one-to-one correspondence
between the parameter $u$ of the gauge fixing curve
and the gauge invariant variable $x_1$. In contrast,
the curve $\gamma\gamma'$ does not intersect each
gauge orbit once. The states labeled by the values of
$u$ from the intervals $(u_1,u_2)$, $(u_2,u_3)$ and 
$(u_3,u_4)$ are gauge equivalent.Two of them must
be discarded to achieve a unique parameterization of
the gauge orbit space. If the first and third intervals
are removed, then, in the $u$--parameterization, the
configuration space would have two ``holes''. There is
no one-to-one correspondence between $u$ and $x_1$.
In fact, the function $u(x_1)$ is multi-valued and
has, in this particular case, three branches.
\newline
{\em b}). Gribov problem in the SO(2) model.
Here the gauge fixing curve $\chi(x_1,x_2)=0$,
specified by a continuous, {\em single-valued}
and everywhere regular
function $\chi$ on the plane
such that $\pl_\theta\chi\neq 0$ at zeros of $\chi$, would intersect
each gauge orbit at least twice, thus making
the Gribov problem unavoidable in any non-invariant
approach. The condition $\pl_\theta\chi\neq 0$ at zeros of $\chi(x_1,x_2)$
is necessary \cite{fd68} to reestablish a canonical
symplectic structure on the physical phase space parameterized by
points of the surface $\sigma=\chi=0$ in the total phase space.}
\end{figure}
It seems that the dynamics of $u$ and the gauge invariant
variable $x_1$ is the same modulo a functional relation
between $u$ and $x_1$.  This is, however, only partially true.
If the gauge fixing curve intersects some gauge orbits more than once,
some {\em distinct} values of $u$ would correspond
to {\em the same} physical states. An example of such
a ``bad'' gauge fixing curve is plotted in Fig.  7a. To achieve
a one-to-one correspondence between physical states and
the values of $u$, one has to remove certain values of $u$
from the real line, thus making ``holes'' in it. These holes
are absent in the gauge invariant description via the variable
$x_1$. The parameter space also has boundaries where dynamics
of $u=u(t)$ may exhibit unusual properties. All these troubles have
been created
just by a ``bad'' choice of the gauge. We observe also
that the function $u(x_1)$ is {\em multi-valued} in
this case. 

An important point to realize is that the Gribov problem in
the above model is fully artificial. The {\em topological structure} of
the gauge orbits is such that it admits a 
gauge fixing that would allow one to construct a system of
Cartesian canonical coordinates on the physical phase space. 
The phase space of the only physical degree
of freedom is a {\em plane}. All the complications of the dynamical
description are caused by  the ``inappropriate'' parameterization
of the physical phase space.

In the SO(N) model studied earlier, the gauge orbits are
spheres centered at the origin. Their topology is not that
of a Euclidean space. This is the reason for the physical
phase space being non-Euclidean. The same applies to the 
Yang-Mills mechanical systems and the 2D Yang-Mills theory
studied above. The gauge orbits in all those models are 
compact manifolds with non-trivial topology, which makes 
the coordinate
singularities in the physical configuration space  {\em unavoidable},
in contrast to the model with the translational gauge symmetry.
The nontrivial topology of gauge orbits is, in general, 
the source of the Gribov obstruction to the 
reduced-phase-space path integral or 
canonical quantization in gauge theories. The reason is
that the physical phase space is not Euclidean in this case,
which, in turn, implies that a conventional representation
of the canonical commutation relations is no longer valid
and should be modified in accordance with the geometry
of the phase space. The artificial Gribov problem, like
in the model with translational gauge symmetry, does not
lead to any difficulty in quantization because the physical
phase space is Euclidean. Note that a bad choice of a gauge is
always possible even in electrodynamics where no one would
expect any obstruction to canonical quantization.
Let us reveal a relation between the Gribov obstruction 
and the topology of the gauge orbits in a more explicit way. 

Suppose we have the constraints $\sigma_a=\sigma_a(p,q)$.
To parameterize the physical phase space, we introduce
supplementary (gauge) conditions $\chi_a(q,p)=0$ such
that $\{\chi_a,\chi_b\}=0$ and 
the matrix $M_{ab}=\{\chi_a,\sigma_b\}$
is not degenerate, i.e., the Faddeev-Popov determinant 
$\Delta_{FP}=\det M_{ab}$ does not vanish.
A symplectic structure on the physical phase space
can locally be reestablished by means of a canonical
transformation \cite{fd68} $p,q\rightarrow p^*,q^*; \tilde{p}_a, 
\tilde{q}_a=\chi_a$. In the new canonical variables
we get $M_{ab}=\{\tilde{q}_a,\sigma_b\}=\pl\sigma_b/\pl \tilde{p}_a$.
The condition $\Delta_{FP}\neq 0$ allows one to solve
the equation $\sigma_a=0$ for the nonphysical canonical
momenta $\tilde{p}_a =\tilde{p}_a(p^*,q^*)$ which, together
with the conditions $\tilde{q}_a=0$ introduces a 
parameterization of the physical phase space by the canonical
coordinates $p^*, q^*$. The condition $\Delta_{FP}\neq 0$
is crucial for establishing a canonical symplectic structure
on the physical phase space. There may {\em not} exist
regular everywhere functions $\chi_a$ such that this condition
is met everywhere on the surface $\sigma_a=\chi_a=0$. 
This turns out to be the case when gauge orbits have
a nontrivial topology.

To illustrate the importance of gauge orbit topology for
the physical phase space geometry, let us consider the
SO(2) model (see also figure 7.b). First we remark that
one can {\em always} find a singe-valued regular function
$\chi({\bf x})$ on the plane such that its zeros form a curve 
that intersects each circle (gauge orbit) {\em precisely once}.
Thus, the condition $\chi=0$ is the {\em global cross section}
of the associated fiber bundle, or global gauge condition.
From this point of view there is no difference between the
model (\ref{tra}) and the SO(2) model. The difference 
appears when one attempts to establish the induced symplectic
structure on the physical phase space parameterized by points
of the surface $\sigma=\chi=0$. The canonical symplectic structure exists
if the Faddeev-Popov determinant $\{\sigma,\chi\}\neq 0$
does not vanish \cite{fd68}. In the model with the translational
gauge symmetry we have $\sigma=p_2$ and, hence, the condition
reads $\pl_2\chi\neq0$, which can easily be achieved with the 
choice $\chi =x_2$ on the entire surface $\sigma=\chi=0$. In the
SO(2) model we have $\sigma=({\bf p},T{\bf x})=p_\theta$, where
$p_\theta$ is the canonical momentum for the angular variable
$\theta$ on the plane. Therefore $\{\chi,\sigma\} =\pl_\theta \chi
\neq 0$. Let the function $\chi({\bf x})$ vanish, say, at the $\theta=0$
(the radial variable $r$ is fixed). 
Since $\pl_\theta\chi$ cannot be zero, the function $\chi$ changes sign 
when its argument passes through the point $\theta=0$. 
The function $\chi$ must be 
a periodic function of $\theta$ because it is single-valued on the
plane. Therefore it has to change sign at least one more time before
$\theta$ approaches $2\pi$, that is,
there exists another point $\theta=\theta_0\neq 0$ on the 
orbit $r=const$
such that $\chi(\theta_0)=0$. Thus, any curve $\chi({\bf x}) =0$,
$\{\sigma,\chi\}\neq 0$ would intersect each gauge orbit at least
twice, and the surface $\sigma=\chi=0$ cannot be isomorphic
to the physical phase space.
The periodicity of $\chi$ along the directions tangent to
the gauge orbits is due to the nontrivial topology of the orbits.

{\em Remark}. If multi-valued gauge conditions are used
to remove the gauge freedom (see, e.g., \cite{neub}), then
the canonical transformation that separates the total phase
space variables into the physical and nonphysical canonical
variables is generally related to curvilinear coordinates
\cite{fd68}. There will be singularities at the
points of the configuration space where the multi-valued $\chi$
is ill-defined. For instance, if we set $\chi =\theta$ in the 
SO(2) model, then the origin is the singular point in the physical
sector described by the radial variable. This singularity is
clearly associated with the conic structure of the physical phase space
as we have seen in section 3.4. Multi-valued gauges have
an additional bad feature: they would, in general, lead to
a multi-valued Faddeev-Popov effective action. 

In the literature one can find another model which has been
intensively studied in an attempt to resolve 
the Gribov obstruction \cite{h2,h3,fujikawa}.
This is the so called helix model \cite{h1}. 
It is obtained by a kind of merging the translational gauge
model (\ref{tra}) and the  SO(2) model.
The Lagrangian reads
\be
L = \frac 12(\dot{x}_3 - y)^2 + \frac 12\left[(\dot{x}_1+ yx_2)^2 
+(\dot{x}_2- yx_1)^2\right] - V\ .
\label{hel}
\ee
It is invariant under simultaneous time-dependent rotations
of the vector $(x_1,x_2)$ and translations of $x_3$:
\ba
y&\rightarrow& y+\dot{\xi}\ ,\\ 
x_3&\rightarrow& x_3+ \xi,\\
x_1&\rightarrow& x_1\cos\xi -x_2\sin\xi\ ,\\ 
x_2&\rightarrow& x_2\cos\xi +x_1\sin\xi .
\ea
The potential $V$ is a function of two {\em independent} Casimir functions
\be
C_1 = x_1\cos x_3 +x_2\sin x_3 \ , \ \ \  
C_2 = x_2\cos x_3 -x_1\sin x_3\ ,
\label{c12}
\ee
which are invariant under the gauge transformations. In fact,
any gauge invariant function is a function of $C_{1,2}$.
After excluding the Lagrange multiplier $y$ from
the configuration space, we find that 
the gauge orbits in the model are helices extended along
the $x_3$ axis.  
The topology of the gauge orbits in the 
model is that of the real line and thus trivial. There is {\em no}\/
topological obstruction to find a regular single--valued gauge
fixing condition that would provide a Cartesian system
of coordinates on the physical phase space. 
For instance, the plane $x_3 = 0$ intersects each
gauge orbit, specified by fixed values of $C_{1,2}$, 
precisely once. No Gribov ambiguity occurs in 
contrast to the models with topologically nontrivial gauge
orbits studied above. The Gribov problem here can only 
be {\em artificially}\/ created by a bad choice of the gauge. 
An example of a bad choice of the gauge is easy to find.
Configurations in the plane
$x_2=0$ would have infinitely many Gribov copies. Indeed, the plane
$x_2=0$ intersects each helix winding around the third axis
at the points related to one another by transformations
$x_1\rightarrow (-1)^n x_1,\ x_3 \rightarrow x_3 + \pi n$ with
$n$ being any integer. The modular domain on the gauge fixing
surface in configuration space is therefore a half--strip $x_1\geq 0,\
x_3\in [-\pi,\pi)$. One can also make the number of copies
depending on the configuration itself by taking, e.g.,
the gauge interpolating the bad and good gauges, $x_3 + a x_2=0$.
When $a=0$ we recover the good gauge, and when $a$ approaches
infinity we get the bad gauge.  
 
Thus, the model exhibits {\em no} obstruction to either the reduced
phase-space canonical or path integral quantization because
the physical phase space in the model is obviously a four
dimensional {\em Euclidean} space. From this point of view
the model has no difference from the translational gauge model
discussed earlier. 

{\em Remark}. In the gauge $x_2=0$, it looks like the physical
phase space is not $\Rs^4$ because of the restrictions $x_1\geq 0$
and $x_3\in [\pi,-\pi)$. This is not the case. As one might see
from the form of the Casimir functions (\ref{c12}), the 
gauge $x_2=0$ corresponds to the parameterization of the physical
phase space by the canonical variables associated with the 
polar coordinates on the $C_{1,2}$-plane, while the gauge $x_3=0$
is associated with the natural 
Cartesian canonical coordinates on the physical
phase space. Both the parameterization are related by
a {\em canonical} transformation.
In section 3.4 it is shown that by going over to 
polar coordinates (as well as to any curvilinear coordinates) one
cannot change the geometrical structure of the phase space.
The artificial Gribov problem in this model is just a question
of how to {\em regularize} the conventional Liouville
path integral measure on the {\em Euclidean} phase space
with respect to general canonical transformations. This,
as a point of fact, can be done in general \cite{kld,kld2}.
As far as the particular gauge $x_2=0$ is concerned, one
knows perfectly well how to change variables in the 
path integral (or in the Schr\"odinger equation) 
from the Cartesian to polar coordinates
in the plane \cite{pcoor,pcoor2,lpcc,book}.

\subsection{Arbitrary gauge fixing in the SO(2) model}

Although a good choice of the gauge 
could greatly simplify the dynamical description 
of the physical degrees of freedom, we often use
bad gauges for the reasons that either the geometry of
gauge orbits is not explicitly known or the variables
parameterizing the gauge orbit space and associated with
a particular gauge (like the Coulomb gauge in Yang-Mills
theory) have a convenient physical interpretation.
Here we take a closer look at some 
dynamical artifacts that may occur through a bad
choice of the gauge. These artifacts would be purely
gauge dependent or, in other words, they are coordinate
dependent, meaning that they can be {\em removed} by changing 
a parameterization of the gauge orbit space. However
the physical interpretation may also considerably change upon
going over to the new variables related to the initial
ones by a {\em nonlinear} transformation, like the transverse
gluons are easy to describe in the Coulomb gauge, while 
it would be a hard task to do so using the gauge invariant
loop variables ${\rm tr}\,{\rm P}
\exp [ig\oint (d{\bf x}, {\bf A})]$ which
can be used to parameterize the gauge orbit space in the Yang-Mills
theory. 

We limit our consideration to the SO(2) model.
The reason is, first of all, that a general case
(meaning a general gauge in a general gauge theory) 
would be rather involved to consider in details,
and it is hardly believed that the artificially
created Gribov-like problem is of great physical
significance. Secondly, the idea
is general enough to be extended to any gauge model. So,
the gauge orbits are circles centered at the origin. The
configuration space is a plane spanned by the vector
variable ${\bf x}$.

Any gauge condition $\chi({\bf x})=0$ determines a curve on a plane 
$\Rs^2$ over which a physical variable ranges. The curve 
$\chi({\bf x}) =0$ must cross each orbit at least 
once because a gauge choice is 
nothing but a choice of a parameterization of the gauge orbit space. In the 
model under consideration, this yields that the curve has to go 
through the origin to infinity. Let us introduce a smooth parameterization 
of the gauge condition curve 
\be 
{\bf x} = {\bf x}(u) = {\bf f}(u)\ ,\ \ \ \ \ u\in \Rs\ , 
\label{gf1}
\ee 
where ${\bf f}(0) = 0$ and $|{\bf f}|\rightarrow \infty$ as 
$u\rightarrow \pm\infty$ so that $u$ serves as a physical variable
which we can always choose  to range the whole real line.
If $f_2=0$ and $f_1=u$, we recover the unitary gauge considered above. 
 
Let the points ${\bf x}$ and 
${\bf x}_s$ belong to the same gauge orbit, then ${\bf x}_s= \Omega_s 
{\bf x},\ \Omega_s\in SO(2)$. Suppose the curve (\ref{gf1}) intersects a 
gauge orbit at points ${\bf x}={\bf f}(u)$ and ${\bf x}_s = {\bf f}(u_s)$. 
We have also $u_s= u_s(u)$ because ${\bf f}(u_s)= \Omega_s {\bf f }(u)$. 
If the structure of gauge orbits is assumed to be unknown, the function 
$u_s(u)$ can be found  by solving the following equations 
\ba 
\chi(\Omega_s{\bf f})&=&0\ \label{gf2},\\ 
\Omega_s(u){\bf f}(u) &=& {\bf f}(u_s(u))\ . 
\label{gf3}
\ea 
Eq. (\ref{gf2}) is to be solved for $\Omega_s$ while $u$
is kept fixed. The trivial solution, 
$\Omega_s =1$, always exists by 
the definition of ${\bf f}$. All the solutions 
form a set $S_\chi$ of discrete residual 
gauge transformations. Eq. (\ref{gf3}) 
determines an induced action of $S_\chi$ on the variable $u$ spanning
the gauge fixing curve, i.e., it specifies the functions $u_s(u)$.
The set $S_\chi$ is not a group 
because for an arbitrary $\chi$ a composition $\Omega_{s}\Omega_{s'}$ of 
two elements from $S_\chi$ might not belong to $S_\chi$ since
it may not satisfy (\ref{gf2}), while for each $\Omega_s$ 
there exists the inverse element 
$\Omega_s^{-1}$ such that $\Omega_s^{-1}\Omega_s = 1$. 
Indeed, suppose we have two different solutions 
$\Omega _s$ and $\Omega _{s'}$ 
to the system (\ref{gf2})--(\ref{gf3}). 
The composition $\Omega 
_s\Omega _{s'}$ is not a solution to (\ref{gf2}), 
i.e. $\chi(\Omega _s\Omega _{s'} 
{\bf f}(u))=\chi(\Omega _sf(u_{s'}))\ne 0$ because,
in general, ${\bf f}(u_{s'})\ne {\bf f}(u)$ whereas  we only have
$\chi(\Omega _s{\bf f}(u))=0$. From the geometrical point of
view, this simply means that, although the configurations ${\bf f}$,
$\Omega_s {\bf f}$ and $\Omega_{s^\prime} {\bf f}$ are in
the gauge fixing curve, the configuration 
$\Omega_s\Omega_{s^\prime} {\bf f}$ is not necessarily in it.

The functions $u_s(u)$ determined by (\ref{gf3}) do not 
have a unique analytic continuation to the covering space $\Rs$
isomorphic to the gauge fixing curve ${\bf x}={\bf f}(u), u\in \Rs$, 
otherwise the composition $u_s\circ u_{s'}=u_{ss'}(u)$ would be uniquely 
defined and, hence, one could always find an element $\Omega _{ss'}= 
\Omega _s\Omega _{s'}$ being a solution to (\ref{gf2}), 
which is not the case. 
Moreover, a number of elements of $S_\chi$ can depend on $u$. 
 
To illustrate our analysis, let us take an explicit function
${\bf f}(u)$, find the functions $u_s(u)$ and investigate
their analytic properties.
Set $f_1= -u_0,\ f_2=-\gamma(2u_0 + u)$ for $u< - u_0$ and $f_1= u,\ 
f_2=\gamma u$ for $u> -u_0$ where
$\gamma $ and $u_0$ are positive constants. The 
curve is plotted in Fig 7b (see the curve $BPB'$ in it).
It touches circles (gauge orbits) of radii $r_1= u_0$ and 
$r_2=u_0\gamma_0,\ \gamma_0=\sqrt{1+\gamma^2}$
(the points $Q$ and $P$ in Fig. 7b, respectively). It intersects twice all 
circles with radii $r< r_1$ and $r> r_2$, whereas any circle 
with a radius from the interval $r\in (r_1,r_2)$ has four 
common points with the gauge condition curve. Therefore, $S_\chi$ has one 
nontrivial element for $u\in \Rs_1\cup\Rs_3,\ \Rs_1= (-u_0/\gamma_0, 
u_0/\gamma_0),\ \Rs_3= (-\infty,-3u_0)\cup (u_0,\infty)$ and three 
nontrivial elements for $u\in \Rs_2= (-3u_0,-u_0/\gamma_0)\cup 
(u_0/\gamma_0, u_0)$. 
In Fig. 7b the point $P^\prime$ correspond to $u=-3u_0$, $P^{\prime\prime}$
to $u=u_0$, $Q^\prime$ to $u=-u_0/\gamma_0$ and $Q^{\prime\prime}$
to $u=u_0/\gamma_0$, i.e., $\Rs_1$ is the segment 
$(Q^\prime Q^{\prime\prime})$,
$\Rs_2= (BP^\prime)\cup (P^{\prime\prime}B^\prime)$ and 
$\Rs_3= (P^\prime PQ^\prime)\cup 
(Q^{\prime\prime}P^{\prime\prime})$.
Since the points ${\bf f}(u_s)$ and ${\bf f}(u)$ belong to the same 
circle (gauge orbit), the functions $u_s$ 
have to obey the following equation 
\be 
{\bf f}^2(u_s)= {\bf f}^2(u)\ . 
\label{gf4}
\ee 
Denoting $S_\chi = S_\alpha$ for $u\in \Rs_\alpha,\ 
\alpha = 1,2,3$, we have $S_1=\Z_2,\ u_s(u)= -u;\ S_2$ is determined by 
the following mappings of the interval $K_2 = (u_0/\gamma_0,u_0)$ 
\ba 
u_{s_1}(u) &=& -u\ ,\ \ \ \ \ \ \ \ \ \ \ \ \ \ \ \ \ \ \ \ \ \  
\ \ \ \ \ \ \ 
u_{s_1}:\  K_2\rightarrow (-u_0,-u_0/\gamma_0)\ 
\label{gf5};\\ 
u_{s_2}(u) &=& -2u_0 + \frac{\gamma_0}{\gamma}\left(u^2 - 
\frac{u_0^2}{\gamma _0^2}\right)^{1/2} ,\ \  
u_{s_2}:\ K_2\rightarrow (-u_0,-2u_0)\ 
\label{gf6};\\ 
u_{s_3}(u)&=&-2u_0 - \frac{\gamma_0}{\gamma}\left(u^2 -
\frac{u_0^2}{\gamma _0^2}\right)^{1/2}
 ,\ \ 
u_{s_3}:\ K_2\rightarrow (-2u_0,-3u_0)\ ; 
\label{gf7}
\ea 
and for $S_3$ we get 
\be 
u_s(u) = - 2u_0 - \frac{\gamma_0}{\gamma}\left(u^2 -
\frac{u_0^2}{\gamma _0^2}\right)^{1/2}\ : 
\ \ \ (u_0,\infty)\rightarrow (-3u_0, -\infty)\ .
\label{gf8} 
\ee 
The functions (\ref{gf6})--(\ref{gf7}) do {\em not} have a 
{\em unique analytic} continuation 
to the whole domain $\Rs_2$ (observe the square root function
in them) and, hence, their composition is ill-defined. 
The mappings (\ref{gf5})--(\ref{gf7}) do not form a group. 
Since they realize a representation 
of $S_\alpha$, $S_\alpha$ is not a group. 
 
The physical configuration space is, obviously, isomorphic to 
 $K= \cup K_\alpha,\ 
K_\alpha= \Rs_\alpha/S_\alpha$, i.e. $K_\alpha$ is a fundamental domain 
of $\Rs_\alpha$ with respect to the action of $S_\chi=
S_\alpha$ in $\Rs_\alpha$, 
$\Rs_\alpha = \cup \hat{R}K_\alpha,\ \hat{R}$ ranges over $S_\alpha$. 
Upon solving (\ref{gf4}) (or (\ref{gf2})--(\ref{gf3})) 
we have to choose a particular interval 
as the fundamental domain where the solutions are
analytic functions. We have set $K_2 = (u_0/\gamma_0,u_0)$ in 
(\ref{gf5})--(\ref{gf7}). 
Another choice  would lead to a {\em different} form of the functions 
$u_s$ (to another representation of $S_\chi$ in $\Rs_2$). Setting, for 
example, $K_2= (-2u_0,-u_0)$ we obtain from (\ref{gf4}) 
\ba 
u_{s_1}(u)&=& - 4u_0-u\ ,\ \ \ \  \ \ \ \ \ \ \ \ \ \ \ \ \ \ \ \ \ \ \ \  
u_{s_1}:\  K_2\rightarrow (-3u_0,-2u_0)\ ;\\ 
u_{s_2}(u)&=& -\frac{1}{\gamma_0}
\left[u_0^2+\gamma^2(2u_0 +u)^2\right]^{1/2} ,\ \ 
u_{s_2}:\  K_2\rightarrow \left(-u_0,-\frac{u_0}{\gamma_0}
\right)\ ;\\ 
u_{s_3}(u)&=& \frac{1}{\gamma_0}
\left[u_0^2+\gamma^2(2u_0 +u)^2\right]^{1/2}
\ ,\   
\ \ u_{s_3}:\  K_2\rightarrow \left(\frac{u_0}{\gamma_0},u_0
\right)\ . 
\ea 
To find the group elements $\Omega_s(u)$ corresponding to $u_s(u)$, one 
should solve Eq.(\ref{gf4}). Setting $\Omega_s =
\exp(-T\omega_s)$, where 
$T_{ij}=-T_{ji},\ T_{12}=1$, the only generator
of SO(2), and 
substituting (\ref{gf5})--(\ref{gf7}) into (\ref{gf3}), we find 
\ba 
\omega_{s_1}(u) &=& \pi\ 
\label{gf12};\\ 
\omega_{s_2}(u) &=& \frac{3\pi}{2} - \sin^{-1}\left( 
\frac{u_0}{\gamma_0u}\right) - \tan^{-1}\gamma\ ;\\ 
\omega_{s_3}(u) &=& \frac{\pi}{2} + \sin^{-1}\left( 
\frac{u_0}{\gamma_0u}\right) - \tan^{-1}\gamma\ , 
\label{gf14}
\ea 
where $u\in K_2=(u_0/\gamma_0,u_0)$. Elements of $S_{1,3}$ are 
obtained analogously. It is readily seen that $\Omega_{s_1} 
\Omega_{s_2}\neq \Omega_{s_3}$, etc., i.e. the elements $\Omega_s$ 
do not form a group. An alternative choice of $K_2$ results in a 
modification of the functions (\ref{gf12})--(\ref{gf14}). 
 
Thus, under an inappropriate gauge fixing, residual gauge transformations 
might not form a group (no composition for elements);
the parameterization of  
${\rm CS}_{\rm phys}$ appears to be complicated. 
One could assume that all the complications of 
the ${\rm CS}_{\rm phys}$ structure, 
${\rm CS}_{\rm phys}\sim K$, found above have been caused by using 
{\em gauge non-invariant} variables 
for describing physical degrees of freedom. Indeed, we have chosen 
a ``bad'' gauge $\chi({\bf x})=0$ and gained a complicated set of 
residual gauge transformations (Gribov-like problem). 
However, one can easily turn the 
variable $u$ into a formally 
{\em gauge-invariant} one by means of a special 
canonical transformation. The set $S_\chi$ will appear again due to 
topological properties of such a canonical transformation rather than 
due to gauge fixing ambiguities. The
coordinate singularities in the physical phase space parameterized by
such gauge-invariant canonical variables will be present again. 
Since local canonical  coordinates on the 
gauge invariant phase space (\ref{1}) can only be specified modulo
canonical transformations, it is natural to expect, and we will see
this shortly, that the arbitrariness of gauge
fixing may always be re-interpreted as the arbitrariness
in choosing local canonical coordinates on the physical
phase space. If one cares only about 
a {\em formal} gauge invariance of canonical variables, i.e.,
vanishing Poisson brackets of the canonical
variables with the constraints,  and ignores a geometrical 
structure of the physical phase space (\ref{1}), 
then the choice of the canonical coordinates might lead to 
some artificial (coordinate dependent) 
singularities in the Hamiltonian formalism which are
similar to those in the non-invariant approach.

\subsection{Revealing singularities in a formally gauge invariant
 Hamiltonian formalism} 
 
The gauge condition $\chi({\bf x})=0$ induces a parameterization 
of the physical phase space by some local canonical variables.
To construct them,
consider the following canonical transformation of ${\bf x}$ and 
${\bf p}$
\ba 
{\bf x}&=& \exp(T\theta){\bf f}(u)\ 
\label{gf15};\\ 
p_\theta &=& {\bf p}T{\bf x}= \sigma\ ,\ \ \ 
p_u = \frac12({\bf p},{\bf x})\frac{d}{du}\ln {\bf x}^2\ , 
\label{gf16}
\ea 
where in (\ref{gf16}) the derivative 
$d{\bf x}/du= \exp(T\theta){\bf f}'(u)$ 
is expressed via $\theta({\bf x})$ and $u({\bf x})$.
We also obtain that 
$\{\theta,p_\theta\}=\{u,p_u\} = 1$ (if $\{x_i,p_j\}= 
\delta_{ij}$) all other Poisson brackets vanish. 
We remark that the case $f_1 = u$ and $f_2=0$
corresponds to the polar coordinates on the plane,
$u^2={\bf x}^2$. The matrix $\exp(T\theta)$
rotates the ray $x_2=0, x_1=u=r>0$ so that it sweeps
the entire plane.  For arbitrary smooth functions $f_i(u)$,
Eq. (\ref{gf15}) defines a generalization of the polar coordinates.
The plane is now swept by segments
of the curve ${\bf x}={\bf f}(u)$ rotated by the matrix
$\exp(T\theta)$, where $\theta\in [0,2\pi)$. 
The segments are traced out by the 
the vector function ${\bf x}={\bf f}(u)$ for those 
values of $u\in K\subset \Rs$ for which Eq. (\ref{gf15})
determines a one-to-one correspondence between the components of 
${\bf x}$ and the new variables $u$ and $\theta$.
For example, if ${\bf x}={\bf f}(u)$ is the curve $\gamma O\gamma'$
plotted in Figure 7b, then a possible choice of $K$ is the union
of the sets $[0,u_2')$ and $[u_2,\infty)$, where $|{\bf f}(u_2')|=
|{\bf f}(u_2)|$, but $u_2'<u_2$.
The parameter $u$ is {\em gauge-invariant} 
since ${\bf f}^2(u)={\bf x}^2$. We shall call such a change
of variable {\em associated} with (or {\em adjusted } to) both
the chosen gauge condition and the gauge transformation law. 
We have already used such curvilinear coordinates. 
These are the spherical coordinates
for the SO(N) model which are naturally associated with
the unitary gauge $x_i=0, i\neq 1$, or the functional
curvilinear coordinates (\ref{aah}) associated with
the Coulomb gauge in the 2D Yang-Mills theory. 

So, given a gauge transformation law and a desired gauge condition,
such curvilinear coordinates can be constructed in any
gauge model by acting by a generic gauge group element
on elements the gauge fixing surface. The latter is subject to
the only condition that each gauge orbits has at least one common 
point with it. The parameters of the gauge
transformation and those spanning the gauge fixing surface are the new 
curvilinear coordinates. Clearly, the parameters of the
gauge fixing surface become gauge invariant in such an 
approach. We postpone for a moment the analysis of topological
properties of this change of variables 
and complete constructing the Hamiltonian formalism.
 
Since $p_\theta$ coincides with the constraint, we conclude that $\theta$ 
is the nonphysical variable in the model; $\sigma =p_\theta$ generates 
its shifts, whereas $\{\sigma, u\}=\{\sigma ,p_u\}= 
0$ and, hence, $u$ and $p_u$ are {\em gauge-invariant}. 
Using the decomposition 
\be 
{\bf p} = p_\theta
\frac{T{\bf x}}{{\bf x}^2} + p_u\frac{{\bf x}}{\mu(u)}\ , 
\ee 
where $\mu (u)=(d{\bf f}/du,{\bf f})$, and the constraint $p_\theta =0$
we derive the physical Hamiltonian
\be
H_{ph}=\left. \left(\frac12{\bf p}^2 +V({\bf x}^2)\right)
\right\vert_{p_{\theta}=0}=
\frac12\frac{{\bf f}^2(u)}{\mu^2(u)}\ p_u^2 +
V({\bf f}^2(u))\ .
\label{gf18}
\ee
Hamiltonian equations of motion generated by 
(\ref{gf18}) provide a formally gauge-invariant
dynamical description. 

Let us find the hidden set of transformations $S_\chi$.
As we have pointed out above, dynamics is sensitive to a phase
space structure. Therefore, to complete the formally
gauge-invariant description, one
should describe the phase space parameterized
by the local canonical variables $u$ and $p_u$.
Let us forget for a moment about the gauge symmetry and
the constraint $p_\theta=0$ induced by it and consider relation 
(\ref{gf15}) as a
change of variables. We will be interested in the topological
properties of the change of variables.
There should be a one-to-one
correspondence between points ${\bf x}\in \Rs^2$
and $\theta,\ u$. The latter yields a restriction on admissible values
of $\theta$ and $u$, $\theta\in [0,2\pi)$ and $u\in K\subset \Rs$. To see
this, we allow the variables $\theta$ and $u$ to have their values on
the whole real axis and consider transformations $\theta,\ u\rightarrow
\theta +\theta_s =\hat{R}\theta,\ u_s=\hat{R} u$ such that
\be
{\bf x}(\hat{R}\theta, \hat{R}u)= {\bf x}(\theta, u)\ .
\label{gf19}
\ee
We assume ${\bf f}(u)$ to be a real {\em analytic} function on $\Rs$.
Points $\hat{R}\theta,\ \hat{R}u $ of the $(u,\theta)$-plane
 are mapped to one
point on the ${\bf x}$-plane. The mapping (\ref{gf15}) becomes
one-to-one, i.e., it determines a change of variables, if one
restricts values of $\theta$ and $u$ by the modular domain $\tilde{K}=
\Rs^2/\tilde{S}$ where transformations from $\tilde{S}$ are defined by
(\ref{gf19}). The set $\tilde{S}$ is decomposed into the product $T_e\times
S_\chi$ where elements of $T_e$ are translations of $\theta$ through the
group manifold period,
\be
T_e\ :\ \ \ \ \ \theta\rightarrow \theta + 2\pi n, \ \ \ \
u\rightarrow u,\ \ \ n\in \Z\ ,
\label{gf20}
\ee
and $S_\chi$ {\em formally} 
coincides with the set of residual gauge transformations
in the gauge $\chi =0$. Indeed, 
let $\Omega_s= \exp(T\omega_s(u))$ satisfy (\ref{gf2})--(\ref{gf3}).
Then we have $x(u,\theta) =\exp(T\theta)\Omega_s^{-1}\Omega_s {\bf f}(u)=
x(\hat{R}_su,\hat{R}_s\theta)$ where
\be
S_\chi\ :\ \ \ \ \theta\rightarrow\hat{R}_s\theta =
\theta - \omega_s(u)\ ,\ \ \ \
u\rightarrow \hat{R}_su= u_s(u)\ .
\label{gf21}
\ee
Thus, $\tilde{K}\sim [0,2\pi )
\cup K$ with $K$ being the 
fundamental modular domain for the gauge 
$\chi=0$. In the case of the polar coordinates, $S_\chi=\Z_2,\ \omega_s
=\pi$ and $u_s= -u$, hence $K\sim \Rs_+$ (a positive semiaxis).

Under the transformations (\ref{gf20}), 
the canonical momenta (\ref{gf16}) remain untouched, while
\be
p_\theta \rightarrow p_\theta\ ,\ \ \ \ \ \ \
p_u\rightarrow \left(\frac{du_s}{du}\right)^{-1}p_u\equiv p_{u_s}
=\hat{R}_sp_u
\label{gf22}
\ee
under the transformation (\ref{gf21}).
In the new canonical variables, a state with given values
of canonical coordinates ${\bf p}$ and ${\bf x}$
corresponds to phase-space points $(p_\theta, \hat{R}_s\theta,\hat{R}_sp_u,
\hat{R}_su)$, $\hat{R}_s$ runs over $S_\chi$, provided $\theta\in [0, 2\pi)$.
Therefore, values of the new canonical variables connected{\bf }
with each other by 
the $S_\chi$-transformations are physically indistinguishable.
 
Consider a phase-space plane, where $p_\theta =0$ and $\theta$ has a 
fixed value, and states $(p_\theta=0,\theta,\hat{R}_sp_u, \hat{R}_su)$ on 
it. These states differ from each other only by values of the angular 
variable $(0,\theta, \hat{R}_sp_u, \hat{R}_su)\sim (0, 
\hat{R}^{-1}_s\theta, p_u, u)$ where $\hat{R}^{-1}_s\theta = \theta + 
\omega_s(u)$. If now we switch on the gauge symmetry, the angular 
variable becomes nonphysical and, hence, the difference between all those 
states disappears. They correspond to the same physical state. Thus, 
the transformations $u,p_u\rightarrow u_s,p_{u_s}$ 
relate {\em distinct} points in the phase space spanned by
$p_u$ and $u$, which correspond to the very {\em same} physical 
physical state of the system. Therefore they should be identified
to describe  $\ph$ in the parameterization chosen.  
For the polar coordinates, we obviously get $\ph 
=cone(\pi)$. The conic singularity is also present in the
new variables (it is non-removable due to the 
nontrivial topology of the gauge orbits), 
but there appear additional singular points
which are pure coordinate artifacts and merely
related to the fact that the function $u=u(r)$ ($r$ is 
the radial variable on the plane) is {\em multi-valued}.
There is {\em no} curvature at those points of the phase
space. The transformations $S_\chi$ are nothing but the transformations
which relate different branches of the function $u(r)$ to
one another as one might see from (\ref{gf4}) since 
${\bf f}^2(u) = r^2$.   
 
One should emphasize that in the approach being developed
the transformations 
$\hat{R}\in S_\chi$ in the $(u,p_u)$-plane cannot be regarded as the ones 
generated by the constraint $\sigma=p_\theta $ since $\{\sigma,u\}= 
\{\sigma,p_u\}=0$ in contrast to the gauge fixing description 
considered above. Physical variables are 
chosen so that the set $S_\chi$ determining their phase space coincides 
formally with the set of residual gauge transformations in the gauge 
fixing approach. 
Thus, all artifacts inherent 
to an inappropriate gauge fixing may well emerge
in a formally gauge-invariant approach.
To see them,  we compare phase-space trajectories 
in the canonical variables $r=|{\bf x}|,\ p_r=({\bf x,p})/r$ and 
$u,\ p_u$. They are connected by the canonical transformation $r=r(u)= 
|{\bf f}(u)|,\ p_r=rp_u/\mu = p_u(dr/du)^{-1}$. We also assume the function 
${\bf f}$ to be differentiable so that $dr/du =0$ only at two points 
$u=u_{1,2}'$ and $dr/du>0$ as $u< u_2'$ and $u>u_1'$, while $dr/du < 0$ if 
$u\in (u_2',u_1')$. 
Our assumptions mean that the curve ${\bf x}={\bf f}(u),\ 
u\geq 0$, goes from the origin, crosses 
the circle $|{\bf x}|= r_1 = r(u_1)$
at ${\bf x}= {\bf f}(u_1)$ and reaches the circle $|{\bf x}| = r_2 
=r(u_2')$, touches it at ${\bf x}= {\bf f}(u_2')$, returns back to the 
circle $|{\bf x}| = r_1$, and, after touching it at the point ${\bf x} 
={\bf f}(u_1')$, tends to infinity, crossing the circle $|{\bf x}| =r_2$ at 
${\bf x}={\bf f}(u_2)$. An example of such a curve is given in
Fig. 7b (the curve $\gamma O\gamma'$) and in Fig. 8 (right)).
 
In a neighborhood of the origin, $\ph$ has the conic structure 
as we have already learned. This local structure is preserved upon the 
canonical transformation to the variables $u, p_u$ because it is 
a smooth and one-to-one mapping of the strip $r\in (0,r_1)$ on $ 
u\in (0,u_1)$. The same holds for the map of the half-plane $r>r_2$ 
onto the half-plane
$u>u_2$. Troubles occur in the domain $r\in (r_1,r_2)$ where the 
inverse function $u=u(r)$ becomes multi-valued; it has three branches 
in our particular case. States belonging to the strips $u\in (u_1,u_2'), 
\ u\in (u_2',u_1')$ and $u\in (u_1',u_2)$ are physically equivalent because 
there are transformations from $S_\chi$ mapping the strips on each other 
and leaving points $p_r, r\in (r_1,r_2)$ invariant. 
  
To investigate what happens to phase-space trajectories in the 
region $u\in (u_1,u_2)$ of the phase space, 
consider a motion with a constant 
momentum $p_r$ and suppose that the particle is outgoing from the 
origin $r=0$. On the $(p_u,u)-$plane, the particle motion corresponds 
to a point running along a curve going from the origin $u=0$. As soon as 
the phase-space point crosses the line $u=u_1$, 
there appear two ``phantom'' 
\begin{figure}
\centerline{\psfig{figure=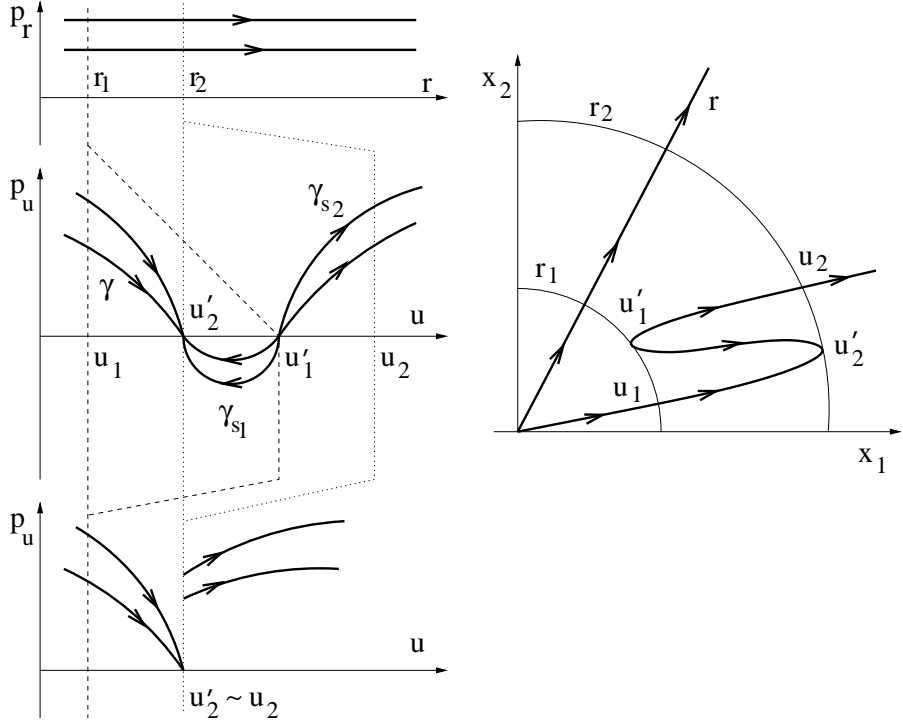}}
\caption{\small Phantom trajectories caused by coordinate
singularities occurring through a bad parameterization
of the physical phase space.}
\end{figure}
phase-space trajectories outgoing from the point $p_u=0, u=u_1'$
because the point $u_1$ is $S_\chi$-equivalent to $u_1'$. Note
also that 
$p_{u_1'}=p_{u_2'}= 0$ since $dr/du = 0$ at $u=u_{1,2}'$.
The process is shown in Fig. 8. The interval $(r_1,r_2)$
is represented by the three intervals $(u_1,u_2')$,
$(u_2',u_1')$ and $(u_1',u_2)$ in the $u$--parameterization.
They are ranges of the three branches of the multi-valued
function $u(r)$. The dashed and dotted lines in the figure
show the ``splitting'' of the points $r_1$ and $r_2$, respectively.
The trajectories at $u=u_1'$ appear right after crossing
the line $u=u_1$ by the system. So a single trajectory
in the $r$--parameterization is represented by the three
trajectories in the $u$--representation on the interval
$(r_1,r_2)$. 

If $u_{s_1}$ and $u_{s_2}$ map 
$(u_1,u_2')$ onto $(u_2',u_1')$ and $(u_1',u_2)$, respectively, so that 
$r(u)=r(u_{s_1})= r(u_{s_2}),\ u\in (u_1,u_2')$, then the ``phantom'' 
trajectories, shown in Fig. 8 as $\gamma_{s1}$
and $\gamma_{s2}$, are described by the 
pairs $\hat{R}_{1,2}p_u, \hat{R}_{1,2}u$ 
(cf. (\ref{gf22})) where the point $p_u,u$ traces out the trajectory 
$\gamma$ in the phase space region $u\in (u_1,u_2')$. 
Since $du_{s_1}/du <0$ and $du_{s_2}/du >0$, the ``phantom'' 
trajectory $\hat{R}_2p_u,\hat{R}_2u$ goes from the origin, while 
the point $\hat{R}_1p_u, \hat{R}_1u$ traces out the 
trajectory in the opposite direction. Note that the 
momentum $\hat{R}_1p_u$ is negative for this trajectory since
$dr/du$ is negative in the interval $(u_1,u_2)$. 
The points $p_u,u$ and $\hat{R}_1p_u,\hat{R}_1u$ arrive at
$p_u=p_{u_2'}=0, u=u_2'$ in the same time 
and annihilate each other, whereas a 
``phantom'' particle moving along the branch $\hat{R}_2p_u,\hat{R}_2u$ 
approaches the line $u=u_2$. In the next moment of time the system 
leaves the interval $r\in (r_1,r_2)$ (or $u\in (u_1,u_2)$). 
 
Such ``branching'' of classical phase-space 
trajectories is a pure artifact of 
an inappropriate parameterization of $\ph$ (or, as we have argued 
above, of a bad gauge fixing). It has to be removed by gluing all the 
``phantom'' trajectories (branches). In so doing, we {\em cannot} however 
{\em avoid breaking} the trajectories at the singular points $u=u_{1,2}'$. 
Indeed, consider trajectories approaching the line $u=u_1$ with 
{\em different} momenta $p_r$ from the origin and crossing it.
Since the motion in the phase-space strips $(u_2',u_1')$ and $(u_1',u_2)$
is physically equivalent to the one in the strip $(u_1,u_2')$,
we can cut out those two strips from the physical domain of the local
canonical variables $u$ and $p_u$. The state $u=u_2', p_u=0$
is equivalent to the state $u=u_2, p_u=0$, so we can glue them
together making just a point-like joint between two phase-space
domains $u<u_2'$ and $u>u_2$. In principle, we could  glue the
edges of the cut 
shown by the dotted line in the bottom of Fig. 8 since the 
phase-space points in the vicinity of $u=u_2'$ are 
$S_\chi$--equivalent to those in the vicinity of $u=u_2$
and, therefore, correspond to the same physical states.
This would restore the original {\em conic} structure of the physical
phase space which certainly cannot depend on the parameterization.
However, the continuity of the phase-space trajectories is lost.
Every trajectory approaching the line $u=u_2'$ from the origin
would fall into the point $p_u=0$ on this line because
$p_u =dr/du p_r$ and $dr/du$ vanishes at $u=u_2'$. So there is no
trajectory that could cross this line with non-zero momentum.
On the other hand, trajectories approaching the line $u=u_2$
from infinity can have a non-zero momentum. Therefore we
always gain the discontinuity by gluing the lines $u=u_2'$
and $u=u_2$. The artificial attractor at the phase-space
point $p_u=0, u=u_2'$ corresponds to one of the zeros of the
Faddeev-Popov determinant $\mu(u_{1,2}')=0$. 
It is, obviously, absent in another gauge or, as we have
just learned, in another parameterization of the physical
phase space.

We conclude that the use of formally gauge invariant 
canonical variables (i.e, those whose Poisson bracket
with the constraints vanishes) may well exhibit the
same type of singularities as the non-invariant approach
based on the gauge fixing. For this reason, it is
of great importance to study the geometrical structure of 
the physical phase space
{\em before} introducing any explicit parameterization of it
either via gauge fixing or by local formally gauge invariant
canonical coordinates in order to avoid unnecessary (artificial)
complications associated with a bad parameterization.

\subsection{Symplectic structure on the physical phase space} 

The existence of the singularities in any parameterization 
of the physical phase space by a set of canonical variables 
naturally leads to the question
whether one could get around this trouble by using local
{\em noncanonical} coordinates. The answer is {\em affirmative}, although 
it does not come for free. The idea is a generalization 
of the approach proposed in section 3.3. Suppose we know
a set of all independent Casimir functions $C_i(q)$ in a 
gauge theory, where $q$ labels points in the total configuration
space. Clearly, the values of the Casimir functions parameterize
the gauge orbit space. We also assume $C_i(q)$ to be regular on
the entire configuration space. Then we can introduce 
another set of variables $\Pi_i(q,p) = \<p,\pl_qC_i\>$,
where $\<,\>$ is an inner product such that the phase-space
functions $\Pi_i$ are invariant under gauge transformations 
on the phase space spanned by $q$ and $p$. The canonical
symplectic structure in the total phase space would induce
a {\em non-canonical} symplectic structure on the physical
phase space spanned by variables $C_i$ and $\Pi_i$
\be
\{C_i,C_j\} = 0\ ,\ \ \ \{C_i,\Pi_j\}= D_{ij}(C)\ ,\ \ \ 
\{\Pi_i,\Pi_j\} = \bar{D}_{ij}(\Pi,C)\ , 
\label{css}
\ee
where the functions $D_{ij}$ and $\bar{D}_{ij}$ depend
on the structure of the constraint algebra.
The Hamiltonian dynamics can be reformulated in terms of these
gauge invariant variables with the symplectic structure (\ref{css})
just as has been done in section 3.3 for the simplest case.
If the Hamiltonian is regular in the total phase space, classical
phase-space trajectories $C_i(t),\Pi_i(t)$ do not have any singularities
because they are regular gauge invariant functions on the total
phase space. In this way one can 
always circumvent the coordinate singularities in classical theory. 

However, the induced symplectic structure
would vanish at certain points like the right-hand side of Eq. (\ref{PQ})
vanishes at $Q=0$. For the model discussed in section 4, $C_i(x)=
\tr x^{\nu_i}$, where $\nu_i$ are degrees of the independent Casimir
polynomials. So, $\Pi_i= \tr (px^{\nu_i-1})$. For groups of rank 2,
these variables are related to $\Phi_i$ and $\pi_i$ introduced in
section 4.6 by a coordinate transformation. It is not hard to be
convinced that, for instance, the function $D_{ij}$ vanishes 
for some values of $C_i$. Using the gauge invariance of $C_i$ one
can show that the singularities of the symplectic structure occur
exactly at those configurations of $C_i$ that correspond to
values of $x=h$ on the boundary of the Weyl chamber, $C_i=
C_i(x)=C_i(h)$ (cf. section 7.4). Similarly, in the SU(2) 
Yang-Mills theory in two dimensions, one can take $C(A)=
\tr{\rm P}\exp(ig\oint dxA)$ and $\Pi=\<E,\delta/\delta A\>C(A)$.
Then the symplectic structure reads $\{C,\Pi\}=1-C^2$
(after an appropriate rescaling $C$ and $\Pi$ by some constants depending
on $g$ and $l$). Thanks to the gauge invariance, 
$C\sim \cos[\pi(a,\omega)/a_0]$,
where $\omega=\tau_3/4$, and
$\<E,\delta/\pl A\>\sim (p_a,\pl/\pl a)$.
Zeros of the symplectic structure are obviously
related to the boundary of the Weyl cell where the Polyakov loop
variable attains its maximal (minimal) values. So, in this 
approach the gauge invariant induced symplectic structure
inherits the information about the physical phase space structure.

In contrast to the simplest case (\ref{PQ}), the symplectic structure
(\ref{css}) may no longer have a Lie algebra structure, which poses
substantial technical difficulties in its quantization because
it is hard to find a representation of the corresponding
commutation relations.
In Yang-Mills theory, with each spatial loop one can associate
a Casimir function, being the trace of the path-ordered exponential
of a generic connection along the spatial loop. These functionals
form an overcomplete set of gauge invariant variables (there are
identities between them \cite{migpr}) that can be used to parameterize
the orbit space. A quantum mechanical description in term of loop
variables can be developed (see, e.g., \cite{migpr} for a review) 
but it is still technically complicated
in practical use. The symplectic structure based on loop variables
has been proposed to quantize gravity \cite{asht,rs} (see 
\cite{baez,ashta,ashta1} for advances in this approach).

\section{Quantum mechanics and the gauge symmetry}
\setcounter{equation}0

Upon going over to a quantum mechanical description
of gauge systems the following main questions are
to be put forward.
First, can one promote first-class constraints into
operator equalities? Second, can the nonphysical variables
be excluded before a canonical quantization? 
Under the canonical quantization we imply the procedure
of promoting canonical symplectic coordinates 
$p_i$ and $q_i$ in the phase space of the system
into self-adjoint operators $\hat{p}_i$ and $\hat{q}_i$
which satisfy the Heisenberg commutation relations
\be
[\hat{q}_j,\hat{p}_k] = i\hbar\{q_j,p_k\}=i\hbar\delta_{jk}\ , 
\label{7.1} 
\ee
\be
[\hat{q}_j,\hat{q}_k] = i\hbar\{q_j,q_k\}=0\ ,\ \ \ 
[\hat{p}_j,\hat{p}_k] = i\hbar\{p_j,p_k\}=0\ ,
\label{7.1a}
\ee
where $\hbar$ is the Planck constant. The canonical
operators can be realized as linear operators in a Hilbert space. 
The states $|\psi\>$ of the
system are vectors of the Hilbert space.
For instance, one can take the representation of the Heisenberg
algebra in the space of square integrable 
complex functions $\<q|\psi\>=\psi(q)$, $\int dq |\psi|^2<\infty$.
Then 
\be 
\<q|\hat{q}_j|\psi\> = q_j\psi(q)\ ,\ \ 
\<q|\hat{p}_j|\psi\> = -i\hbar \pl_j\psi(q)\ ,
\label{7.2}
\ee
where $\pl_j$ stands for the partial derivative $\pl/\pl q_j$.
One should emphasize that the self-adjointness of the canonical
operators $\hat{p}_i$ and $\hat{q}_i$ is guaranteed by that the
phase space is a Euclidean space and $p,q$ refer to the Cartesian
system of coordinates on it.
The time evolution of the system is described by the Schr\"odinger
equation
\be 
i\hbar\pl_t|\psi(t)\> = \hat{H}|\psi(t)\>\ .
\label{7.3}
\ee
Here $\hat{H}$ is the Hamiltonian operator which is obtained
from the classical Hamiltonian by replacing the canonical variables
by the corresponding operators. 
The quantum Hamiltonian obtained in such a way is by no
means unique. Since the canonical operators are noncommutative,
there is, in general, a great deal of operator ordering ambiguity.
The condition of hermiticity of $\hat{H}$ is not generally sufficient
to uniquely specify the operator ordering. In addition, one
should also keep in mind that any quantization recipe is only
a guess for the right theory. Nature is quantum.
One should start, in fact, from quantum mechanics and derive
all the properties of our classical world from it by means
of the classical approximation, i.e., when the effects of
the noncommutativity of the canonical operators are negligible.
This can be achieved by studying the formal limit in which
the Planck constant vanishes. Unfortunately, we do not
have enough experience to postulate  the quantum laws 
prior to the classical ones. For this reason we use
various quantization procedures and believe that by
means of them we guess the quantum physics right. 
So, the quantum Hamiltonians are,
in principle, allowed to have any quantum corrections (of 
higher
orders of $\hbar$) which disappear in the classical limit.
This corrections can either be decided experimentally by observing
the energy spectrum of the system, or, sometimes, theoretically
by analyzing self-consistency of quantum theory, meaning that
the quantum theory obtained by means of a certain quantization
rule does not contain any internal contradiction, nor does it 
contradict some fundamental theoretical principles which
we believe to be true and superior. 

Canonical quantization fulfills the correspondence principle. 
This can be most easily seen from 
the Heisenberg representation of the time evolution
\be
i\hbar\, \frac{d}{dt}\,\hat{F}=[\hat{F},\hat{H}]\ ,
\label{7.4}
\ee
where $\hat{F}$ is any operator constructed out of the canonical
operators. In the formal limit $\hbar\rightarrow 0$, $(i\hbar)^{-1}
[,] \rightarrow \{,\}$ as follows from the canonical 
commutation relations, the Heisenberg equations turn into
the Hamilton equations of classical mechanics. The 
Schr\"odinger and Heisenberg representations of the time evolution
are related through the unitary transformation 
\be
|\psi(t)\> = \hat{U}_t|\psi\>\ , \ \ 
\hat{F}(t) = \hat{U}_t^\dagger \hat{F}\hat{U}_t\ ,\ \ 
\hat{U}_t = e^{-it\hat{H}}\ .
\label{7.5}
\ee
Here the states with the time label and the operators without
it refer to the Schr\"odinger picture, while the states without
the time label and the operators with it belong to the Heisenberg
picture. The numerical values of the amplitudes 
$\<\psi|\hat{F}(t)|\psi'\> = \< \psi(t)|\hat{F}|\psi'(t)\>$ do not depend
on the picture in which they are computed.

Having made all the above definitions and reservations about them,
we can now proceed to answer the questions about quantization
of gauge systems.
The answer to the first
question can be anticipated through the analysis of the 
simplest gauge model where the gauge symmetry is just
a translation of one of the Cartesian coordinates spanning
the configuration space of the system. The constraint coincides
with one of the canonical momenta $\sigma=p=0$. We cannot promote
this classical equality into the operator equality $\hat{p}=0$
because this would be in conflict with the canonical commutation
relation (\ref{7.1}): $\hat{q}\hat{p}- \hat{p}\hat{q}=i\hbar$.
To circumvent this problem and, nevertheless, to have a quantum theory whose 
classical limit complies with the existence of 
the first class constraints, one should restrict the physical
states by those annihilated by the operator version of the 
constraints
\be
\hat{\sigma}_a|\psi\>=0\ .
\label{7.6}
\ee
This recipe has been proposed in the works of
 Dirac \cite{can} (see also \cite{diraclec})
 and Bergmann \cite{berg}. Its consistency is
guaranteed by the properties of the first class constraint algebra
\be
[\hat{\sigma}_a,\hat{\sigma}_b]=\hat{f}_{ab}^c\hat{\sigma}_c\ ,\ \
[\hat{\sigma}_a,\hat{H}]=\hat{f}_a^b\hat{\sigma}_b\ ,
\label{7.7}
\ee
where $\hat{f}_{ab}^c$ and $\hat{f}_a^b$ are some functions
of canonical operators. One should remark that the constraints
may also exhibit the operator ordering ambiguity upon promoting
them into operators. Therefore one of the conditions which 
should be imposed on  constraints is that
the constraints remain in involution (\ref{7.7}) upon quantization.
This is {\em necessary} for the consistency of the Dirac rule (\ref{7.6}).
Sometime it turns out to be impossible to fulfill this condition.
This is known as the quantization anomaly of first-class constraints.
An example of such an anomaly is provided by Yang-Mills theory 
with chiral massless fermions \cite{fads}. 
In other theories, e.g., the string theory, the condition
of the absence of the anomaly may impose restriction on physical
parameters of the theory (see, e.g., \cite{hamachi}).

In what follows we shall always deal with gauge theories where
the constraints generate {\em linear} gauge transformations
in the configuration space: $q \rightarrow \Omega(\omega) q$.
In the Schr\"odinger picture, the Dirac condition means the 
gauge invariance of the physical states
\be
e^{i\omega^a\hat{\sigma}_a}\psi(q) = \psi(\Omega(\omega)q)=\psi(q)\ .
\label{7.8}
\ee
The norm of the Dirac states is proportional to the volume
of the gauge orbit through a generic point $q$
because the wave function (\ref{7.8}) is constant along
the gauge orbit. An 
apparent difficulty within the Dirac quantization  scheme
is a possible
non-renormalizability of the physical states. If the gauge orbits
are noncompact, then the norms are divergent. Even if the gauge
orbits
are compact, the norm can still be divergent if the number of
nonphysical degrees of freedom is infinite, like in gauge field
theories. This means, in fact, that the physical states do {\em not}
belong to the original Hilbert space. 

In the simple case, when the constraint coincides with a canonical
momentum, the problem can be resolved by discarding the 
corresponding degree of freedom. This does not lead to any 
contradiction because 
the wave function does not depend on
one of the {\em Cartesian} coordinates. 
This  coordinate can be excluded
at the very beginning, i.e., before the canonical quantization
(\ref{7.1})--(\ref{7.1a}).
The existence of the constraint means that the corresponding
variable is nonphysical. It belongs to the nonphysical
configuration space which is orthogonal to the physical one.
The nonphysical degrees of freedom cannot affect any physical
process. The divergence of the norm, on the other hand, is
exactly caused by the integration over the nonphysical space.
Therefore in Cartesian coordinates the integral over nonphysical
variables can be omitted without any effect on the physical
amplitudes.  
 This procedure may {\em not} be consistent if the nonphysical
degrees of freedom are described by {\em curvilinear} coordinates.
In this case the problem  amounts
to our second question about excluding the nonphysical variables
{\em before} quantization. 

If the number of nonphysical
degrees of freedom is finite and the gauge orbits are compact,
there is no problem with the implementation of the Dirac rule.
In the case of gauge field theory, 
the number of nonphysical degrees of freedom is infinite.
For compact gauge groups the norm problem can, for instance, 
be resolved by introducing
a finite lattice regularization. After factorizing the volume
of the gauge orbits in the scalar product, one removes the 
regularization.

Now we turn to the second question. Here the crucial observation
made by Dirac, is that the canonical quantization is, in general, consistent
when applied with the dynamical coordinates and momenta
referring to a Cartesian system of axes and not to more
general curvilinear coordinates \cite{dirac}. We have seen that in gauge
theories physical phase space coordinates are 
typically {\em not} Cartesian coordinates, and the physical
phase space is often a non-Euclidean space. 
So the canonical
quantization of the reduced phase space might have internal
inconsistencies.
Another important observation, which follows from our
analysis of the physical phase space in gauge models,  
is that the parameterization 
of the physical phase space
is defined {\em modulo general canonical transformations}. Quantization
and canonical transformations are {\em non-commutative} operations,
in general. On the other hand, there are infinitely many ways to
remove nonphysical variables before quantization.  Various 
parameterization of the physical phase space obtained in such 
a way are related to one another by canonical transformations.
Thus, the canonical quantization after the elimination of nonphysical
variables may lead to a quantum theory which depends on the
parameterization chosen \cite{prokhorov,ooa}. 
Clearly, this indicates a possible theoretical
inconsistency of the approach since quantum mechanics
of the physical degrees of freedom cannot depend on the way
the nonphysical variables have been excluded, i.e., on the
chosen gauge. We shall illustrate our general preceding remarks
with explicit examples of gauge models.

{\em Remark}. The noncommutativity of the canonical quantization
and canonical transformations does not mean that it is impossible
to develop a parameterization independent (coordinate-free)
quantum theory on the physical phase space (\ref{1}). Actually,
it can be done for constrained systems in general \cite{cfq1,cfq2,cfq3}.
A naive application of the canonical quantization,
which is often done in physical models, is subject to
this potential problem, while other methods may still 
work (e.g., the Bohr-Sommerfeld semiclassical
quantization applies to non-Euclidean phase spaces).

\subsection{Fock space  in gauge models}

The Bohr-Sommerfeld semiclassical quantization has led us to the 
conclusion that the  geometry of the physical phase space  
affects the spectrum of the harmonic oscillator.
Let us now verify whether our semiclassical analysis is compatible
with the gauge invariant approach due to Dirac. Consider first
the SO(N) model. We shall not quantize the Lagrange multipliers
since they represent pure nonphysical degrees of freedom. 
Only the canonical variables ${\bf x}$
and ${\bf p}$ are promoted to the self-adjoint operators  $\hat{\bf x}$
and $\hat{\bf p}$ satisfying the Heisenberg commutation relations.
In what follows we shall also assume units 
in which the Planck constant $\hbar$
is one. When needed it can always be restored from dimensional arguments.
Let us introduce a new set of operators 
\be
\hat{\bf a} = (\hat{\bf p} - i\hat{\bf x})/\sqrt{2}\ ,\ \ \ 
 \hat{\bf a}^\dagger = (\hat{\bf p} + i\hat{\bf x})/\sqrt{2}\ ,
\label{711}
\ee
which are called the destruction and creation operators, respectively.
The dagger stands for the hermitian conjugation.
The operators (\ref{711}) satisfy the commutation relations
\be 
[\hat{a}_j,\hat{a}_k^\dagger] = \delta_{jk}\ ,\ \ \ 
[\hat{a}_j,\hat{a}_k] = [\hat{a}_j^\dagger,\hat{a}_k^\dagger ] =  0\ .
\label{712}
\ee
An orthonormal basis of the total Hilbert space is given by the states
\be
|n_1,n_2,...,n_N\> \equiv |{\bf n}\> = \prod_{k=1}^{N}\, 
\frac{\left(\hat{a}_k^\dagger\right)^{n_k}}{\sqrt{n_k!}}
|0\>\ ,\ \ \  \hat{a}_k|0\> \equiv 0\ ,\ \ \ \<0|0\>=1\ ,
\label{713}
\ee
where $n_k$ are non-negative integers.
In this representation the Hamiltonian of an isotropic
harmonic oscillator have the form
\be
\hat{H} = \frac 12\left(\hat{\bf a}^\dagger\hat{\bf a}+\hat{\bf a}
\hat{\bf a}^\dagger\right) = \hat{\bf a}^\dagger\hat{\bf a}+
\frac N2\ .
\label{714}
\ee
The Dirac physical subspace is defined by the condition that
the operators of constraints annihilate any state from it:
\be 
\hat{\sigma}_a |\Phi\> = (\hat{\bf a}^\dagger, T_a\hat{\bf a})
|\Phi\> =0\ .
\label{715}
\ee
There is no operator ordering ambiguity in the constraints
thanks to the antisymmetry of the matrices $(T_a)_{jk}$.

The vacuum state $|0\>$ belongs
to the physical subspace since it is annihilated by the 
constraints. Hence,  any physical state can be constructed 
by applying an operator $\hat{\Phi}$
that commutes with the constraints,
$[\hat{\Phi}, \hat{\sigma}_a]=0$, to the vacuum state.
In fact, it is sufficient to assume that the commutator
vanishes weakly, i.e., $[\hat{\Phi}, \hat{\sigma}_a]\sim
\hat{\sigma}_a$, to guarantee that $\hat{\sigma}_a\hat{\Phi}
|0\>=0$. However, it is clear that any state can be obtained
by applying a function  {\em only} of the creation 
operators to the vacuum state. Therefore $\hat{\Phi}$
may also be a function only of the creation operators.
Since the constraints are linear in the destruction
operators, their commutator with $\hat{\Phi}$ cannot
depend on the constraints and, therefore, has to vanish.

To describe all possible operators that commute with
the constraints, we observe that the constraints generate
SO(N)-rotations of the destruction and creation operators. 
This follows from the commutation relations
\be
[\hat{\sigma}_a, \hat{\bf a}]= -T_a\hat{\bf a}\ ,\ \ \ 
[\hat{\sigma}_a, \hat{\bf a}^\dagger ]= 
-T_a\hat{\bf a}^\dagger\ . 
\label{716}
\ee
Thus, the operator $\hat{\Phi}$ must be a gauge invariant
function of the creation operators. This holds in general.
Operators that commute with the operators of constraints
are gauge invariant.  This is a quantum version of the analogous
statement in classical theory: The Poisson bracket of gauge
invariant quantities with the constraints vanishes. The
correspondence principle is fulfilled for observables.

Returning to the model, one can say that $\hat{\Phi}$
is a function of independent Casimir operators built of
${\bf a}^\dagger$. For the
fundamental representation of the group SO(N) 
there is only one independent Casimir operator
which is $(\hat{\bf a}^\dagger)^2$. Note that the
system has only one physical degree of freedom.
The powers of this
operator applied to the vacuum state form a basis
in the physical subspace \cite{ufn}
\be
|\Phi_n\> = \left(
\frac{4^nn!\Gamma(n +N/2)}{\Gamma(n/2)}
\right)^{-1/2} \left[(\hat{\bf a}^\dagger)^2
\right]^n |0\>\ .
\label{717}
\ee
The coefficients have been chosen so that $\<\Phi_k|\Phi_n\>=\delta_{kn}$.

The basis  vectors (\ref{717}) are also eigenvectors of
the oscillator Hamiltonian.  From the commutation 
relation 
\be 
[\hat{\bf a}^\dagger\hat{\bf a}, (\hat{\bf a}^\dagger)^2]
=2(\hat{\bf a}^\dagger)^2\ ,
\label{717a}
\ee
the eigenvalues follow
\be
E_n = 2n + N/2\ ,
\label{718}
\ee
that is, the distance between energy levels is doubled.
This effect has been observed in the semiclassical quantization
of the system. It has been caused by the conic structure
of the physical phase space. Here we have established it again
using the explicitly gauge invariant approach. The vacuum
energy depends on $N$, while in the Bohr-Sommerfeld approach
it does not because the physical phase space structure and the 
physical classical Hamiltonian do not depend on $N$.

Let us now turn to gauge systems with many physical
degrees of freedom. In classical theory we have seen 
that the non-Euclidean structure of the physical
phase space causes a specific kinematic coupling
between the physical degrees of freedom, 
because of which
only collective excitations of the physical degrees of
freedom occur. The kinematic coupling has also been shown
to have a significant 
effect on the semiclassical spectrum of the physical 
excitations.  Now we can verify whether our conclusion
is consistent with the Dirac approach.
We take first the gauge model
where the total configuration space is a Lie algebra
and the action of the gauge group in it is the adjoint
action of the group in its Lie algebra. 
Introducing the operators $\hat{a}=\hat{a}_b\lambda_b$
the Dirac condition for the gauge invariant states
can be written in the form
\be
\hat{\sigma}_b|\Phi\> = 
f_{bcd}\hat{a}_c^\dagger\hat{a}_d|\Phi\>=0\ .
\label{ag1}
\ee
Thanks to the antisymmetry of the structure 
constants $f_{bcd}=-f_{bdc}$, there is no operator
ordering ambiguity in the constraints. So they remain
in involution after quantization.

To solve the equation (\ref{ag1}) for the physical states,
we can use the same method as for the SO(N) model.
Since the vacuum state belongs to the physical Hilbert
space, any physical state can be obtained by applying 
a gauge invariant operator built out of $\hat{a}^\dagger$ 
to the vacuum state. The problem is reduced to seeking
all independent Casimir polynomials that can be constructed
from $\hat{a}^\dagger$. From the commutation relation
$[\hat{\sigma}_b,\hat{a}_c^\dagger ] = 
f_{bcd}\hat{a}_d^\dagger$ we infer that the operator
$\hat{a}^\dagger$  is transformed by the adjoint action
of the gauge group. Therefore the independent Casimir
polynomials are 
\be
P_{\nu_j}(\hat{a}^\dagger) = \tr \left(\hat{a}^\dagger\right)^{\nu_j}\ ,
\label{ag2}
\ee
where the trace is related to a matrix basis $\lambda_b$ in the
Lie algebra; the integers $\nu_j,\ j=1,2,..., r={\rm rank}\, G$,
are degrees of the independent Casimir polynomials, $\nu_1=2$
for all  groups. For the groups of rank 2, we have $\nu_2=
3,4,6$ for SU(3), Sp(4)$\sim$SO(5) and G${}_2$, respectively \cite{zhel}. 
We remark also that the use of a matrix
representation is not necessary to construct the gauge
invariant polynomials of $\hat{a}^\dagger$. In general,
gauge invariant operators are polynomials of $\hat{a}_b^\dagger$
whose coefficients are {\em invariant symmetric} tensors 
in the adjoint representation of the Lie algebra. Alternatively
the operators (\ref{ag2}) can be written via the {\em irreducible}
invariant symmetric tensors $d^{(\nu)}_{b_1b_2\cdots b_\nu}$, where
ranks $\nu$ of the tensors  equal corresponding degrees  
of the independent Casimir polynomials. The irreducible 
invariant symmetric tensors form a basis for all invariant symmetrical
tensors \cite{zhel}. Accordingly, the operators
\be
P_{\nu_j}(\hat{a}^\dagger) =
d^{(\nu_j)}_{b_1b_2\cdots b_{\nu_j}}\hat{a}_{b_1}^\dagger
\hat{a}_{b_2}^\dagger\cdots\hat{a}_{b_{\nu_j}}^\dagger\ 
\label{ag3}
\ee
form a basis of gauge invariant polynomials of the creation operators.
The irreducible invariant tensors can be obtained from the
commutation relations of the basis elements of the Lie algebra. 
For instance,
for SU(3) the invariant symmetrical tensors are $\delta_{ab}$
and $d_{abc}$ which are proportional to traces of two and
three Gell-Mann matrices, respectively.

A basis in the physical Hilbert space is
given by the states \cite{plb89}
\be
|n_1,n_2,...,n_r\> =
\left[P_{\nu_1}(\hat{a}^\dagger)\right]^{n_1}
\left[P_{\nu_2}(\hat{a}^\dagger)\right]^{n_2}
\cdots
\left[P_{\nu_r}(\hat{a}^\dagger)\right]^{n_r}
|0\>\ ,
\label{ag4}
\ee
where $n_j$ are non-negative integers.
These states are eigenstates of the oscillator Hamiltonian.
The eigenvalues follow from the commutation relation
$[\hat{a}_b^\dagger\hat{a}_b, P_{\nu}(\hat{a}^\dagger) ]
=\nu P_{\nu}(\hat{a}^\dagger)$ and have the form
\be
E_n = \nu_1n_1 +\nu_2n_2 +\cdots +\nu_rn_r + N/2\ .
\label{ag5}
\ee
Up to the ground state energy this is the spectrum
of the $r$--dimensional harmonic oscillator with 
frequencies equal to ranks of the irreducible symmetric
tensors in the adjoint representation of the Lie algebra.
We have anticipated this result from the semiclassical
quantization of the $r$--dimensional {\em isotropic}
harmonic oscillator with a hyperconic
structure of its physical phase space described in
section 3. 

In the matrix gauge model discussed in section 4.8
we take $V_q= \omega_q^2{\bf x}_q^2/2$ and $\omega_1
\neq\omega_2$ \cite{ufn,book}. The destruction and
creation operators (\ref{711}) carry an additional
index $q=1,2$. The constraint
(\ref{mat2b}) and the Hamiltonian (\ref{mat2a}) assume the form
\ba
\hat{\sigma} &=& (\hat{{\bf a}}_1^\dagger, T\hat{{\bf a}}_1) +
(\hat{{\bf a}}_2^\dagger, T\hat{{\bf a}}_2)\ ,\\
\hat{H}&=& \omega_1(\hat{{\bf a}}_1^\dagger,\hat{{\bf a}}_1)
+ \omega_2(\hat{{\bf a}}_2^\dagger,\hat{{\bf a}}_2)
+\omega_1 +\omega_2 \ ,\label{hammat}
\ea
where the term proportional to the constraint in the Hamiltonian
(\ref{mat2a}) has been omitted because it vanishes on the physical
states. Since the vacuum is annihilated by the constraint 
operator, $\hat{\sigma}|0\>=0$,
the physical states are generated by the independent 
invariants of the orthogonal group SO(2) which are composed 
of the vectors $\hat{\bf a}_q^\dagger$:
\be
\hat{b}_q^\dagger = \left(\hat{\bf a}_q^\dagger\right)^2\ ,\ \ \
\hat{b}_3^\dagger = \left(\hat{\bf a}_1^\dagger,
\hat{\bf a}_2^\dagger\right)\ ,\ \ \ 
\hat{b}_4^\dagger= \varepsilon_{ij}\hat{a}_1^{(i)\dagger}
\hat{a}_2^{(j)\dagger}\ ,
\label{inv1}
\ee
where $\varepsilon_{ij}=-\varepsilon_{ji}$ is a totally
antisymmetric tensor, $\varepsilon_{12}=1$. Recall that
the group SO(2) has two invariant irreducible tensors $\delta_{ij}$
and $\varepsilon_{ij}$. The operators (\ref{inv1}) are
all {\em independent} operators which can be composed of
the two vectors $\hat{\bf a}_q^\dagger$ and the two
invariant tensors.

Here the following should be noted. All the invariant
operators (\ref{inv1}), except $\hat{b}_4^\dagger$ are invariant under
the larger group O(2) = SO(2)$\otimes \Z_2$ (the nontrivial
element of $\Z_2$ corresponds to the reflection of one of the coordinate
axes, which changes sign of $\hat{b}_4^\dagger$). Should the 
operator $\hat{b}_4^\dagger$ be included among the operators
that generate the basis of the physical Hilbert space?
In other words what is the gauge group of the model: SO(2)
or O(2)? We remark that the similar question exists
in gauge theories {\em without} fermions: What is the gauge
group G or G$/Z_G$, where $Z_G$ is the center of G \cite{polbook}?
Yet, we have already encountered this question when studying
the physical phase space in the 2D Yang-Mills theory in section 5.
Following the arguments given there we point out that formally 
all information about the dynamics is contained
in the Lagrangian. In the Hamiltonian formalism, 
any finite gauge group transformation is an iteration of
infinitesimal gauge transformations generated by the constraints.
Therefore only the transformations which
can be continuously deformed towards the group unity have to be
included into the gauge group. The existence of the discrete
{\em gauge} group cannot be established for the Lagrangian
(\ref{mat2}). The group O(2) can be made a gauge
group of the model only by a supplementary condition
that the physical states are invariant under the transformations
from the center of O(2). Another possibility would be
to consider a larger gauge group where O(2) is a subgroup,
e.g., SO(3). In view of these arguments, we include
the operator $\hat{b}_4^\dagger$ into the set of physical
operators.

Because of the identity $\varepsilon_{ij}\varepsilon_{kn}
=\delta_{ik} \delta_{jn}-\delta_{in}\delta_{jk}$,
the operator $(\hat{b}_4^\dagger)^2$ can be expressed via
the other operators $ \hat{b}_a^\dagger, a=1,2,3$ so that
the basis of the physical Hilbert space is given by the 
states \cite{ijmp91}
\begin{equation}
\left(b^\dagger_1\right)^{n_1}
\left(b^\dagger_2\right)^{n_2}
\left(b^\dagger_3\right)^{n_3}|0\>\ ,\ \ \ 
\left(b^\dagger_1\right)^{n_1}
\left(b^\dagger_2\right)^{n_2}
\left(b^\dagger_3\right)^{n_3}b^\dagger_4|0\>\ ,
\end{equation}
where $n_a$ are non-negative integers. 
The physical states acquire a phase factor $\pm 1$
under the transformations from the center of O(2).
Similarly, the physical states of the 2D Yang-Mills theory 
get  a phase factor under homotopically nontrivial gauge
transformations as will be shown in section 7.6.
The spectrum of the Hamiltonian
(\ref{hammat}) reads
\be
E_{\bf n}= 2n_1\omega_1 + 2n_2\omega_2 +
n_3(\omega_1 +\omega_2) + n_4(\omega_1 +\omega_2)
+\omega_1 +\omega_2\ ,
\ee
where $n_4=0,1$. Here we see again that the oscillators
are excited in pairs, the same effect we have anticipated from
the analysis of the physical phase space of the model in section 4.8.
The physical frequencies are $2\omega_{1,2}$ and $\omega_1
+ \omega_2$, while the original frequencies of the 
{\em uncoupled} oscillators (cf. (\ref{mat2a})) are just $\omega_{1,2}$.

The lesson one could learn from the above analysis
is that, when describing a quantum gauge theory in term
of only physical degrees of freedom (e.g. the 
Hamiltonian path integral), it is of great importance 
to take into account the true structure of the physical
phase space in order to establish the equivalence 
with the Dirac gauge invariant operator formalism.

\subsection{Schr\"odinger representation of physical states}

In the path integral formalism one uses an explicit parameterization
of the physical configuration space (the Lagrangian path integral)
or that of the physical phase space (the Hamiltonian path integral).
It is often the case that the structure of gauge orbits is
so complicated that a parameterization is chosen 
on the basis of a physical ``convenience'' which may not
be the best choice from the mathematical point of view.
To develop the path integral formalism which uniquely
corresponds to the Dirac gauge invariant approach,
it seems useful to investigate, within the operator formalism, 
the role of coordinate singularities, that unavoidably occur in any
parameterization of a non-Euclidean physical phase space
by canonical variables.

In the case of the SO(N) model the total Hilbert space
is the space of square integrable functions 
$\psi({\bf x})$ in the $N$--dimensional
Euclidean space.  The gauge invariance condition
means that the physical wave functions must be
invariant under the SO(N) rotations of the argument.
So the physical motion is the radial motion. Recall
that the constraints in the model are nothing but
the components of the angular momentum of the
particle. The motion with zero angular momentum is
radial. Physical wave functions $\Phi$ depend only on the radial
variable $r=|{\bf x}|$. Therefore a natural way to solve
the Schr\"odinger equation for eigenfunctions of the
Hamiltonian is to make use of spherical coordinates.
In the equation 
\be 
\left[-\frac 12 \Delta_N + V({\bf x}^2)\right]\Phi_E =E\Phi_E\ ,
\label{sa1}
\ee
where $\Delta_N$ is the N-dimensional Laplace operator,
we introduce the spherical coordinates and omit all
the terms of the corresponding Laplace-Beltrami operator
containing the derivatives with respect to the angular
variables because the physical wave functions are
independent of them. 
The radial part of the Laplace-Beltrami operator
is the physical kinetic energy operator. The equation
assumes the form
\be
\left[-\frac{d^2}{dr^2} - \frac{N-1}{r}\,\frac{d}{dr} +V(r^2)
\right]\Phi_E(r) = 2E\Phi_E(r)\ .
\label{sa2}
\ee
We shall solve it for the oscillator potential $V= r^2/2$.

To this end, we make the substitution $\Phi=r^2\exp(-r^2/2)
\phi(r)$ and introduce a new variable $z=r^2$ so that
the function $f(z)=\phi(r)$ satisfies the equation
\be
zf'' + (a-z)f' -bf=0\ ,
\label{sa3} 
\ee
in which $a=N/2$ and $b=(a-E)/2$. The solution of 
this equation that is {\em regular} at the origin $z=r^2=0$
is given by the confluent hypergeometric function
\be
f(z)= {}_1F_1(b,a;z)\ .
\label{sa4} 
\ee
From the condition that $\Phi_E(r)$ decreases as
$r$ approaches infinity, which means that the 
function $f(z)$ must be a polynomial, i.e., $b=-n$,
we find the spectrum (\ref{718}).  The distance 
between the oscillator energy levels is doubled.
Making use of the relation between the function
${}_1F_1$ and the Laguerre polynomials $L_1^a$,
${}_1F_1(-n,a+1;z)=L_n^a(z)\Gamma(n+1)
\Gamma(a+1)/\Gamma(n+a+1)$, we can represent
the eigenfunctions as follows \cite{prokhorov,ufn,book}
\be
\Phi_n(r) = c_nL_n^{-1+N/2}(r^2)\, e^{-r^2/2}\ ,
\label{sa5}
\ee
with $c_n$ being normalization constants.
The physical wave functions are 
normalizable with the scalar product
\be
\int d^Nx |\Phi_n|^2 = \Omega_N \int_0^\infty
dr r^{N-1}|\Phi_n|^2 \rightarrow 
\int_0^\infty dr r^{N-1}|\Phi_n|^2 \ ,
\label{sa6}
\ee
where $\Omega_N$ is the total solid angle in
$\Rs^N$ (the volume of the nonphysical configuration
space) which we can include into the
norm of physical states. 

Let us compare our results with those we would have
obtained, had we quantized the system {\em after}
eliminating all nonphysical degrees of freedom, say,
by imposing the unitary gauge $x_i=0, i\neq 1$.
The gauge-fixed classical Hamiltonian can be obtained
by solving the constraints for $p_i, i\neq 1$, substituting
the solution into the original Hamiltonian and then
setting all $x_i$, except $x_1$, to zero. It would have
the form
\be
H_{\rm phys} = \frac 12 (p_1^2 + x_1^2)\ .
\label{sa7}
\ee
Clearly, the canonical quantization of this  Hamiltonian
would lead to the spectrum $E_n = n+1/2$ which gives
the energy level spacing different from that found in the gauge
invariant approach. 

The reason of failure of the canonical quantization is 
obviously that the phase space spanned by the variables
$p_1$ and $x_1$ is not a plane, but a cone unfoldable
into a half-plane. If the cone is cut along the momentum
axis, then we have to impose the restriction on the
admissible values of $x_1$: It has to be non-negative.
The operator $\hat{p}_1 = -i\pl/\pl x_1$ is not self-adjoint
on the half-axis in the space of square integrable
functions. Therefore $\hat{p}_1$ cannot be identified
with the physical observables, while the Hamiltonian  (\ref{sa7})
can be made self-adjoint. 
A possible way is to quantize the theory 
in the {\em covering} space, i.e., on 
the full real line spanned by $x_1$, and then to implement
the condition that the physical states must be invariant
under the parity transformation $x_1 \rightarrow -x_1$
\be 
\phi(x_1) = \phi(-x_1)\ .
\label{sa8}
\ee
In so doing, the right energy level spacing of the 
oscillator is restored.  Recall that the wave
function of the one-dimensional harmonic oscillator
are
\be
\phi_k (x_1) = c_n' H_k(x_1) e^{-x_1^2/2}\ , 
\label{sa9}
\ee
where $H_k$ are Hermite polynomials. They have the
property that $H_k(-x_1) = (-1)^kH_k(x_1)$. So
the physical values of $k$ are even integers, $k=2n$.

Although the invariance under the residual (discrete)
gauge transformations of the physical wave functions
has led us to the right energy level spacing, the quantum
theory still differs from that obtained by the gauge invariant 
Dirac procedure. The physical eigenstates in both theories
are different. This, in turn, means that the amplitudes for
the same physical processes, but described
within the two quantum theories, will {\em not} be the same.
 
Thus, in general, the canonical quantization of a gauge fixed
theory with an additional condition of the invariance of
the physical states with respect to the residual gauge
transformations may lead to a {\em gauge dependent}
quantum theory, which is not acceptable for a physical theory.
Yet, though the variable $x_1$ is assigned to describe the 
physical degree of freedom, the state $\hat{x}_1\phi(x_1)$,
where $\phi$ is a physical state satisfying (\ref{sa8}),
is {\em not} a physical state. The action of the operator $\hat{x}_1$
throws the states out of the physical subspace
because it does not commute with the parity transformation,
which is a rather odd property of a ``physical'' variable.  This
is not the case for the radial variable $r$ used in the Dirac
approach. It is still not the whole story. Here we have
been lucky  not to have had an ordering problem in the physical
Hamiltonian after eliminating the nonphysical degrees
 of freedom, thanks to the simplicity of the constraints
 and the appropriate choice of the gauge. In general, the elimination
 of the nonphysical variables would lead to the operator
 ordering problem in the physical kinetic energy. A solution
 to the ordering problem is generally not unique. On the other
 hand, an explicit form of the classical kinetic energy
 depends on the chosen gauge. Therefore it might be difficult
 to find a {\em special} ordering of the operators in the 
 physical Hamiltonian such that the spectrum would be
 independent  of the parameterization of the physical
 configuration space, or of the chosen gauge. If any
 operator ordering is assumed, say, just to provide
 hermiticity of the Hamiltonian, the spectrum would
 generally be {\em gauge-dependent}.
An explicit example is discussed in section 7.7.  
 This observation seems especially important for gauge
 theories where the structure of gauge orbits is unknown
 (or hard to describe, like in the Yang-Mills theory), 
and, hence, no ``appropriate'' gauge fixing condition exists. 

Let us analyze the singular point $r=0$ in the Dirac
approach. This point can be
thought as the Gribov horizon since $r=|x_1|$ 
in the unitary gauge $x_2=0$.  We  recall
that any gauge invariant parameterization of the
physical configuration space  can be
related to a special gauge fixing condition through curvilinear
coordinates associated with the gauge transformation
law and the chosen gauge, as has been shown in 
section 6.2. In the non-invariant approach the
singular points form the Gribov horizon; in the invariant
approach the singular points appear as the singular
points of the change of variables, like the origin in the
the spherical coordinates, i.e., as zeros of the Jacobian.
This is also the case for the Yang-Mills theory (see section 10.1).

For the sake of simplicity
let us take the group SO(3).  By means of the substitution $\Phi(r)
=\phi(r)/r$ Eq.(\ref{sa2}) can be transformed to the 
ordinary one-dimensional Schr\"odinger equation
$-\phi'' /2 +V\phi= E\phi$. Since the potential is an
even function of $r$ (a consequence of the gauge invariance),
the solutions to this equation  have certain parity.
The Hamiltonian  commutes with the parity transformation, so some of the
eigenvalues would correspond to odd eigenfunctions, some
to even ones. For example, we can take the harmonic
oscillator, $\phi_k(r) = c_kH_k(r) \exp(-r^2/2)$. For odd
$k$, the wave functions $\Phi_k(r) = \phi_k(r)/r$ are even,
while for even $k$ they are odd. We have eliminated 
the solutions that are  {\em not} invariant under the 
parity transformation $r\rightarrow -r$. The reason is
that these solution are {\em not regular} at the origin $r=0$.
Indeed, $H_{2n} (0) \neq 0$ so there is a singularity
$1/r$. Although this singularity is integrable since
the scalar product has the density $r^2$, the singular
solution to the Schr\"odinger equation must be excluded.
As has been pointed out by Dirac \cite{dirac},  singular solutions
of the Schr\"odinger equation with a regular potential
obtained in curvilinear coordinates are not solutions
in the original Cartesian coordinates. Indeed, the wave
functions with the singularity $1/r$ would not satisfy the
Schr\"odinger equation in the vicinity of the origin because
$\Delta_{(3)}(1/r)= 4\pi \delta^3(x) $.  

Regular  even functions of $r$ are regular functions
of $r^2={\bf x}^2$ and, hence, they have a unique 
{\em gauge-invariant} analytic continuation into the 
whole original configuration space. We conclude that 
the {\em regularity} condition for wave functions at the singular points 
in a chosen parameterization  of the physical configuration  space 
eliminates nonphysical states and provides one-to-one
correspondence with the explicitly gauge invariant approach
that does not rely on any parameterization of the physical
configuration space. This conclusion is
rather general and can be extended to 
all gauge theories (see section 8.7). 
Thus, the Gribov obstruction in the Schr\"odinger
representation of quantum gauge theories can be solved
 in the following way. 
Given a gauge condition, construct 
the curvilinear coordinates associated 
with it and the gauge transformation law. Solve the constraint
equations in the new coordinates and find the physical Hamiltonian.
Solve the Schr\"odinger equation under the condition that 
the physical wave functions are regular at the points
where the Jacobian of the change of variable vanishes.

\subsection{The Schr\"odinger representation in the case of
many physical degrees of freedom}

To obtain the Dirac gauge invariant wave functions in
gauge models with many physical degrees of freedom, we
will follow the general scheme formulated at the very end
of the preceding section. We take the model where the 
configuration space is a Lie algebra and the gauge group
acts in the adjoint representation in it. A natural parameterization
of the physical configuration space is provided by the gauge
$x=h$, where $h$ belongs to the Cartan subalgebra.
The associated curvilinear coordinates have been constructed
in Section 4.3 (see (\ref{huz})). The physical wave function 
are functions of $h$ because the constraints generate
shifts of the variables $z$. The Laplace-Beltrami operator
in general curvilinear coordinates has the form
\be 
\Delta_{LB} = \frac{1}{\sqrt{g}}\pl_j\,
\left( g^{jk}\sqrt{g}\, \pl_k\right)\ ,
\label{sag1}
\ee
where $g=\det g_{jk}$, $g_{jk}$ is the metric in the curvilinear
coordinates and $g^{jk}$ is the inverse of $g_{jk}$.
The metric (\ref{zdz}) is block-diagonal so the Laplace-Beltrami
operator is a sum of the physical  and the nonphysical
terms. Since the physical wave function are independent of $z$,
we omit the second term containing the derivatives $\pl_z$.
The metric in the physical sector is Euclidean, but the Jacobian
$\kappa^2$ is not trivial (cf. (\ref{hhh})). 
The physical part of the Laplace-Beltrami operator reads
\be
\frac{1}{\kappa^2}\,(\pl_h,\kappa^2\pl_h) = \frac{1}{\kappa}\,
\Delta_{(r)}\,\kappa - \frac{\Delta_{(r)}\kappa}{\kappa}
=\frac{1}{\kappa}\,\Delta_{(r)}\,\kappa\ ,
\label{sag2}
\ee
where $\Delta_{(r)}=(\pl_h,\pl_h)$ is the $r$-dimensional 
Laplace operator. The vanishing of the second term 
in the right-hand side of the first equality can be 
demonstrated by the explicit computation
\ba
\frac{\Delta_{(r)}\kappa}{\kappa} &=&  \ \sum_{\alpha\neq\beta >0}\,
\, \frac{(\alpha,\beta)}{(h,\alpha)(h,\beta)}\\
&=&\sum_{P_{\alpha\beta}}\ \ \sum_{\alpha\neq\beta >0 \in
 P_{\alpha\beta}} \ \
\frac{(\alpha,\beta)}{(h,\alpha)(h,\beta)} =0\ .
\label{sag3} 
\ea
Here the sum over the positive roots $\alpha\neq\beta>0$
has been divided into the sum over the positive roots contained
in a plane $P_{\alpha\beta}$ and a sum over all planes.
The sum in one plane is calculated explicitly. The  relative
directions of the roots in one plane are specified by the 
matrix $\cos^2 \theta_{\alpha\beta} = (\alpha,\beta)^2/[(\alpha,
\alpha)(\beta,\beta)]$ whose elements may only have
the values $0,1/4,1/2$ and $3/4$. That is, the quantity 
(\ref{sag3}) for a group of rank $r$ is determined by
that for the groups of rank 2. By an explicit computation
one can convince oneself that it vanishes for SU(3),
Sp(4)$\sim$SO(5),  and G${}_2$ \cite{hel}.

The Schr\"odinger equation in the physical configuration
space is written as
\be
\left(-\,\frac{1}{2\kappa}\,\Delta_{(r)} \,  \kappa + V\right)
\Phi = E\Phi\ .
\label{sag4}
\ee
Its solutions must be normalizable with respect to the 
scalar product 
\be
\int d^Nx |\Phi|^2 = {\cal V}_G/{\cal V}_H\int_{K^+}
d^rh \kappa^2|\Phi|^2\rightarrow 
\int_{K^+}d^rh\kappa^2|\Phi|^2\ ,
\label{sag4a}
\ee
where ${\cal V}_G$ is the volume of the group
manifold and ${\cal V}_H=(2\pi)^r$ is the volume of the 
stationary subgroup of a generic element $x=h$ which is
the Cartan group $G_H\sim [\times{\rm U(1)}]^r$.
The ratio of these factors is the result of the integration
over the variable $z$ (cf. (\ref{nrz}) and the paragraph after
(\ref{hhh})). The gauge
orbits are compact in the model, so their volume
can be included into the norm of physical
states, which is shown by the arrow in (\ref{sag4a}).

Eq. (\ref{sag4}) can be transformed to 
the standard Schr\"odinger equation
in the $r$--dimensional Euclidean space
by the substitution $\Phi = \phi/\kappa$. Let $\phi$
be a solution in the Euclidean space. The physical 
wave functions $\Phi$ must be regular at the singular
points where the Jacobian
(or the Faddeev-Popov determinant) $\kappa^2$
vanishes. To obtain the physical solutions, we observe
that the Hamiltonian $\hat{H}_{(r)}= -\Delta_{(r)}/2 +V$
commutes with the operators
$\hat{R}$ that transform the argument of the wave functions
by the Weyl group. This follows from the invariance of the 
Laplace operator and the potential under the Weyl transformations.
The Weyl group can be regarded  as the group of residual
gauge transformations in the gauge $x=h$. As  
the potential $V$ is gauge invariant, it must be invariant 
under the Weyl group transformations. 
Thus, if $\phi_E(h)$ is an eigenfunction
of $\hat{H}_{(r)}$, then $\phi_E(\hat{R}h)$ is also its eigenfunction
with the same eigenvalue $E$.
Let us take a ray through a generic point on the hyperplane
$(\alpha, h)=0$ and perpendicular to it, 
and let the variable $y$ span the ray so that
$y=0$ at the point of intersection of the ray with the hyperplane.
The potential $V$ is assumed to be a regular function 
everywhere. Therefore the eigenfunctions $\phi_E$ are
regular as well. 
Since $\kappa\sim y$ as $y$ approaches zero,
the function $\Phi_E$ has the singularity
$1/y$ in the vicinity of the hyperplane $(\alpha,h)=0$, which
is a part of the boundary of the Weyl chamber $K^+$.
Consider an element $\hat{R}_\alpha$ of the Weyl group 
which is a reflection in the hyperplane $(\alpha,h)=0$,
i.e., $\hat{R}_\alpha \alpha = -\alpha$.
Then, $\hat{R}_\alpha y =-y$. The function $\phi(h)/
\kappa(h) +\phi(\hat{R}_\alpha h)/\kappa(\hat{R}_\alpha h)$
satisfies Eq.(\ref{sag4}) and regular in the vicinity of the 
hyperplane $(\alpha,h)=0$.  This analysis can be done
for any positive root $\alpha$, which may lead us to the guess
that the functions
\be
\Phi_E(h) = \sum_W \left[\kappa(\hat{R}h)\right]^{-1}\phi_E(\hat{R}h)
\label{sag5}
\ee
are regular solutions to Eq.(\ref{sag4}) on the entire
Cartan subalgebra. Let us show that
this is indeed the case. 

First of all we observe that 
\be
\kappa(\hat{R}h) = \pm \kappa(h)
\label{sag6}
 \ee
because $\mu(h)=\kappa^2(h)$ is an invariant of the Weyl
group since the Weyl group is the group of permutations
and reflections of the roots which preserve the root pattern
(cf. section 4.3).
The negative sign in (\ref{sag6}) corresponds to an odd number
of reflections in the group element $\hat{R}$. Any reflection 
in a hyperplane through the origin can be viewed as an orthogonal
transformation. In the matrix representation $\det \hat{R} =\pm 1$
because $\hat{R}$ is a composition of reflections in the hyperplanes
orthogonal to simple roots. Next, we invoke the following theorem 
from group theory \cite{hel,zhel}. Any polynomial $p(h)$ in the Cartan
subalgebra with the property $p(\hat{R}h) = \pm p(h)$ can be represented
in the form
\be
p(h) = \kappa(h)C(h)\ ,
\label{sag7} 
\ee
where a polynomial $C(h)$ is invariant under the Weyl group.
By construction the function (\ref{sag5}) is invariant under
the Weyl transformations. Making use of the relation (\ref{sag6}),
the physical wave function can also be 
represented in the form $\Phi_E(h) = [\kappa(h)]^{-1}
\tilde{\phi}_E(h)$ where $\tilde{\phi}_E(\hat{R}h) = \pm\tilde{\phi}_E(h)$.
Let us decompose  $\tilde{\phi}_E(h)$ into a power series
and re-group the latter into a sum of terms of the same order in $h$:
\be
\tilde{\phi}_E(h)= \sum_{n=0}^\infty \tilde{\phi}_n^{(E)} p_n(h)\ .
\label{sag8}
\ee
The polynomials $p_n(h)$ of order $n$ satisfy the condition
of the above theorem $p_n(\hat{R}h)=\pm p_n(h)$. Therefore
\be
\Phi_E(h) = \sum_{n=0}^\infty \tilde{\phi}^{(E)}_n C_n(h)\ ,
\label{sag9}
\ee
where $C_n(h)$ are polynomials invariant under the Weyl
group. We remark that not for every order $n$ there exists
an invariant polynomial $C_n$. For instance, there is no
invariant polynomial of order one. Therefore some of the coefficients
$\tilde{\phi}^{(E)}_n$ necessarily vanish.

Now let us prove the converse that any regular solution
of the Schr\"o\-din\-ger equation (\ref{sag4}) 
is invariant under the Weyl group.
In the total configuration space the solutions of the Schr\"odinger
equation can be written in the form
\be
\phi_E(x) = \phi_E(h,z) = \Phi_E^{(k)}(h) Y_{(k)}(z)\ , 
\label{sag10}
\ee
where  $Y_{(k)}(z)$ are eigenfunctions of the Casimir operators
in the algebra generated by the operators $\hat{\sigma}_a$
of constraints, and the index $(k)$ stands for a set of corresponding
eigenvalues, $E=E(k)$. The functions $\Phi_E^{(0)}(h)$
form a basis in the physical subspace, $Y_{(0)}(z)=const$.
Consider the symmetry transformation of the new
variables $h$ and $z$ in  (\ref{huz}) under which the old
variables $x$ are invariant. These transformations
contain translations of $z$ on the periods of the manifold
$G/G_H$ ($h$ is not changed) and the Weyl group
\be
x\rightarrow x\ ,\ \ \ h\rightarrow \hat{R}h=\Omega h\Omega^{-1}\ ,
\ \ \ S(z) \rightarrow \Omega S(z) =S(z_\Omega)\ .
\label{sag11}
\ee
The functions (\ref{sag10}) must be invariant under these
transformations. Hence $\Phi_E^{(0)}(h)$ must be invariant
under the Weyl group. The functions $\Phi_E^{(0)}(h)$ are
also regular because the functions $\Phi_E(x)$ are regular.

We have established the one-to-one correspondence
between analytic gauge invariant functions $\Phi_E(x)$
in the total configuration space  and analytic functions
$\Phi_E(h)$ invariant under the Weyl group in the reduced
theory. In group theory this statement is known as the
theorem of Chevalley which asserts \cite{zhel} that any polynomial
in the Cartan subalgebra invariant under the  Weyl group
has a unique analytic continuation to the Lie algebra
that is invariant under the adjoint action of the group.
Since polynomials form a dense set in the space of
analytic functions, the  statement is also valid for
analytic functions. 
The regularity condition of the physical wave functions
at the Gribov horizon (the boundary of the Weyl chamber)
on the gauge fixing surface has
been crucial to prove the equivalence of the gauge fixed
formalism to the explicitly gauge invariant
approach due to Dirac. Attention should be drawn to the
fact that this boundary condition does not allow 
separation of variables in the Schr\"odinger equation,
even if the potential would allow it,
i.e., the physical wave functions {\em cannot} be factorized into
a product of wave functions for each component of 
$h$. This is the evidence of the kinematic coupling
in the quantum theory, the effect we have observed in the
classical theory. The above example also provides us with the 
key idea for how to deal with the coordinate singularities
in quantum theory: The physical amplitude must be regular
at the singular points in any particular coordinate system
assumed on the orbit space. There is no {\em need} to postulate
the invariance of the physical states under the residual gauge
(Gribov) transformation. It is ensured by the regularity condition.

\subsection{The theorem of Chevalley and the Dirac states
for groups of rank 2}

Although the theorem of Chevalley establishes the one-to-one
correspondence between the gauge invariant Dirac states and
the states invariant under the residual gauge transformations
in the non-invariant approach, an explicit construction of the
analytic continuation might be tricky. A general idea is to
find an explicit form of the physical wave functions
in the new variables $P_\nu(h)=\tr h^\nu$ 
instead of the components of $h$, 
where $P_\nu(x)=P_\nu(h)$ are the independent Casimir polynomials
(or functions in a general case) in the chosen gauge.
We fulfill this program for groups of rank 2 in the case
of the oscillator potential \cite{tmp89}, just to give an idea
of how hard it might be to realize in general.
We take the variables $\Phi_{1,2}$ introduced in section 4.6.
To calculate the density $\kappa(h)$ in the new variables,
we make use of Lemma III.3.7 in \cite{helgason2} which asserts
that
\be
\det\left(\frac{\pl P_{\nu_k}}{\pl h^j}\right)
=c'\,\kappa(h)\ ,\ \ \ c'=const\ ,\ \ \ k,j=1,2,...,{\rm rank}\, X\ .
\label{sad1}
\ee
Applying (\ref{sad1}) to groups of rank 2, we find that
$\kappa^2\sim \Phi_1^{2\nu}(c_2 +c_1\Phi_2 - \Phi_2^2)$
where $\Phi_{1,2}$ are defined by (\ref{phi12}) for $x=h$
and the coefficients are specified after Eq. (\ref{tre}).
The variable $\Phi_2$ is then replaced by $(\Phi_2-b)/\sqrt{a}$
(cf. (\ref{phi2})). As a result we obtain
\be
\mu(h)=\kappa^2(h) = c\,\Phi_1^{2\nu}(1 - \Phi_2^2)\ ,
\label{cd1}
\ee
where $c$ is a constant. The Chevalley's theorem
applies to the density $\mu(h)$. Eq. (\ref{cd1}) determines
the analytic gauge invariant continuation of $\mu$
to the whole configuration space. It is a polynomial of rank $2\nu$
constructed out of two independent Casimir polynomials
$P_2(x)$ and $P_\nu(x)$. A gauge invariant function
$\Psi_E(x)$ is a regular function of $\Phi_{1,2}$. So substituting
$\Psi_E(x)= [\kappa(\Phi_1,\Phi_2)]^{-1}\varphi_E(\Phi_1,\Phi_2)$
into the Schr\"odinger equation in the total configuration space
$\hat{H }\Psi_E =E\Psi_E$, we find the equation
\ba
\hat{H}_{ph}\varphi_E&=&E\varphi_E; \label{cd2}\\
\hat{H}_{ph}&=&
-\frac{1}{2\Phi_1}\,\pl_1\,\Phi_1\,
\pl_1 -\frac{\nu^2}{2\Phi_1^2}\,
\left[1-\Phi_2^2\right]^{1/2}\pl_2\,
\left[1-\Phi_2^2\right]^{1/2}
\pl_2 + \frac{\Phi_1^2}{2}\ ,
\nonumber
\ea
where $\pl_{1,2}$ are partial derivatives with respect
to $\Phi_{1,2}$.
Solutions are sought in the form $\varphi_E=g(\Phi_1)F(\Phi_2)$.
Observe that just as in the classical theory discussed in section 4.6,
the new variables allow us to separate independent oscillator modes
and thereby to solve the kinematic coupling problem.
Equation (\ref{cd2}) is equivalent to two equations
\ba
&-&\left[1-\Phi_2^2\right]F'' +\Phi_2 F' + \gamma F=0\ ;
\label{cd3}\\ 
&-&g'' - \frac{1}{\Phi_1}\,g' -\left(\frac{\gamma \nu^2}{\Phi_1^2}
-\Phi_1^2 +2E\right) g = 0\ ,
\label{cd4}
\ea
where $\gamma$ is a constant of separation of the variables.
Since the function $\Psi_E(x)=\Psi_E(h)$ has to be finite
at the boundaries of the Weyl chamber ($\mu=0$
when $\Phi_2=\pm 1$), the following 
boundary conditions are to be imposed on $F$
\be
F(\pm 1)=0\ .
\label{cd5}
\ee 
The solution of (\ref{cd3}) satisfying this condition
is given by 
\be
F_m(\Phi_2)= \sin[(m+1)\cos^{-1}\Phi_2] 
= \left(1-\Phi_2^2\right)^{1/2}U_m(\Phi_2)\ ,
\label{cd6}
\ee
where $U_m(\Phi_2) $ are
the Chebyshev polynomials, $m=0,1,2,...$, and $\gamma =-(m+1)^2$.
Equation (\ref{cd4}) is transformed to the standard form
(\ref{sa3})
by the substitution $g= \Phi_1^{\nu(m+1)}e^{-\Phi_1^2/2}
f(\Phi_1)$ and by introducing a new variable $z=\Phi_1^2$.
In Eq. (\ref{sa3}) one should set $a= \nu(m+1)+1$ and
$b= -(E-a)/2$. Thus, the spectrum and the gauge invariant eigenfunctions
are
\ba
E_{nm}&=& 2n +\nu m + N/2\ ,\\
\Psi_{nm}&=&c_{nm}\Phi_1^{\nu m}U_m(\Phi_2)L_n^{\nu(m+1)}(\Phi_1^2)
e^{-\Phi_1^2/2}\ ,
 \label{cd7}
\ea
where $c_{mn}$ are normalization constants. The dimension $N$
of the gauge group specifies the ground state energy, as
in the Fock space approach. To establish this within
the Schr\"odinger picture, we have used the relation \cite{zhel}
$N= \nu_1\nu_2\cdots\nu_r +r$, i.e., $N=2\nu +2$ for the
groups of rank 2.

From the expression (\ref{cd7}) we infer that $\Psi_{nm} $
depends only on the Casimir polynomials $P_{2,\nu}$.
For the groups Sp(4)$\sim$SO(5) and G${}_2$, the factor of the 
exponential in (\ref{cd7}) is a polynomial of $P_{2,\nu}$
since $\nu$ is an even integer ($\nu=4,6$, respectively),
and, therefore, $\Phi_1^{\nu m}$ is a polynomial for
any positive integer $m$. In the case of SU(3), $\nu=3$,
and for odd $m$, $\Phi_1^{3m}$ is proportional
to the nonpolynomial factor 
$[P_2]^{1/2}$. However in section 3.6, it has been
pointed out that the coefficient $b$ in (\ref{tre}) vanishes
for SU(3), thus leading to $\Phi_2 = \sqrt{6}P_3\Phi_1^{-3}$.
Hence, the nonpolynomial factor in $\Phi_1^{3m}U_m(\Phi_2)$ 
is canceled out.
Since $P_{2,\nu}(x)=P_{2,\nu}(h)$, 
the wave functions (\ref{cd7}) are invariant under the Weyl 
group, and have a unique gauge invariant continuation
to the Lie algebra (the total configuration space). 

{\em Remark}. The approach can also be applied to obtain
explicitly gauge invariant wave functions for the SO(2)
gauge matrix model with the oscillator potential.
The idea is to write first the Schr\"odinger equation
in the curvilinear coordinates (\ref{mat5}). The physical
wave functions do not depend on $\theta$, so the corresponding
derivative should be omitted in the Laplace-Beltrami operator.
Next, one introduces a new set of curvilinear coordinates
to separate the variables in the Schr\"odinger equation
(to remove the kinematic coupling): $q^2 = r\cos\varphi,
q^3=r\sin\varphi$. We refer to the works \cite{ijmp91,heinzl}
for the details.

\subsection{The operator approach to quantum 
Yang-Mills theory on a cylinder}

Here we analyze coordinate singularities 
in the Schr\"odinger picture
for a soluble gauge system 
with infinitely many degrees of freedom \cite{plb91,plb93}.
Following the Dirac method we replace the canonical variables 
$E(x)\rightarrow -i\hbar 
\delta/\delta A(x)$, $A(x)\rightarrow A(x), A(x)\in 
{\cal F}$, 
by the corresponding operators and get the quantum theory in 
the Schr\"odinger functional representation \cite{loos,loos2},
\ba 
\hat{H}\Phi_n[A] = -\frac{\hbar^2}{2}\left\langle\frac{\delta}{\delta A}, 
\frac{\delta}{\delta A}\right\rangle\Phi_n[A]&=&E_n\Phi_n[A]\ ,
\label{2ym1}\\ 
\hat{\sigma }\Phi _n[A]=-i\hbar\nabla(A)\frac{\delta}{\delta A}\Phi_n[A] 
&=&0\ .
\label{2ym2} 
\ea 
The states are now given by {\em functionals} on the space ${\cal F}$.
In accordance with the general method proposed in the 
end of Section 7.4, to solve Eq. (\ref{2ym2}) 
and to project the Hamiltonian in (\ref{2ym1}) onto 
the gauge orbit space, one should introduce 
curvilinear coordinates associated with both a gauge transformation 
law and a chosen gauge condition. These coordinates are given in
 (\ref{aah}). In the new variables the orbit space
is parameterized by homogeneous connections $a$ from the Cartan
subalgebra. Following the analysis of the moduli space in section
5.1, we also impose the condition $a\in K^+_W\sim {\cal F}/{\cal G}$ 
to ensure 
a one-to-one correspondence between the ``old'' and ``new'' variables. If 
we assume that $a$ ranges over the entire Cartan subalgebra, then
the values of the new variables 
$\Omega\Omega^{-1}_s,\ \hat{R}a$, for all $\hat{R}$  
from the affine Weyl group $W_A$, are mapped 
to the same configuration $A(x)$ by (\ref{aah}). 
Therefore the admissible values 
of $a$ in (\ref{aah}) must be restricted by the Weyl cell $K_W^+$. 
We show below that
the constraint operator $\hat{\sigma }$  commutes with
the curvilinear variable $a$ and, therefore, 
$a$ is a  formally gauge-invariant variable. 

The norm of the physical states is defined according to
the rule (\ref{vym})
\be 
\int_{{\cal F}}\prod\limits_{x\in {\bf s}^1}dA(x)\Phi^*_n[A] 
\Phi_{n'}[A]\rightarrow \int_{K_W^+}da\kappa^2(a) 
\Phi_n^*(a)\Phi_{n'}(a)=\delta_{nn'}\ , 
\label{2ym3}
\ee 
where the infinite constant $C(l)=\int_{{\cal G}/G_H}\prod_x dw(x)$ 
is removed by a renormalization of the physical states, which we denote 
by the arrow in (\ref{2ym3}). 
The integration over the nonphysical variables $w$ yields an infinite
factor, thus making the physical states non-normalizable in the 
original Hilbert space, even though the gauge orbits are compact.
The origin of the divergence is the infinite number of nonphysical
degrees of freedom. Frankly speaking, at each space point $x$  
nonphysical degrees of freedom
contribute a finite factor, proportional to the volume of $G$, to
the norm of a physical state. As one might see from (\ref{geo6}),
the way to get around of this difficulty is to make the number
of Fourier modes finite, renormalize the physical states and then
remove the regularization.  We have implied this procedure
done in (\ref{2ym3}). 

The independence of the physical state from $w(x)$ as well as
the formal gauge invariance of $a$ can be demonstrated
explicitly by solving the Gauss law in the new curvilinear 
coordinates (\ref{aah}). We assert that 
\be 
\hat{\sigma }\Phi _n[a,\omega ]= 
-i\hbar g\hat{\Omega}^T\frac{\delta }{\delta w}\Phi_n[a,w] =0\ ; 
\label{ym4a}
\ee 
here we have used the notation $(\hat{\Omega}y)_a=
(\Omega y\Omega^{-1})_a= 
(y\hat{\Omega}^T)_a\equiv \hat{\Omega}_{ab}y_b
,\ \hat{\Omega}^T\hat{\Omega}= \hat{\Omega}\hat{\Omega}^T 
= 1$ for any element $y\in X$. Since $\det\hat{\Omega}\neq 0$, 
the physical states are  functionals independent of $w(x)$.
To prove (\ref{ym4a}), we first derive the 
following relations from (\ref{dif}): 
\ba 
\delta a&\equiv& da={\cal P}_0^H\hat{\Omega}^T\delta A\ ,
\label{ym5a}\\ 
\delta w&=&-g\nabla^{-1}(a)(1-{\cal P}_0^H)\hat{\Omega}^T\delta A 
\label{ym6a}
\ea 
with ${\cal P}_0^H$ being a projector on the subspace
${\cal F}_0^H$ of spatially homogeneous functions taking their values
in the Cartan subalgebra. Recall that 
the operator 
$\nabla(a)$ is invertible on $(1-{\cal P}_0^H){\cal F}$. The
following simple computation leads us to the desired result 
\ba 
\nabla(A)\frac{\delta}{\delta A}&=&\nabla(A)\left[ 
\left(\frac{\delta a}{\delta A},\frac{\pl}{\pl a}\right)_a + 
\left\langle\frac{\delta w}{\delta A},\frac{\delta }{\delta w}\right\rangle_w 
\right] 
\label{ym7a}\\ 
&=&\nabla(A)\left[\left({\cal P}_0^H\hat{\Omega}^T\right)^T\frac{\pl}{\pl a} 
+\left(-g\nabla^{-1}(a)(1-{\cal P}_0^H)\hat{\Omega}^T\right)^T 
\frac{\delta}{\delta w}\right]
\label{ym8}\\ 
&=&\hat{\Omega}\nabla(a){\cal P}_0^H\frac{\pl}{\pl a} 
+g\hat{\Omega}\nabla(a)(1-{\cal P}_0^H)\nabla^{-1}(a)\frac{\delta}{\delta w} 
\nonumber\\
&=& g\hat{\Omega}^T\frac{\delta}{\delta w}\ . 
\label{ym9a}
\ea 
In (\ref{ym7a}), the subscript of the scalar product brackets denotes 
variables over whose indices the scalar product is taken, i.e. 
all indices  labeling independent degrees of freedom 
described by $A(x)$ (the Lie algebra ones and $x\in {\bf S}^1$) 
 in the scalar products in (\ref{ym7a}) 
are left free. Equality (\ref{ym8}) 
is obtained by the substitution of $\delta a/\delta A(x)$ and 
$\delta w(y)/\delta A(x)$ which are taken from (\ref{ym5a})
and (\ref{ym6a}), respectively. To get  (\ref{ym9a}), 
we have used the identities 
$\nabla(a){\cal P}_0^H\pl/\pl a$ $\equiv 0$ and $\nabla(A) 
\hat{\Omega}=\hat{\Omega}\nabla(a)$. 
 
Thus, the operator of multiplication on the variable $a$ 
commutes with the constraint operator 
 $[\hat{\sigma },\hat{a}]=0$.   In this 
approach, the Gauss law (\ref{2ym2}) can formally be solved even in 
four dimensions \cite{plb91}. 
As has already been argued in section 6.3, 
such a formally gauge invariant approach 
is {\em not}, in general, free of  coordinate singularities. 
We now turn to analyze the role of these singularities
in quantum theory.
 
To project the functional Laplace operator in (\ref{2ym1}) on the gauge 
orbit space spanned by the variable $a$,  
one should calculate the Laplace-Beltrami 
operator in the new functional variables (\ref{aah}) and omit in it
all terms containing the variational derivative
$\delta/\delta w$. The metric (\ref{metric}) is block diagonal.
The physical and nonphysical parts of the kinetic energy
operator are decoupled (cf. (\ref{sag1})). After a transformation
similar to (\ref{sag2}), we arrive at the quantum mechanical problem 
\be 
\hat{H}_{ph}\Phi_n(a)=\left[-\frac{\hbar^2}{4\pi l}\frac{1}{\kappa(a)} 
\Delta_{(r)}\,\kappa(a) -E_C\right]\Phi_n(a)=E_n\Phi_n(a)\ , 
\label{ym10a}
\ee 
where we have taken into account 
that the metric on the physical space is flat: $g^{aa}= (2\pi l)^ 
{-1}$, and that the function $\kappa(a)$ is an eigenfunction
of the Laplace operator  (cf. (\ref{c.11})),
\be 
E_C=-\frac{\hbar^2}{4\pi l}\,
\frac {\Delta_{(r)}\kappa }{\kappa}= 
\frac{\pi\hbar^2}{a_0^2l}\left(\rho,\rho\right)=
\frac{\pi\hbar^2 N}{24a_0^2l} \ .
\label{ym10b} 
\ee  
Substituting $\Phi_n = \kappa^{-1}\phi_n $ into (\ref{ym10a}) we 
find that $\phi_n$ is an $r$-dimensional plane wave, 
$\exp(2\pi i(\gamma _n,a)/a_0)$. However, not all values of 
the momentum vector $\gamma_n\in H$ are admissible
because only regular solutions to (\ref{ym10a}) 
have a physical meaning. 
The regularity condition requires that the functions
$\phi_n(a)$ should vanish on the hyperplanes orthogonal to positive 
roots, $(\alpha ,a)=n_\alpha a_0$, $n_\alpha$ an integer, 
as the factor $\kappa^{-1}(a)$ has 
simple poles on them. Since $(\hat{R}\gamma _n,\hat{R}\gamma _n)=(\gamma 
_n,\gamma _n),\ \hat{R}$ is from the Weyl group $W$, the
superposition of the plane waves
\begin{equation} 
\Phi _n(a)\sim [\kappa(a)]^{-1} \sum\limits_{\hat{R}\in W}^{}
\det \hat{R} \exp 
\left\{\frac{2\pi 
i}{a_0}(\hat{R} \gamma _n,a) \right\}\equiv[\kappa(a)]^{-1}\phi_n(a) 
\label{ym11}
\end{equation} 
is an eigenstate of the physical Hamiltonian
with the eigenvalue
\begin{equation} 
E_n=\frac{\pi \hbar ^2}{a_0^2l}\left[(\gamma _n,\gamma _n)-(\rho ,\rho 
)\right]\ . 
\label{ym12}
\end{equation} 
The function (\ref{ym11}) is also regular at the hyperplanes
$(a,\alpha )=n_\alpha a_0$, provided the momentum $\gamma_n$
attains {\em discrete} values such that the number
\begin{equation} 
\frac{2(\gamma _n,\beta )}{(\beta ,\beta )}\in \Z
\label{ym13}
\end{equation} 
is an integer for any root $\beta $. Thus, the {\em regularity} condition
has a dramatic effect on the physical spectrum: It appears to be
{\em discrete}, rather than continuous as one might naively expect after
removing all nonphysical degrees of freedom by a gauge
fixing because the system has no potential. Moreover,
the  regular eigenfunctions (\ref{ym11}) have a unique
{\em gauge invariant} analytic continuation to the
whole configuration space  ${\cal F}$. They are
characters of all irreducible representation of
the Polyakov loop ${\rm P}\exp[ ig\oint dx A(x)]$. Therefore
the wave functions as well as the eigenvalues (\ref{ym12}) we have obtained
do {\em not} depend on the particular parameterization
of the gauge orbit space we have chosen to solve
the Gauss law and the Schr\"odinger equation \cite{plb93,2d}. 
 
To prove the regularity of the functions (\ref{ym11}), let us 
decompose $a$ into two parts $a=a^{||}+a^{\perp}$, such that 
$(a^\perp,\alpha)=0$ for a root $\alpha $ and let $W^{(\alpha)}$ be the 
quotient $W/\Z_2^{(\alpha)},\ \Z_2^{(\alpha)}$
$=\{1,\hat{R}_\alpha\}$, 
where $\hat{R}_\alpha \alpha= - \alpha $ and, therefore, $\hat{R}_\alpha 
a^\perp = a^\perp,\ \hat{R}_\alpha a^{||} = -a^{||},\ \det\hat{R}_\alpha $
$= -1$. 
Then the sum in (\ref{ym11}) can be rewritten as follows 
\begin{eqnarray} 
\phi _n(a) &\sim & \sum\limits_{\hat{R}\in W^{(\alpha )}}^{}
\left[
\det 
\hat{R}\exp \left\{\frac{2\pi i}{a_0}(\hat{R}\gamma 
_n,a)\right\}\right.+\nonumber\\
&\ &+\left.\det (\hat{R}_\alpha 
\hat{R})\exp \left\{\frac{2\pi i}{a_0}(\hat{R}_\alpha \hat{R}\gamma 
_n,a)\right\}\right]=  \label{ym12a}\\ 
&= & \sum\limits_{\hat{R}\in W^{(\alpha )}}^{}\det \hat{R}
\exp\left\{ \frac{2\pi 
i}{a_0} (\hat{R}\gamma _n,a)\right\}\left[1-\exp \left\{-\frac{4\pi 
i}{a_0}(\hat{R}\gamma _n,a^{||})\right\}\right]\ . 
\nonumber
\end{eqnarray}
Here we have  used the identities $\det(\hat{R}_\alpha\hat{R})
=-\det\hat{R}$ and
\be
\left(\hat{R}_\alpha\hat{R}\gamma_n, a\right)=
\left(\hat{R}\gamma_n,a^\perp-a^{||}\right) =
\left(\hat{R}\gamma_n,a\right)- 2\left(\hat{R}\gamma_n,a^{||}\right)\ .
\ee
In a neighborhood of the hyperplane $(a,\alpha )=n_\alpha a_0$
with a nonvanishing integer $n_\alpha$,
we have  $a^{||}=n_\alpha a_0 \alpha /(\alpha ,\alpha )+ 
\epsilon 
\alpha $ where $\epsilon \rightarrow 0$. 
The sum in the third line of Eq. (\ref{ym12a}) vanishes as 
$\epsilon \rightarrow 0$ if the factor in the brackets vanishes. This
yields the condition that 
$\ 2(\hat{R}\gamma _n,\alpha )/(\alpha ,\alpha 
)$ must be  an integer. Since $\hat{R}\alpha 
=\beta$ 
is a root, we conclude that the function (\ref{ym11})
 is regular, provided  the momentum $\gamma_n$
satisfies the condition (\ref{ym13}).
 
For any $\gamma _n$  satisfying (\ref{ym13}),  
a vector $\hat{R}_0\gamma _n,\ \hat{R}_0\in W$, 
also satisfies (\ref{ym13}) 
and corresponds to the {\em same} energy level 
(\ref{ym12}) because the Killing form is 
$W$-invariant. Replacing $\gamma _n$ by $\hat{R}_0\gamma _n$ 
in (\ref{ym11}) we 
have $\phi _n(a)\rightarrow \det \hat{R}_0\phi _n(a)
=\pm \phi_n(a)$, which means that 
{\em linearly independent} wave functions corresponding to each energy level 
(\ref{ym12}) are determined only by 
$\gamma_n$ modulo the Weyl transformations, that is,
$\gamma _n \in H/W\sim K^+$, thus leading to the condition 
$(\omega ,\gamma _n)> 
0,\ \omega $ ranging over simple roots. Moreover, if $\gamma _n\in \pl K^+$, 
meaning that $(\gamma _n,\omega )=0$ for a certain simple root $\omega $, 
then the corresponding wave function vanishes because
$\phi _n(a)=0$. Indeed, changing the summation in (\ref{ym11}) $\hat{R} 
\rightarrow \hat{R}\hat{R}_\omega $, $\hat{R}_\omega\omega=-\omega$ and 
making use of the relations $\det \hat{R}_\omega =-1$ 
and $\hat{R}_\omega \gamma _n=\gamma _n$ 
(since $\hat{R}_\omega$ is a reflection
in the hyperplane perpendicular to $\omega$ and  $(\gamma _n,\omega )=0$) we 
get $\phi _n(a)=-\phi _n(a)$ and, hence, $\phi _n(a)=0$. 
 
The regular solutions of the Schr\"odinger equation
(\ref{ym10a}) are invariant with respect of the affine Weyl 
group, 
\be
\Phi _n(\hat{R}_{\alpha,m}a)=
\left[\kappa (\hat{R}_\alpha a)\right]^{-1}
\phi _n(\hat{R}_\alpha a)
=\Phi _n(a)\ .
\ee
It is a simple consequence of the property (\ref{c.9}) of the
function $\kappa(a)$.
Thus, we observe again that in the   Dirac quantization scheme
there is no need to postulate
the invariance of physical states with respect to 
residual gauge transformations.  
The {\em regularity} condition for the wave functions at the 
{\em singular points} of the chosen orbit space parameterization 
ensures this invariance. 

Now we obtain an explicitly gauge invariant analytic continuation
of the physical wave functions into the total functional configuration
space ${\cal F}$. Recall that we solved a similar problem for the
mechanical gauge models by means of the theorem of Chevalley.
Here we invoke other remarkable facts from group theory
to achieve the goal: The relation (\ref{c.2}) between the function
$\kappa$ and the Weyl determinant, and the Weyl formula
for the characters $\chi _{\Lambda _n} $ of 
irreducible representations of Lie groups
\cite{ch}, p.909. We get 
 \begin{equation} 
\Phi _n(a)=c_n\, \frac{\sum\limits_{\hat{R}\in W}^{}\det \hat{R}\exp 
\left\{\frac{2\pi i}{a_0}(\rho +\Lambda _n,\hat{R}a)\right\} }
{\sum\limits_{\hat{R}\in 
W}^{}\det \hat{R}\exp \left\{\frac{2\pi i}{a_0}(\rho ,\hat{R}a)\right\} }= 
c_n\chi _{\Lambda _n}\left(e^{2\pi ia/a_0}\right) \ ,
\label{ym14}
\end{equation} 
where $c_n$ are normalization constants and 
$\gamma _n=\rho +\Lambda _n$. The lattice formed by vectors 
$\Lambda _n$ labels the irreducible representations
of the Lie group. The sum over the Weyl group in 
(\ref{ym11}) should vanish for all $\gamma _n$ 
such that $(\gamma _n,\gamma _n 
)<(\rho ,\rho )$ because the function (\ref{ym11}) must be regular,
which is, in turn, possible 
only if $(\gamma _n,\gamma _n )\ge (\rho ,\rho )$
as one can see from the explicit form (\ref{c.2}) 
of the function $\kappa(a)$.
This latter condition on the norm of $\gamma_n$ ensures also 
that the spectrum (\ref{ym12}) is non-negative.
For the character 
$\chi _{\Lambda _n}$ we have the following representation 
\begin{equation} 
\chi _{\Lambda _n}\left(\exp \frac{2\pi ia}{a_0}\right)=\tr\left(\exp 2\pi 
igla\right)_{\Lambda _n}= \tr\left({\rm P}\exp ig 
\oint_{S^1}^{}Adx\right)_{\Lambda _n}\ , 
\label{ym15}
\end{equation} 
where by $(e^y)_{\Lambda _n}$  we imply 
the group element  $e^y$ in the irreducible 
representation $\Lambda _n$. 
The last equality in (\ref{ym15}) follows from the fact that 
the variable $a$ is related to a generic connection $A(x)$
by a gauge transformation.
Formula (\ref{ym15}) establishes the gauge invariant analytic
continuation of the eigenstates (\ref{ym14}) to the total configuration
space. Thus, the 
solutions to the system of functional equations (\ref{2ym1}) and
(\ref{2ym2}), which are independent of any parameterization of the gauge
orbit space,  are given by the characters of the Polyakov loop 
in all irreducible representations 
of the gauge group. 
 
The wave functions (\ref{ym14}) are orthogonal 
with respect to the scalar product (\ref{2ym3}). 
This follows from the orthogonality of the characters (\ref{ym15}). 
For normalization coefficients $c_n$ we obtain 
\ba
\delta _{nn'} &=&\int_{K^+_W}^{}da\kappa ^2(a)\Phi _n(a)\Phi ^*_n(a)
\label{ym16}\\
&=& 
2^{2N_+}c_nc^*_{n'}\int_{K^+_W}^{}da\sum\limits_{\hat{R},\hat{R}{}^\prime\in 
W}^{} \det \hat{R}\hat{R}{}^\prime
\exp \left\{\frac{2\pi i}{a_0}(a,\hat{R}\gamma _n 
-\hat{R}{}^\prime\gamma _{n'})\right\}\ .
\nonumber 
\ea 
The integrand in (\ref{ym16}) is a periodic function on 
the Cartan subalgebra. Its periods are determined by 
the geometry of the Weyl cell. 
Therefore, the integral over the periods vanishes for all $ 
\hat{R}\ne \hat{R}{}^\prime$ because 
$\gamma _n$ and $\gamma _{n'}$ belong to the 
Weyl chamber and the Weyl group acts simply and transitively on the set of 
the Weyl chambers. Hence, there is no Weyl group element $\hat{R}$ such that 
$\hat{R}\gamma _n=\gamma _{n'}$ if $\gamma _{n,n'}\in K^+$. 
For $\hat{R}=\hat{R}{}^\prime$ the integral differs from zero 
only for $\gamma _n=\gamma _{n'}$, i.e., when the periodic
exponential equals one. Thus, 
\begin{equation} 
\vert c_n\vert =2^{-N_+}(N_W\cdot V_{K^+_W})^{-1/2} \ ,
\label{ym17}
\end{equation} 
where $N_W=\nu_1\nu_2\cdots\nu_r=\dim G -{\rm rank}\, G$ 
is a number of elements in the Weyl group, $V_{K^+_W}$ is the 
volume of the Weyl cell. 
 
The energy spectrum (\ref{ym12}) seems to depend on 
normalization of the roots in the Lie algebra. Recall, however, 
that the norms of the roots are fixed by the structure constants in the 
Cartan-Weyl basis (see section 4.2 for
details). If the roots are  rescaled 
by a factor $c$, which means, in fact, rescaling the structure constants 
in the Cartan-Weyl basis by the factor $c^{-1}$, 
the invariant scalar product $(x,y)=\tr({\rm ad}x,{\rm ad}y)$
gets rescaled accordingly, i.e., by $c^{-2}$. 
Therefore the spectrum (\ref{ym12}) does not depend on the rescaling factor
because the factor $c^{-2}$ in the scalar
product is canceled against $c^2$ resulting from 
rescaling $\gamma_n$ and $\rho$ by $c$.

{\em Remark}. The Coulomb gauge can be fixed prior to canonical
quantization. Such an approach has been considered by Hetrick
and Hosotany \cite{het}. Some boundary conditions at the Gribov
horizon must be assumed. The choice of the boundary conditions
is not unique and depends on a self-adjoint extension of
the Laplace operator in the Weyl cell.
The spectrum also depend on the self-adjoint extension and 
differs from (\ref{ym12})
to order $O(\hbar)$. The model can be solved without any explicit
parameterization of the gauge orbit space via a gauge fixing.
According to the earlier work of Migdal \cite{mig}
devoted to the lattice version of the model, all physical
degrees can be described by the Polyakov loop extended
around the compactified space (the circle). Rajeev
formulated the Schr\"odinger equation in terms of the Polyakov
loop and solved it \cite{raj}. Our conclusions \cite{plb93} coincide with 
those obtained in \cite{mig,raj,mic}. Although we have used an 
explicit parameterization of the gauge orbit space via the Coulomb
gauge which has singularities, 
all the eigenstates found are explicitly gauge
invariant and regular 
in the total configuration space. The technique developed
is important 
for establishing a gauge invariant path integral formalism and
resolving the Gribov obstruction within it.

It is noteworthy that despite the fact that the physical configuration
space has an orbifold structure, the quantum theory obtained
from the Dirac formalism differs from a general quantum mechanics
on orbifolds \cite{qobi}, where wave functions are, generally,
allowed to have singularities at the singular points of the 
configurations space. In our approach the regularity condition
plays the major role in maintaining the gauge invariance if
an explicit parameterization of the orbit space is used. 

\subsection{Homotopically nontrivial Gribov transformations}

Having found the physical wave functions in the parameterization 
of the gauge orbit space, which is associated with the Coulomb
gauge, we can investigate their properties under homotopically
nontrivial residual gauge transformations. The wave functions
(\ref{ym14}) can be regarded as gauge invariant functions
(\ref{ym15}) {\em reduced} on the gauge fixing surface $\pl A=0$
in the functional space ${\cal F}$.  The Coulomb gauge is not
complete and, therefore, there are Gribov copies on the gauge
fixing surface. They are related, in general, 
either by homotopically trivial
or nontrivial gauge transformations \cite{baal92}. 
We have excluded the latter,
when calculating the physical phase space, because they cannot
be generated by the constraints, that is, two classical states related
by homotopically nontrivial transformations are, in fact, two
different physical states, so they have to correspond to two
different points in the physical phase space. Here we demonstrate
that the physical wave functions are {\em } not invariant under
homotopically nontrivial residual gauge transformations. 
The analogy can be made with instanton physics \cite{jackiw,van}
in Yang-Mills theory.
An instanton is a classical solution of Euclidean equations of motion that
connects two distinct classical vacua related 
by a homotopically nontrivial gauge transformation. Physical
wave functions acquire a phase factor under such a transformation.

Consider the group SU(2) first. 
The algebra has one positive root $\omega$. 
Solutions to (\ref{ym13}) are given by $\gamma_n=\omega n/2$ where 
$n$ ranges {\em positive} 
integers because $K^+=\Rs_+$ and $\pl K^+$ coincides with the origin 
$\gamma_n=0$. The spectrum and wave functions respectively read 
\begin{eqnarray} 
E_n &= &\frac{\pi \hbar^2}{4a_0^2l}\ (n^2-1) (\omega,\omega)\ ,\ \ \ 
n=1,2,...;
\label{hm1}\\ 
\Phi_n &= &c_n\,\frac{\sin\left[\pi n(a,\omega)/a_0\right]}
{\sin\left[\pi(a,\omega)/a_0\right]}\ . 
\label{hm2}
\end{eqnarray} 
Substituting $n=2j+1,\ j=0,1/2,1,...$, into (\ref{hm1}) we observe that 
$E_n$ is proportional to eigenvalues of the quadratic Casimir operator 
of SU(2); $E_n\sim j(j+1)$ where the spin $j$ labels the irreducible 
representations of SU(2). 

Let us introduce a new variable $\theta$ such that 
$a=a_0\omega\theta/(\omega,\omega)$.  When $a$ ranges the  Weyl cell
$K_W^+$, the variable $\theta$ spans the open interval $(0,1)$. 
The measure $da$ is defined in the orthonormal 
basis in $H$ (meaning that $H\sim \Rs^r$). For the SU(2) case we have 
$da \equiv da_3,\ a=\sqrt{2}\omega a_3$ so that $(a,a)=a_3^2,\ a_3\in \Rs$.
Here we have used $(\omega,\omega)=1/2$ for SU(2). 
Hence, the normalization coefficients $c_n$ in (\ref{hm2}) are
\begin{equation} 
c_n=\left(\sqrt{2}a_0\int_0^1d\theta\sin^2\pi n\theta\right)^{-1/2} 
=\left(\frac{a_0}{\sqrt{2}}\right)^{-1/2}\ . 
\label{hm3}
\end{equation}
The action of 
homotopically non-trivial elements (\ref{center}) of
an arbitrary simple compact gauge group on the argument
of the wave functions are determined by the shifts 
$a\rightarrow a+i/g\Omega_s\pl\Omega^{-1}_s$ where $\Omega_s
=\exp(ix\eta/l)$ and (cf. (\ref{unit})) 
\begin{equation} 
\exp(2\pi i\eta) = z\in Z_G\ . 
\end{equation} 
The lattice $\eta$ is given by integral linear combinations of elements 
$\alpha/(\alpha,\alpha)$, with $\alpha$ 
ranging over the root system, because \cite{hel}
\begin{equation} 
\exp\frac{2\pi i\alpha}{(\alpha,\alpha)}\in Z_G\ 
\label{zg}
\end{equation} 
for any root $\alpha$.
Thus, homotopically non-trivial gauge transformations 
are generated by shifts (cf. (\ref{eta}) and the example
of SU(3) given in Figure 5) 
\begin{equation} 
a\rightarrow a+\frac{n\alpha a_0}{(\alpha,\alpha)}\ ,\ \ \ n\in \Z\ . 
\label{hm4}
\end{equation} 
In the matrix representation of SU(2) the only 
positive root is $\omega=\tau_3/4$ (see section 3.2). Then $\exp 
(2\pi i\omega/(\omega,\omega)) = \exp i\pi \tau_3 = -e \in \Z_2 
=Z_{su(2)}$. 
Therefore in the case of SU(2) we get the following transformation
of the wave functions (\ref{hm2})
\begin{equation} 
\Phi_n\left(a+\frac{a_0\omega n}{(\omega,\omega)}\right) 
=(-1)^{n+1}\Phi_n(a)\ , 
\end{equation} 
i.e., the  physical states acquire a phase factor under homotopically 
nontrivial gauge transformations. 

The analysis can easily be extended to an arbitrary group 
by using the properties of the root pattern and the Weyl
representation of the characters (\ref{ym14}) under the
transformations (\ref{hm4}). However the use of the explicit form of
the gauge invariant wave functions (\ref{ym15}) in the total configuration
space ${\cal F}$ would lead to the answer faster. Making a homotopically
nontrivial gauge transformation of the Polyakov loop and taking 
into account the twisted periodicity condition (\ref{center}), we find
\be
\tr\left({\rm P}\exp ig\oint Adx\right)_{\Lambda_n} \rightarrow  
\tr\left(z{\rm P}\exp ig\oint Adx\right)_{\Lambda_n}\ .
\ee
Thus, the Gauss law (\ref{2ym2}) provides 
only the invariance of physical states with respect to gauge 
transformations which can be continuously deformed towards the 
identity. 

\subsection{Reduced phase-space quantization versus the Dirac
 approach}

The key idea to include a gauge condition chosen for parameterization
of the physical configuration space into the Dirac scheme is to use 
the curvilinear coordinates associated with the gauge condition and
the gauge transformation law to solve the constraints and find
the physical quantum Hamiltonian. We have also seen that this
approach can be applied in classical theory. Here we will compare
quantum theories obtained by the Dirac procedure and by what is
known as the reduced phase-space quantization. By the latter one usually
implies that nonphysical degrees of freedom are removed by a suitable
canonical transformation such that the constraints 
are fulfilled if some
of the new canonical momenta. Due to the gauge invariance, the 
corresponding canonical coordinates are cyclic, i.e., the Hamiltonian
does not depend on them. So the physical Hamiltonian is obtained
by setting the nonphysical momenta to zero. Finally, the theory
is canonically quantized. The point we would like to stress in 
the subsequent analysis is the following. All quantum theories
obtained by the Dirac procedure with various parameterizations
of the physical configuration space (i.e., with various gauges)
are unitarily equivalent. Thus,  physical quantities like
the spectrum of the Hamiltonian are independent of the parameterization
of the physical configuration space. In contrast, the reduced 
phase-space quantization involves ambiguities which, when
not taken care of, may lead to a  {\em gauge dependent} 
quantum theory.  Here we discuss gauge systems in rather
general settings and turn to examples only to illustrate
general concepts. 
 
Let operators $\Omega$ acting  in a space isomorphic to $\Rs^N$ 
realize a linear representation of a compact group G. 
Consider a quantum theory determined by the Schr\"odinger equation 
\be 
\left(-\frac 12\left\langle\frac{\pl}{\pl x},\frac{\pl}{\pl x}\right\rangle 
+ V(x)\right) 
\psi_E = E\psi_E\ .
\label{b.1}
\ee 
where $x\in \Rs^N$, and the group G acts on it as
$x\rightarrow \Omega(\omega)x,\ \Omega(\omega)\in G$; 
$\langle x,y\rangle = \sum_{p=1}^Nx_py_p = \langle 
\Omega x,\Omega y\rangle$ is 
an invariant 
scalar product in the representation space.
We also assume that the potential is invariant under
G-transformations $V(\Omega x)=V(x)$.
 The eigenfunctions $\psi_E$ are normalized by the condition 
\be 
\int_{\Rs^N}dx\psi_E^*(x)\psi_{E'}(x)= \delta_{EE'}\ . 
\label{b.2}
\ee 
The theory turns into a gauge theory 
if we require that physical states are 
annihilated by operators $\hat{\sigma}_a=\hat{\sigma}_a^\dagger$ 
generating $G$-transformations of $x$, 
$\hat{\sigma}_a\Psi(x)=0$. These conditions determine a physical 
subspace in the Hilbert space. By definition, we have 
\be
\exp(i\omega_a \hat{\sigma}_a)\psi(x) = \psi(\Omega(\omega)x)\ .
\label{b.3}
\ee
Therefore, the physical states are $G$-invariant 
\be 
\Psi(\Omega(\omega)x)= \Psi(x)\ . 
\label{b.4}
\ee 
 Let the number of physical degrees of freedom in the system  
equal $M$, then a number of independent constraints is $N-M$. 
One can also admit that $N$ or $M$, or both of them, are infinite.
Like in the 2D Yang-Mills theory, we can always introduce
a {\em countable} functional orthogonal 
basis in any gauge field theory and regard
the coefficients of the decomposition of the fields
over the basis functions as independent degrees of freedom \cite{kl99}. 

Suppose we would like to span the physical configuration space 
$\Rs^N/G$ by local coordinates satisfying  a gauge condition 
$\chi(x)=0$. The gauge condition fixes the gauge arbitrariness
modulo possible discrete gauge transformations, that is,   
there is no nonphysical degree of freedom left. Let $u\in \Rs^M$ be 
a parameter of the gauge condition surface; $x=f(u)$ such that 
$\chi(f(u))$ identically vanishes for all $u\in \Rs^M$. By analogy 
with (\ref{gf15}) we introduce curvilinear coordinates 
associated with the chosen gauge and the gauge transformation
law
\be 
x=x(\theta,u)=\Omega(\theta)f(u)\ , 
\label{b.5}
\ee 
where variables $\theta$ ran over the manifold $G/G_f$ with $G_f$ 
being a stationary group of the vector $x=f,\ G_ff=f$. The subgroup
$G_f$ is nontrivial if the constraints are reducible like in the mechanical
model discussed in section 4. 
The metric tensor in the new coordinates reads 
\be 
\langle dx,dx \rangle = \langle df, df \rangle + 2\langle df,d\theta f 
\rangle + \langle d\theta f, d\theta f\rangle \equiv g_{AB}dy^Ady^B\ , 
\label{b.6}
\ee 
where we have put $d\theta =
\Omega^\dagger d\Omega $ and $dy^1\equiv du,\ dy^2\equiv 
d\theta$. An integral in the new variables assumes the
form
\be 
\int_{\Rs^N}dx\psi(x) = \int_{G/G_f} \wedge d\theta 
\int_K d^M u\mu(u)\psi\left(\Omega(\theta)f(u)\right)\ ; 
\label{b.7}
\ee 
here $\mu(u)= (\det g_{AB})^{1/2},\ K$ is a domain in $\Rs^M$ such 
that the mapping (\ref{b.5}), $K\oplus G/G_f\rightarrow 
\Rs^N$, is one-to-one, i.e., $K\sim \Rs^N/G$ modulo possible
boundary identifications. To determine  the modular domain $K$, 
one should find transformations $\theta,u 
\rightarrow \hat{R}\theta, \hat{R}u,\ \hat{R}\in \tilde{S}_\chi $ which 
leave $x$ unchanged, $x(\hat{R}\theta, \hat{R}u)=x(\theta,u)$. 
Obviously, $\tilde{S}_\chi =T_e\times S_\chi $ where $T_e$ is a 
group of translations of $\theta$ through periods of the manifold 
$G/G_f$, while the set $S_\chi $ is obtained by solving Eqs.(\ref{gf2})
and (\ref{gf3}) with  
${\bf f}$ replaced by $f\in \Rs^N,\ u\in R^M,\ \Omega_s\in G$, so 
$K\sim \Rs^M/S_\chi $. Indeed, if
Eq.(\ref{gf2})  has non-trivial solutions 
(the trivial one $\Omega _s=1$ always
exists by the definition of $f(u)$), then all points
$\Omega _sf$ belong to the gauge condition surface and, 
hence, there exists a function $ u_s=u_s(u)$ such that 
$\Omega _sf(u)= f(u_s)$. 
The transformations $\Omega_s$ determine
the Gribov copies on the gauge fixing surface.
Consider transformations of $\theta$ generated 
by the group shift $\Omega (\theta)\rightarrow \Omega (\theta)\Omega ^{-1}_s 
=\Omega (\theta_s),\ 
\theta_s=\theta_s(\theta,u)$. Setting $\hat{R}u=u_s$ and $\hat{R} 
\theta= \theta_s$ we see that the transformations $\hat{R}\in S_\chi $ 
leave $x=x(\theta,u)$ unchanged. To avoid a ``double'' counting in the 
integral (\ref{b.7}), 
one has to restrict the integration domain for $u$ to the 
quotient $\Rs^M/S_\chi \sim K$. The modular domain $K$ can be specified 
as a portion of the gauge condition surface $x=f(u),\ 
u\in K\subset \Rs^M$, which has just one common point with any gauge orbit. 
 
A choice of the modular domain is not unique 
as we have already seen in section 6.2. 
In (\ref{b.7}), we assume the choice of $K$ such that
$\mu >0$ for $u\in K$. Having chosen the parameterization 
of $K$, we fix a representation of $S_\chi $ by functions 
$\hat{R}u=u_s(u),\ u\in K,\ u_s\in K_s$, i.e., $K$ is the 
domain of the function $u_s(u)$ and $K_s$ is its range.
The intersection  $K_s\cap K_{s'}=\emptyset$ is an empty set for 
any $\hat{R}\neq\hat{R}{}^\prime$. Then $\Rs^M= \cup_s K_s$ up to a set of zero 
measure being a unification of the boundaries $\pl K_s$. We define 
an orientation of $K_s$ so that for all $\hat{R}\in S_\chi $, 
$\int_{K_s}du\phi(u)\geq 0$ for any $\phi(u)\geq 0$, and the following 
rules hold 
\ba 
\int_{\Rs^M}du \phi(u) &=& 
\sum\limits_{S_\chi }\int_{K_s}du\phi(u)\ ,
\label{b.8}\\ 
\int_Kdu |J_s(u)|\phi(u)&=& \int_{K_s}du\phi(u^{-1}_s(u))\ , 
\label{b.9}
\ea 
where $J_s(u) $ is the Jacobian of the change of variable 
$u\rightarrow u_s(u)$, the absolute value of 
$J_s$ has been inserted into the right-hand side of (\ref{b.9}) to 
preserve the positive orientation of the integration domain. 

{\em Remark}. A number of elements in $S_\chi $ can depend on $u$. 
We follow the procedure described 
in section 6.2.
We define a domain $\Rs^M_\alpha\subseteq \Rs^M $ such that 
$S_\chi =S_\alpha$ has a fixed number of elements for all $u\in 
\Rs^M_\alpha$. 
Then $K=\cup_\alpha K_\alpha,\ K_\alpha 
=\Rs^M_\alpha/S_\alpha,\ \Rs^M= \cup_\alpha\Rs^M_\alpha$. The 
sum in (\ref{b.8}) means $\sum_{S_\chi }=\sum_\alpha\sum_{S_\alpha}$ 
and $K_s$  carries an additional subscript 
$\alpha$. In what follows we will omit it and use the simplified 
notations (\ref{b.8})--(\ref{b.9}) 
to avoid piling up  subscripts in formulas. The subscript 
$\alpha$ can be easily restored  by means of the rule 
just explained.
 
Let us illustrate some of the concepts introduced with the 
example of the SO(2) model of section 6.2. We have $G=SO(2), 
G_f=1,\ \det g_{AB} = {\bf f}'^2{\bf f}^2 -({\bf f}',T{\bf f})^2 
= ({\bf f}',{\bf f})^2 = \mu^2(u)$. We take a particular 
form of ${\bf f}$ considered in section 6.2 as an example.
Set $K=\cup_\alpha K_\alpha,\ 
K_1=(0,u_0/\gamma_0),\ K_2=(u_0/\gamma_0, u_0),\ K_3$\ 
$ = (u_0,\infty)$, 
i.e. $K=\Rs_+$, then $\int_{-\infty}^\infty du\phi = 
\sum_\alpha\int_{\Rs_\alpha}du\phi$ and (\ref{b.8}) means that the upper 
integral limit is always greater than the lower one, for example, 
$$
\int_{\Rs_2}du\phi(u) =\left(\int_{-3u_0}^{-2u_0}+ 
\int_{-2u_0}^{-u_0} + 
\int_{-u_0}^{-u_0/\gamma_0}+ 
\int_{u_0/\gamma_0}^{u_0}\right) du\phi(u)\ , 
$$ 
where the terms of the sum correspond to integrations over 
$\hat{R}_3K_2,\ \hat{R}_2K_2$, $\hat{R}_1K_2$ and $K_2$, 
respectively. The explicit form of functions $u_s(u)$
is given by (\ref{gf5})--(\ref{gf6}). 
The following chain of equalities 
is to illustrate the rule (\ref{b.9})
\be 
\!\!\int_{\hat{R}_3K_2}\! du_{s_3}\phi =\!\! 
\int_{-3u_0}^{-2u_0} 
\!\! du_{s_3}\phi =\!\! \int_{u_0}^{u_0/\gamma_0}\! 
 du J_{s_3}\phi =\! 
-\! \int_{u_0/\gamma_0}^{u_0}\!\! du J_{s_3}\phi =\!\!
 \int_{K_2} \!
du |J_{s_3}|\phi\ ; 
\ee 
the last equality results from $J_{s_3}= du_{s_3}/du < 0$ 
(cf. (\ref{gf7})). 
 
Solutions to the constraint equations $\hat{\sigma}_a{\Psi}(x)=0$ are 
given by functions independent of $\theta$, 
\be 
{\Psi}(x)=\Psi(\Omega (\theta)f(u))= 
{\Psi}(f(u))\equiv \Phi(u)\ , 
\label{b10}
\ee 
because $\hat{\sigma}_a$ generate only shifts of $\theta$, while 
$u$ is invariant. To obtain a physical Hamiltonian, one has to write 
the Laplacian in (\ref{b.1}) via the new variables (\ref{b.5}),
pull all the derivatives with respect to $\theta$
to the right and then set them to zero. In doing so, 
we get 
\be 
\hat{H}^f_{ph}\Phi_E(u)= 
\left(\frac12\hat{p}_ig^{ij}_{ph}\hat{p}_j + V_q^f(u) + 
V(f(u))\right)\Phi_E(u) = E\Phi _E(u)\ ; 
\label{b11}
\ee 
here we have introduced hermitian momenta $\hat{p}_j = -i\hbar\mu^{-1/2} 
\pl_j\, \mu^{1/2},\ \pl_j =\pl/\pl u^j$; 
the induced inverse metric $g^{ij}_{ph}$ on the physical 
configuration space is the $11$-component 
(see (\ref{b.6})) of a tensor $g^{AB}$ inverse 
to $g_{AB},\ g^{AC}g_{CB}=\delta^A_B,\ g^{ij}_{ph}= (g^{11})^{ij},\ 
i,j=1,2,...,M$. The quantum potential, 
\be 
V_q^f=\frac{\hbar^2}{2\sqrt{\mu}}\,(\pl_ig^{ij}_{ph})(\pl_j\sqrt{\mu}) + 
\frac{\hbar^2}{2\sqrt{\mu}}\,g^{ij}_{ph}(\pl_i\pl_j\sqrt{\mu})\ ,
\label{b12}
\ee 
occurs  after an appropriate re-ordering of 
the operators $\hat{u}^i$ and $\hat{p}_i$ in the original Laplace-Beltrami 
operator to transform it to the form of
the kinetic energy operator in
the Hamiltonian in (\ref{b11}). 
The scalar product is reduced to 
\be 
\int_{\Rs^N}dx\Phi^*_E(u)\Phi_{E'}(u)\rightarrow 
\int_Kd^Mu \mu(u)\Phi^*_E(u)\Phi_{E'}(u)=\delta_{EE'}\ , 
\label{b13}
\ee 
where the integral over $G/G_f$ has 
been included into the norm of physical states.
The renormalization procedure is denoted by 
the arrow in (\ref{b13}). The construction of an operator 
description of a gauge theory in a given gauge condition is completed. 
 
 In this approach the variables $u$ appear to be gauge-invariant; 
they parameterize the physical configuration space 
${\rm CS}_{\rm phys}=\Rs^N/G$. 
Two different choices of $f(u)$ (or the gauge condition $\chi$) 
correspond two different parameterizations 
of ${\rm CS}_{\rm phys}$ related to one  another by a change of variables 
$u=u(\tilde{u})$ in (\ref{b11})--(\ref{b13}) because
$x=f(\tilde{u}(u))=\tilde{f}(u)$.
Therefore quantum theories constructed with different gauges 
are unitary equivalent in the Dirac approach
because the Hamiltonian in (\ref{b11}) is invariant under
general coordinate transformations $u\rightarrow \tilde{u}(u)$. 
The physical
quantities like the spectrum of the Hamiltonian (\ref{b11})
are independent of the choice of $\chi$ or $f$. This
holds despite  that the explicit form of the physical Hamiltonian
{\em depends} on the concrete choice of $f$. We emphasize
that  the form (\ref{b12}) of the quantum potential
 is crucial for establishing  the unitary equivalence
of quantum theories in different gauges \cite{mont}.  
 
To illustrate this statement, consider the simplest example 
$G=SO(2),\ M=1,\ g_{ph}=r^2(u)/\mu^2(u)$, and compare descriptions 
in the coordinates (\ref{gf15}) and in the polar ones ($f_1=r, f_2=0$). 
With this purpose we change variables $r=r(u)$ in (\ref{b11})--(\ref{b13}). 
For  $u\in K$ the function $r(u)$ is invertible, $u=u(r), r\in \Rs_+$. 
Simple straightforward calculations lead us to the 
following equalities 
$\hat{H}^f_{ph}=1/2\hat{p}_r^2 + V_q(r) 
+ V,\ \hat{p}_r=-i r^{-1/2}\pl_r \, r^{1/2},\ V_q 
=-\hbar^2(8r^2)^{-1},\ \int_K du\mu\phi = \int_0^\infty drr\phi$. 
It is nothing 
but quantum mechanics of a radial motion on a plane. All theories 
with different $f$'s are unitarily equivalent to it and, therefore, 
to each other. A specific operator ordering  
obtained in the Dirac method 
ensures the unitary equivalence. Had $V_q$ been different from 
(\ref{b12}), the spectrum of the physical Hamiltonian in (\ref{b11})
would generally have depended on the gauge. This statement
can also be verified in general 
by an explicit computation of the Hamiltonian
in (\ref{b11}) in the new parameterization $\tilde{u} =\tilde{u}(u)$:
The Hamiltonian remains {\em invariant} under general
coordinate transformations if the quantum
potential has the form (\ref{b12}). Thus, the operator ordering
appears to be of great importance for the gauge invariance of the 
theory in a chosen parameterization of the physical configuration
space. The Dirac method leads to the operator ordering
that guarantees the unitary equivalence of all representations
of a quantum gauge theory with various parameterizations of
the gauge orbit space (see also the remark
at the very end of this section). 

Now let us take a formal classical limit of the Hamiltonian 
in (\ref{b11}), meaning that $\hbar=0$ and the operators
$\hat{p}$ and $\hat{u}$ are replaced by commutative canonical
variables $p$ and $u$. The classical Hamiltonian is
\be
H_{ph} = \frac 12 g^{ij}_{ph}p_ip_j + V(f(u))\ .
\label{b14}
\ee
This Hamiltonian can also be obtained by the canonical
transformation associated with the change of variables
(\ref{b.5}) just like we derived the Hamiltonian (\ref{gf18})
for the SO(2) model in an arbitrary gauge.  The constraints
$\sigma_a$ become linear combinations of the momenta 
conjugated to the variables $\theta$ that span the gauge
orbits. Thanks to the gauge invariance,
the Poisson bracket of the total Hamiltonian and the 
constraints is zero. The canonical momenta conjugated
to the $\theta$'s are integrals of motion, and, therefore,
the variables $\theta$ are cyclic: The Hamiltonian does
not depend on them. The Hamiltonian (\ref{b14})
is a reduction of the total Hamiltonian on the physical
phase space.

Had we eliminated the nonphysical degrees of freedom
in the classical theory, the Hamiltonian (\ref{b14}) 
would have been the starting point to develop a quantum 
theory.
The difficulties arising in this approach are twofold.
First, the physical phase space may not be
Euclidean. In particular, local canonical coordinates 
$u$ may not take their values in
the full Euclidean space $\Rs^M$. Therefore a 
canonical quantization runs into a notorious problem of the
self-adjointness of the corresponding momentum operators. Second,
the kinetic energy exhibits an operator ordering ambiguity.
The hermiticity condition for the quantum Hamiltonian 
is not generally sufficient to fix the operator ordering uniquely.
The physical Hamiltonian (\ref{b14}) describes a motion in a curved 
space with the metric $g^{ph}_{ij}$. What quantization 
procedures for motion in curved spaces are on the market? 
The most popular one is to replace the kinetic energy by the 
corresponding Laplace-Beltrami operator (\ref{sag1})
for the physical metric \cite{pod}. Let us see
what quantum theory emerges when this approach is
applied to the Hamiltonian (\ref{gf18}) which is a one-dimensional
version of (\ref{b14}). A general consideration would be slightly
more involved, but leads to the same conclusion.  Comparing
(\ref{gf18}) and (\ref{b14}) we see that $g(u)= r^2/\mu^2$,
where $r^2(u) = {\bf f}^2(u)$, plays
the role of the {\em inverse} metric. Therefore the density in
the volume element,
being the square root of the determinant of the metric, is 
$\gamma(u) = \mu/r$. According to (\ref{sag1}), the kinetic
energy is quantized by the rule
\be
g(u)p^2 \rightarrow -\hbar^2\frac{1}{\gamma}\,\pl_u\,  g\gamma\pl_u
= -\hbar^2 \frac{r}{\mu}\, \pl_u \,  \frac{r}{\mu}\,\pl_u =
 -\hbar^2 \pl_r^2\ ,
\ee
where we have used the relation $dr/du = \mu/r$. The scalar product
measure is transformed accordingly
\be
\int_K du \gamma\phi
 = \int_K du \frac {\mu}{r}\phi = \int_0^\infty dr\phi\ .
\ee
The operator $\pl_r^2$ is not essentially self-adjoint on the half-axis.
Its self-adjoint extensions form a one-parametric
family characterized by a real number $c=(\pl_r\psi/\psi)_{r=0}$. 
Thus, the naive replacement of the kinetic energy by the Laplace-Beltrami
operator does not lead, in general, to a self-adjoint Hamiltonian and its
self-adjoint extension may not be unique. The boundary
$\pl K$ may have a complicated geometrical structure, which 
could make a self-adjoint extension of the kinetic energy
a tricky problem in the case of many physical degrees of freedom, 
needless to say about the field theory case.

One of the reasons  that the above method fails
is inherent to any gauge theory with a non-Euclidean
orbit space.
The density $\mu(u)$ on the gauge orbit space does {\em not}
coincide with the
square root of the determinant of the induced physical metric
$g^{ph}_{ij}$. One could therefore abandon
the above quantization recipe and require that the volume
element of the orbit space 
should be calculated by the reduction of the volume 
element $d^Nx$ onto the gauge fixing surface. 
The canonical momenta $\hat{p}= -i\hbar 
\mu^{-1/2}\pl_u \,  \mu^{1/2}$ are hermitian 
with respect to the scalar
product $\int_K du \mu \phi^*_1\phi_2=\<1\vert 2 \>$. 
Hence, hermiticity of 
the physical Hamiltonian can be achieved by an appropriate 
operator ordering, say, by a symmetrical one
\be
g(u)p^2 \rightarrow \hat{p} g(u) \hat{p} + O(\hbar)\ .
\ee
But now we face another problem. The $\hbar$--corrections
to the quantum kinetic energy operator should be precisely of
the form (\ref{b12}), otherwise the spectrum of the physical
Hamiltonian would depend on the chosen gauge to parameterize
the physical configuration space. In the Dirac approach 
the necessary operator ordering has been generated automaticly,
while in the reduced phase-space quantization approach we have
to seek a resolution of this problem separately. Thus, the Dirac
approach has advantages in this regard. 

{\em Remark}. The operator ordering ambiguities in the reduced
phase-space quantization might be resolved in the sense that 
the spectrum of the quantum Hamiltonian would not depend
on the parameterization of the gauge orbit space. One can require
that a physically acceptable operator ordering should provide
an {\em invariance} of the physical Hamiltonian under general coordinate
transformations $u \rightarrow u(\tilde{u})$. This condition
would lead to the physical Hamiltonian that coincides with 
that obtained in the Dirac approach modulo quantum
corrections containing the Riemann {\em curvature} tensor of the gauge
orbit space (any scalar potential that can be built out
of the physical metric tensor). This type of corrections is known
in quantization on curved manifolds (without a gauge symmetry)
\cite{dewitt,cheng,dekker,cs1,cs2,cs3}.
In gauge theories such an addition would mean a modification
of the canonical Hamiltonian by corrections to order $\hbar^2$.
The curvature of the gauge orbit space does not
depend on the choice of local coordinates and, hence, is gauge
invariant (cf. the example of the gauge matrix model
in section 4.8). Thus, an addition of curvature terms to a
quantum Hamiltonian would be consistent with the gauge invariance.
So far there seem to be no theoretical reason to forbid
such terms, unless they affect the Yang-Mills perturbation theory,
which seems unlikely because the perturbation theory deals with
field fluctuations that are much smaller in amplitude than
the inverse curvature of the orbit space. Possible nonperturbative
effects of such terms are unknown.
  
\section{Path integrals and the physical phase space structure}
\setcounter{equation}0

In this section we develop the path integral formalism 
for quantum gauge systems. The goal is
to take into account the geometrical structure of  either the 
physical configuration space in the Lagrangian path
integral or the physical phase space in the Hamiltonian
path integral. A modification of the conventional path
integral formalism stems from the very definition of the 
sum over paths. So we first give a derivation of the path
integral in a Euclidean space and then look for what 
should be modified in it in order to reproduce the Dirac operator
formalism  for gauge theories. 

\subsection{Definition and basic properties of the path integral}

Let us take a quantum system with one degree of freedom.
Let $|q,t\rangle$ be an eigenstate of the Heisenberg position operator
\begin{equation}
\hat{q}(t)|q,t\rangle =q|q,t\rangle\ .
\label{pi.1}
\end{equation}
The operator $\hat{q}(t) $ depends on time and so do its eigenstates.
Making use of the relation (\ref{7.5}) 
between the Heisenberg and Schr\"odinger pictures we find
\begin{equation}
|q,t\rangle =e^{it\hat{H}/\hbar}|q\rangle\ .
\label{pi.2}
\end{equation}
The probability amplitude that a system which was in the eigenstate
$|q'\rangle$ at the time $t=0$ will be found to have the value $q$ of the 
Heisenberg position operator $\hat{q}(t)$ at time $t>0$ is
\begin{equation}
\langle q,t|q'\rangle =\langle q|e^{-it\hat{H}/\hbar}|q'\rangle =
U_t(q,q')\ .
\label{pi.3}
\end{equation}
The amplitude (\ref{pi.3}) is called the evolution operator kernel, 
or the transition
amplitude. It satisfies the Schr\"odinger equation
\begin{equation}
i\hbar \pl _tU_t(q,q')=\hat{H}(q)U_t(q,q')
\label{pi.4}
\end{equation}
with the initial condition
\begin{equation}
U_{t=0}(q,q')=\langle q|q'\rangle =\delta (q-q')\ .
\label{pi.5}
\end{equation}
Any state $|\Psi\rangle $ evolving 
according to the Schr\"odinger equation can
be represented in the following form
\be
\Psi _t(q)=  \< q,t|\Psi\> =
\int_{}^{}dq'U_t(q,q')\Psi _0(q')\ ,
\label{pi.6}
\ee
where $\Psi _0(q')=\<q'|\Psi\> $  is the initial wave function.

The kernel $U_t(q,q')$ contains all information about dynamics
of the quantum system. There exists a representation of it as 
a Feynman sum over paths weighted by the exponential of the
classical action \cite{feyn}.
We derive it following the method proposed by Nelson \cite{nelson}
which is based on  the Kato-Trotter product formula. The derivation
can easily be extended to gauge systems. For this reason
we reproduce its details.
For any two  self-adjoint operators  $\hat{A}$ and $\hat{B}$,
in a separable Hilbert space such that the operator  
$\hat{A} +\hat{B}$ is self-adjoint on the intersection of the
domains of the operators  $\hat{A}$ and $\hat{B}$
the following relation holds \cite{kt,kt1,schul}
\begin{equation}
e^{i(\hat{A}+\hat{B})}=\lim_{N\rightarrow
\infty}\left(e^{i\hat{A}/N}e^{i\hat{B}/N}\right)^N\ .
\label{pi.7}
\end{equation}
Assume the Hamiltonian $\hat{H}$ to be a sum of kinetic and potential
energies $\hat{H}=\hat{H}_0+V(\hat{q}),\ H_0=\hat{p} ^2/2$, and set
$\hat{A}=-t\hat{H}_0/\hbar$ and $\hat{B}=-t\hat{V}/\hbar$ in the
Kato-Trotter formula. We have
\begin{equation}
e^{-it\hat{H}/\hbar}=\lim_{N\rightarrow\infty}\left[e^
{-i\epsilon\hat{H}_0/\hbar} e^{-i\epsilon \hat{V}/\hbar}\right]^N\equiv
\lim_{N\rightarrow\infty}(\hat{U}_\epsilon)^N\ ,
\label{pi.8}
\end{equation}
where $\epsilon =t/N$. It is easy to verify that the evolution
operator kernel for the Hamiltonian being just the kinetic energy has the form
\begin{equation}
U^0_t(q,q')=\< q|e^{-it\hat{H}_0/\hbar}|q'\>=\left(2\pi i\hbar t
\right)^{-1/2} \exp \left\{\frac{i(q-q')^2}{2\hbar t}\right\}\ ,
\label{pi.9}
\end{equation}
i.e., it is a solution to the Schr\"odinger equation (\ref{pi.4}) with
$\hat{H}=\hat{H}_0=-\hbar ^2\pl _q^2/2$ and the initial condition
(\ref{pi.5}). Consider the matrix element of the operator
$\hat{U}_\epsilon=\hat{U}_\epsilon^0\exp(-i\epsilon\hat{V}/\hbar)$ 
in (\ref{pi.8}). We have
\be
 \< q_{j+1}|\hat{U}_\epsilon|q_j\> =
\left(2\pi i\hbar \epsilon\right)^{-1/2}\exp
 \left\{\frac{i}{\hbar} \epsilon\,\left(\frac{(q_{j+1}-q_j)^2}{2\epsilon
^{2}}-V(q_j)\right)\right\}\ .
\label{pi10}
\ee
Inserting the resolution of unity
\begin{equation}
1=\int_{-\infty}^{\infty}dq|q\>  \<q|
\label{pi11}
\end{equation}
between the operators $\hat{U}_\epsilon$ in the product
$(\hat{U}_\epsilon)^N=\hat{U}_t$ we find
\begin{equation}
U_t(q,q')=\lim_{N\rightarrow\infty}\left(\frac{2\pi i \hbar t}{N}
\right)^{-N/2}\int_{}^{}dq_1dq_2\ldots dq_{N-1}e^{i
S(q,q_{N-1},\ldots ,q_1,q')/\hbar}\ ,
\label{pia12}
\end{equation}
where
\begin{equation}
S(q,q_{N-1},\ldots ,q_1,q')=\sum\limits_{j=0}^{N-1}\frac{t}{N}\left[
\frac{1}{2}(q_{j+1}-q_j)^2\left(\frac{t}{N}\right)^{-2}-V(q_j)\right]
\label{pi13}
\end{equation}
with $q_0\equiv q'$ and $q_N\equiv q$.

Let $q(\tau)$ be a polygonal path going through points
$q_j=q(j\epsilon)$ and connecting points $q(\tau =0)=q'=q_0$ and $q(\tau
=t)=q=q_N$ so that
on each interval $\tau \in [j\epsilon,(j+1)\epsilon]$ it is a linear
function of $\tau$ 
\begin{equation}
q(\tau)=(q_{j+1}-q_j)\left(\epsilon\tau -j\right)+q_j\ .
\label{pi14}
\end{equation}
The classical action of this path is
\begin{equation}
S[q]= \int_{0}^{t}d\tau\left[\frac{1}{2}
\left(\frac{dq(\tau)}{d\tau}\right)^2-
V(q(\tau)) \right]\approx S(q,q_{N-1},\ldots ,q_1,q')\ .
\label{pi15}
\end{equation}
Thus, for sufficiently large $N$, integrating with respect to $q_1,\ldots
, q_{N-1}$ in (\ref{pia12}) is like integrating over all polygonal paths
having $N$ segments. In the limit $N\rightarrow \infty$, polygonal paths
turn into continuous paths.
The continuity of the paths contributing to the Feynman integral
follows from the fact that  (\ref{pi13}) is a {\em Gaussian} distribution
of $\Delta_j = q_{j+1}-q_j$ so that the expectation value
of $\Delta^{2n}$ is proportional to $\epsilon^n$. Therefore the
main contribution to the discretized
integral (\ref{pia12}) comes from $|\Delta_j| \sim \sqrt{\epsilon}
\rightarrow 0$ as $\epsilon$ approaches zero, i.e., the distance
between neighboring points of the path vanishes as $\sqrt{\epsilon}$.
It should be noted however that for a generic
action the distance between neighboring points of paths in the Feynman
sum may have a different dependence on the time slice $\epsilon$,
and the paths will not be necessarily continuous.

In the continuum limit, the integral (\ref{pia12}) looks
like a sum over all continuous paths
connecting $q$ and $q'$ and weighted by the
exponential of the classical action
\begin{equation}
\< q,t|q'\> =
U_t(q,q')=\sum\limits_{paths}^{}e^{iS[q]/\hbar}=
\int\limits_{q(0)=q'}^{q(t)=q}{\cal D}
qe^{iS[q]/\hbar}\ ,
\label{pi16}
\end{equation}
where
\begin{equation}
{\cal D}q=\lim_{N\rightarrow\infty}\left(\frac{2\pi i\hbar
t}{N}\right)^{-N/2} dq_1dq_2\cdots dq_{N-1}\equiv {\cal Z}^{-1}_0
\prod_{\tau =0}^{t}dq(\tau)\ .
\label{pi17}
\end{equation}
The integral (\ref{pi16}) is called the Lagrangian path integral.

The transition amplitude of a free particle (\ref{pi.9}) can be
written as the Gaussian integral
\be
U_t^0(q,q') = (2\pi\hbar)^{-1}\int_{-\infty}^\infty dp
\exp\frac{i}{\hbar}\left\{p(q-q')/t - p^2t/2\right\}\ .
\ee
The expression in the exponential is the action of a free particle
moving with the momentum $p$.
Making use of this representation in each stage of the Lagrangian
path integral derivation we obtain the Hamiltonian path integral
representation of the transition amplitude
\be
U_t(q,q') = \int {\cal D}p{\cal D}q \, e^{\frac{i}{\hbar}\int_0^t d\tau
[p\dot{q} - H(p,q)]}\ ,
\label{pi18}
\ee
where $H(p,q)$ is the classical Hamiltonian of the system, and
the measure is defined as the formal time product of the Liouville
measures on the phase space
\be
 {\cal D}p{\cal D}q = \lim_{N\rightarrow\infty} \frac{dp_N}{2\pi\hbar}
 \prod_{j=1}^{N-1}\frac{dp_jdq_j}{2\pi\hbar} \equiv
 \prod_{\tau =0}^{t}\frac{dp(\tau)dq(\tau)}{2\pi\hbar}\ .
\ee
Observe one extra integration over the momentum and the normalization
of the phase space measure by $2\pi\hbar$.

Let $\Psi_E(q)$ be normalized eigenfunctions of the Hamiltonian $\hat{H}$
with the eigenvalues $E$. Then we can derive the spectral decomposition
of the transition amplitude
\be
U_t(q,q')= \sum_{E,E'}\<q|E\>\<E|\hat{U}_t|E'\>\<E'|q'\>=
 \sum_{E} e^{-itE/\hbar}\Psi_E(q)\Psi^*_E(q')\ .
\label{pi19}
\ee
This decomposition will be useful to establish
the correspondence between the Dirac operator formalism and the 
path integral formalism for gauge theories.

A generalization of the path integral formalism to systems with
many degrees of freedom is straightforward. The kernel of 
$\hat{U}_t^0$ is a product of the kernels (\ref{pi.9}) for each Cartesian
degree of freedom. The rest of the derivation remains the same.
In the field theory, a
lattice regularization of the functional integral is usually
assumed. The analysis of the continuum limit leads
to the conclusion that the support of the functional integral
measure is in the space of distributions rather than 
continuous field configurations (see, e.g., \cite{zinn,kl99}).
Yet, the removal of
the lattice regularization in strongly interacting field theory is
not simple and, in general, may pose a problem \cite{zinn}.   

\subsection{Topology and boundaries of the configuration space
in the path integral formalism}

The configuration space of a system may have a non-trivial topology.
It can be, for instance, due to constraints. 
Consider a planar motion constrained to a circle. The
system is known as a rigid rotator. Its quantum mechanics is
described by the Hamiltonian
\begin{equation}
\hat{H}_0=-\frac{\hbar ^2}{2}\frac{\pl ^2}{\pl \varphi ^2}\ ,
\label{tp1}
\end{equation}
where the angular variable $\varphi$ spans the configuration space
being a circle of unit radius, $\varphi\in [0,2\pi)$. The entire difference
between the quantum motion on the line and circle lies in the topologies
of these spaces. 
The topology of the  rotator configuration space -- 
the fact that it is a circle rather
than a line -- is accounted for by the periodicity condition imposed on
state vectors
\begin{equation}
\< \varphi +2\pi |\Psi\>  =\< \varphi |\Psi\>
\label{tp2}
\end{equation}
for any $|\Psi\> $.  Accordingly,  the resolution of unity for the 
rotator differs from that for the free particle (\ref{pi11})
\begin{equation}
1=\int_{0}^{2\pi}d\varphi |\varphi\>  \< \varphi |\ .
\label{tp3}
\end{equation}
Observe that the integral is taken over a {\em finite} interval.

The transition amplitude $\< \varphi ,t|\varphi '\> $  must satisfy
the Schr\"odinger equation and the periodicity condition (\ref{tp2}) for
both arguments $\varphi $ and $ \varphi '$. 
Since the Hamiltonians for free particle and the rotator have the same
form, the solution to the
Schr\"odinger equation for the free motion is given by (\ref{pi.9}),
where the variable $q$ is replaced by $\varphi$, and can also
be written as the path integral
\be
\<\varphi,t|\varphi\> = \int\limits_{\varphi(0)=\varphi'}
^{\varphi(t)=\varphi} {\cal D}\varphi 
\exp\left[\frac{i}{2\hbar}\int_0^t
d\tau \dot{\varphi}^2\right]\ . 
\label{tp4}
\ee
Clearly, the transition amplitude (\ref{pi.9}) does not
satisfy the periodicity condition
(\ref{tp2}) and neither does the path integral (\ref{tp4}). 
The measure of the path integral in (\ref{tp4}) is the
standard one, that is, in every 
intermediate moment of time it is integrated
over the entire real line $\varphi (\tau) \in (-\infty, \infty), \
0< \tau < t$.
Looking at the resolution of unity
(\ref{tp3}) one could argue that
the integration in the infinite limits is the source of the trouble
because it seems to be in conflict with the path integral definition
(\ref{pia12}), where the resolution of unity has been used, and  
the replacement of
$\int_{-\infty}^{\infty}d\varphi(\tau)$
by $\int_{0}^{2\pi}d\varphi(\tau)$
in the path integral measure (\ref{tp4}) 
(in accordance with the folding (\ref{pia12}))
would have to improve the situation. However,
making the time-slicing regularization of the path integral
measure (\ref{tp4}) and restricting the integration to the 
interval $[0,2\pi)$ we immediately see
that we are unable to calculate the Gaussian integrals in the 
folding (\ref{pia12}) due to the
finiteness of the integration limits. 
Thus, such a modification of 
the folding (\ref{pia12}) would fail 
to reproduce the solution (\ref{pi.9}) of the Schr\"odinger equation.
This leads to the conclusion that 
the formal restriction of the integration domain in the path
integral {\em contradicts} the operator formalism.
An important point is that
even for an infinitesimal interval of time
$t\rightarrow 0$, the amplitude (\ref{pi.9}) does not satisfy
the periodicity condition and, therefore,
{\em cannot be used to construct
the path integral as the limit of the folding (\ref{pia12})}
that stems from the Kato-Trotter product formula.

To find a right relation between the transition amplitudes
on a line and circle, we invoke the superposition
principle in quantum mechanics.  Let $\varphi\in [0,2\pi)$ and
the initial point $\varphi'$ may take its values on 
the whole real line which is the {\em covering} space of the 
circle. The circle can be regarded as
a quotient space $\Rs/T_e$ where $T_e$ is a group of translations
$\varphi\rightarrow \varphi+2\pi n$.
If $\varphi'$
describes the states of the rotator, then the states $\varphi' +2\pi n$,
where $n$ runs over integers, corresponds to the same physical
state. Therefore the Feynman sum over paths should  include
paths outgoing from  $\varphi' +2\pi n$ and ending at $\varphi$
in accordance the superposition principle.
Thus, the transition amplitude for the rotator has the form \cite{schul}
\begin{equation}
\< \varphi ,t|\varphi '\>_{c}=\sum\limits_{n=-\infty}^{\infty}
\left({2\pi i\hbar t}\right)^{-1/2}\exp \frac{i(\varphi -
\varphi'+2\pi n)^2}{2\hbar t}\ .
\label{tp5}
\end{equation}
Here by the suffix $c$ we imply ``circle''.
The sum over $n$  can be
interpreted as a sum over winding numbers of a classical trajectory
around the circle.
The function (\ref{tp5}) satisfies the Schr\"odinger equation and
the periodicity condition. Let us take the limit of zero time:
\begin{equation}
\lim_{t\rightarrow 0}\< \varphi ,t|\varphi '\>_{c}=
\sum\limits_{n=-\infty}^{\infty} \delta (\varphi -\varphi '+2\pi n)
\label{tp6}
\end{equation}
which coincides with $\delta (\varphi -\varphi ')$ for physical values of
$\varphi ,\varphi '\in [0,2\pi)$ and defines a {\em continuation} of 
the unit operator kernel into the covering space of the physical
configuration space.  The notion of the covering space
as well as the continuation of the unit operator kernel to the covering
space will be useful in the path integral formalism for gauge theories.  
The concept of the covering space has been used to construct the 
path integral over non-Euclidean {\em phase} spaces, e.g., the sphere
\cite{anton}. A similar structure of the path integral occurs
when passing to curvilinear coordinates in the measure \cite{lpcc,shja91}
and in quantum dynamics on compact group manifolds \cite{mar}.

The kernel (\ref{tp5}) can be used in the folding 
(\ref{pia12}) with the resolution of unity (\ref{tp3}) without any
contradiction. Indeed, we have
\begin{eqnarray}
\< \varphi ,t|\varphi '\>_{c} &= &
\prod\limits_{j=1}^{N-1}\left(
\int_{0}^{2\pi} d\varphi_j\right)
\< \varphi ,\epsilon|\varphi_{N-1} \>_{c}
\cdots
\< \varphi_1 ,\epsilon|\varphi '\>_{c}= 
\nonumber\\
 &= &
\prod\limits_{j=1}^{N-1}\left(
\int_{-\infty}^{\infty} d\varphi_j\right)
\< \varphi ,\epsilon|\varphi_{N-1} \>
  \cdots
\< \varphi_1 ,\epsilon|\varphi '\>_{c}= \nonumber\\
&=&
 \sum\limits_{n=-\infty}^{\infty}
\prod\limits_{j=1}^{N-1}\left(
\int_{-\infty}^{\infty} d\varphi_j\right)
\< \varphi ,\epsilon|\varphi_{N-1} \>
\cdots
\< \varphi_1 ,\epsilon|\varphi'+2\pi n \>\ .
\label{tp7}
\end{eqnarray}
Here in the first equality we used the sum over the winding
number to extend the integration to the whole real line and
to replace the infinitesimal transition amplitude on the circle
by that on the line.
In the limit $N\rightarrow \infty$, the 
expression (\ref{tp7}) yields the path integral
\ba
\< \varphi ,t|\varphi '\>_{c} &=&
\sum\limits_{n=-\infty}^{\infty}
\int\limits_{\varphi(0)=\varphi'+2\pi n}^{\varphi(t)=\varphi}
{\cal D}\varphi e^{i\int_0^t d\tau \dot{\varphi}^2/2\hbar }
\label{tp8}\\
&=&\int_{-\infty}^\infty d\varphi'' 
\< \varphi ,t|\varphi ''\> Q(\varphi '',\varphi ')\ ,
\label{tp9}
\ea
where the kernel ${Q}$ is given by  (\ref{tp6}). It 
defines a {\em periodic} continuation
of any function  on the interval $[0,2\pi)$ to the covering space:
\be
\Psi^Q(\varphi+2\pi)=\Psi^Q(\varphi)=\int_0^{2\pi}d\varphi'
Q(\varphi,\varphi')\Psi(\varphi')\ ,
\label{tp10}
\ee
and, thereby, ensures that
the action of the evolution operator constructed by the sum over
paths in the {\em covering} space preserves the periodicity
of the physical states $\Psi_t(\varphi +2\pi) = \Psi_t(\varphi)$,
where
\be
\Psi_t(\varphi)=\int_0^{2\pi}d\varphi'
\<\varphi,t|\varphi'\>_c\Psi_0(\varphi')=
\int_{-\infty}^{\infty}
d\varphi''\<\varphi,t|\varphi''\>\Psi^Q_0(\varphi'')\ .
\label{star}
\ee
A similar representation can also be established for the path integral
of a free particle in the infinite well. In this case the transition
amplitude should satisfy the zero boundary conditions
\begin{equation}
\< q=0,t|q'\>  =\< q=L,t|q'\>=\< q,t|q'=0\>=\< q,t|q'=L\>  =0 \ ,
\label{bo1}
\end{equation}
where $L$ is the size of the well. The resolution of unity reads 
\begin{equation}
\int_{0}^{L}dq|q\>\< q|=1\ .
\label{bo2}
\end{equation}
The formal restriction of the
integration domain in the path integral would yield an incorrect answer 
because the kernel of $\hat{U}_\epsilon^0$ in the Kato-Trotter product
formula does not have the standard form (\ref{pi.9}). The right
transition amplitude compatible with the Kato-Trotter operator
representation of the evolution operator
is obtained by the superposition principle
\cite{pauli,jk}. It can be written as follows \cite{217}
\ba 
 \< q,t|q'\>_{box} =
 \int_ {-\infty}^\infty dq''\,
 \int\limits_{q(0)=q''}^{q(t)=q}{\cal D}q 
 \exp\left\{\frac{i}{2\hbar}\int_0^t
 d\tau \dot{q}^2\right\}\, Q(q'',q')\ ;
\label{bo3}\\
Q(q,q')= \sum_{n=-\infty}^\infty \left[
\delta(q-q' + 2Ln)-\delta(q+q' +2Ln)\right]\ .
\label{bo4}
\ea
The contributions of trajectories
going from $x'+2Ln$ to $x$ and of those going from $-x'+2Ln$ to $x$
have opposite signs, which is necessary to provide 
the zero boundary conditions (\ref{bo1}).
A straight trajectory $x'+2Ln\rightarrow x$ can be interpreted as a
continuous trajectory {\em inside} the well which connect $x',x\in(0,L)$
and have $2n$ reflections
from the well walls because it has the same
action. Contributions of the trajectories
$-x'+2Ln\rightarrow x$ are
equivalent to contributions of trajectories inside the well
with an odd number of reflections $2n+1$.
The problem of zero boundary conditions in the path integral 
formalism in general has been studied in \cite{pde1}. 

The lesson following from our analysis is that 
the restriction of the integration domain in the path integral,
which might seem to be motivated
by the prelimit expression (\ref{pia12}), is ruled out because the
infinitesimal transition amplitude, that is used in the 
Kato-Trotter product formula for the path integral, 
has no ``standard'' form (\ref{pi.9})
if the system configuration space has either a nontrivial topology,
or boundaries, or both of them.
This is the key observation for constructing the path integral
formalism {\em equivalent} to the Dirac operator quantization
of gauge systems.

\subsection{Gribov obstruction to the path integral quantization
of gauge systems}

The Feynman representation of quantum mechanics has led
to a {\em new} quantization postulate which is known as the
path integral quantization \cite{feyn}. Given a classical Hamiltonian 
$H=H(p,q)$ and the canonical symplectic structure on
the phase space, the transition amplitude of the corresponding
classical system is given by the Hamiltonian path integral
(\ref{pi18}). The correspondence principle is guaranteed by 
the stationary phase approximation of the path integral (\ref{pi18})
in the formal limit $\hbar\rightarrow 0$, i.e., in the dynamical
regime when the classical action is much greater than the Planck
constant. For many physically interesting models this postulate
is valid. It is natural to extend it as quantization postulate
to general Hamiltonian systems, and, thereby, to avoid the
use of {\em noncommutative} variables (operators) to describe
quantum systems. 
This attractive idea has, unfortunately, some shortcomings,
which, as we will see, appear to be relevant for the path integral
formalism in gauge theories. 

The action functional of systems with gauge symmetry is constant
along the directions traversed by gauge transformations in the path 
space. Therefore the Feynman sum (\ref{pi16}) would diverge.
In the Hamiltonian formalism the gauge symmetry leads to constraints
and the appearance of nonphysical variables. 
The physical motion occurs in the physical phase space, the 
quotient of the constraint surface by the gauge group. In his pioneering
work \cite{fd68}, Faddeev proposed the following modification  
of the path integral measure for systems with first-class constraints:
\begin{equation}
{\cal D}p{\cal D}q \rightarrow {\cal D}p{\cal D}q
\delta(\sigma_a)\delta(\chi_a)\Delta_{FP}
= {\cal D}p^*{\cal D}q^*{\cal D}\tilde{p}
{\cal D}\tilde{q}\delta(\tilde{p}_a)\delta(\tilde{q}_a)\ . 
\label{fp1}
\end{equation}
Here $\delta(\sigma_a)$ reduces the Liouville measure onto the constraint
surface at every moment of time, 
while the {\em supplementary} (or gauge) conditions $\chi_a=0$ are to
select a representative from the gauge orbit through
a point $q$. The function 
\be
\Delta_{FP} = \det\{\chi_a,\sigma_b\}\ ,
\label{fp2}
\ee
known as the Faddeev-Popov determinant \cite{fp,fd68}, effects a local 
reestablishment of the {\em local} Liouville measure on the ${\rm
 PS}_{\rm phys}$; it is assumed that $\{\chi_a,\chi_b\}=0$.
 If $\Delta_{FP}\neq 0$, then one can show \cite{fd68} that there exists
 a canonical transformation $p,q\rightarrow p^*,q^*; \tilde{p},
 \tilde{q}$ such that
the variables  $p^*, q^*$ form a set of local canonical
coordinates on ${\rm PS}_{\rm phys}$,
and $\tilde{p},\tilde{q}$ are nonphysical phase-space variables 
(see also section 6.1 for details). 
Assuming that the Liouville
path integral measure remains invariant under canonical
transformations,
the equality (\ref{fp1}) is readily established
(after solving the constraints
$\sigma_a=0$ for the nonphysical momenta 
$\tilde{p}_a=\tilde{p}_a(p^*,q^*)$,
the shift of the integration variable $\tilde{p}_a
\rightarrow \tilde{p}_a -\tilde{p}_a(p^*,q^*)$ has to be done).
The method has successfully been applied to the
perturbation theory for quantum gauge fields \cite{fp} and provided
a solution to two significant problems: 
the unitarity problem in perturbative path integral quantization
of Yang-Mills fields \cite{feynym} and the problem of
constructing a local gauge fixed effective action \cite{dew}.

If the physical phase space is non-Euclidean, the transformation
(\ref{fp1}) is no longer true. 
 The evidence for this obstruction  
is the impossibility to introduce a set of supplementary conditions that
provide a global parameterization of the physical phase space
by the canonical coordinates $p^*,q^*$ without singularities,
a situation which is often rendered concrete in the vanishing or even sign 
changing of the Faddeev-Popov determinant
\cite{gribov}. In section 6.1 we have shown that the 
condition $\Delta_{FP}\neq 0$ cannot be met everywhere for
any single-valued functions $\chi_a$ if gauge orbits have
a nontrivial topology.
The surface $\sigma_a=\chi_a=0$ may not
be isomorphic to ${\rm PS}_{ph}$, i.e., 
it still has gauge-equivalent configurations
(Gribov copies). Assuming that the local coordinates $p^*,q^*$ span
the surface $\sigma_a=\chi_a=0$, the physical phase space will be
isomorphic to a certain (gauge-dependent) domain within it
(modulo boundary identifications).

We have seen that the formal restriction of the integration domain, say,
to the modular domain, to remove the contribution of physically
equivalent configurations is not consistent and  contradicts 
the operator formalism. 
Another remark is that the parameterization of the physical phase space
is defined modulo general canonical transformations. Different choices
of the supplementary condition $\chi$ 
would lead to different parameterizations
of the physical phase space which are related by canonical transformations. 
Physical amplitudes 
cannot depend on any particular parameterization, i.e., they have
to be independent of the choice of the supplementary condition.
However, the formal Liouville measure ${\cal D}p^*{\cal D}q^*$ 
does not provide any genuine covariance of the path integral
under general canonical transformations as has
been argued in the Introduction. Thus, the measure
(\ref{fp1}) must be modified to take into account the non-Euclidean
geometry of the physical phase space, which is natural,
given the fact that path integral quantization of the phase
space geometries different from the Euclidean one leads to
quantizations different from the canonical one based on the canonical
Heisenberg algebra \cite{kld,kld2,kl88,klmc97}.

In the Yang-Mills theory, the Coulomb gauge turns out to
be successful for a consistent path integral quantization in
the high energy limit where the coupling constant is small,
and the geometry of the physical phase space does not affect
the perturbation theory. In the infrared limit, where the coupling
constant becomes large, the coordinate singularities associated with
the Coulomb gauge invalidate the path integral quantization
based on the recipe (\ref{fp1}) as has been first observed by
Gribov \cite{gribov}. Therefore a successful non-perturbative
formulation of the path integral in Yang-Mills theory is impossible
without taking into account the (non-Euclidean) 
geometry of the physical phase space.

\subsection{The path integral on the conic phase space}

To get an idea of how the Faddeev-Popov recipe should 
be modified if the physical phase space is not Euclidean,
we take the isotropic oscillator in three-dimensional space
with the gauge group $SO(3)$ \cite{vest88,tmp89}. 
The reason of taking the group
$SO(3)$ is that the quantum Hamiltonian (\ref{sa2}) constructed
in the Dirac formalism can be related to the corresponding
one-dimensional problem by rescaling the wave functions,
$\Phi=\phi/r$. So one can compare the path integrals for
oscillators with flat and conic phase spaces. For an arbitrary
orthogonal group, the Dirac quantum Hamiltonian contains
the quantum potential $V_q = \hbar^2(N-3)(N-1)/(8r^2)$
as compared with the Hamiltonian of 
the corresponding one-dimensional system
(see (\ref{b12})). A general technique
to construct the path integral over a non-Euclidean
phase space, which takes into
account the operator ordering problem, is given in section 8.7. 

Making the substitution $\Phi_n=\phi_n/r$ in (\ref{sa2}) for $N=3$
and the oscillator potential, we find that the functions
$\phi_n$ are eigenfunctions of the one-dimensional harmonic
oscillator ($\hbar =1$)
\be
\phi_n(r) = c_n H_n(r) e^{-r^2/2} \ . 
\label{con1}
\ee
The physical eigenstates are given by the regular functions
\be
\Phi_{2k+1} (r) = \tilde{c}_{2k +1}\frac{H_{2k+1}(r)}{r} \, e^{-r^2/2}\ .
\label{con2}
\ee
The singular functions for $n=2k$ do not satisfy the Schr\"odinger
equation at the origin $r=0$ (cf. the discussion 
in section 7.2).
To compare the transition amplitude of the oscillator with
flat and conic phase spaces, we 
compute $c_n$ and $\tilde{c}_n$ and then make use of the spectral
decomposition (\ref{pi19}). The normalization
constants $c_n$ of the wave functions of the ordinary oscillator
are calculated with respect to the standard measure
$\int_{-\infty}^\infty dr|\phi_n|^2=1$. The normalization constants
$\tilde{c}_n$ of the Dirac states (\ref{con2}) are evaluated
with respect to the measure (\ref{sa6}), 
$\int_0^\infty dr r^2|\Phi_n|^2=1$.
This leads to the relation between the normalization constants
\be
\tilde{c}_{2k+1} = \sqrt{2}\, c_{2k+1}\ .
\label{con3}  
\ee
The transition amplitude for the harmonic oscillator is given
by the spectral decomposition (\ref{pi19})
\be
U_t(r,r') = \sum_{n=0}^\infty c_n^2H_n(r)H_n(r')e^{-(r^2+r'^2)/2}\,
e^{-itE_n}\ ,
\label{con3a}
\ee
where $E_n= n+1/2$ is the energy spectrum.
Let us apply the spectral decomposition (\ref{pi19}) to the 
system with the eigenfunctions (\ref{con2}). The result is \cite{vest88}
\ba
U_t^c(r,r') &=& \frac{1}{rr'}\,\left[ U_t(r,r') - U(r,-r')\right]
\label{con4}\\
&=&\int_{-\infty}^\infty \frac{dr''}{rr''}\, U(r,r'')Q(r'',r')\ ,
\label{con5}
\ea
where the kernel $Q$ is given by
\be
Q(r'',r) = \delta(r''-r') + \delta(r''+r')\ ,
\label{con5a}
\ee
for $r''\in \Rs$ and $r'>0$. The equality (\ref{con4}) follows
from both the symmetry property of the Hermite polynomials  
under the parity transformation $H_n(-r)= (-1)^nH_n(r)$
(so that even $n$ do not contribute to the right-hand side
of (\ref{con4})) and the relation (\ref{con3})
between the normalization constants. Note that there is
no extra factor $1/2$ in the right-hand side of (\ref{con4}).
The kernel (\ref{con3a}) has the standard path integral
representation
\be
U_t(r,r')= \int\limits_{r(0)=r'}^{r(t)=r} {\cal D}r 
\exp\left\{ \frac{i}{2}\int_0^t d\tau \left(\dot{r}^2
-r^2\right)\right\}\ .
\label{con6}
\ee
The measure involves integrations in {\em infinite} limits
over $r$.

Introducing the integration over a momentum variable
in (\ref{con6}), we can see that it coincides with the 
Faddeev-Popov integral in the unitary gauge $x_2=x_3=0$. Indeed,
taking, for example, $p_1x_2 -x_1p_2=0$ and $p_1x_3 -x_1p_3=0$
as independent constraints we find the Faddeev-Popov
determinant $\Delta_{FP}= x_1^2$ which vanishes at the origin
$x_1=0$ indicating the singularity of the unitary gauge. Integrating
then over the nonphysical variables the Faddeev-Popov measure
turns into the Liouville measure for the variables $x_1$ and $p_1$,
thus leading to the path integral (\ref{con6}) $x_1=r$ after
integrating out the momentum variable $p_1$. 
The Faddeev-Popov determinant is canceled against the corresponding
factor resulting from the delta functions of the independent constraints.
Thus, the transition amplitude obtained by the Faddeev-Popov integral
differs from the transition amplitude derived in the Dirac approach.
The reason is that
the unitary gauge is not complete. We have the Gribov transformations
$x_1 \rightarrow -x_1$. The modular domain is the positive half-axis.
In section 6.1 it has been shown that due to a nontrivial topology
of the gauge orbits, there is no smooth single-valued supplementary
condition in the model which would provide a parameterization of
the physical phase space (a cone) by canonical coordinates without
singularities. This is the Gribov obstruction to the Hamiltonian
path integral quantization. The physical reason behind it is the 
non-Euclidean structure of the physical phase space.

The solution to the Gribov obstruction given by the formula (\ref{con5})
implies a simple procedure. First construct the path integral 
in the covering space, i.e., on the whole line; then  symmetrize the 
result with respect to the residual gauge transformations. The operator
$\hat{Q}$ does this job. The transition amplitude on the covering space
does not, in general, coincide with the Faddeev-Popov 
phase-space path integral. Observe the factor $(rr')^{-1}$ in (\ref{con5}).
The deviation would stem from the fact that the insertion of the 
delta-functions of constraints and supplementary conditions into
the path integral measure means the 
elimination of nonphysical degrees of freedom
{\em before} canonical quantization (canonical quantization of
$p^*$ and $q^*$),  while in the Dirac
approach the nonphysical degrees of freedom are excluded 
{\em after} quantization, which is not generally the same. 
The physical degrees of freedom
are frequently described by curvilinear coordinates. That is why
we get the factor $(rr')^{-1}$ which is related to the density $r^2$
in the scalar product. In general, there could also be an operator
ordering correction to the Faddeev-Popov effective action (cf. 
the discussion at the end of section 7.7).  The nonphysical
variables do not disappear without a trace as a consequence 
of the fact that they are associated with curvilinear coordinates.

The sum (\ref{con4}) can be interpreted as the sum
over trajectories inside the modular domain $r>0$.
Due to the gauge invariance, the action of the trajectories
outgoing from $-r'$ to the origin is the same as the action
of the reflected trajectory from $r'$ to the origin. 
This may be interpreted as contributions to the Feynman integral
of trajectories $r(\tau)\geq 0$ outgoing from $r'>0$ and reflecting
from the origin before coming to the final point $r>0$.
The amplitude (\ref{con4}) does not vanish at
$r=0$ or $r'=0$, i.e., the system has a non zero
probability amplitude to reach the horizon. 
The situation is similar to the  path integrals discussed 
in section 8.2. 

\subsection{The path integral in the Weyl chamber}

Let us illustrate the Kato-Trotter product formula (\ref{pi.8}) 
by constructing the 
path integral for the model in the adjoint representation discussed
in Section 3.  Although a direct analysis of the evolution
(Schr\"odinger) equation for a generic potential would  lead to the answer
faster \cite{tmp89,plb89}, it is instructive 
to apply the Kato-Trotter product formula. 
The aim is to show how the integration over the {\em modular domain},
being the Weyl chamber, in the {\em scalar product} turns into the
integration over the {\em entire covering space} (a gauge fixing surface)
in the {\em path integral}. This has already been demonstrated 
when deriving the path integral
on the circle (see (\ref{tp7})--(\ref{tp9})) 
and, as we will show, holds for gauge theories as well.  

The key observation we made in the very end of
section 8.2 is that the kernel (\ref{pi.9}) of the evolution
operator for a free motion
should be modified in accordance with the true geometry of the
physical configuration space. 
To find the right evolution kernel for the free motion, 
we have to solve the Schr\"odinger equation (see (\ref{sag4})) 
in the Dirac operator formalism
\be
- \frac{1}{2\kappa}\Delta_{(r)}\left(\kappa U_t^{0D}(h,h')\right) = 
i\pl_tU_t^{0D}(h,h') \ ,
\label{wc1}
\ee
where the superscript $D$ stands to emphasize that the 
amplitude is obtained via the Dirac operator formalism, and
the superscript $0$ means the free motion as before.
The solution must be regular for all $t>0$ and turn into 
 the unit operator kernel with respect to the scalar
product (\ref{sag4a}) at $t=0$.  According to the analysis of
section 7.3, we make the substitution
$U_t^{0D}(h,h') =[\kappa(h)\kappa(h')]^{-1}
U_t^0 (h,h')$, solve the equation and symmetrize the 
result with respect to the Weyl group. The kernel $U_t^0(h,h')$ satisfies
the free Schr\"odinger equation in $H\sim \Rs^r$. So it is a product
of the kernels (\ref{pi.9}) for each degree of freedom. Thus, we get
\ba
U_t^{D0}(h,h')&=& (2\pi it)^{-r/2} \sum_W \left[\kappa(h)
\kappa(\hat{R}h')\right]^{-1}
\, \exp\left\{\frac{i(h-\hat{R}h')^2}{2t}\right\}
\label{wc2}\\
&=&\int_{H}\frac{dh''}{\kappa(h)\kappa(h'')} 
U_t^0(h,h'')Q(h'',h')\ ;
\label{wc3}\\
Q(h'',h')&=&\sum_W \delta(h''-\hat{R}h'),\ \ \ h''\in H\ ,\ \ \
 h'\in K^+\ .
\label{wc4}
\ea
As $t$ approaches zero, the kernel (\ref{wc2}) turns into the unit
operator kernel
\be 
\<h|h'\> = \sum_W\left[\kappa(h)\kappa(\hat{R}h')\right]^{-1}
\delta(h-\hat{R}h')  
\label{wc5}
\ee
which equals the unit operator kernel 
$[\kappa(h)]^{-2}\delta (h-h')$ for $h,h'$ from the Weyl chamber $K^+$.  
It is noteworthy that by taking the limit $t\rightarrow 0$ 
in the regular solution to the Schr\"odinger equation we have obtained
a W-invariant continuation of the unit operator kernel to the covering
space (the Cartan subalgebra) of the fundamental modular domain
(the Weyl chamber).
 
Due to the W-invariance of the potential $V(\hat{R}h)=V(h)$
(the consequence of the gauge invariance), we also find
\be
\<h|e^{-it\hat{V}}|h'\> = e^{-itV(h)}\<h|h'\>\ .
\label{wc6}
\ee
Thus, the infinitesimal evolution operator
kernel reads
\be
U^D_\epsilon(h,h') = \int_{H}\frac{dh''}{\kappa(h)\kappa(h'')}\,
U_\epsilon(h,h'')Q(h'',h')\ ,
\label{wc7}
\ee
where $U_\epsilon(h,h')$ is the r-dimensional version of
the kernel (\ref{pi10}). We will also write the integral relation
(\ref{wc7}) in the operator form $\hat{U}_\epsilon^D =
\hat{U}_\epsilon\hat{Q}$.
To obtain the folding (\ref{pia12}) that converges
to the path integral, one has to calculate the folding of the kernels 
(\ref{wc7}). The difference from the standard path integral derivation 
of section 8.1
is that the integration domain is restricted to the Weyl chamber and
the $\hat{U}_\epsilon^D$ does not have the standard form (\ref{pi10}).
Next we prove that 
\be
U_t^D=\left(\hat{U}_\epsilon^D\right)^N = 
\left(\hat{U}_\epsilon\hat{Q}\right)^N =
\left(\hat{U}_\epsilon\right)^N\hat{Q}=\hat{U}_t\hat{Q}\ ,
\label{wc8}
\ee
where the folding 
$\left(\hat{U}_\epsilon\right)^N$ is given by the standard 
expression (\ref{pi12}), i.e., {\em without} the restriction
of the integration domain.

To this end we calculate the action of the kernel (\ref{wc7})
on any function $\phi(h)$. We get 
\ba
\hat{U}_\epsilon^D\phi(h)&=&
\int_{H}dh''\int_{K^+}\frac{dh'\kappa^2(h')}
{\kappa(h)\kappa(h'')}\, U_\epsilon(h,h'') Q(h'',h')\phi(h')\\
&=&\int_{H}dh''\, \frac{\kappa(h'')}{\kappa(h)}\,
U_\epsilon(h,h'') \sum_W\Theta_{K^+}(\hat{R}h'') \phi(\hat{R}h'')
\ ,
\ea
where $\Theta_{K^+}(h)$ is the characteristic function of the 
Weyl chamber, i.e., it equals one for $h\in K^+$ and vanishes 
otherwise. To do the integral over $h'$, we use the invariance
of $\kappa^2$ relative to the Weyl group and $\delta(h-\hat{R}h')
=\delta (\hat{R}^{-1}h-h')$ (recall $\det\hat{R}= \pm 1$).
If the function $\phi$ is invariant under the Weyl 
group, then
\be
\hat{U}_\epsilon^D\phi(h)= 
\int_{H}dh''\, \frac{\kappa(h'')}{\kappa(h)}\,U_\epsilon(h,h'')
\phi(h'')\ ,
\ee
because $\sum_W\Theta_{K^+}(\hat{R}h)=1$ except for a set of
zero measure formed by the hyperplanes orthogonal to positive
roots where the Faddeev-Popov determinant $\kappa^2(h)$
in the gauge $x=h$ vanishes.
Taking the W-invariant kernel (\ref{wc7}) as $\phi$, we find that
the kernel $U_{2\epsilon}^D(h,h')$ of the operator $(\hat{U}_\epsilon^D)^2$
has the form (\ref{wc7}) where $\epsilon$ is replaced by $2\epsilon$
and 
\be 
U_{2\epsilon}(h,h'') =\int_{H} d\bar{h}\, 
U_{\epsilon}(h,\bar{h})U_{\epsilon}(\bar{h},h'')\ .
\ee
The proof of (\ref{wc8}) is accomplished by a successive
repeating of the procedure in the folding
$\hat{U}_\epsilon^D\cdots\hat{U}_\epsilon^D$ from left to
right. Thus the path integral has the form \cite{plb89,tmp89}
\be 
U_t^D(h,h') = \sum_W\left[\kappa(h)\kappa(\hat{R}h')\right]^{-1}
\int\limits_{h(0)=\hat{R}h'}^{h(t)=h}{\cal D}
h\ e^{i\int_0^t d\tau [\dot{h}^2/2 - V(h)]}\ .
\label{wc10}
\ee
The exponential contains the gauge-fixed action. Due to the 
Weyl invariance of the action, the sum over the Weyl group
can be interpreted as contributions of trajectories reflected
from the boundary of the Weyl chamber (cf. the analysis
of the harmonic oscillator in Fig. 4 and the discussion
at the end of section 4.5). 
\begin{figure}
\centerline{\psfig{figure=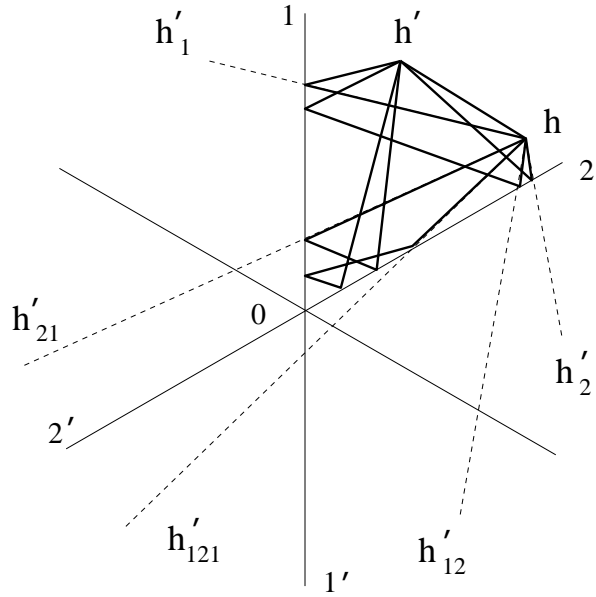}}
\caption{\small The case of SU(3). The modular domain is
the Weyl chamber being the sector $1O2$ with the angle
$\pi/3$. In addition to a trajectory
connecting the initial configuration $h'$ and the final
configuration $h$,
there are five trajectories connecting them and containing
reflections from the Gribov horizon (the Weyl chamber boundary).
All these trajectories contribute to the ``free'' transition
amplitude. The points $h'_{1,2,12,21,121}$ are the Weyl
images of $h'$ obtained by all compositions of 
the mirror reflections with respect to the lines $1O1'$ and $2O2'$. 
For instance, $h'_{12}$ is obtained by the reflection
in $1O1'$ and then in $2O2'$, etc.
Such an interaction with the horizon induces the kinematic
coupling of the physical degrees of freedom. The transition
amplitude cannot be factorized, even though the Hamiltonian
would have {\em no} interaction between the physical degrees of freedom.}  
\end{figure}
That is, a trajectory
outgoing from $\hat{R}h'$ and ending at $h\in K^+$
has the same action as the trajectory outgoing
from $h'\in K^+$, reflecting from the boundary $\pl K^+$
(maybe not once) and ending at $h\in K^+$. An example
of the group SU(3) is plotted in Figure 9. 

We stress again that the reflections
are {\em not} caused by any force action (no infinite
potential well as in the case of a particle in a box).
The physical state of the system is not changed at the 
very moment of the reflection. 
Thanks to the square root of the Faddeev-Popov determinant
at the initial and final points
in the denominator of Eq. (\ref{wc10}), 
the amplitude does {\em not} vanish when either the
initial or final point lies on the boundary of the Weyl
chamber, that is, the system can reach the horizon
with nonzero probability. This is in contrast to
the infinite well case (\ref{bo1}).
The occurrence of the reflected
trajectories in the path integral measure
is the price we have to necessarily pay when 
cutting the hyperconic physical
phase space to unfold it into a part of a Euclidean
space spanned by the canonical coordinates $h$ and $p_h$
and, thereby, to establish the relation between the 
the path integral measure on the hypercone and the 
conventional Liouville phase space measure.
A phase space trajectory $p_h(\tau),
h(\tau), \tau\in [0,t]$ that contributes 
to the phase-space path integral, obtained from (\ref{wc10}) 
by the Fourier transform (\ref{pi18}),
may have {\em discontinuities } since the momentum
$p_h(\tau)$ changes abruptly when the trajectory
goes through the cut on the phase space. Such trajectories
are absent in the support of the path integral measure
for a similar system with a Euclidean phase space. 
The Weyl symmetry of the probability amplitude guarantees
that the physical state of the system does not change
when passing through the cut, which means that the system
does not feel the discontinuity of the phase space trajectory
associated with particular canonical coordinates on the hypercone.

{\em Remark}. The path integral (\ref{wc10}) is invariant
relative to the Weyl transformations. Therefore it has
a {\em unique, gauge-invariant} analytic continuation
into the total configuration space in accordance with the 
theorem of Chevalley (cf. section 7.4). 
It is a function of the independent
Casimir polynomials $P_\nu(x)$ and $P_\nu(x')$, which
can also be anticipated from the spectral decomposition
(\ref{pi19}) over {\em gauge} invariant eigenstates.
Thus, the path integral (\ref{wc10}) does {\em not} depend on any
particular parameterization of the gauge orbit space. 

\subsection{Solving the Gribov obstruction in the 2D Yang-Mills
theory}

The Jacobian (\ref{geo4}) $\kappa^2(a)$ calculated in section 5.2 is
the Faddeev-Popov determinant in the Coulomb gauge
$\pl A=0$ (with the additional condition that $A\in H$). 
Indeed, the Faddeev-Popov operator is
$\{\chi,\sigma \}= \{\pl A,\nabla(A) E\}=-
\pl\nabla(A) $. Since the Coulomb gauge is not complete
in two dimensions (there are homogeneous continuous 
gauge transformations left), the determinant  $\det[- \pl\nabla(A) ]$
should be taken on the space ${\cal F}\ominus {\cal F}_0$,
i.e., homogeneous functions should be excluded
from the domain of the Faddeev-Popov operator (these are
zero modes of the operator $-\pl\nabla(A)$). The residual
continuous gauge arbitrariness generated by the constraints
(\ref{homc}) is fixed by the gauge $A=A_0=a$, where $a$ is
from the Cartan subalgebra. On the surface 
$A_0=a$, the Faddeev-Popov operator
in the space ${\cal F}_0$ of constant functions 
has the form $\{\sigma_0, A_0\} =   {\rm ad}\, A_0=
{\rm ad}\, a$. It vanishes identically on the subspace
of constant functions taking their values in the Cartan
subalgebra ${\cal F}_0^H=H$. This indicates that 
there is still a continuous gauge arbitrariness left. These
are homogeneous transformations from the Cartan subgroup.
They cannot be fixed because they leave the connection
$A=a$ invariant. As we have already remarked, this is due
to the reducibility of the constraints (the Gauss law) 
in two dimensions
(not all the constraints are independent). 
In the reducible case the Faddeev-Popov determinant 
should be defined only
for the set of independent constraints (otherwise it identically
vanishes). Thus, the Faddeev-Popov
operator acts as the operator $-\pl\nabla(a)$
in the space ${\cal F}\ominus{\cal F}_0$ and as
${\rm ad}\, a$ in ${\cal F}_0\ominus {\cal F}_0^H\sim
X\ominus H$.

An additional simplification, thanks to two dimensions, is that 
the determinant  $\det[- \pl$ $\nabla(A)] 
=\det i\pl \det i\nabla (A)$ is factorized, and the infinite constant
$\det i\pl$ can be neglected. On the constraint surface we have
$\nabla(A)=\nabla(a)$.
The operator  $-ig{\rm ad}\, a$ acting in $X\ominus H=
{\cal F}_0\ominus {\cal F}_0^H$ coincides 
with $\nabla(a)$ acting in the same space of constant
functions taking their values in the orthogonal supplement
to the Cartan subalgebra. Thus, the Faddeev-Popov determinant is
$\det i\nabla(a)$, where the operator $i\nabla(a)$
acts in ${\cal F}\ominus {\cal F}_0^H$. 
The determinant $\det i\nabla(a)$
is the Jacobian $\kappa^2(a)$ computed in section 5.4 
modulo some (infinite) constant.
We see again that the Jacobian of the change
of variables associated with the chosen gauge and the gauge
transformation law is proportional to the Faddeev-Popov determinant 
in that gauge.

The Faddeev-Popov determinant vanishes if $(a,\alpha)=n_\alpha a_0$
for any integer $n_\alpha$ and a positive root $\alpha$.
What are the corresponding zero modes of the Faddeev-Popov operator?
Let us split the zero modes into those which belong to the 
space ${\cal F}\ominus{\cal F}_0$ and those from ${\cal F}_0
\ominus {\cal F}_0^H$, i.e., the spatially 
nonhomogeneous and homogeneous ones. They satisfy the equations
\ba
\nabla(a)\xi &=& \pl \xi -ig[a,\xi]=0\ , \ \ \ \ \  
\xi\in {\cal F}\ominus{\cal F}_0\ ; \label{zerm1}\\
\nabla(a)\xi_0&=&  -ig[a,\xi_0]=0\ , \ \ \ \ \ 
\xi_0\in {\cal F}_0\ominus {\cal F}_0^H\ .\label{zerm2}
\ea
A general  solution to Eq. (\ref{zerm1}) reads
\be
\xi(x)= e^{igax}\bar{\xi}e^{-igax}
=\exp[igx({\rm ad}\, a)]\bar{\xi}
\ , \ \ \ \ \bar{\xi}\in {\cal F}_0\ominus {\cal F}_0^H\ .
\label{zerm3}
\ee
The zero modes must be periodic functions $\xi(x+2\pi l)=\xi(x)$
because the space is compactified into a circle of radius $l$.
This imposes a restriction on the connection $A=a$ under which
zero mode exist, and accordingly 
the Faddeev-Popov determinant vanishes at
the connection satisfying these conditions. 
Let us decompose the element $\bar{\xi}$ over the Cartan-Weyl
basis: $\bar{\xi}=\sum_{\alpha>0}(\bar{\xi}_\alpha^{+}e_\alpha
+\bar{\xi}_\alpha^{-}e_{-\alpha})$. 
The constant $\bar{\xi}$ cannot contain a Cartan subalgebra
component, otherwise $\xi(x)$ would have a component from
${\cal F}_0^H$. Making use of the commutation
relation (\ref{A.2}) we find
\be
\xi(x) = \sum_{k=0}^\infty \left[igx({\rm ad}a)\right]^k\bar{\xi}=
\sum_{\alpha>0}\left[e^{igx(a,\alpha)}\bar{\xi}_\alpha^{+}e_\alpha
+ e^{-igx(a,\alpha)}\bar{\xi}_\alpha^{-}e_{-\alpha}
\right]\ .
\label{zerm4}
\ee
Each coefficient in the decomposition (\ref{zerm4}) must
be periodic, which yields
\be 
(a,\alpha)= a_0 n_\alpha \ ,\ \ \ n_\alpha\neq 0\ .
\label{zerm5} 
\ee
We conclude
that the Faddeev-Popov operator has an infinite number
of independent nonhomogeneous zero modes
labeled by all roots $\pm\alpha$ and integers $n_{\pm\alpha}\neq 0$
if the connection is in any of the hyperplanes (\ref{zerm5}).
Each term in the sum (\ref{zerm4}) satisfies Eq. (\ref{zerm1})
and, therefore, can be regarded as an independent zero mode.
The zero modes are orthogonal with respect to the scalar
product $\int_0^{2\pi l}dx (\xi^*_1,\xi_2)$ where $(e_{\alpha})^*
=e_{-\alpha}$ (cf. (\ref{A.5})).
The condition $n_\alpha\neq 0$ ensures that $\xi(x)$ is {\em not}
homogeneous. However, the Jacobian $\kappa^2(a)$ vanishes
on the hyperplanes $(a,\alpha)=0$. 
Where are the corresponding zero modes? They come
from Eq. (\ref{zerm2}). Let us decompose $\xi_0$ over
the Cartan Weyl basis: $\xi_0=
\sum_{\alpha>0}(\xi^+_\alpha e_\alpha +\xi^-_\alpha e_{-\alpha})$.
We recall that $\xi_0$ does not have a Cartan subalgebra component.
From the commutation relation (\ref{A.2}) it follows that
Eq. (\ref{zerm2}) has $\dim G-\dim H$ (the number of all
roots) linearly independent solutions proportional to $e_{\pm\alpha}$, 
provided the connection satisfy the condition
$(a,\alpha)=0$ (cf. also the analysis in section 4.3 between 
(\ref{ifi}) and (\ref{hhh})). So the Faddeev-Popov determinant
should vanish on the hyperplanes $(a,\alpha)=0$ as well. 

The Gribov copies are found by applying the affine Weyl transformations
to configurations on the gauge fixing surface. The fundamental
modular domain is compact and isomorphic to the Weyl cell.
The Faddeev-Popov determinant vanishes on its boundary.
Note also that there are copies {\em inside} of the Gribov region
(i.e., inside the region bounded by zeros of the Faddeev-Popov
determinant), but they are related to one another  by 
{\em homotopically nontrivial} gauge transformations 
which are not generated by the constraints
(see Figure 5 where the case of SU(3) is illustrated).

Now we construct a modified path integral that solves the 
Gribov obstruction in the model \cite{plb93}. Let us take first the simplest
case of SU(2).
We will use the variable $\theta=(a,\omega)/a_0$ introduced 
in section 7.6. The Weyl cell is the open interval $\theta\in (0,1)$
and $\kappa(\theta)= \sin\pi\theta$. The affine Weyl
transformations are 
\be
\theta \rightarrow \theta_{p,n} = p\theta + 2n\ ,\ \ \ p=\pm 1\ ,
\label{go1}
\ee
where $n$ ranges over all integers.
The interval $(0,1)$ is the quotient of the real line by the
affine Weyl group (\ref{go1}). A transition
amplitude is a solution to the Schr\"odinger
equation $(\hbar=1)$,
\be
\left[-\frac{1}{2b\sin(\pi\theta)}\,\frac{\pl^2}{\pl\theta^2}\,
\,\sin(\pi\theta) -
E_C\right]U_t^{D}(\theta,\theta')= i\pl_tU_t^{D}(\theta,\theta')\ ,
\label{sin1}
\ee
that is {\em regular} at the boundaries $\theta=0,1$ and
satisfies the initial condition, 
\be
U_{t=0}^{D}(\theta,\theta')=\<\theta|\theta'\>=
\left[\sin(\pi\theta)\sin(\pi\theta')\right]^{-1}\delta(\theta-\theta')\ ,
\label{sin2}
\ee
where $\theta,\theta'\in (0,1)$. It has the form
\ba
U_t^D(\theta,\theta') &=& \left(2\pi i t b\right)^{-1/2}
\sum_{p=-1}^{1}
\sum_{n=-\infty}^\infty 
\frac{\exp\left\{\frac{i(\theta-\theta_{p,n}')^2}{2tb}+iE_Ct\right\}}
{\sin(\pi\theta)\sin(\pi\theta_{p,n}')}\ .
\label{go2}\\
&=&
\frac{e^{iE_Ct}}{(2\pi itb)^{1/2}}\sum_{n=-\infty}^\infty
\frac{\exp\left\{\frac{i(\theta-\theta'+2n)^2}{2bt}\right\}
-\exp\left\{\frac{i(\theta+\theta'+2n)^2}{2bt}\right\}}
{\sin \pi\theta\,\sin\pi\theta'}\ .      
\label{go2aa}
\ea
We have included all parameters of the kinetic energy 
in the Hamiltonian (\ref{ym10a}) into the constant 
$b=4\pi la_0=4\pi/(lg^2)$. The sum in (\ref{go2}) is 
extended over the residual gauge transformations
(the affine Weyl group), or, in other words, over 
the Gribov copies of the initial configuration $\theta'$
in the gauge fixing surface. 
The regularity of the transition amplitude at $\theta =n$
or $\theta' =n$ is easy to verify. The numerator and
the denominator in the sum
in (\ref{go2aa}) vanish if either $\theta$ or $\theta'$
attains an integer value, but the ratio remains finite
because the zeros are simple. 
The exponential in (\ref{go2}) is 
nothing, but the evolution operator kernel of a free particle
on a line.  It can be written as the path integral with
the standard measure which involves {\em no}
restriction of the integration region to the modular domain.
The action of a free particle coincides with the Yang-Mills
action in two dimensions in the Coulomb gauge. That is,
we have found the way to modify the Faddeev-Popov
reduced phase-space path integral to resolve the
Gribov obstruction. The sum over the Gribov copies 
of the initial configuration $\theta'$ in the covering
space (the gauge fixing surface) can again be interpreted
as contributions of the trajectories that reflect from the 
Gribov horizon several times before they reach the final
point $\theta$. The amplitude does not vanish if the initial
or final point is on the horizon.

A generalization to an arbitrary compact group is straightforward
\cite{plb93}
\be
U_t^D(a,a') = \sum_{W_A}\left[\kappa(a)\kappa(\hat{R}a')\right]^{-1}
\int\limits_{a(0)=\hat{R}a'}^{a(t)=a}{\cal D}
a \ e^{i\int_0^t d\tau [\pi l\dot{a}^2 + E_C]}\ .
\label{go3}
\ee
The path integral for a free particle in $r$ dimensions has the
standard measure. The transition amplitude obviously
satisfies the Schr\"odinger evolution equation. One can also
verify the validity of the representation (\ref{go3}) by the
direct summation of the spectral representation (\ref{pi19})
of the transition amplitude because we know the explicit
form of the eigenstates (\ref{ym14}). However we will give
another derivation of (\ref{go3}) which is more general
and can be used for obtaining a Hamiltonian path integral
for any gauge theory from the Dirac operator formalism.

Consider a spectral decomposition of the unit operator kernel
\be
\<a|a'\> = \sum_E \Phi_E(a)\Phi_E^*(a') = [\kappa(a)\kappa(a')]^{-1}
\delta(a-a')\ , \ \ \ a,a' \in K_W^+\ .
\label{go4}
\ee
The eigenfunctions $\Phi_E(a)$ are
the gauge invariant eigenfunctions (\ref{ym14}),
(\ref{ym15}) reduced  on the gauge fixing surface. 
Therefore the kernel (\ref{go4}) is, in fact, a genuine
unit operator kernel on the gauge orbits space, which does not
depend on any particular parameterization of the latter.
Clearly, $\Phi_E(a)$ are invariant under
the residual gauge transformations, under the affine Weyl
transformations. Let us make use of this fact to obtain a continuation
of the unit operator kernel to the nonphysical region $a\in H$,
i.e., to the whole covering space of the modular domain $K_W^+$.
The following property should hold $\<a|\hat{R}a'\>= \<a|a'\>$
because $\Phi_E(\hat{R}a)=\Phi_E(a)$. Therefore
\ba
\<a|a'\> &=& \sum_{W_A}[\kappa(a)\kappa(\hat{R}a')]^{-1}
\delta(a -\hat{R} a')
\label{go5}\\
&=&\int_H \frac{da''}{\kappa(a)\kappa(a'')}\,\delta(a-a'')Q(a'',a')
 \label{go6}\\
&=& \int_H \frac{da''}{\kappa(a)\kappa(a'')}\,
 \int_H \frac{dp}{(2\pi)^r}\, e^{ip(a-a'')}Q(a'',a')\ ,
\label{go7}
\ea
where $a\in H$ and $a'\in K^+_W$, $pa\equiv (p,a)$; the 
kernel of the operator $\hat{Q}$ is defined as before
\be
Q(a,a') = \sum_{W_A}\delta(a-\hat{R}a')\ .
\label{go7a}
\ee 
The extended unit operator kernel coincides with
the transition amplitude in the limit $t\rightarrow 0$
as we have learned in section 8.2
(see (\ref{tp6})). Now we can construct the infinitesimal transition
amplitude by means of the relation
\be
U_\epsilon^D (a,a') = \left(1-i\epsilon \hat{H}_{ph}(a)\right)
\<a|a'\> + O(\epsilon^2)\ ,
\ee
where the physical Hamiltonian is taken from the Dirac 
quantization method (\ref{ym10a}) (see also (\ref{b11})
for a general case). Applying the physical Hamiltonian
to the Fourier transform of the unit operator kernel
(\ref{go7}), we obtain the following representation
\ba
U_\epsilon^D(a,a') &=&\int_H \frac{da''}{\kappa(a)\kappa(a'')}\,
U_\epsilon(a,a'') Q(a'',a') +O(\epsilon^2)\ ,
\label{go8}\\
U_\epsilon(a,a'') &=& 
 \int_H \frac{dp}{(2\pi)^r}\, \exp\left\{ 
ip(a-a'')- i\epsilon\left(\frac{p^2}{4\pi l} - E_C\right)\right\}
\label{go9}\\
&=&
(i\epsilon/l)^{-r/2} \exp\left\{\frac{i\pi l (a-a'')^2}{\epsilon} +
i\epsilon E_C\right\}\ .
\label{go10}
\ea
The function $p^2/(4\pi l)=H_{ph}$ in (\ref{go9}) 
is the classical gauge-fixed
Hamiltonian, the addition $E_C$ is a quantum correction to it
resulting from the operator ordering.
To obtain (\ref{go10}), we did the Gaussian integral over the
momentum variable. 

The folding $(\hat{U}_\epsilon\hat{Q})^N$ can be calculated
in the same fashion as it has
been done in the preceding section. The only difference
is that the integration in the scalar product is extended 
over the Weyl cell which is {\em compact}. 
Due to the invariance of the amplitude 
$U_\epsilon^D(a,a')$ relative to the affine Weyl transformations,
all the operators $\hat{Q}$ in the folding can be pulled
over to the right with the result that the integration over the Weyl cell
is replaced by the integration over the whole Cartan
subalgebra (the covering space) in the folding
$(\hat{U}_\epsilon)^N$ (thanks to the sum over the 
affine Weyl group generated by $\hat{Q}$).
Thus, the formula (\ref{go3}) is recovered again.

The amplitude (\ref{go3}) has a unique analytic continuation
into the original functional configuration space ${\cal F}$,
which results from the spectral decomposition (\ref{pi19})
and the representation (\ref{ym15}) for the eigenfunctions.
It is a function of two
Polyakov loops for the initial and final configurations of the 
vector potential. 
Therefore the probability amplitude  
does not depend on any particular parameterization
of the gauge orbit space, which has been used to compute
the corresponding path integral. It is a genuine
{\em coordinate-free} 
transition amplitude on the gauge orbit space ${\cal F}/{\cal G}$. 

Replacing the time $t$ by the imaginary one $t\rightarrow -i\beta$, 
one can calculate the partition function 
\be
Z(\beta) =\tr\, e^{-\beta \hat{H}}=
\int_{K_W^+} da \kappa^2(a)U_\beta^D(a,a)=
\sum_{\Lambda_n}e^{-\beta E_n}\ ,
\label{pf}
\ee
where the sum is extended over the irreducible representations
$\Lambda_n$ (see (\ref{ym12})).
Thanks to the sum over the affine Weyl group in (\ref{go3}), the integral
in (\ref{pf}) can be done explicitly. 
The result coincides with the earlier calculation
of the partition function of the 2D Yang-Mills theory on the 
lattice where no gauge fixing is needed since the path integral
is just a finite multiple integral \cite{mig,t3}. The partition function can
also be calculated directly from the spectral decomposition 
(\ref{pi19}) and the orthogonality of the characters of the
irreducible representations (\ref{ym15}) which are eigenfunctions
in the model.

As a conclusion, we comment that the formalism developed above
provides us with a necessary modification
of the Faddeev-Popov Hamiltonian path integral which takes into
account the non-Euclidean geometry of the physical phase space
and naturally resolves the Gribov obstruction. It determines 
an explicitly gauge invariant transition amplitude on the gauge
orbit space. Next we will develop a general method of constructing
such a path integral formalism in gauge theories directly
from the Kato-Trotter product formula {\em without} 
any use of the Schr\"odinger equation. Moreover, the 
new path integral formalism will allow us to {\em deduce}
the corresponding Schr\"odinger equation on the orbit space.

{\em Remark}. 
The method of constructing the Hamiltonian path integral, 
based on the continuation of the unit
operator kernel to the whole gauge fixing surface, 
can be applied to a generic gauge model of the Yang-Mills
type discussed in section 7.7 \cite{plb91,mont}. The effective 
Hamiltonian that emerges in the Hamiltonian path integral
will not coincide with the classical gauge-fixed Hamiltonian
(\ref{b14}). It will contain additional terms corresponding to
the operator ordering corrections
that appear in the Dirac quantum Hamiltonian in (\ref{b11}).
In this way, one can construct the path integral that takes
into account both the singularities of a particular coordinate
parameterization of the orbit space and the operator ordering
which both are essential for the gauge invariance of the
quantum theory as has been argued in section 7.7.

\subsection{The projection method and a modified
Kato-Trotter product formula for the evolution operator in
gauge systems }

The path integral quantization is regarded as an independent
quantization recipe from which the corresponding operator
formalism is to be derived. So far we have explored the other way
around. It is therefore of interest to put forward the following
question. Is it possible to develop a self-contained path
integral quantization of gauge systems that does not rely
on the operator formalism? The answer is affirmative \cite{klsh4}.
The idea is to combine the Kato-Trotter product formula
for the evolution operator in the total Hilbert space and
the projection on the physical (Dirac) subspace. In such
an approach no gauge fixing is needed in the path integral
formalism \cite{pr1,pr2,kl97}. The gauge invariant 
path integral can then be reduced onto any gauge fixing
surface.

Let the gauge group G be compact in a generic gauge model of
the Yang-Mills type discussed in section 7.7.
Consider the projection operator \cite{pr1,pr2,kl97}
\be 
\hat{\cal P} = \int_G d\mu_G(\omega) 
e^{i\omega_a\hat{\sigma}_a}\ ,
\label{pr1}
\ee
where the measure is normalized on unity, $\int d\mu_G =1$,
e.g., it can be the Haar measure of the group G. 
The operators of constraints are assumed to be hermitian.
So, $\hat{\cal P}=\hat{\cal P}^\dagger =\hat{\cal P}^2$.
The Dirac gauge invariant states (\ref{b.4}) are obtained by 
applying the projection operator (\ref{pr1}) to all states
in the total Hilbert space
\be
\Psi(x) = \hat{\cal P}\psi(x) = \int_G d\mu_G(\omega)
\psi(\Omega(\omega)x)\ .
\label{pr2}
\ee 
If the group
is not compact, one can take a sequence of the rescaled projection
operators $c_\delta\hat{\cal P}_\delta$ where $\hat{\cal P}_\delta$
projects on the subspace $\sum_a\hat{\sigma}_a^2\leq \delta$.
In the limit $\delta \rightarrow 0$ a Hilbert space isomorphic to
the Dirac physical subspace is obtained. 
To make the procedure rigorous, the use
of the coherent state representation is helpful \cite{kl97,klles}.
An explicit form of the projection operator kernel in the coherent
state representation for some gauge models has been obtained in
\cite{pr1,ufn,kl97,221}.

From the spectral representation of the evolution operator
kernel (\ref{pi19}) it follows that 
the physical evolution operator is obtained by the projection
of the evolution operator onto the physical subspace in the total
Hilbert space
\be
\hat{U}_t^D = \hat{\cal P}\hat{U}_t\hat{\cal P}\ ,
\label{pr3}
\ee
where the superscript ``D'' stands for ``Dirac''. 
The path integral representation of the physical
evolution operator kernel is then derived by taking
the limit of the folding sequence
\be
\hat{U}_t^D=(\hat{U}_\epsilon^D)^n =
(\hat{\cal P}\hat{U}_\epsilon\hat{\cal P})^n=
\left(\hat{\cal P}\hat{U}_\epsilon^0\hat{\cal P}e^{-i\epsilon \hat{V}}
\right)^n \equiv 
\left(\hat{U}_\epsilon^{0D}e^{-i\epsilon \hat{V}}
\right)^n\ ,
\label{pr4}
\ee
where the gauge invariance of the potential is assumed,
$[\hat{V},\hat{\sigma}_a]=0$.  Equation (\ref{pr4}) is the modified
version of the Kato-Trotter product formula (\ref{pi.8}) 
for the path integral construction
in gauge systems \cite{klsh4}.

Let us see how the main features of the modified reduced phase-space
path integral, like the sum over copies and operator ordering corrections
to the classical action, emerge from this representation.
First of all we reduce the theory on the gauge fixing surface by introducing
new curvilinear coordinates (\ref{b.5}) associated 
with the chosen gauge condition
 and the gauge transformation law. For wave functions we get
\be
\Psi(f(u)) = \int_G d\mu_G(\omega)\psi(\Omega(\omega)f(u))
\equiv \Phi(u)\ .
\label{pr5}
\ee
The invariance of the physical states  (\ref{pr5}) 
with respect to the Gribov
transformations $u\rightarrow \hat{R}u=u_s(u)$ follows
from the relation $f(u) = \Omega_s^{-1}(u)f(u_s)$, which defines the 
Gribov transformations, and the right-shift invariance of the measure
on the group manifold. 
To make use of the modified Kato-Trotter formula (\ref{pr4}), we have
to construct the kernel of $\hat{U}_\epsilon^{0D} =\hat{U}_\epsilon^0
\hat{\cal P}$. Applying the projection operator to the infinitesimal
 evolution operator kernel of a {\em free} motion
in the total configuration space we find
\be
U_\epsilon^{0D}(x,x')= (2\pi i\epsilon)^{-N/2} \int_G d\mu_G(\omega)
\exp\left\{\frac{i\< x-\Omega(\omega)x'\>^2}{2\epsilon}\right\}\ ,
\label{pr6}
\ee
where by $\<x\>^2$ we imply the invariant scalar product
$\<x,x\>$. The kernel (\ref{pr6}) is explicitly gauge invariant.
Reducing it on the gauge fixing surface by the change of
variables (\ref{b.5}) we find
\be
U_\epsilon^{0D}(u,u') = (2\pi i\epsilon)^{-N/2} \int_G d\mu_G(\omega)
\exp\left\{\frac{i\<f(u)-\Omega(\omega)f(u')\>^2}{2\epsilon}\right\}\ ,
\label{pr7} 
\ee
where $u$ and $u'$ belong to the fundamental modular domain
$K$. Formula (\ref{pr7}) determines an analytic continuation
of the transition amplitude to the 
entire gauge fixing surface (the covering space of the 
modular domain $K$). The analytic 
continuation is invariant under the Gribov transformations
\be 
U_\epsilon^{0D}(u, \hat{R}u') =U_\epsilon^{0D}(u, u')\ .
\label{pr7a}
\ee
The evolution of the physical states governed just by
the free Hamiltonian is given by
the equation
\be
\Phi_\epsilon(u) = 
\int_K du'\mu(u')U_\epsilon^{0D}(u,u')\Phi_0(u')\ ,
\label{pr7aa}
\ee
where the density $\mu(u)$ is the Faddeev-Popov
determinant on the gauge fixing surface \cite{babelon}. 
Formulas (\ref{pr6})--(\ref{pr7aa})
are obviously valid for a finite time, $\epsilon\rightarrow t$. This
follows from  the modified Kato-Trotter formula for zero potential $V=0$.

By construction, the kernel (\ref{pr7}) turns into a unit operator
kernel as $\epsilon\rightarrow 0$. Moreover, thanks to the invariance
property (\ref{pr7a}), we get a unique continuation of the 
unit operator kernel to the covering space of the modular domain
\ba
\<u|u'\>&=& \int_Gd\mu_G(\omega)
\delta^N\left(f(u)-\Omega(\omega)f(u')\right)\\
&=&
\sum_{S_\chi} [\mu(u)\mu(\hat{R}u')]^{-1/2}\delta^M(u-\hat{R}u')\ ,
\label{pr8}\\
&=&\int\frac{du''}{[\mu(u)\mu(u'')]^{1/2}}\,\delta^M(u-u'')Q(u'',u')\ ,
\label{pr9}
\ea
where $u$ is a generic point on the gauge fixing surface,
$u'$ belongs to the modular domain and $Q(u'',u')
=\sum_{S_\chi}\delta^M(u''-\hat{R}u')$; the integration in 
(\ref{pr9}) is extended over the whole gauge fixing surface. 
Recall that the functions
$\hat{R}u' = u_s(u')$ are well defined after the modular domain
is identified (see sections 6.2 and 7.7).  
Due to the gauge invariance of the potential
we obviously have
\be 
\<u|e^{-i\epsilon \hat{V}}|u'\>=e^{-i\epsilon V(f(u))}\<u|u'\>\ .
\label{pr10}
\ee
Thus, the basic idea is to project the infinitesimal transition
amplitude of a free motion onto the gauge orbit space, rather than
to reduce the formal local measure $\prod_{\tau =0}^{t}
dx(\tau)$ onto the gauge fixing
surface by means of the Faddeev-Popov identity \cite{fp}
\be
1= \Delta_{FP}(x)\int_Gd\mu_G(\omega)
\delta^{N-M}\left(\chi(\Omega(\omega)x)\right)\ .
\label{fpid}
\ee
From the mathematical point of view, the folding
(\ref{pr4}) of the kernels (\ref{pr7}) and
(\ref{pr10}) leads to a certain measure for the averaging {\em
functions} $\omega=\omega(t)$ in the continuum limit. By making
use of the classical theory of Kolmogorov, one can show that this
measure is a countably additive {\em probability} measure for $\omega(t)$
such that {\em any} set of values of $\omega(t)$ at {\em any} set 
of distinct times is equally likely \cite{cfq1}. 

Our next step is to calculate the averaging integral {\em explicitly} 
by means of
the stationary phase approximation as $\epsilon\rightarrow 0$.
It would be technically rather involved to do this in our general settings.
We shall outline the strategy and turn to concrete examples in next
section to illustrate the procedure.

The stationary phase approximation can be applied
before the reduction of $U_\epsilon^D(x,x')$ on a gauge
fixing surface. No gauge fixing is needed a priori.
A {\em deviation} from the conventional
gauge-fixing procedure results from the fact that there
may be more than just one stationary point.

{\em Remark}. As a point of fact, the averaging integral in
the Faddeev-Popov identity (\ref{fpid}) may also have 
contributions from several points in the gauge parameter space
\cite{th71}. To characterize the path integral measure,
one needs to know the effect of the gauge group averaging
on correlators between neighboring points of a path contributing
to the path integral. Because of the locality of the Faddeev-Popov identity,
such information cannot be obtained from (\ref{fpid}), while
the amplitude (\ref{pr7}) does determine all correlators
between neighboring points on paths on the gauge orbit space.
 
We can always shift the origin of the averaging variable $\omega$ so that
one of the stationary points is at the origin $\omega=0$.
Let $\hat{T}_a$ be operators generating gauge transformations of $x$. 
Decomposing the distance $\<(x-\Omega(\omega)x')\>^2$
in the vicinity of the stationary
point, we find 
\be
\<x-x',\hat{T}_ax'\>=0 \ .\label{dgl}
\ee 
In the formal
continuum limit, $x-x'\approx \epsilon \dot{x}$, we get the condition 
$\sigma_a(\dot{x},x)\equiv 
(\dot{x},\hat{T}_ax)=0$ induced by the averaging procedure.
This is nothing, but the Gauss law enforcement for trajectories 
contributing to the path integral for the folding (\ref{pr4}). 
Suppose there exists a gauge condition
$\chi_a(x)=0$, which involves no time derivatives, such that a generic
configuration $x=f(u)$ satisfying it also fulfills identically 
the discretized Gauss law (\ref{dgl}), i.e., $\<f-f',\hat{T}_af'\>\equiv 0$,
where $f=f(u)$ and $f'=f(u')$.
We will call it a {\em natural} gauge.
In this case all other stationary points in the integral (\ref{pr7})
are $\omega_c=\omega_s$ where $\Omega (\omega_s)f(u)=
f(u_s)$. That is, the transformations $\Omega(\omega_s)$ generate
Gribov copies of the configuration $x=f(u)$ on the gauge fixing
surface. 
Therefore we get a {\em sum} over the stationary points in the 
averaging integral (\ref{pr7}) if the Gribov problem is present.

Still, in the continuum limit we have to control all terms of
order $\epsilon$. This means that we need not only the leading
term in the stationary phase approximation of (\ref{pr7}) but also
the next two corrections to it. Therefore the group element $\Omega(\omega)$
should be decomposed up to order $\omega^4$ because $\omega^4/\epsilon
\sim \epsilon$ as one is easily convinced by rescaling the 
integration variable $\omega \rightarrow \sqrt{\epsilon}\,\omega$. 
The averaging measure should also be decomposed up to the necessary order
to control the relevant $\epsilon$-terms. The latter would yield 
{quantum corrections} to the classical potential associated
with the operator ordering in the kinetic energy operator
on the orbit space. We stress that the averaging procedure
gives a {\em unique} ordering so that the integral is invariant
under general coordinate transformations on the orbit space, i.e.,
does not depend on the choice of $\chi$. Thus, 
\ba
U_\epsilon^{0D}(u,u')&=&(2\pi i\epsilon)^{-M/2}\sum_{S_\chi}
D^{-1/2}(u,\hat{R}u')\nonumber\\
 &\times&\left\{\exp\left[i\< f(u)-f(\hat{R}u')\>^2/2\epsilon - 
i\epsilon \bar{V}_q(u,\hat{R}u')\right] + O(\epsilon^2)\right\}
\label{ad0a}\\
&\equiv& \sum_{S_\chi} D^{-1/2}(u,\hat{R}u') 
\tilde{U}_\epsilon(u,\hat{R}u')\ , 
\label{ad0b}
\ea
where $D(u,u')$ is the conventional determinant arising in the
stationary phase approximation, $\hat{R}u' =u_s(u'), u'\in K$,
and by $\bar{V}_q$ we denote a contribution
of all relevant corrections to the leading order.
The amplitude
$U_\epsilon^D(u,u')$ is obtained by adding $-i\epsilon V(f(u))$
to the exponential in (\ref{ad0a}). Note that 
$V(f(u))=V(f(\hat{R}u))$ thanks
to the gauge invariance of the potential. We postpone for a moment
a discussion of the quantum corrections $\bar{V}_q$.

In general, the equations $\sigma_a(\dot{x},x)=0$ are not
integrable, therefore the natural gauge 
does not always exist. In this case we consider two possibilities.
Let $\Omega_c(u,u')$ be the group element at a stationary
point in (\ref{pr7}). Decomposing the distance in the vicinity
of the stationary point we get $\<f-\Omega_cf',\hat{T}_a
\Omega_cf'\>=0$, $f'=f(u')$. Let $\chi$ be such that the latter condition
is also satisfied if $f(u')$ is replaced by $f(\hat{R}u^\prime)$
where $u'\in K$. Then the sum over the stationary points is again
a sum over the Gribov residual transformations. In Eq. (\ref{ad0a})
we have to replace 
\be
f(\hat{R}u')\rightarrow \Omega_c(u,\hat{R}u') f(\hat{R}u')
\ ,\ \ \  u'\in K\ .
\ee
In the most general case, the sum over stationary points may
not coincide with the sum over Gribov copies in a chosen gauge. However 
for sufficiently small $\epsilon$, the averaged
short-time transition amplitude can always be represented in
the form (\ref{ad0b}) for some $\tilde{U}_\epsilon$. 
Indeed, as $\epsilon$ approaches 
zero, the amplitude $U_\epsilon^{0D}(u,u')$ tends to the unit
operator kernel (\ref{pr8}) that contains the sum over
the Gribov copies. Each delta function in the sum (\ref{pr8})
can be approximated by the corresponding amplitude of a free motion
up to terms of order $\epsilon$. Thus, the sum over copies should
always emerge in the short-time transition amplitude
as $\epsilon$ gets sufficiently small.
A general method to obtain it is to make an asymptotic expansion
of the left-hand side of Eq. (\ref{pr7aa}) as  $\epsilon\rightarrow 0$
after taking the averaging integral in (\ref{pr7}) in the stationary
phase approximation.

The folding of two infinitesimal evolution operator kernels is 
given by
\be
U_{2\epsilon}^{D}(u,u') = \int_K du_1 \mu(u_1) U_\epsilon^D(u,u_1)
U_\epsilon^D(u_1,u')\ .
\label{ad0c}
\ee
Let us replace 
$U_\epsilon^D(u,u_1)$ in (\ref{ad0c}) by the sum (\ref{ad0b}) and
make use of (\ref{pr7a}) applied to the second kernel in (\ref{ad0c}):
$U_\epsilon^D(u_1,u')=U_\epsilon^D(\hat{R}u_{1},u')$. 
Note that $\hat{R}u_1=u_s(u_1)$ and the functions $u_s$ are
well defined because $u_1\in K$. Since the measure
on the orbit space does not depend on a particular choice of
the modular domain, $du_s\mu(u_s)=du\mu(u)$,
we can extend the integration to the entire covering space
by removing the sum over ${S}_\chi$ (cf. (\ref{b.8}))
\be
U_{2\epsilon}^{D}(u,u')= \int du_1 |\mu(u_1)|
D^{-1/2}(u,u_1)
\tilde{U}_{\epsilon}(u,u_1)U_{\epsilon}^{D}(u_1,u')\ .
\label{ad0d}
\ee
The absolute value bars account for a possible sign change
of the density $\mu(u)$ (the Faddeev-Popov determinant on 
the gauge fixing surface).
The procedure can be repeated from left to right 
in the folding (\ref{pr4}), thus removing the restriction 
of the integration domain and the sum over copies in all
intermediate times $\tau\in (0,t)$. The sum over ${S}_\chi$ for the 
initial configuration $u'$ {\em remains} in the integral.

Now we can formally take a continuum limit with the result
\ba
U_t^D(u,u') &=& \sum_{S_\chi}\left[
\mu(u)\mu(\hat{R}u')\right]^{-1/2}\!\!
\!\!\int\limits_{u(0)=\hat{R}u'}^{u(t)=u}\!\!\!{\cal D}u\sqrt{\det g^{ph}}\,
e^{iS_{eff}[u]}\label{ad0e}\\
S_{eff}&=&
\int_0^td\tau \left[(\dot{u},g^{ph}\dot{u})/2 - V_q(u) -V(f(u))\right] \ ,
\label{ad0ee}
\ea
where $g^{ph}_{ij}=g^{ph}_{ij}(u)$ is the induced metric on the orbit space
spanned by local coordinates $u$ (cf. (\ref{b11})). The local 
density $\prod_{\tau=0}^t\sqrt{\det g^{ph}}$ should be understood 
as the result of the integration over the momenta in the 
corresponding time-sliced phase-space path integral where
the kinetic energy is $p_jg_{ph}^{jk}p_k/2$ with $p_j$
being a canonical momentum for $u^j$. A derivation of (\ref{ad0e})
follows the standard technique in the path integral formalism.
One has to set $u' = u -\Delta$ for $u=u(\tau)$ and $u'=u(\tau-\epsilon)$
in each intermediate moment of time $\tau$ and make a decomposition
into the power series over $\Delta$ in every infinitesimal
evolution operator kernel in the folding (\ref{pr4}). 
According to the relation 
between the volume of a gauge orbit through $x=f(u)$, the induced
metric $g^{ph}$, and the Faddeev-Popov determinant \cite{babelon}, we
get 
\be
D(u'+\Delta,u')=\Delta_{FP}^2(f(u'))/\det g^{ph}(u') + O(\Delta)\ , 
\label{8.100}
\ee
where $\Delta_{FP}(f(u)) =\mu(u)$.
Relation (\ref{8.100}) explains the cancellation of the absolute value
of the Faddeev-Popov determinant in the folding (\ref{pr4}) computed 
in accordance with the rule (\ref{ad0d}). The term $\Delta^2/\epsilon$
in the exponential (\ref{ad0a}) gives rise to the kinetic
energy $(\Delta,g^{ph}\Delta)/2\epsilon + O(\Delta^3)$. The metric
$g^{ph}$ can be found from this quadratic form. 

A technically most involved part to calculate is the operator ordering
corrections $V_q(u)$ in the continuum limit. Here we remark that
$D(u,u')$ has to be decomposed up to order $\Delta^2$, while
the exponential in (\ref{ad0a}) up to order $\Delta^4$ because 
the measure has support on paths for which $\Delta^2\sim \epsilon$
and $\Delta^4 \sim \epsilon^2$. There is a technique,
called the equivalence rules for Lagrangian path integrals
on manifolds, which allows one to convert terms $\Delta^{2n}$
into terms $\epsilon^n$ and thereby to calculate
$V_q$ \cite{lpi1,lpi2,lpi3,lpi4} (see also \cite{book} for a detailed review): 
\be    
\Delta^{j_1}\cdots\Delta^{j_{2k}} \rightarrow (i\epsilon\hbar)^k
\sum_{p(j_1,\ldots , j_{2k})}
g_{ph}^{j_1j_2}\cdots g_{ph}^{j_{2k-1}j_{2k}}\ ,
\label{ad0f}
\ee
in the folding of the short-time transition amplitudes,
where the sum is extended over all permutations of the 
indices $j$ to make the right-hand side of (\ref{ad0f})
symmetric under permutations of the $j$'s. 
Following the (\ref{ad0f}) one can derive the Schr\"odinger 
equation for the physical amplitude (\ref{ad0e}).
The corresponding Hamiltonian operator on the orbit space has the form
($\hbar =1$)
\be
\hat{H}_{ph} = -\frac{1}{2 \mu}\,\pl_j\left(\mu\, g_{ph}^{jk}\,
\pl_k\right) + V(f(u))\ ,
\label{ad011}
\ee
where $\pl_j=\pl/\pl u^j$. It can easily be transformed to $\hat{H}^f_{ph}$
in (\ref{b11}) by introducing the hermitian momenta $\hat{p}_j$.
Observe that the kinetic energy in (\ref{ad011}) does {\em not}
coincide with the Laplace-Beltrami operator on the orbit space
because $[\det g^{ph}_{ij}]^{1/2}\neq \mu$.
The operator (\ref{ad011}) is invariant
under general coordinate transformations on the orbit space, i.e., its
spectrum does {\em not} depend on the choice of local coordinates $u$
and, therefore, is gauge invariant. 

Thus, we have developed a self-contained path integral quantization in gauge
theories that takes into account both the coordinate
singularities associated with a parameterization of the 
non-Euclidean physical phase or configuration space and the operator
ordering corrections to the effective
gauge fixed action, which both are important for
the gauge invariance of the path integral. The essential
step was to use the projection on the Dirac physical
subspace directly in the Kato-Trotter representation of the 
evolution operator. It guarantees the unique correspondence
between the path integral and gauge invariant operator formalisms.
The equivalence of the path integral quantization
developed here to the Dirac operator approach discussed
in section 7.7 follows from the simple fact that the projection
operator (\ref{pr1}) commutes with the total Hamiltonian
in (\ref{b.1}) (due the gauge invariance of the latter).
Therefore the evolution (Schr\"odinger) equation in the physical
subspace should have the form
\be
i\pl_t\hat{U}_t^D =
i \pl_t\hat{\cal P}\hat{U}_t\hat{\cal P}=\hat{H}
\hat{\cal P}\hat{U}_t\hat{\cal P} =\hat{H}_{ph}\hat{U}_t^D\ ,
\ee
where $\hat{H}_{ph}$ is given by (\ref{b11}), because the
projection eliminates the dependence on the nonphysical
variables $\theta$ (cf. (\ref{b.5})) in the transition
amplitude in the total configuration space as has been
shown in (\ref{pr7}).

{\em Remark}. If gauge orbits are not compact, the 
integral over the gauge group in (\ref{pr6}) may still
exist, although the measure $d\mu_G(\omega)$ is
no longer normalizable; the Riemann measure
on the gauge orbit  can be taken as the measure
$d\mu_G$. For example, if the gauge group acts as a
translation of one of the components of $x$, say, 
$x_1\rightarrow x_1 +\omega$, the integration over
$\omega$ in the infinite limits with the Cartesian measure
$d\omega$ would simply eliminate $x_1-x_1'$ from the 
exponential in (\ref{pr6}). A general procedure 
of constructing the coordinate-free phase-space path integral
based on the projection method  
in gauge theories has been developed in \cite{cfq1,cfq2}.

\subsection{The modified Kato-Trotter formula for
gauge models. Examples.}

Let us illustrate the main features of the path integral 
quantization method based on the modified Kato-Trotter
formula. We start with the simplest example of the SO(N) model. 
For the pedagogical reasons, 
we do it in two ways. First we calculate the averaging
integral {\em exactly} and then use the result to develop
the path integral. Second, we obtain the same result
using the stationary phase approximation in the average
integral. The latter approach is more powerful and general
since it does not require doing the averaging integral exactly. 
As has been
mentioned in section 8.4, for $N\neq 3$ the kinetic
energy would produce a quantum potential of the form
$V_q = (N-3)(N-1)/(8r^2)$ if the unitary gauge, 
$x_1=r, x_i=0, i\neq 1$, is used to parameterize the orbit space. 
Assuming a spherical
coordinate system (as the one associated with the chosen gauge
and the gauge transformation law) we get for the 
infinitesimal amplitude (\ref{pr7})
\ba
U_\epsilon^{0D}(r,r') &=& 
\frac{{\cal V}_{N-1}e^{i(r^2 + r'^2)/2\epsilon}}
{{\cal V}_N(2\pi i\epsilon)^{N/2}
}\, \int_0^\pi d\theta
\sin^{N-2}\theta \exp\left\{-\frac{irr'}{\epsilon} 
\cos\theta\right\}\label{bes1}\\
&=&
\frac{{\cal V}_{N-1}}{{\cal V}_N(\pi i)^{N/2}}
\frac{\Gamma(\nu +1/2)\Gamma(1/2)}{2\epsilon(rr')^{\nu/2}}\,
J_\nu\left(\frac{rr'}{\epsilon}\right)\, 
e^{i(r^2 + r'^2)/2\epsilon}\ ,
\ea
where ${\cal V}_N$ is the volume of the N-sphere of unit radius,
$\theta$ is the angle between ${\bf x}$ and ${\bf x}'$, $J_\nu$ is the 
Bessel function where $\nu = N/2 -1$. 
The factor ${\cal V}^{-1}_N$ is inserted to normalize the 
averaging measure on unity. As a side remark we
note that for $N=3$ and a {\em finite} time, $\epsilon\rightarrow t$,
Eq. (\ref{bes1}) turns into (\ref{con4}). 
As $\epsilon$ is 
infinitesimally small, we should take the asymptotes of the
Bessel function for a large argument, keeping only the 
terms of order $\epsilon$. Making use of the asymptotic expansion
of the Bessel function \cite{ryz}
\ba
J_\nu(z) &=& 
\sqrt{\frac{2}{\pi z}}\left\{
\cos z_\nu - 
\sin z_\nu\,\,
\frac{\Gamma(\nu +3/2)}{2z\Gamma(\nu-1/2)}\right\}\ ,
\label{bes}\\
z_\nu &=& z- \pi\nu/2 -\pi/4\ ,
\ea
we find, up to terms of order $O(\epsilon^2)$,
\ba
U_\epsilon^{0D}(r,r')&=& (2\pi i\epsilon)^{-1/2}
\int_{-\infty}^\infty
\frac{dr''}{(rr'')^{(N-1)/2}}\,
e^{i(r-r'')^2/(2\epsilon)
-i\epsilon V_q(r)}
Q(r'',r') \ ,\label{bss1}\\
Q(r'',r')&=&\delta(r''-r')+\delta(r''+r')\ ,
\ea
where $V_q= (N-1)(N-3)/(8r^2)$ is the quantum potential.
The first term in the exponential is identified
with the kinetic energy $\dot{r}^2/2$ in the 
effective gauge fixed action, while the second
one is the quantum potential (\ref{b12}). 
The projection method automatically reproduces
the density $r^{N-1}$ of the scalar product measure as a prefactor
of the exponential (the Faddeev-Popov determinant
in the unitary gauge). The existence of the quantum potential
barrier near the Gribov horizon $r=0$ would change the
phase with which trajectories reflected
from the horizon contribute to the sum over paths.  
Observe that there are no absolute value bars in the 
denominator of the integrand in (\ref{bss1}).
The phase is determined by the phases of the two exponentials
in the asymptote of the Bessel function (\ref{bes}).

In the stationary phase approximation of the average
integral (\ref{bes1}), we have to control
the corrections of order $\epsilon$ in the
exponential. The stationary points are $\theta=0$
and $\theta =\pi$. In the vicinity of $\theta=0$,
we decompose $\cos\theta \approx 1-\theta^2/2 +\theta^4/24$
and in the measure $\sin\theta \approx \theta -\theta^3/6$.
The cubic and quartic terms give the contribution of order
$\epsilon$. This can be immediately seen after rescaling the 
integration variable $\theta\rightarrow \theta/\sqrt{\epsilon}$.
Keeping only the $\epsilon$ and $r$ dependencies
and the phase factors, the contribution of the first
stationary point $\theta=0$ to the averaging 
integral can be written in the form 
\be
\frac{e^{i(r -r')^2/2\epsilon}}{(i\epsilon)^{N/2}}
\!\int_0^\infty \! d\theta
\epsilon^{N/2-1}\left(\theta -\frac{\epsilon\theta^3}{6}\right)^{N-2}\!
\left(1-\frac{irr'}{24}\,\epsilon \theta^4\right)e^{-irr'\theta^2/2}\ .
\label{exam1}
\ee
Contributions of the averaging measure and the group element
$\Omega(\omega)$ in (\ref{pr7}) to the next-to-leading
order of the stationary phase approximation 
are given, respectively, by  the  $\theta^3$- and
$\theta^4$-terms in the parenthesis. All the quantum
corrections are determined by them. Indeed, doing the integral
we get 
\be
\frac{e^{i(r -r')^2/2\epsilon}}{(2\pi i\epsilon)^{1/2}(rr')^{(N-1)/2}}
\left(1-\frac{i\epsilon (N-1)(N-3)}{8rr'}\right) + O(\epsilon^2)\ ,
\label{bss2}
\ee
where all numerical factors of (\ref{bes1}) have
now been restored. The expression in the parenthesis is 
nothing but the exponential of the quantum potential
up to terms of order $\epsilon^2$. Similarly, we can
calculate the contribution of the second stationary
point $\theta=\pi$. The result has the form (\ref{bss2})
where $r'\rightarrow -r'$ because $\cos\pi =-1$. Thus,
we have recovered the result (\ref{bss1}) again.

The lesson we can learn from this exercise is the following.
When doing the stationary phase approximation in the average
integral in (\ref{pr7}), the group element $\Omega(\omega)$
and the averaging measure must be decomposed up to such order
in the vicinity of every stationary point that 
the integral would assume the form $\epsilon^{-M/2}A$
where $A$ is decomposed up to $O(\epsilon^2)$ and
$M$ is the number of physical degrees of freedom.  

For the Yang-Mills theory in (0+1) spacetime, we set
$f(u)\rightarrow h\in H$. 
The distance in the exponential (\ref{pr7}) is taken with respect
to the Killing form: $(h-\exp(i{\rm ad}{\omega})h')^2$.
The equation for stationary points is
\be 
i\left(h-e^{i{\rm ad}\,{\omega_c}}h', 
{\rm ad}\,e_{\pm\alpha}( e^{i{\rm ad}{\omega_c}}h')\right) =0\ ,
\label{new1}
\ee
for any positive root $\alpha$. The operators ${\rm ad}\, e_{\pm\alpha}$
generate the adjoint action of $G/G_H$, $G_H$ is the Cartan subgroup,
on the Lie algebra. Equation (\ref{new1}) 
has a trivial solution $\omega_c=0$ because ${\rm ad}e_{\pm\alpha}(h')=
[e_{\pm\alpha}, h']$ is orthogonal to any element of the Cartan subalgebra
$H$. That is, $x=h$ is the natural gauge.
All nontrivial solutions are exhausted by the elements of
the Weyl group: $\exp[i{\rm ad}\,{\omega_c}]=\hat{R}\in W$.
The averaging measure has the form \cite{hel}
\be
d\mu(\omega) = d\omega \det\left\{ (i{\rm ad}\,\omega)^{-1}
\left[e^{i{\rm ad}\,\omega}-1\right]\right\}\ .
\ee
The determinant has to be decomposed up to the second
order in  ${\rm ad}\,\omega$, while the exponential in the 
distance formula up to the fourth order, in the fashion
similar to (\ref{exam1}). The second variation 
of the distance at the stationary point follows from the 
decomposition 
\ba
\left(h-e^{i{\rm ad}\,\omega}h'\right)^2 &=&
\left(h-h' 
-i{\rm ad}\,\omega h' +\frac 12({\rm ad}\,\omega)^2h'\right)^2 + O(\omega^3)
\nonumber\\
&=& \left(h-h'\right)^2 
-\left(\omega, {\rm ad}\,h\,{\rm ad}\,h'(\omega)\right) +O(\omega^3)\ .
\label{dist}
\ea
 Therefore 
\be
D^{1/2}(h,h') =\det\left[ -{\rm ad}\,h\,{\rm ad}\,h'\right]^{1/2} =
\kappa (h)\kappa(h')\ .
\label{dhh}
\ee
For another stationary point, the configuration $h'$ has to be
replaced by $\hat{R}h'$, $\hat{R}\in W$. If we do such a replacement
{\em formally} in (\ref{dhh}), we get an ambiguity. 
Indeed, $\det[i{\rm ad}\,h']=\kappa^2(h') =\kappa^2(\hat{R}h')$,
while the function $\kappa(h')$ can change sign under the Weyl
transformations (cf. (\ref{hhh}) and (\ref{sag6}). 
The question is: How do we define the square root
in (\ref{dhh})? This is a quite subtle and important question
for the formalism being developed in general.  
If we put the absolute value bars in the right-hand side of Eq. (\ref{dhh}),
as it seems formally correct,
the corresponding short-time transition amplitude would not coincide
with the one obtained in section 8.5 by solving the Schr\"odinger
equation. That is, the {\em phase} with which the trajectories
reflected from the Gribov horizon contribute to the sum over paths would
be incorrect. To determine a correct phase, we note that, 
if $\hat{D}$ is a strictly positive operator, then
\be
\int d\omega \exp[-(\omega,\hat{D}\omega)] \sim D^{-1/2}\ ,\ \ \ 
D=\det\hat{D}\ .
\label{gi}
\ee
If $\hat{D}\rightarrow \pm i\hat{D}$, the integral (\ref{gi}) is obtained
by an {\em analytic} continuation of the left-hand side of (\ref{gi}).
In our case $\hat{D}\omega = i{\rm ad}\,h\,{\rm ad}\,h'(\omega)=
i[h,[h',\omega]]$, $\omega\in X\ominus H$, 
because the distance (\ref{dist}) is multiplied by $i$ in the 
exponential (\ref{pr7}). 
Making use of the Cartan-Weyl basis, the quadratic
form can be written as
\be
(\omega,\hat{D}\omega) =i\sum_{\alpha >0}(h,\alpha)(h',\alpha)
\left[(\omega^\alpha_c)^2 +(\omega^\alpha_c)^2\right]\ .
\label{qf}
\ee
Here we have used the commutation relations (\ref{adda6}).
The operator $-i\hat{D}={\rm ad}\, h{\rm ad}\, h'$ is strictly
positive if $h,h'\in K^+$ because $(h,\alpha)>0$ for any
positive root $\alpha$. Recall that a positive root $\alpha$
is a linear combination of simple roots with {\em non-negative}
integer coefficients, and, by definition, a scalar product of
$h\in K^+$ and any simple root is strictly positive.    
The replacement of $h$ or $h'$  by $\hat{R}h$ or $\hat{R}h'$, respectively,
induces permutations and reflections of the roots in the 
product $\prod_{\alpha>0}(h,\alpha)(h^\prime,\alpha)$ which
emerges after the integration over $\omega_{c,s}^\alpha$
since $(\hat{R}h,\alpha)=(h,\hat{R}^{-1}\alpha)$ and the Weyl group
preserves the root pattern. Permutations do not change the product.
A reflection in the hyperplane perpendicular to a positive
root $\alpha$, $\hat{R}\alpha =-\alpha$, 
changes sign of an {\em odd} number of
factors in it and may make some permutations among other 
positive roots, too. Indeed, for any two positive roots, $\beta$ and
$\gamma$, distinct from $\alpha$, the reflection can only occur
pairwise: $\hat{R}\beta =-\gamma$ and $\hat{R}\gamma =-\beta$ because
$\hat{R}^2=1$.  According
to the analytic continuation of (\ref{gi}), each of the integrals
over $\omega^\alpha_{s,c}$ ($\alpha$ fixed) 
would contribute the phase factor $\exp(-i\pi/2)$
when $h'$ is replaced by $\hat{R}h'$ and $\det{\hat{R}}=-1$ (reflection),
thus making together the phase $\exp(-i\pi)=-1=\det\hat{R}$, while
the pairwise reflections give rise to the total phase
$[\exp(-i\pi/2)]^{4k}=1, k=0,1,...$. Therefore,
an analytic continuation of (\ref{dhh}) assumes the form
\ba
D^{1/2}(h,\hat{R}h') &=&\det{\hat{R}}
\det\left[ -{\rm ad}\,h\,{\rm ad}\,\hat{R}h'\right]^{1/2} =
\det\hat{R}\,\,|\kappa (h)\kappa(\hat{R}h')|\nonumber\\
&=&\kappa (h)\kappa(\hat{R}h')\  .
\label{dhha}
\ea

This is the power and the beauty of the new path integral formalism.
A change of
the probability amplitude phase 
after hitting the horizon by the system
is uniquely determined
whatever parameterization of the orbit space is used. In contrast,
in the reduced phase-space quantization the phase change is not unique
and depends on a self-adjoint extension of the kinetic energy operator
in the modular domain. Needless to say, 
the very construction of a self-adjoint 
extension may be an extremely hard technical problem, given the
fact that the modular domain depends on the gauge choice. 

The number of stationary points in the averaging integral can be
infinite. This would indicate that the physical configuration space
may be {\em compact} in certain directions. Feynman conjectured
that a compactification of the configuration space in certain
directions due to the gauge symmetry might be responsible for
the mass gap in the spectrum of (2+1) Yang-Mills theory \cite{feyn21}
(a finite gap between the ground state energy and 
the first excited state energy). We have
seen that such a conjecture is indeed true for (1+1) Yang-Mills theory.
Now we can establish this within our path integral quantization
of gauge theories without solving the Schr\"odinger equation.
The averaging integral is now a {\em functional} integral over
the gauge group ${\cal G}/G_H$. A rigorous definition of the normalized
averaging measure can be given via a lattice regularization of the
theory (see section 10.5). 
To achieve our goal, it is sufficient to calculate
the leading order of the stationary phase approximation for
the averaging integral, for which no lattice regularization is needed. 
The key observation is that  
the sum over an infinite number of stationary
points has a similar effect on the spectrum of a {\em free} 
motion (there is no magnetic field in 2D Yang-Mills theory) 
as the sum over the winding numbers in the 
free particle transition amplitude discussed in section 8.2: 
The spectrum becomes discrete.  

Let us turn to the details. 
The quadratic form in the  exponential
in (\ref{pr6}) assumes the form $\<A -A'{}^{\Omega}\>^2$
for any two configurations $A(x)$ and $A'(x)$. It is the 
distance between two configurations
$A(x)$ and $ A^\prime{}^\Omega(x)$ introduced by
Feynman \cite{feyn21}.  The scalar 
product has the form $\<\, ,\,\>=\int_0^{2\pi l}dx(\, ,\, )$.
According to our general analysis,  
the gauge group average enforces the Gauss law 
$\sigma(\dot{A},A) =\nabla(A)\dot{A}=0$.
The orbit space 
can be parameterized by constant connections $A(x) = a$ taking
their values in the Cartan subalgebra $H$. An infinitesimal
gauge transformation of $a$ has the form $\delta a=\nabla(a)\omega$
where $\omega(x)\in {\cal F}\ominus{\cal F}^H_0$ (cf. sections 5.1
and 8.6). The gauge $A(x)=a$
is the natural gauge because the Gauss law is satisfied identically
$\sigma(\dot{a},a)=\nabla(a)\dot{a}\equiv 0$. Thus, if $\omega(x) =0$
is a stationary configuration, $\Omega(0)=e$, then all other
stationary configurations $\omega_c(x)$ in the functional averaging integral
(\ref{pr7}) must be given by the Gribov transformations of the 
gauge fixed potential $A(x)=a$, i.e., $\Omega(\omega_c)$
generate transformations from the affine Weyl group. 
To find the function $D(a,a')$, we decompose the distance up to
the second order in the vicinity of the stationary point
\ba
\left\<a- a'{}^\Omega\right\>^2 &=&
\left\<a-a'-\nabla(a')\omega +\frac 12[\nabla(a')\omega,\omega]\right\>^2
+ O(\omega^3)\nonumber\\
&=&\left\<a-a'\right\>^2 - \left\<\omega,\nabla(a)\nabla(a')\omega\right\>
+ O(\omega^3)\ .
\ea
The Gaussian functional integration over $\omega$ yields
\be
D^{1/2}(a,a') =\det[-\nabla(a)\nabla(a')]^{1/2}\sim \kappa(a)\kappa(a')\ ,
\label{daa}
\ee
where $\kappa^2(a)\sim \det[i\nabla(a)]$ 
is the Faddeev-Popov determinant in the chosen gauge (cf. section 8.6). 
One should be careful when taking the square root in (\ref{daa}) for
other stationary points, i.e., when $a'\rightarrow \hat{R}a'$, $\hat{R}$ is
from the affine Weyl group, or when $a$ or $a'$ is outside the modular
domain being the Weyl cell.  By making use of the
representation (\ref{geo311}) -- (\ref{geo3}) and the analyticity arguments
similar to those given above to prove (\ref{dhha}), it is not hard
to be convinced that the absolute value bars must be omitted
when taking the square root in (\ref{daa}). Formula (\ref{gi})
is applied to the operator $-\nabla(a)\nabla(a')$ which is 
strictly positive in ${\cal F}\ominus{\cal F}_0^H$ if $a$ and $a'$
are in the modular domain $K_W^+$.

The folding of the short-time transition amplitudes can be computed
along the lines of section 8.5 and leads to the result (\ref{go3}).
To calculate the Casimir energy $E_C$, the higher-order corrections
must be taken into account in addition to the leading term of the
stationary phase approximation as has been explained with the example
of the SO(N) model.   The effects on the energy spectrum
caused by the modification of the path integral (due to the sum
over Gribov copies) can be found
from the pole structure of the trace of the resolvent
\ba
\tr\,\hat{R}(\tau) &=& \tr\,\left(\tau - i\hat{H}\right)^{-1}=
\int_0^\infty dt e^{-\tau t}\,{\rm tr}\,\hat{U}_t^D\ ,
\label{14}\\
{\rm tr}\,\hat{U}_t^D &=& \int_K du\mu(u) U_t^D(u,u)\ .
\label{15}
\ea
In particular, thanks to the sum over infinite number of Gribov copies,
the resolvent for (1+1) Yang-Mills theory has discrete poles
(cf. (\ref{pf}).  Thus, we have verified
Feynman's conjecture for (1+1) Yang-Mills theory 
without any use of the operator formalism.

To illustrate the effects of curvature of the orbit space, we
consider a simple gauge matrix model of section 4.8. 
Let $x$ be a real 2$\times$2
matrix subject to the gauge transformations $x\rightarrow
\Omega(\omega)x$ where $\Omega\in SO(2)$. An invariant
scalar product reads $(x,x')={\rm tr}\,x^Tx'$ with $x^T$ being
a transposed matrix $x$. The total configuration space is $\Rs^4$.  
Let $T$ be a generator of SO(2).  Then $\Omega(\omega)
=\exp(\omega T)$. 
The Gauss law enforced by the projection, $\sigma=
(\dot{x},Tx) =0$, is not integrable.
We parameterize the orbit space by triangular matrices $\rho$, 
$\rho_{21}\equiv 0$ (the gauge $x_{21}=0$). The residual gauge 
transformations form the group $S_\chi=\Z_2$: 
$\rho\rightarrow \pm \rho$.
The modular domain is a positive half-space
$\rho_{11}>0$. According to the analysis of section 4.8, we have
$\mu(\rho)=\rho_{11}$ (the Faddeev-Popov determinant). 
The plane $\rho_{11}=0$ is the Gribov horizon.
The averaging measure in (\ref{pr7}) reads $(2\pi)^{-1}
d\omega$ and the integration is extended over the interval $[0,2\pi)$.
The quadratic form in the exponential in (\ref{pr7}) reads
\be
\left(\rho-e^{\omega T}\rho^\prime\right)^2 =
(\rho,\rho) +(\rho^{\prime },\rho^\prime)
-2(\rho,\rho')\cos\omega  -2(\rho,T \rho^\prime)\sin\omega\ .
\label{matpi1}
\ee
A distinguished feature of this model from those considered above
is that the stationary point is a function of $\rho$ and $\rho^\prime$.
Taking the derivative of (\ref{matpi1}) with respect to $\omega$ and 
setting it to zero, we find 
\be
\omega_c  =\tan^{-1}\frac{(\rho,T \rho^\prime)}
{(\rho,\rho^\prime)}\ ,\ \ \ \omega_c^s=\omega_c +\pi\ .
\ee
The second stationary point $\omega_c^s$ is associated with 
the Gribov transformation $\rho\rightarrow -\rho$. 

A geometrical
meaning of the transformation $\rho^\prime\rightarrow \exp(\omega_c
T)\rho^\prime$ is transparent. 
The distance $[(\rho-\rho^\prime)^2]^{1/2}$
between
two points on the gauge fixing plane  is greater
than the minimal distance between the two gauge orbits
through $x=\rho$ and $x^\prime =\rho^\prime$.
By shifting $x^\prime$ along the gauge orbit to 
 $x^\prime_c = \exp(\omega_c T)\rho^\prime$,
a minimum of the distance between the orbits is achieved.
 In such a way the metric on the orbit
space emerges in the projection formalism. To find its explicit form,
we substitute
$\omega =\omega_c(\rho,\rho')$ into (\ref{matpi1}), set $\rho^\prime =
\rho-\Delta$ and decompose (\ref{matpi1}) in a power series over
$\Delta$. The quadratic term (the leading term) determines
the metric. We get 
\be
(\Delta, g^{ph}(\rho)\Delta)=
(\Delta,\Delta) +(\Delta,T \rho)
(T\rho, \Delta)/(\rho,\rho) \ , 
\label{mmm}
\ee
which coincides with the metric (\ref{mat4}).

In the stationary phase approximation 
the cosine and sine
in (\ref{matpi1}) should be decomposed up to fourth order
in the vicinity of the stationary point. In this model 
quantum corrections do not vanish. The short-time transition
amplitude on the orbit space is
\ba
U_\epsilon^D (\rho,\rho') &=& D^{-1/2}(\rho,\rho')
\tilde{U}_\epsilon(\rho,\rho') +D^{-1/2}(\rho,-\rho')
\tilde{U}_\epsilon(\rho,-\rho')\label{mm1}\\
\tilde{U}_\epsilon(\rho,\rho')&=&
(2\pi i\epsilon)^{-3/2} 
e^{iS_\epsilon(\rho,\rho')} \ ,\label{mm1a}\\
S_\epsilon(\rho,\rho')&=&\frac{1}{2\epsilon} 
\left[(\rho,\rho) +(\rho^\prime,\rho^\prime) -
2D(\rho,\rho')\right]
-\frac{\epsilon}{8D(\rho,\rho')} -\epsilon V(\rho)\ ,
\label{matpi2}
\ea 
where $-2D(\rho,\rho')$ is given by the two last terms in Eq. (\ref{matpi1})
at the stationary point $\omega=\omega_c$. Up to order $\Delta^2$
it can be written in the form
\be
D(\rho,\rho') = \mu(\rho)\mu(\rho')\,\det{}^{-1} 
g^{ph}(\rho')+ O(\Delta^2)\ .
\label{mmm1}
\ee   
Here we have used an explicit form of the metric (\ref{mmm})
and $\mu =\rho_{11}$ to compute $\det g^{ph}=\mu^2/(\rho,\rho)$. 
As before, an analytic continuation of the Gaussian integral (\ref{gi})
must be applied to obtain $D^{-1/2}$ outside the modular domain $\rho_{11}>0$.
The result, expanded into a power series over $\Delta$, is obtained by
taking the square root of the right-hand side of
(\ref{mmm1}) even
though $ \rho$ and $\rho'$ range over the entire gauge fixing surface.
The phase of $D^{-1/2}$ is determined only by the sign of the Faddeev-Popov
determinant $\mu$ at the points $ \rho$ and $\rho'$ because
the determinant of the physical metric is positive. The phase is
invariant under permutations of $ \rho$ and $\rho'$ in
(\ref{mmm1}) because terms $O(\Delta^2)$ in $\det g^{ph}$ do not
affect it. The leading term in (\ref{mmm1}) specifies the phase of
$D^{-1/2}(\rho,\rho')$ in the continuum limit. 

According to (\ref{ad0c})--(\ref{ad0d}), the folding of $N+1$ kernels
(\ref{mm1}) contains the following density
\be
\frac{|\mu_N|\cdots |\mu_1|\,|\mu_0|}{\left[D_{N+1,N}\cdots
D_{2,1}D_{1,0} \right]^{1/2}}
=\frac{\prod_{k=0}^N\det^{1/2} g^{ph}_k}
{\left[\mu_{N+1}\mu_{0}\right]^{1/2}}
+ O(\epsilon)\ ,
\label{mm2}
\ee
with $N$ being the number of integrations in the folding;
$\mu_k=\mu(\rho_k)$, $D_{k,k-1}=D(\rho_k,\rho_{k-1})$ etc,
$k=0,1,...,N+1$, and  $\rho_{0,N+1}$ are initial and final
configurations, respectively. All terms $O(\Delta^2)$ are
assumed to have been converted into $O(\epsilon)$ by means
of the equivalence rule (\ref{ad0f}). In the numerator of
the right-hand side of (\ref{mm2}), the density at
the initial state $\det^{1/2}g^{ph}_0$ can  be
replaced by the density at the final state $\det^{1/2}g^{ph}_{N+1}$.
The choice depends on the base point (pre-point or post-point)
in the definition of the path integral on a curved space.
In other words, the short-time action (\ref{matpi2}) 
in the amplitude (\ref{mm1a})
can be decomposed in powers of $\Delta$ either  at the point $\rho$
(post-point) or at the point $\rho'$ (pre-point). 
Both representations differ in terms of order $\epsilon$.
We have chosen
the pre-point decomposition in $D$ (cf. (\ref{mmm1})) and $S_\epsilon$. 
The base point can be changed by means of the equivalence rules
(\ref{ad0f}).
If we make a Fourier transformation for $\Delta$ in each kernel
(\ref{mm1a}) involved in the folding, the $N+1$ factors in
the numerator of (\ref{mm2}) would cancel against
the same factors resulting from the integrals over momentum variables,
thus producing a local Liouville measure in the formal continuum limit.
The number of momentum integrals should be exactly $N+1$
because it exceeds by one
the number of integrals over configurations (see section 8.1). 

\subsection{Instantons and the phase space structure}

Here we discuss the simplest consequences of the modification
of the path integral for instanton calculus in gauge quantum mechanics.
The instantons are used in quantum theory to calculate tunneling
effects \cite{ins1,van}. Consider a one-dimensional quantum
systems with a periodic potential \cite{ins1}. The ground
state in the vicinity of each potential minima is degenerate.
The degeneracy is removed due to the tunneling effects, and 
the ground state turns into a zone. It appears that knowledge of
the solutions
of the Euclidean equations of motion (the equations of motion
in the imaginary time $t\rightarrow -i\tau$) allows one to
approximately calculate the energy levels in the zone and find
the corresponding wave functions (the $\theta$-vacua).

Let us take the SU(2) model from section 3 with the periodic
potential $V(x) =1-\cos[(x,x)^{1/2}]$. Since the 
cosine is an even function, the potential is a regular function
of the only independent Casimir polynomial $P_2(x)=(x,x)$.
The analogous one-dimensional model has been well studied 
(see, e.g., \cite{ins1} and references therein). In our case
the phase space of the only physical degree of freedom is a cone.

Consider the Euclidean version of the theory. In the Lagrangian
(\ref{ym1}) we replace $t\rightarrow -i\tau$ and $y\rightarrow iy$.
Recall that $y$ is analogous to the time component of the Yang-Mills
potential which requires the factor $i$ in the Euclidean 
formulation \cite{ins1}.
The Lagrangian assumes the form $L\rightarrow L_E=
(D_\tau  x)^2/2 + V(x)$. The dynamics of the only 
physical degree of freedom is described by the element of the Cartan
subalgebra $x=h\lambda_1\in H$ ($\lambda_1$ is the only basis element
of $H\sim\Rs$, $(\lambda_1,\lambda_1)=1$).  Solutions
of the Euclidean equations of motion 
\be
\frac{d}{d\tau}\frac{\pl L_E}{\pl \dot{x}}=\frac{\pl L_E}{\pl x}\ ,\ \ 
\ \ \ \frac{\pl L_E}{\pl \dot{y}}=0\ ,
\ee
where the overdot denotes the Euclidean time derivative $\pl_\tau$,
depend on the arbitrary functions $y=y(\tau)$ whose variations 
generate the gauge transformations of the classical solutions $x(\tau)$
(see section 4.1 for details). Removing the gauge arbitrariness
by imposing the condition $y=0$, we get the following equation
for $h$ (cf. (\ref{ym9}))
\be
\ddot{h}=\sin h\ .
\label{i1}
\ee 
The instanton solution of Eq. (\ref{i1}) has the form \cite{ins1}
\be
h(\tau) = h_{inst}(\tau) = 4 \tan^{-1}\exp(\tau-\tau_c) + 2\pi m\ ,
\ \ \ \tau_c =const\ .
\ee
It connects the local minima of the potential:
$x^2_{inst}\rightarrow (2\pi m)^2$ as $\tau\rightarrow -\infty$,
and $ x^2_{inst}\rightarrow [2\pi (m+1)]^2$ as 
$\tau\rightarrow \infty$, where $x_{inst}(\tau) = h_{inst}(\tau)
\lambda_1$ in the chosen gauge.

Equation (\ref{i1}) is the same as in the analogous one-dimensional
model $L_E = \dot{h}^2/2 + 1 - \cos h$, $h\in\Rs$, i.e.,
with the Euclidean phase space $\Rs^2$. For this model
the wave function of the $\theta$-vacuum is calculated 
as follows \cite{ins1}. First, one finds the amplitude
$U_\tau(2\pi m,2\pi m')$ in the semiclassical approximation
of the corresponding path integral. The instanton solution
serves as the stationary point. In the limit $\tau\rightarrow
\infty$, the main contribution comes from the states of the 
lowest zone (the contributions of higher levels are exponentially
suppressed):
\ba
U_\tau (2\pi m,2\pi m')&=& \<2\pi m|e^{-\tau \hat{H}}|2\pi m'\>
\nonumber\\
&\approx& \int_0^{2\pi}d\theta
\<2\pi m|\theta\>\<\theta|2\pi m'\> e^{-\tau E_\theta}\ ,
\label{i2}
\ea
as $\tau\rightarrow\infty$, where $\theta$ parameterizes
the energy levels $E_\theta$ in the lowest zone. The 
amplitude $\<2\pi m|\theta\>$ is extracted from
the path integral in the semiclassical approximation for
the instanton solution (\ref{i1}). The details can be found
in \cite{ins1} where it is shown that ($\tau\rightarrow \infty$)
\ba
U_\tau (2\pi m,2\pi m')&\approx& \int_0^{2\pi}\frac{d\theta}
{2\pi^{3/2}}\, e^{-i(m-m')\theta } e^{-\tau E_\theta}\ ,
\label{i3}\\
E_\theta &=& \frac 12 - e^{-S_0}S_0 K \cos\theta\ ;
\ea
here $S_0$ is the instanton action, $K$ a constant independent
of $\theta$ (the instanton determinant \cite{ins1}).
The amplitude $\<2\pi m|\theta\>\sim \exp(-im\theta)$
follows from the comparison of (\ref{i2}) and (\ref{i3}).
It specifies the value of the vacuum wave function
$\<h|\theta\>$ in the local minima of
the potential, $h=2\pi m$. Therefore the wave function
$\<h|\theta\>$ can be approximated by the superposition
\be
\<h|\theta\> \approx 
c \sum_{m=-\infty}^\infty e^{-im\theta}\<h|2\pi m\>\ ,
\label{i4}
\ee
where $\<h|2\pi m\> \sim \exp[-(h-2\pi m)^2/2]$ is the ground
state wave function in the oscillator approximation 
in the vicinity of each potential minima.

To find how the above calculations are modified in the case
when the physical degree of freedom has the conic phase space,
one has to take the amplitude $U^D_{\tau}(2\pi m,2\pi m')$
instead of $U_{\tau}(2\pi m,2\pi m')$ in (\ref{i3}). 
In Eq. (\ref{wc10}) we take $W=\Z_2$ (in the SU(2) case)
and replace $t$ by $-i\tau$.
Since the algebra su(2) is isomorphic to so(3), the amplitude
is also given by (\ref{con4}) (where $r\rightarrow h$).
Making use of this relation we find  \cite{boyan}
\be
U^D_\tau(2\pi m,2\pi m')\approx \int_0^{2\pi}\frac{d\theta}
{\pi^{3/2}}\frac{\sin(m\theta)\sin{m'\theta}}{(2\pi)^2mm'}
\, e^{-\tau E_\theta}\ .
\ee
Therefore the change of the phase space structure does not
affect the distribution of the energy levels in the lowest
zone. However, it does affect the amplitudes $\<2\pi m|\theta\>$,
thus leading to the modification of the wave function of the 
$\theta$-vacuum:
\be
\<h|\theta\>^D = c\sum_{m=-\infty}^\infty \frac{\sin m\theta}
{2\pi m}\, \<h|2\pi m\>\ .
\label{i5}
\ee
From the obvious relation $\<-h|2\pi m\>=\<h|-2\pi m\>$
we infer that the function (\ref{i5}) is even,
$\<-h|\theta\>^D=\<h|\theta\>^D$, i.e., invariant under
the residual Weyl transformations, while (\ref{i4})
does not have a definite parity. 

That the energy level distribution in the lowest zone
is not sensitive to the conic phase space structure
holds, in general, only for the continuum spectrum.
One can make the analogy with a free particle. The
change of the phase space structure from the plane
to the cone has no effect on the spectrum. The latter
would not be the case for systems with a discrete 
spectrum, like the harmonic oscillator. A similar
phenomenon might be expected for the instantons.
Consider, for example, the double well potential
$V= ( x^2 -v^2)^2$ and the gauge group SU(2).
The corresponding one-dimensional system has
been well studied \cite{van} where it was shown that 
the lowest zone contains only two levels because the classical
ground state is doubly degenerate. Returning to the gauge model,
we take the gauge $x=h\lambda_1$ so that $h=\pm v$ are classical
minima of the potential. Therefore the lowest zone would also seem 
to contain two levels. 
The lower level has an odd wave function, while
the upper one has an even wave functions. Such an
unusual parity property (typically one expects
the lowest level to have an even wave function) 
is a consequence of the fact that wave functions of the corresponding 
one-dimensional system have to be multiplied by the odd density
factor $(h)^{-1}$ (cf. section 7.3) to get the wave
functions of the gauge system. The reduction
of the phase space from the plane to cone implies that
the odd functions are to be excluded. 

The analysis of more complicated gauge systems would not
add essentially new features. Given a classical
solution, one should evaluate the path integral
in the semiclassical approximation, multiply it
by the Faddeev-Popov determinant at initial and
final configurations raised to the negative $1/2$ power 
as prescribed by (\ref{ad0e}), and symmetrize the
result relative to the residual gauge transformations.

\subsection{The phase space of gauge fields in the 
minisuperspace cosmology}

Another simple effect due to the non-Euclidean
structure of the physical phase space in gauge theories
can be found in the minisuperspace (quantum) cosmology.
Consider the Einstein-Yang-Mills theory. The theory
is complicated for a general analysis, but one can 
introduce a set of simplifying assumptions and consider
closed cosmologies with an $\Rs\times S^3$ topology.
These are known as minisuperspace cosmological models
\cite{hartle}. They are used to study the wormhole
dynamics \cite{verbin}. Wormholes are Riemannian manifolds
which have two or more asymptotically Euclidean regions.
They are believed to play an important role in quantum
gravity \cite{hartle,lav,col}. It is known however that
there is no wormhole solutions of the Einstein equations
in vacuum \cite{eev1,eev2,eev3}. The presence of matter
changes the situation \cite{eev1}. We consider the case
when only gauge fields are present.
The gauge fields on a 
homogeneous space are described by the SO(4)-invariant
Ansatz \cite{verbin,orfeu,orfeuze}. The reduced 
system contains only a finite number of degrees of
freedom of gravitational and gauge fields.  
Our primary interest will be to find  
effects caused by the non-Euclidean geometry of
the physical phase space of
the Yang-Mills fields.

In the minisuperspace approach to the Einstein-Yang-Mills
system, the prototype of a four-dimensional
wormhole may be described by the SO(4) symmetric metric. 
The most general form of such a metric, i.e., a metric 
which is spatially homogeneous and isotropic in the 
spacetime of the $\Rs\times S^3$ topology, is given
by the Friedmann-Robertson-Walker Ansatz \cite{frw}
\be
g_{\mu\nu}dx^\mu dx^\nu = 
\frac{2G_g}{3\pi}\left[ -N^2(t) dt^2 + \rho^2(t)
\theta^i\theta^i\right]\ ,
\label{mini1}
\ee
where $N(t)$ and $\rho(t)$ are arbitrary nonvanishing
functions of time, $G_g$ is the gravitational constant
and $\theta^i$ are the left-invariant one-forms
($i=1,2,3$) on the three-sphere $S^3$ satisfying
the condition $d\theta^i = -\varepsilon_{ijk}\theta^j
\wedge \theta^k$. 
The Ansatz for gauge fields in 
the metric  (\ref{mini1}) has been proposed by 
Verbin and Davidson \cite{verbin} for the group SU(2) and
generalized to an arbitrary group in works \cite{orfeuze,orfeu}.
The gauge fields with the SO(n) group, $n>3$,
are described by a scalar $z(t)\in \Rs$,
a vector ${\bf x}(t) \in \Rs^l, l=n-3$ and a
real antisymmetric $l\times l$ matrix $y=y^a\lambda_a$
with $\lambda_a$ being generators of SO($l$).
The effective Einstein-Yang-Mills action reads
\be
S= \frac 12\int dt\,\frac{N}{\rho}\left\{
-\left(\frac{\rho}{N}\dot{\rho}\right)^2+
\left(\frac{\rho}{N}\dot{z}\right)^2+
\left(\frac{\rho}{N}D_t{\bf x}\right)^2-2V\right\}\ ,
\label{mini2}
\ee
where $D_t$ is the covariant derivative introduced
for the SO(n) models in section 3.
The potential has the form
\be
V= \frac{\alpha_g}{3\pi}\left[
\left(z^2 +{\bf x}^2 -\frac{3\pi}{2\alpha}\right)^2
+ 4z^2{\bf x}^2\right]
-\frac 12 \rho^2 +\frac{\lambda}{2}\rho^4\ ,
\label{wormpot}
\ee
with $\alpha_g=g^2/(4\pi)$ being the Yang-Mills 
coupling constant, $\lambda =
2G_g\Lambda/(9\pi)$, and $\Lambda$ the cosmological
constant.

The action is invariant under the gauge transformations
(\ref{so.2}) and time reparameterizations
\be
t\rightarrow t'(t)\ ,\ \ \ N(t) \rightarrow N(t')
\frac{dt'}{dt}\ .
\ee
Therefore our analysis of the phase space structure
of the gauge fields applies here. The gauge fields
have two physical degrees of freedom. As $z$ is 
gauge invariant, it  has a planar phase space, while
the other physical degree of freedom $|{\bf x}|$ 
has a conic phase space just like in
the model discussed in section 3. 
A quantum theory can be developed by the methods discussed
in sections 7.2 and 8.7. The corresponding
path integral has been obtained
in \cite{plb91m}. It has the same structure as the one
derived in section 8.7, and, hence, leads to a modification
of the semiclassical approximation 
where wormhole solutions play the role of a stationary point. 
Here, however, we study only the classical effects
caused by the non-Euclidean structure of the physical
phase space on the wormhole dynamics, in particular,
on the wormhole size quantization. The wormhole size quantization
was first observed by Verbin and Davidson \cite{verbin}
for Yang-Mills fields with the group SU(2). In this case
$SU(2)\sim SO(3)$, i.e., $l=0$ in the minisuperspace model.
So, the physical phase space is a plane. We need 
the gauge groups of higher ranks to see the effect
of the non-Euclidean
structure of the physical phase space. 

The wormholes are solutions to the Euclidean equations
of motion ($t \rightarrow -i\tau, y\rightarrow iy$)
for the action (\ref{mini2}) with a particular 
behavior for $\rho(\tau)$: $\rho^2(\tau)\sim \tau^2$
as $\tau\rightarrow \pm \infty$. The simplest example
of the wormhole is known as the Tolman wormhole \cite{eev1}.
This is a closed radiation-dominated universe, and
\be
\rho^2(\tau) = 4b^2 + \tau^2\ .
\label{size}
\ee  
The positive constant $b$ is identified as the wormhole
radius (or size). The idea is to find solutions
of the minisuperspace Einstein-Yang-Mills system
which have an asymptotic behavior as (\ref{size}).
It turns out that such solutions exist,
provided the constant $b$ is quantized \cite{verbin}:
\be
b=b_n\sim \Lambda^{-1/2} \exp(-\pi n/\sqrt{2})\ .
\label{whq}
\ee
In the gauge sector the solutions are determined modulo
gauge transformations associated by various choices of
the Lagrange multiplier $y(\tau)$. So we are free to 
fix the gauge so that $x_i(\tau)=\delta_{i1}x(\tau)$
(cf. section 3.2). The time reparameterization gauge
freedom is fixed by going over to the conformal time 
$d\eta = d\tau/\rho(\tau)$.
The use of the conformal time has advantage that the equations
of motion for $\rho$ and gauge fields are decoupled. 
From the action principle we find
\be
\frac{d^2x}{d\eta^2}=\frac{\pl V}{\pl x}\ ,\ \ \ 
\frac{d^2z}{d\eta^2}=\frac{\pl V}{\pl z}\ .
\label{eem1}
\ee
On any line $x=az$ the potential (\ref{wormpot})
has the form of the double well. Therefore the Euclidean
equations of motion (\ref{eem1}) should have periodic
solutions oscillating around the local minima $x=z=0$ of
the Euclidean potential $-V$. For every periodic solution
in the gauge sector, one can find a periodic solution
for $\rho$ \cite{orfeu}.
 The solution
$\rho(\eta)$ is interpreted as a wormhole connecting
two points in the {\em same} space. Therefore the gauge
fields should be the same at both sides of the wormhole. 
Since $z(\eta)$ and $x(\eta)$ are periodic
(with the periods $T_{z,x}$), the period
$T_\rho$ (the Euclidean time between two $\rho$-maxima)
should be an integer multiple of their periods \cite{verbin}
\be
T_\rho = nT_z=m T_x\ .
\label{mini3}
\ee
The relation (\ref{mini3}) leads to the exponential
quantization of the wormhole size \cite{verbin,orfeu}.
For the gauge group $SU(2)$, the integer $n$ determines
the wormhole size quantization (\ref{whq}). For the group
SO(n), $n>3$, the wormhole size depends on both the integers $n,m$
\cite{orfeu}. 

The phase space of the $x$-degree of freedom
is a cone unfoldable into a half-plane. 
Since the $x(\eta)$ oscillates around the origin $x=0$,
the corresponding phase-space trajectory winds about
the phase-space origin.
Therefore
the physical degree of freedom  $x$ needs twice less
time to return to the initial state (see section 3),
that is,
$T_x^{ph}= \frac 12 T_x$, thus leading
to the modification of the wormhole size quantization
rule
\be
T_\rho = nT_z=m T_x^{ph}= \frac m2 T_x\ .
\label{mini3a}
 \ee
If the theory contains fields realizing different
representations of the gauge group, the periods of their
physical oscillations would be determined by degrees
of the independent Casimir operators for a given
representation \cite{plb91m}. The modification
of the wormhole size quantization would have
an effect on quantum tunneling in quantum gravity
involving wormholes.
The minisuperspace quantum theory with gauge fields
and fermions is discussed in \cite{evora}.

\section{Including fermions}
\setcounter{equation}0

So far we have investigated the effects of the non-Euclidean
geometry of the physical phase space 
on classical and quantum dynamics of bosonic systems
with gauge symmetry. In realistic models, gauge and fermionic fields 
are typically coupled in a gauge 
invariant way.  The fermionic degrees of freedom are also
subject to gauge transformations.  However they are described
by Grassmann (anticommutative) variables, so one cannot
eliminate nonphysical degrees of freedom in a gauge theory
by imposing a gauge in the fermionic sector.
A total configuration or phase space of the system 
can be regarded as a superspace spanned by some
number of bosonic and Grassmann variables \cite{super1,super2,super3}.
The definitions (\ref{1}) and (\ref{2}) of the physical
phase and configuration spaces apply in this case too.
If, when calculating the quotient spaces (\ref{1})
or (\ref{2}), one eliminates nonphysical degrees of freedom by
fixing a gauge in the bosonic sector, then the residual
gauge transformations, that might occur, provided the
topology of the gauge orbits is nontrivial, would act
on {\em both} physical  bosonic and fermionic variables
of the corresponding superspace, thus changing its
structure significantly after identifying gauge equivalent configurations. 
The aim of the subsequent analysis
is to investigate the effects of non-Euclidean geometry
of the physical configuration and phase spaces in 
gauge models with fermionic degrees of freedom.
We will see that the kinematic coupling of bosonic
degrees of freedom, which occurs because of a non-Euclidean
geometry of the physical phase space, exists
also for fermionic degrees of freedom, and this, in turn, 
has a significant effect on their quantum dynamics.

\subsection{2D SUSY oscillator with a gauge symmetry}

Consider a simple supersymmetric extension of the SO(2)
gauge model of the isotropic oscillator. The Lagrangian reads
\be
L = \frac 12 \left(\dot{\bf x} - yT{\bf x}\right)^2 
+i \mbf{\psi}^*\left(\dot{\mbf{\psi}} -iy\Gamma\mbf{\psi}
\right) - \frac 12{\bf x}^2 - \mbf{\psi}^*\mbf{\psi}\ .
\label{fer1}
\ee
Here $\mbf{\psi}$ is a two dimensional vector with complex
Grassmann components, $\psi_i$, $ i=1,2$. 
If $\theta_{1,2}$ are two (real) Grassmann elements, $\theta_{1,2}^2=0$,
then we can define a complex Grassmann element by $\psi=
\theta_1 +i\theta_2$ and $\psi^*=\theta_1 -i\theta_2$.
The complex conjugation obeys the following rule $(c\psi_1\psi_2)^*=
c^*\psi_2^*\psi_1^*$ where $c$ is a complex number. 
The matrix $T$ is a generator of SO(2) as before, and $\Gamma$ is 
diagonal matrix, $\Gamma_{11}=-\Gamma_{22}=1$.
The Lagrangian is invariant under the gauge transformations
\be
{\bf x}\rightarrow e^{\omega T}{\bf x}\ ,\ \ \
\mbf{\psi}\rightarrow e^{i\omega \Gamma}\mbf{\psi}\ ,\ \ \
y\rightarrow y +\dot{\omega}\ .
\label{fer2}
\ee
To construct the Hamiltonian formalism for this model, we
have to deal with the second class constraints in the fermionic
sector because the Lagrangian is linear in the velocities 
$\dot{\mbf{\psi}}$ and $\dot{\mbf{\psi}}^*$. The usual way is
to introduce the Dirac bracket and solve the second class
constraints \cite{diraclec}. We observe however that in any
first-order Lagrangian the term linear in velocities,
like  $i\mbf{\psi}^*
\dot{\mbf{\psi}}$, can be regarded as a symplectic one-form.
So the corresponding symplectic structure is obtained by
taking the exterior derivative of it. The same symplectic
structure emerges if one proceeds along the lines of the 
Dirac treatment of the second class constraints. Therefore
we simply assume that the variables $\mbf{\psi}$ and
$\mbf{\psi}^*$ are canonical variables in the fermionic
sector and $\{\psi_j,\psi_k^*\}=\{\psi_k^*,\psi_j\}=-i\delta_{jk}$
by definition.
That is, the action with the Lagrangian (\ref{fer1}) should be
regarded as the {\em Hamiltonian} action for the fermionic degrees
of freedom. 
On a phase space being a supermanifold
the symplectic structure has the following parity transformation
property \cite{martin,martin2}:
\be
\{A,B\}= -(-1)^{p_Ap_B}\{ B,A\}\ ,
\label{par}
\ee
where $p_A$ is the Grassmann parity of the function $A$, i.e.,
$p_A$ is zero, if $A$ is an even element of the Grassmann algebra, 
and one, if $A$ is odd. 
The Poisson bracket for odd functions is symmetric, while
for even functions it is antisymmetric. A generic element of
the Grassmann algebra can always be represented as a sum of
odd and even elements. The Poisson bracket on the superspace
is bilinear and satisfies the Leibnitz rule and the Jacobi
identity which are, respectively, $\{ A,BC\} =
\{ A,B\} C + (-1)^{p_Bp_A}B\{ A,C\}$ and 
$(-1)^{p_Ap_C}\{\{ A,B\},C\} + {\rm cycle\ perm.} =0$.

The Hamiltonian of the model reads
\be
H= \frac 12 {\bf p}^2 +\frac 12 {\bf x}^2 +\mbf{\psi}^*\mbf{\psi}
-y\sigma\ ,
\label{fer3}
\ee
where the secondary constraint 
\be
\sigma = {\bf p}T{\bf x} + \mbf{\psi}^*\Gamma \mbf{\psi} =0
\label{fer4}
\ee 
generates {\em simultaneous} gauge transformation 
in the bosonic and Grassmann sectors of the phase space.
In classical theory, solutions to the equations of motion
are elements of the superspace, i.e., ${\bf x}={\bf x}(t)$
is a general even element of the Grassmann algebra generated
by the initial values of $\mbf{\psi}^*_0=\mbf{\psi}^*(0)$
and $\mbf{\psi}_0=\mbf{\psi}(0)$. In fact, a  generic
interaction $V=V({\bf x},\mbf{\psi},\mbf{\psi}^*)$ between 
fermions and bosons would require such an interpretation
of the classical dynamics on the superspace \cite{book}
because the time derivatives $\dot{\bf x}$ and $\dot{\mbf{\psi}}$
are, respectively, generic even and odd functions on the superspace.
Since there is no preference in the choice of the initial
moment of time, the initial configurations of the bosonic
coordinates and momenta should also be regarded as generic
even elements of the Grassmann algebra. Therefore the 
constraint (\ref{fer4}) is {\em not} 
"decoupled" into two independent
constraints in the bosonic and fermionic sectors.
 
If the nonphysical degrees of freedom are eliminated by
imposing the unitary gauge $x_2=0$, then the residual
gauge transformations would act on {\em both} the bosonic
and fermionic variables
\ba
x_1\rightarrow -x_1&\ ,&\ \ \ \ \mbf{\psi}\rightarrow -\mbf{\psi}\ ,
\label{fer5}\\
p_1\rightarrow -p_1&\ ,& \ \ \ \ 
\mbf{\psi}^*\rightarrow -\mbf{\psi}^*\ ,
\label{fer5aa}
\ea
thus making the corresponding points of the configuration
or phase space physically indistinguishable. Therefore
the physical phase (super)space would not
have a Euclidean structure. One should stress again that
the gauge fixing has been used only to get  local canonical
coordinates on the physical phase space. The geometrical
structure of the physical phase (super)space is certainly
gauge independent. 

To see the effects caused by the non-Euclidean structure
of the physical phase space, let us turn to the Dirac
quantization of gauge systems and compare it with the gauge
fixed description. 
All the degrees of freedom (except the Lagrange
multiplier $y$) are canonically quantized 
by the rule $\{,\}\rightarrow -i[,]$ 
\be
[\hat{x}_j,\hat{p}_k]=i\delta_{jk}\ ,\ \ \ \ 
[\hat{\psi}_j,\hat{\psi}^\dagger_k]_+=\delta_{jk} \ ,
\ee
where $[,]_+$ stands for the anticommutator. 
The Poisson bracket (\ref{par})
is symmetric for odd variables, therefore upon quantization
it should be turned into the anticommutator to maintain
the correspondence principle. Introducing creation
and destruction operators for the bosonic degrees of freedom
(see section 7.1), we write the Dirac constraint equation 
for the physical gauge invariant states in the form
\be
\hat{\sigma}|\Phi\>= \left[
\hat{\bf a}^\dagger T\hat{\bf a} + \hat{\mbf{\psi}}^\dagger
\Gamma\hat{\mbf{\psi}}\right]|\Phi\> =0 \ .
\label{fer6}
\ee
Let $|0\>$ be the vacuum state in the Fock representation, i.e.,
$\hat{\bf a}|0\>=\hat{\mbf{\psi}}|0\>=0$. 
It is a physical state because $\hat{\sigma}|0\>=0$.
Then any physical state
can be obtained by acting on the vacuum by a gauge
invariant function of the creation operators. Thus, to
construct the physical subspace, one has to find all
independent gauge invariant polynomials built out of
$\hat{\bf a}^\dagger$ and $\hat{\mbf{\psi}}^\dagger$.
These are
\ba
\hat{b}_1^\dagger &=& (\hat{\bf a}^\dagger)^2\ ,\ \ \ 
\hat{b}_2^\dagger = \hat{\psi}_1^\dagger\hat{\psi}_2^\dagger\ ,
\label{fer7}\\
\hat{f}_1^\dagger &=& (\hat{a}_1^\dagger+i\hat{a}_2^\dagger)
\hat{\psi}^\dagger_1\ ,
\ \ \ 
\hat{f}_2^\dagger = (\hat{a}_1^\dagger-i\hat{a}_2^\dagger)
\hat{\psi}^\dagger_2\ .
\label{fer7a}
\ea
The operators $\hat{\bf b}^\dagger$ create states with the
bosonic parity, while $ \hat{\bf f }^\dagger$ create 
fer\-mio\-nic
states. Since bosonic and fermionic degrees of freedom can
only be excited in pairs, as one might see from (\ref{fer7})
and (\ref{fer7a}),
we conclude that the spectrum of the supersymmetric
oscillator is
\be
E_n = 2(n_1+n_2+n_3 + n_4)\ ,
\ee
where $n_1$ runs over all non-negative integers, while
$n_{2,3,4} =0,1$ as a consequence of the nilpotence
of the fermionic operators $(\hat{\psi}_1^\dagger)^2=
(\hat{\psi}_2^\dagger)^2=0$. The physical eigenstates are
\be
|{\bf n}\> = c_{\bf n}\, \left(\hat{b}_1^\dagger\right)^{n_1}
\left(\hat{b}_2^\dagger\right)^{n_2}
\left(\hat{f}_1^\dagger\right)^{n_3}
\left(\hat{f}_2^\dagger\right)^{n_4}
|0\>\ ,
\ee
where $c_{\bf n}$ is a normalization constant.
Observe the doubling of the spacing between the oscillator
energy levels in both the fermionic and bosonic sectors, whereas
the Hamiltonian (\ref{fer3}) has the unit oscillator frequency
in the potential, even after the removal of all nonphysical
canonical variables. Our next task is to establish this important
fact in the coordinate (and path integral) approach. 
The goal is to show that the effect is due to the invariance
of the physical states under  
the residual gauge transformations
(\ref{fer5}) acting {\em simultaneously}
on {\em both} bosonic and fermionic degrees of freedom.
We interpret this effect as a consequence of a non-Euclidean
structure of the physical phase  superspace which emerges
upon the identification (\ref{fer5}) and (\ref{fer5aa}).
Had we eliminated the nonphysical variable 
by imposing the unitary gauge and then formally canonically 
quantized the
reduced phase-space system, we would have obtained a 
{\em different}  spectrum which would have the unit spacing between the 
energy levels. 

\subsection{Solving Dirac constraints in curvilinear supercoordinates} 

Consider the Schr\"odinger picture for the above quantum 
supersymmetric oscillator with the gauge symmetry. For the
fermionic degrees of freedom we will use the coherent state
representation as usual \cite{super1}. The states are functions of ${\bf x}$
and a complex Grassmann variable $\mbf{\theta}$ so that
\be
\hat{\mbf{\psi}}^\dagger \Phi = \mbf{\theta}^*  \Phi\ ,\ \ \ 
\hat{\mbf{\psi}}\Phi = \frac{\vec{\pl}}{\pl\mbf{\theta}^*  }\Phi\ ,
\ee
where $\vec{\pl}$ denotes the left derivative with respect to
the Grassmann variables. The scalar product reads \cite{super1}
\be
\<\Phi_1|\Phi_2\> = \int_{\Rs^2}d{\bf x}\int 
d\mbf{\theta}^* d\mbf{\theta} \exp(-\mbf{\theta}^*\mbf{\theta})
\left[\Phi_1({\bf x},\mbf{\theta}^* )\right]^*
\Phi_2({\bf x},\mbf{\theta}^* )
\ . \label{fer8}
\ee
The physical states are invariant under the gauge transformations
generated by the constraints $\hat{\sigma}$
\be
e^{i\omega \hat{\sigma}}\Phi({\bf x}, \mbf{\theta}^* )=
\Phi(e^{\omega T}{\bf x}, e^{-i\omega \Gamma}\mbf{\theta}^* ) 
=\Phi({\bf x}, \mbf{\theta}^* )\ .
\ee
To solve the constraint in the Schr\"odinger representation,
we use again curvilinear coordinates associated with a 
chosen gauge and the gauge transformation law. A new
feature is that the change of variables should be done
on the total superspace since the gauge transformations
act on both commutative and anticommutative coordinates
of the superspace \cite{shja91}. The unitary gauge is the natural one
for the this model. So we introduce the new curvilinear
supervariables $r,\varphi$ and $\mbf{\xi}$ by the relations \cite{mpla91}
\be
{\bf x} = e^{\varphi T}{\bf f}(r) \ ,\ \ \ 
\mbf{\theta}^*  = e^{-i\varphi \Gamma} \mbf{\xi}^*\ ,
\label{fer9}
\ee
where the vector ${\bf f} $ has only one component
$f_i =\delta_{1i}r$. In the bosonic sector, the new variables
are nothing but the polar coordinates. However the angular
variable $\varphi$ also appears in the Grassmann sector as
a parameter of the change of variables.
The variables $r$ and $\mbf{\xi}$ are gauge invariant because
the gauge transformations are translations of $\varphi$. 
Indeed, following the rules of changing variables on the superspace
\cite{super2} we find
\be
\frac{\pl}{\pl\varphi}=
\frac{\pl {\bf x}}{\pl\varphi}\frac{\pl}{\pl {\bf x}}
+\frac{\pl \mbf{\theta}^*}{\pl\varphi}\frac{\pl}{\pl \mbf{\theta}^*}
=(T{\bf x})\,\frac{\pl}{\pl {\bf x}}-i(\Gamma\mbf{\theta}^*)
\,\frac{\pl}{\pl \mbf{\theta}^*} = -i\hat{\sigma}\ .
\label{fer10}
\ee
In the new variables the constraint operator is just the momentum
conjugated to $\varphi$. We stress the importance of changing
variables on the total configuration {\em superspace} to achieve
this result. In this sense the idea of solving the constraints
via the curvilinear coordinates associated with the chosen gauge and
the gauge transformation law has a straightforward generalization
to gauge systems with fermions.

Next, we have to find a physical Hamiltonian. This requires 
a calculation of the Laplace-Beltrami operator in the curvilinear
{\em supercoordinates}. Let us derive it for a special case
 when the change of variable is {\em linear} in the generators
 of the Grassmann algebra \cite{shja91}. 
This would be sufficient to analyze
any gauge model with fermions because the gauge transformations
are usually linear transformations in the fermionic sector. 
Let ${\bf x}$
be a vector from $\Rs^N$ and $\mbf{\theta}$ is an $M$-vector
with components being complex Grassmann variables. Consider
a change of variables
\be
{\bf x} = {\bf x}({\bf y})\ ,\ \ \ \ 
\mbf{\theta}^*= \Omega({\bf y})\mbf{\xi}^*\ ,
\label{fer11}  
\ee
where $\Omega$ is an $M\times M$ matrix. Let ${\bf q}$
and ${\bf Q}$ be  collections
of the old and new supercoordinates, respectively. 
Then taking the differential of the relations (\ref{fer11})
we find the supermatrix $A=A({\bf Q})$ such that
$d{\bf q} = A({\bf Q}) d{\bf Q}$. From this relation follows
the transformation law of the partial derivatives 
$\pl/\pl{\bf q} = A^{-1T}({\bf Q})\pl/\pl{\bf Q}$. In particular,
we find
\ba
\frac{\pl}{\pl x^k}&=& B_k^j({\bf y})\left(\frac{\pl}{\pl y^j }+ 
i\pi_j \right)\  ,
\label{fer12}\\
\pi_j &=& i\mbf{\xi}^*\left(\frac{\pl \Omega}{\pl y^j}\right)
\Omega^{-1T}\, \frac{\pl}{\pl \mbf{\xi}^*}\ ,
\label{fer13}
\ea
where $B^j_k = [(\pl {\bf x}/\pl{\bf y})^{-1}]^j_k$. 
The second term in the right-hand side of Eq. (\ref{fer12})
occurs through the dependence of the new Grassmann variables
on the bosonic variables. 
Making use of the relations (\ref{fer12}) and (\ref{fer13})
we can write the kinetic energy operator in the new curvilinear
supercoordinates
\ba
-\frac 12 \Delta_{(N)} &=&\frac12 \hat{P}_j g^{jk}\hat{P}_k + V_q\ ,
\label{fer14}\\
\hat{P}_k &=& -i\mu^{-1/2}\left(\pl_k + i\pi_k\right)\,  \mu^{1/2}\ , 
\label{fer15}
\ea
where $\pl_k =\pl/\pl y^k$, 
and the quantum potential $V_q$ has the form (\ref{b12}) where
the Jacobian is given by the Berezian (or superdeterminant)
$\mu = {\rm sdet}\, A$ and $g^{jk} = \delta^{mn}B^j_mB^k_m$.
The Jacobian depends only on ${\bf y}$ since the change of
variables is linear in the Grassmann sector.

If the change of variables ${\bf q}={\bf q}({\bf Q})$ is invariant under
the discrete transformations ${\bf q}={\bf q}({\bf Q})=
{\bf q}(\hat{R}{\bf Q})$. Then domain of the new bosonic variables
should be restricted to the modular domain $K\sim \Rs^N/S$ where 
$S$ is formed by all transformations $\hat{R}$, that is,
\be
\int d{\bf x}\phi = \int_K d{\bf y} \mu({\bf y})\phi\ .
\ee
For example, the change of variables (\ref{fer9}) is invariant
under the transformations
\be
r\rightarrow (-1)^nr\ ,\ \ \ \varphi\rightarrow \varphi+ \pi n\ ,
\ \ \ \mbf{\xi}\rightarrow (-1)^n\mbf{\xi}\ .
\label{fer15a}
\ee
The modular domain is $r\in [0,\infty)$ and $\varphi
\in [0,2\pi)$, and the Jacobian is $\mu = r$.

Rewriting the Laplace operator in the quantum Hamiltonian 
in the new variables (\ref{fer9}) and omitting all the derivatives
$\pl/\pl\varphi$ in it we find the Schr\"odinger equation
in the physical subspace
\be
\left(-\frac 12\pl_r^2 -\frac{1}{2r}\,\pl_r
+\frac{1}{2r^2}\, \hat{\sigma}_F + \frac 12\,r^2
+\hat{\mbf{\xi}}^\dagger\hat{\mbf{\xi}} - 1\right)\Phi_E=E\Phi_E\ .
\ee
Here $\hat{\sigma}_F= \hat{\mbf{\xi}}^\dagger\Gamma
\hat{\mbf{\xi}}$.
In the fermionic sector we used the symmetric ordering of
the operators $\mbf{\psi}^*\mbf{\psi}\rightarrow
\hat{\mbf{\psi}}^\dagger \hat{\mbf{\psi}} -1=
\hat{\mbf{\xi}}^\dagger \hat{\mbf{\xi}} -1$.
To solve the Schr\"odinger equation, we split the physical subspace
into four orthogonal subspaces which are labeled by 
quantum numbers of fermions in the corresponding states,
i.e.,  we take $\Phi^{(0)}_E= \Phi^{(0)}_E(r)$, 
$\Phi^{(k)}_E = \xi_k^* F_E^{(k)}(r)$ and 
 $\Phi^{(3)}_E= \xi_1^*\xi_2^* F_E^{(3)}(r)$. These states 
are orthogonal with respect to the scalar product 
\be
\<\Phi_1|\Phi_2\> = \int_0^\infty dr r \int d\mbf{\xi}^*
d\mbf{\xi} e^{-\mbf{\xi}^*\mbf{\xi}}
\left[\Phi_1(r,\mbf{\xi}^*)\right]^*\Phi_2(r,\mbf{\xi}^*)\ .
\ee
The volume  $2\pi$ of  the nonphysical
configuration space spanned by $\varphi$
is included into the norm of the physical
states. The operator $\hat{\sigma}_F$ is 
diagonal in each of the subspaces introduced,
$\hat{\sigma}_F1= \hat{\sigma}_F\xi_1\xi_2=0$
and $\hat{\sigma}_F\xi_k=(\Gamma\mbf{\xi})_k$.
The bosonic wave functions can be found by the same
method used in section 7.2. The regular normalized 
eigenstates and the corresponding eigenvalues are \cite{mpla91}
\ba
\Phi_n^{(0)}&=&\frac{\sqrt{2}}{n!} L_n(r^2) e^{-r^2/2}\ ,
\ \ \ \ \ \ \ \ \ \ \ \ \ E_n^{(0)} = 2n\ ,\label{es1}\\
\Phi_n^{(k)}&=&\frac{\sqrt{2}}{n!\sqrt{n+1}}\,
r\xi_k^* L_n^1(r^2)\, e^{-r^2/2}\ ,\ \ \ 
E_n^{(k)} = 2n+2\ ,\\
\Phi_n^{(3)}&=&\xi_1^*\xi_2^*\Phi_n^{(0)}\ ,\ \ \ 
\ \ \ \ \ \ \ \ \ \ \ \ \ \ E_n^{(3)} = 2n+2 \ ,
\label{es2}
\ea
where $n=0,1,2,...$.
The spectrum is the same as in the Fock representation. 

The wave functions have a unique
gauge invariant continuation into the total configuration
superspace. This follows from the fact that they are
{\em regular} functions of the independent gauge invariant
polynomials $r^2 ={\bf x}^2$, $\xi_1^*\xi_2^* = 
\psi_1^*\psi_2^*$, $r\xi_1^*= z\psi_1^*$ and  $r\xi_2^*= z^*\psi_2^*$,
where $z= x_1 +ix_2$. This is an analog of the theorem
of Chevalley for mixed systems \cite{shja91}.
Now we can see that in the unitary
gauge $x_2=0$ the physical states are {\em invariant}
under the residual gauge transformations (\ref{fer5}),
and, eventually, this symmetry is responsible for
pairwise excitations of the bosonic and fermionic degrees
of freedom since only the compositions $x_1^2,
x_1\mbf{\xi}^*$ and $\xi_1^*\xi_2^*$ are invariant under this
symmetry. Thus, the kinematic coupling
of physical bosonic degrees of freedom, which occurs through
the non-Euclidean structure of their physical
phase space, is also inherent to gauge systems
with bosonic and fermionic degrees of freedom.
This kinematic coupling may considerably affect 
quantum dynamics of the fermionic degrees of freedom
as we proceed to demonstrate. 

\subsection{Green's\  functions\  and\ the\ configuration \
(or phase)\ space\ structure}

In quantum field theory, dynamics of physical excitations
is usually described by Green's functions
which are vacuum expectation values of time ordered
products of the Heisenberg field operators. In gauge theories,
they are calculated in a certain gauge (e.g., a propagator). 
In turn, the gauge
may not be complete, thus leading to some residual gauge
transformations left which reduce the configuration
space of bosonic physical degrees of freedom to a modular
domain on the gauge fixing surface. 
An interesting question is: What happens to fermionic
Green's function? Will they be affected if the configuration
space of bosonic variables is reduced to the modular
domain? The answer is affirmative. 

We illustrate this statement with the example of the 
supersymmetric oscillator with the SO(2) gauge symmetry.
The model is soluble. So all the Green's functions can be
explicitly calculated. We will consider the simplest Green's
function $D_t=\<T(\hat{q}(t)\hat{q}(0))\>_0$, being the 
analogy of the quantum field propagator; $T$ stands for
the time ordered product. Here $\hat{q}(t)$
is the Heisenberg position operator. Taking the Hamiltonian
of a harmonic oscillator for bosonic and fermionic degrees
of freedom
\be
\hat{H} = \hat{b}^\dagger \hat{b} +\hat{f}^\dagger \hat{f} \ ,
\ee
where $[\hat{b},\hat{b}^\dagger]=[\hat{f},\hat{f}^\dagger]_+
=1$, we set $\hat{q}= (\hat{b}^\dagger + \hat{b})/\sqrt{2}$.
Then we find \cite{ramond}
\ba
D_b(t)&=&\<0|T(\hat{q}(t)\hat{q}(0))|0\>= \frac 12\theta(t)e^{-it}
+\frac 12\theta(-t)e^{it}\ ,
\label{gri1}\\
D_f(t)&=&\<0|T(\hat{f}(t)\hat{f}^\dagger(0))|0\>=\theta(t)e^{-it}\ ,
\label{gri2}
\ea
where $\theta(t)$ is the  Heaviside step function.  
It is easy to verify that they satisfy the classical equations
of motion with the source
\be
(-\pl_t^2 - 1)D_b(t)=(i\pl_t -1)D_f(t)=i\delta(t)\ ,
\label{gri2a}
\ee
which define, in fact, the classical Green's functions of the 
Bose- and Fermi-oscillators. The Fourier
transforms, $D(\omega) = \int_{-\infty}^\infty dt\exp(-i\omega t)
D(t)$,  of  the Green's functions have a more familiar form
\be
D_b(\omega) = i\left(\omega^2 -1+i\epsilon\right)^{-1}\ ,\ \ \ \ 
D_f(\omega) = -i\left(\omega -1+i\epsilon\right)^{-1}\ ,
\ee
where $\epsilon >0$ and $\epsilon\rightarrow 0$. The poles
of $D_{b,f}(\omega)$ are determined by the energy of the first 
excited state of the corresponding degree of freedom.

In the unitary gauge,  the SUSY oscillator is described
by the variable $r$ which ranges over the positive semiaxis.
The eigenstates and eigenvectors are given
in Eqs. (\ref{es1})--(\ref{es2}). We 
can investigate the effect of the restriction of the integration
domain in the scalar product on the Green's functions by their
explicit calculation through the spectral decomposition
of the vacuum expectation values
\be
\<0|\hat{q}(t)\hat{q}|0\>=
\sum_{E}e^{-it(E-E_0)}\left|\<0|\hat{q}|E\>\right|^2\ .
\ee 
For the Fourier transforms of 
the two-point functions $D_b^c(t)= \<T(\hat{r}(t)\hat{r})\>_0$
and $D_{fjk}^c(t)= \<T(\hat{\xi}_j(t)\hat{\xi}_k)\>_0$, we obtain
\ba
D_b^c(\omega)&=& \sum_{n=0}^\infty
\frac{\Gamma^2(n-1/2)}{4n!^2}\, \frac{in}{\omega^2 - 4n^2+i\epsilon} \ ,
\label{gri3}\\
D_{fjk}^c(\omega) &= &\delta_{jk}
\sum_{n=0}^{\infty}
\frac{\Gamma^2(n+1/2)}{4n!^2(n+1)}\, \frac{-i}{\omega -2n-2+i\epsilon}\ .
\label{gri4}
\ea
In accordance with the theorem of De Morgan \cite{ww}, the series
(\ref{gri3}) and (\ref{gri4}) are absolutely convergent and define
analytic functions on the complex plane of $\omega$ with simple
poles. Their Fourier transforms 
do {\em not} satisfy the classical equations (\ref{gri2a}).

The reason for such a drastic modification 
of the oscillator
Green's functions is the {\em restriction} of the integration domain
in the {\em scalar product}. 
 In contrast to the ordinary oscillators with a flat phase space
the amplitudes $\<0|\hat{r}|\Phi_n^{(0)}\>$ and 
$\<0|\hat{\xi}_k|\Phi_n^{(k)}\>$ do not vanish for
all $n$, i.e., for all energy levels. In other words, the action
of the operators $\hat{r}$ or $\hat{\xi}_k$ on the ground state
does not excite only the next energy level, but all of them.
One can also say that the variables $r$ and $\mbf{\xi}$ do
{\em not} describe {\em elementary} excitations, but rather
composite objects.
This unusual feature deserves further study. 

To this end we recall the residual symmetry (\ref{fer15a})
of the eigenstates (\ref{es1})--(\ref{es2}).
Making use of it we can continue the physical wave 
functions into the nonphysical domain $r<0$ as well as
extend the integration domain to the whole real line
$\int_0^\infty dr r\phi(r^2) = 
1/2\int_{-\infty}^\infty dr |r|\phi(r^2)$ keeping
the orthogonality of the eigenfunctions. However the
states $\hat{r}\Phi_E=r\Phi_E$ and 
$\hat{\mbf{\xi}}^\dagger\Phi_E= \mbf{\xi}\Phi_E$ occurring in
the Green's functions are
{\em not} invariant under the transformations (\ref{fer15a}).
If we take an {\em analytic} continuation of these 
functions into the covering space, we get the 
obvious result $D_b^c=D_{fjk}^c=0$. This means that
the action of the operators $\hat{r}$ and $\hat{\mbf{\xi}}^\dagger$
throws the states out of the physical subspace. The correspondence
with (\ref{gri3}) and (\ref{gri4}) is achieved when the states
$\hat{r}\Phi_E$ and $\hat{\mbf{\xi}}^\dagger\Phi_E$ are continued
into the covering space to be invariant under the transformations
(\ref{fer15a}), i.e., as $|r|\Phi_E$ and $\varepsilon(r)\mbf{\xi}^*\Phi_E$,
respectively, where $\varepsilon(r)$ is the sign function.
Excitations described by 
the functions $|r|$ and $\varepsilon(r)\mbf{\xi}^*$ would
contain all the powers of the {\em elementary} gauge invariant 
polynomials $r^2$ and $r\mbf{\xi}$, which obviously
describe elementary physical pairwise excitations of the oscillators
in the gauge model. This is why the corresponding
Green's functions contain the sum over the entire spectrum. 

The Green's functions can be calculated in the {\em covering}
space (i.e., on the total gauge fixing surface), provided all
operators in question are replaced by their $S$-invariant
continuations into the covering space $\hat{O}\rightarrow
\hat{Q}\hat{O}=\hat{O}_Q$, where 
\ba
\hat{Q}\Phi(r,\mbf{\xi}^*) =&\,& \int_0^\infty dr'\int
d\mbf{\xi}'^*d\mbf{\xi}' e^{-\mbf{\xi}'^*\mbf{\xi}'}\left[
e^{\mbf{\xi}^*\mbf{\xi}'}\delta(r-r') +\right. \nonumber\\
&+&\left.
e^{-\mbf{\xi}^*\mbf{\xi}'}\delta(r+r')\right]\Phi(r',\mbf{\xi}'{}^*)
\ .\ea
Here the expression in the brackets is the kernel of the 
extending operator $\hat{Q}$. The function $\exp(\mbf{\xi}^*\mbf{\xi}'{})$
is the unit operator kernel in the Grassmann sector. 
It is noteworthy that the kernel of $\hat{Q}$ has the same
structure as, e.g., in (\ref{con5a}), (\ref{wc4}) or (\ref{go7a})
with one natural addition that the residual group acts on both
fermionic and bosonic degrees of freedom in the unit operator kernel.
The kernel
of $\hat{Q}$ is invariant under the transformations (\ref{fer15a})
for its first argument and so is the function $\hat{Q}\Phi$.
In particular, we find
\ba
r&\rightarrow& \hat{Q}r = r_Q=\sum_S \Theta_{\hat{R}K}(r)\hat{R}r
=\varepsilon(r)r= |r|\ ,\\
\mbf{\xi}^* &\rightarrow& \hat{Q}\mbf{\xi}^*
= \mbf{\xi}_Q^*=\sum_S \Theta_{\hat{R}K}(r) \hat{R}\mbf{\xi}^*
=\varepsilon(r)\mbf{\xi}^*\ ,
\ea
where $\Theta_K(r)$ is the characteristic function of the modular
domain $K$ (a half axis in this case) and the sum is extended over
the residual symmetry transformation 
$S$ such that the quotient of the gauge
fixing surface by $S$ is isomorphic to $K$.
The extending  operator $\hat{Q}$ has been introduced 
when studying the path integral formalism for bosonic
gauge theories with a non-Euclidean phase space. 
Here we have a generalization of this concept 
to the simple model with 
fermionic degrees of freedom. Our analysis 
of the Green's functions is also compatible with the
path integral formalism. The Heisenberg operator
$\hat{O}(t)$ is determined by the evolution operator $\hat{U}_t$,
but the latter is modified as $\hat{U}_t\rightarrow 
\hat{U}_t\hat{Q}=\hat{Q}\hat{U}_t\hat{Q}=\hat{U}_t^D$.
Therefore
\be
\<\hat{O}(t)\> = \<\hat{U}_t^{D\dagger}\hat{O}\hat{U}_t\>=
\<\hat{U}_t^\dagger \hat{O}_Q\hat{U}_t\>\ ,\ \ \ \hat{Q}|0\>=|0\>\ ,
\label{fekb}
\ee
where $\hat{U}_t$ is the evolution operator on the covering
space.  If the operator $\hat{O}$ is a reduction
of a {\em gauge invariant} operator on the gauge fixing surface
(like $\hat{O}=\hat{r}^2=\hat{\bf x}^2$ in the above model),
then $\hat{O}_Q=\hat{O}$, and {\em the modular domain
has no effect on its Green's function}, which might have been
anticipated since the dynamics of gauge invariant 
quantities {\em cannot depend on the gauge fixing or on the
way we parameterize the gauge orbit space to regularize
the path integral}.
All we still have to prove is that the evolution operator
has the form $\hat{U}_t\hat{Q}$ when the fermions are added into
the gauge system. 

{\em Remark}. Under certain conditions 
perturbative Green's functions may not
be sensitive to a non-Euclidean structure of 
the phase space. A simple
example is the double well potential discussed at the 
end of section 8.8 and in section 3.5. The potential
has a minimum at $r=v$, so the perturbative 
Green's functions of the operator ${\rho}$ that describes
small fluctuations around the classical vacuum, 
${\rho}={r}-v$, are not affected by the conic 
singularity of the phase space. Indeed, we get 
$\hat{Q}{\rho}= |r|-v\approx r-v$ as long as
$(\< r\>-v)/v<\!\!< 1$ for the states close to the 
perturbative (oscillator) ground state (cf. also
the Bohr-Sommerfeld quantization of the system
discussed in section 3.5). 
The coordinate singularities in the Coulomb
gauge in the 4D Yang-Mills theory seems to be ``far away'' from
the classical vacuum so that the perturbative
Green's functions of gluons are not affected by them
(see section 10). The notion ``far away'' requires 
a dimensional scale in the physical configuration 
space. In 2+1 dimensions it might be constructed out
of a gauge coupling constant (which is dimensional in this 
case) \cite{feyn21}. In the four dimensions, such a scale
could be associated with the curvature of the gauge orbit
space \cite{jak2,baal97}. 
A nonperturbative analysis of Green's functions can be done 
in the $1+1$ QCD on the cylindrical
spacetime (the 2D Yang-Mills theory with fermions in the fundamental
representation). 
The residual gauge transformations from the affine Weyl
group would lead to a specific anomaly because the
Dirac sea (the fermionic vacuum) 
is not invariant under them \cite{lan2}.
The Gribov problem in the Yang-Mills theory with adjoint 
fermions has been studied in  \cite{adjf}.

\subsection{A modified Kato-Trotter formula for gauge
systems with fer\-mions}

Consider a generic gauge system with bosonic and fermionic
degrees of freedom described by commutative variables
$x\in \Rs^N$ and complex Grassmann variables $\psi_k,\ 
k=1,2,..., M$. Let the gauge group act linearly in the 
configuration superspace $x\rightarrow \Omega_b(\omega)x$
and $\psi\rightarrow \Omega_f(\omega)\psi$, where the subscripts
$b$ and $f$ denote the corresponding representations of the
gauge group in the bosonic and fermionic sectors, respectively.
To develop a gauge invariant path integral formalism associated
with the Dirac operator method, we use the projection method
proposed in section 8.7 for the path integral defined via
the Kato-Trotter product formula.

In the coherent state representation of the fermions \cite{super1},
$\hat{\psi}|\psi\> = \psi|\psi\>$, we have
\be
\<\psi|\hat{H}|\psi'\>=
H(\psi^*,\psi')\<\psi|\psi'\>=H(\psi^*,\psi')e^{\<\psi^*,\psi'\>}\ ,
\ee
where $\<\, ,\,\>$ stands for the invariant scalar product in the 
representation space of the gauge group. 
The classical Hamiltonian  $H$ (with possible quantum
corrections due to the operator ordering) is assumed to
be invariant under the gauge transformations. 
For this reason in the Kato-Trotter product formula (\ref{pi.8}), 
the kernel of the ``free'' evolution operator is a product
of the ``free'' evolution operator kernel for the bosonic degrees
of freedom and the {\em unit} operator kernel for the fermionic degrees
of freedom.  By analogy with (\ref{pr6}) we construct the gauge
invariant short-time transition amplitude on the gauge orbit superspace 
\ba
&\ &U^{0D}_\epsilon(x,\psi^*;x',\psi') \label{fek1}\\
&=&(2\pi i\epsilon )^{-N/2} \int_G d\mu_G(\omega)
\exp\left\{\frac{i\<x-\Omega_b(\omega)x\>^2}{2\epsilon}\right\}\,
\exp\<\psi^*,\Omega_f(\omega)\psi'\> \ .
\nonumber
\ea
Due to the explicit gauge invariance of this amplitude,
one can reduce it on any gauge fixing surface, say, $x=f(u)$,
parameterized by a set of variables $u$, just by
changing the variables $x=\Omega_b(\varphi)f(u)$
and $\psi=\Omega_f(\varphi)\xi$ in the superspace.
If the gauge is incomplete and there are discrete residual
transformations determined by the equation 
$f(u_s(u))=\Omega_b(\omega_s(u))f(u)$, then in the limit
$\epsilon\rightarrow 0$ the integral (\ref{fek1}) gets contributions from
several stationary points of the exponential, 
just as in the pure bosonic case (\ref{pr7}) because
the entire time dependence of the kernel (\ref{fek1}) is in its bosonic
part. Therefore the stationary phase
approximation of the gauge group averaging integral
is the same as in the pure bosonic case. 
One should however be aware of the possibility that
the same function $u_s(u)$ may, in general, be generated by
{\em distinct} group elements $\Omega_s$. In the pure
bosonic case the existence of such a degeneracy of
the stationary points in the averaging integral would
lead to a numerical factor in the amplitude. Since
the representations of the physical bosonic and fermionic
variables may be different, the group elements $\Omega_s$
that have the {\em same} action on the bosonic variables
may act {\em differently} on the fermionic variables \cite{shja91}.
The above degeneracy is removed by different contributions
of the fermions in (\ref{fek1}). For instance, if we add 
a multiplet of fermions in the adjoint representation
to the (0+1) SU(2) Yang-Mills model, then the Weyl reflection
$\tau_3\rightarrow -\tau_3$ can be induced by two different
group elements $(-i\tau_1)\tau_3(i\tau_1)=(-i\tau_2)\tau_3(i\tau_2)=
-\tau_3$. As the fermion multiplet has all the components
(no gauge can be imposed on fermions), these groups elements
acts differently on it. Yet, in this particular model,
the stationary group U(1) of $\tau_3$ will act as a continuous
gauge group on the fermionic multiplet, while leaving the boson
variable unchanged. The average over this Cartan group would have
no effect on the ``free'' bosonic amplitude, while it will have
an effect on the fermionic unit operator kernel in (\ref{fek1}).
 
Thus, 
the sum over the residual transformations associated with Gribov
copying on the gauge fixing surface would appear again,
and the residual transformations act on both the bosonic 
and fermionic variables simultaneously. The operator ordering
corrections to the physical kinetic energy of free bosons would
emerge from the pre-exponential factor in the stationary
phase approximation for the gauge group averaging integral
in the limit $\epsilon\rightarrow 0$ as we have illustrated with
the example in the end of section 8.8. 

By analogy with (\ref{pr8}) one can obtain a continuation of 
the unit operator kernel to the total covering space of the 
modular domain
\ba
\<u,\xi|u',\xi'\>&=&\int_Gd\mu_G(\omega)
\delta\left(x -\Omega_b(\omega)x\right)
e^{\<\psi^*,\Omega_f(\omega)\psi'\>} \label{pruo}\\
&=&\int_Gd\mu_G(\omega)
\delta\left(f(u) -\Omega_b(\omega)f(u')\right)
e^{\<\xi^*,\Omega_f(\omega)\xi'\>}\\
&=&\int\frac{du''}{[\mu(u)\mu(u'')]^{1/2}}
\delta(u-u'')e^{\<\xi^*,\xi''\>}Q(u'',\xi''^*;u',\xi')\\
Q(u,\xi^* ;u',\xi')&=&
\sum_{S_\chi} \delta(u-\hat{R}u')e^{\<\xi^*,\hat{R}\xi'\>}\ ,
\label{qfer}
\ea
where $u'$ is from the modular domain, $\hat{R}u'=
u_s(u')$ and $\hat{R}\xi = \Omega_f(\omega_s(u'))\xi$
(observe the $u'$-dependence of the residual gauge
transformations in the fer\-mio\-nic sector).
The kernel (\ref{pruo}) is nothing but the kernel of the
projection operator (\ref{pr1}) for gauge systems with
fermions. It has been used in \cite{corfu} to develop the path integral
formalism in gauge models with a non-Euclidean phase space and
in Yang-Mills theory with fermions, in particular. A general structure
of the kernel (\ref{qfer}) has been analyzed in \cite{shja91}
(see also \cite{book}).
Recent developments of the projection formalism for fermionic
gauge systems can be found in \cite{klj}. 

Since the  bosonic potential, fermionic Hamiltonian
and terms describing coupling between bosons and
fermions are gauge invariant by assumption, we conclude that
the gauge invariant infinitesimal transition amplitude 
{\em reduced} on the gauge fixing surface has the form
\be
U_\epsilon^D(q^*;q') =
\int\frac{dq'' e^{-\<\xi''^*,\xi''\>}}{[\mu(u)\mu(u'')]^{1/2}}\,
U_\epsilon(q^*;q'')Q(q''^*;q')\ ,
\label{fek2}
\ee
where, to simplify the notations, we have introduced 
the supervariable $q$ to denote the collection of 
the bosonic coordinates $u$ and the Grassmann variables $\xi^*$;
accordingly $q^*$ means the set $u$, $\xi$;  $dq\equiv 
dud\xi^* d\xi$,
and  $\mu(u)$ is the Jacobian of the change of variables
on the superspace, or the Faddeev-Popov determinant on
the gauge fixing surface. Here we assume that $\det{\Omega_f}=1$,
which is usually the case in gauge theories of the Yang-Mills type
(otherwise the Jacobian is a product of
the Faddeev-Popov determinant and $\det{\Omega_f}$).
This is no restriction on the formalism being developed.
When necessary, $\det{\Omega_f}$ can be kept in all the formulas,
and the final conclusion that $\hat{U}_\epsilon^D=
\hat{U}_\epsilon\hat{Q},\ \hat{Q}\neq 1$,
is not changed.

To calculate the folding of two kernels (\ref{fek2}),
we first prove the following property of the integration
measure for the modular domain
\be
\int_{K}du\mu(u) \phi = \int_{K_s}du_s(u) \mu(u_s(u))\phi\ ,
\label{fek3}
\ee
where $K_s$ is the range of $u_s(u), u\in K$.
Indeed, since $\det\Omega_f=1$, the Jacobian is fully
determined by the Jacobian in the bosonic sector.
We have $dx=d\mu_G(\omega)du\mu(u)$. Under the 
transformations $u\rightarrow u_s(u)$ and $\Omega(\omega)
\rightarrow \Omega(\omega)\Omega^{-1}(\omega_s(u))$,\
the\ original\ variables\ $x$\ are\  not\  changed,\ so\ 
$dx(u,\Omega)\ =$
$dx(u_s(u),\Omega\Omega^{-1}_s)$. Equation (\ref{fek3}) 
follows from the invariance of the 
measure $d\mu_G$ on the group manifold with respect to the right 
shifts. Eq. (\ref{fek3}) merely expresses the simple fact that
when integrating over the orbit space the choice of a modular
domain is not relevant, any $K_s$ can serve for this purpose.
Consider the action of the infinitesimal evolution operator
(\ref{fek2}) on a function $\Phi(u,\xi^*)$ on the modular domain. 
We have
\ba
\hat{U}_\epsilon^D\Phi  =&\ &
\int dq'' e^{-\<\xi''^*,\xi''\>}
\int_K\frac{dq'\mu(u')e^{-\<\xi'^*,\xi'\>}}{[\mu(u)\mu(u'')]}
\times\nonumber\\
&\ &U_\epsilon(q^*;q'')\sum_{S}
\delta(u''-u_s(u'))e^{\<\hat{R}\xi''^*,\xi'\>}\Phi(q')
\label{feka}
\ea
In the integral over the modular domain we change 
the variables $u'\rightarrow u_s(u')$ and 
$\xi'\rightarrow \hat{R}\xi'$ in each term of the sum
over $S$ (see also section 7.7 for details about the orientation 
of the integration domain in the bosonic sector). 
Making use of (\ref{fek3})
and the relation $\<(\Omega_f\xi)^*,\Omega_f\xi\>=
\<\xi^*,\xi'\>$
we can do the integral over the new variables since
it contains the corresponding delta functions, 
thus obtaining the relation
\be
\hat{U}_\epsilon^D\Phi=
\int dq'' e^{-\<\xi''^*,\xi''\>}
\left(\frac{\mu(u'')}{\mu(u)}\right)^{1/2}
U_\epsilon(q^*;q'') \Phi_Q(q''^*)\ ,
\label{fek3a}
\ee
where the function $\Phi_Q$ is the $S$-invariant 
continuation of the function $\Phi$ outside of the 
modular domain to the whole gauge fixing surface
(or the covering space)
\ba
\Phi_Q(u,\xi^*)&=&\sum_S\Theta_{\hat{R}K}(u)\Phi(\hat{R}^{-1}u,
\hat{R}^{-1}\xi) 
\label{fek4}\\
&=&
\int_Kdu'd\xi'^*d\xi' e^{-\<\xi'^*,\xi'\>}
Q(u,\xi^*;u',\xi')\Phi(u',\xi'^*)\ .
\ea
Here by $\hat{R}^{-1}u$ we imply the function $u_s^{-1}:
K_s=\hat{R}K\rightarrow K$. Recall that the function
$u_s(u)$ determines a one-to-one correspondence between 
the domain $K$ and the range $K_s=\hat{R}K$,
so the inverse function has the domain 
$\hat{R}K$ and the range $K$. 
 The physical wave function are gauge invariant
and therefore they are well defined on the entire gauge
fixing surface and invariant under under the $S$-transformations.
Thus, the action of $\hat{Q}$ 
does {\em not} change physical Dirac states {\em reduced}
on the gauge fixing surface since $\sum_S\Theta_{\hat{R}K}(u)
=1$ just like in the example right after (\ref{wc8}). 
Taking instead of $\Phi$ the gauge invariant infinitesimal evolution operator
kernel (\ref{fek1}) reduced on the gauge fixing surface 
(see (\ref{fek2})), we immediately conclude
that the relation (\ref{feka}) holds for  the folding 
$\hat{U}_{2\epsilon}^D=\hat{U}_{\epsilon}^D\hat{U}_{\epsilon}^D
=\hat{U}_{2\epsilon}\hat{Q}$ where the folding $
\hat{U}_{2\epsilon}=\hat{U}_{\epsilon}\hat{U}_{\epsilon}$
is taken with the standard measure $ du d\xi^* d\xi
\exp(-\<\xi^*,\xi\>)$ and the integration over $u$ is extended over
the whole gauge fixing surface.  
Indeed, when $\Phi_Q$ is replaced by the kernel (\ref{fek2})
in (\ref{fek3a}), then $\hat{Q}\hat{U}_\epsilon^D=\hat{U}_\epsilon^D$,
thanks to the gauge invariance of the projected kernel (\ref{fek2}),
and the factor $[\mu(u'')]^{1/2}$ in (\ref{fek3a}) is canceled
against the corresponding factor $[\mu]^{-1/2}$ in the evolution
operator kernel (\ref{fek2}). 
The path integral representation of $\hat{U}_t$ is given
by the Faddeev-Popov reduced phase space integral
modulo the operator ordering corrections whose exact
form can be calculated from the stationary phase approximation
of the group averaging integral (\ref{fek1}) as has been explained
in section 8. 
This accomplishes the proof of the formula (\ref{fekb})
which was essential for an understanding of the effects
of the modular domain on the gauge fixed Green's functions.

{\em Remark}. To calculate the operator ordering terms,
it is sufficient to decompose $\Omega_f$ up to second order in 
the vicinity of the stationary point, just as the measure
$d\mu_G(\omega)$, because the fermionic exponential in 
(\ref{fek1}) does not contain $\epsilon^{-1}$. 
The second order terms will contribute to the quantum potential,
and therefore the latter may, in general, depend on fermionic
variables.
 
 \section{On the gauge orbit space geometry and gauge fixing in
realistic gauge theories}
\setcounter{equation}0

The non-Euclidean geometry of the physical phase space may significantly
affect quantum dynamics. In particular, a substantial modification
of the path integral formalism is required. 
This should certainly be expected 
to happen in realistic gauge theories. Unfortunately, a 
mathematically rigorous
generalization of the methods discussed so far to
realistic four dimensional gauge field theories can only be 
done if the number of degrees of freedom is drastically reduced
by assuming a finite lattice instead of continuous space,
or by compactifying the latter into torus and considering
small volumes of the torus so that high-momentum states
can be treated perturbatively, and only the lowest (zero-momentum)
states will be affected by the nonperturbative corrections.
The removal of the regularizations is still a major problem
to achieve a reliable conclusion about the role of the configuration
or phase space geometry of the physical degrees of freedom in
realistic gauge theories. For this reason we limit the discussion 
by merely a review of various approaches rather than going into
the details. At the end of this section we apply the projection
method to construct the path integral for the Kogut-Susskind
lattice gauge theory, which seems to us to be a good starting point,
consistent with the gauge invariant operator formalism, 
for studying the effects of the physical phase space geometry
in quantum Yang-Mills theory.

\subsection{On the Riemannian 
geometry of the orbit space in classical Yang-Mills
theory}

The total configuration space of the classical Yang-Mills theory consists
of smooth square integrable gauge potentials (connections) ${\bf A}
={\bf A}({\bf x})\in C^\infty$ 
on the space being compactified into a sphere \cite{singer}
(meaning that the potentials decrease sufficiently fast to zero
at spatial infinity). Potentials take their values in a Lie
algebra of a semisimple compact group $G$ (the structure group).
As before, we use the Hamiltonian formalism
in which the time component $A_0$ of the four-vector $A_\mu$ 
is the Lagrange multiplier for the constraint (the Gauss law)
\be
\sigma ({\bf x}) = \nabla_j({\bf A}) E_j  = 
\pl_jE_j -ig[A_j,E_j]=0\ ,
\label{rel1}
\ee
where the components of the color electrical field ${\bf E} $ are canonical
momenta for ${\bf A}$. We omit the details of constructing 
the Hamiltonian formalism. They are essentially the same as for
the two-dimensional case discussed in section 5.

Gauge transformations are generated by the constraint (\ref{rel1}):
$\delta F=\{\<\omega,\sigma\>,$ $ F\}$ for any functional $F$
of the canonical variables and infinitesimal $\omega$. Finite
gauge transformations are obtained by successive iterations
of infinitesimal transformations. One can show that each
gauge orbit in the configuration space ${\cal A}$ of all 
(smooth) connections intersects at least once 
the hyperplane $\pl_iA_i=0$ \cite{zw2,zw82}. The Coulomb
gauge does not fix constant gauge transformations
because $\pl_iA^{\Omega}_i=\Omega\pl_iA_i\Omega^{-1}=0$ if $\pl_i\Omega=0$.
One can remove this gauge arbitrariness by reducing 
the gauge group ${\cal G}$ to the so called pointed
gauge group ${\cal G}_0$ whose elements satisfy the condition
$\Omega({\bf x}_0)=e$ (group unity) for some fixed point ${\bf x}_0$.
For example, one can identify ${\bf x}_0$ with spatial infinity
by requiring that $\Omega({\bf x})\rightarrow e$ as $|{\bf x}|
\rightarrow \infty$ (the space is compactified into a three-sphere).

Local effects of the orbit space geometry on dynamics of physical
degrees of freedom are caused by a non-Euclidean metric because
the kinetic energy depends on the metric. To construct the metric
on the orbit space ${\cal A}/{\cal G}_0$, we need local coordinates.
The space ${\cal A}$ is an affine space, while the orbit space
has a nontrivial topology \cite{singer}. To introduce local
coordinates on the orbit space, we identify a suitable region
of ${\cal A}$ that upon dividing out the gauge group projects 
bijectively on some open subset of the orbit space. There 
always exists a subset $K$ of ${\cal A}$ which is isomorphic
to the orbit space modulo boundary identifications. The subset
$K$ is called a (fundamental) modular domain. To construct $K$,
one uses a gauge fixing, i.e., the modular domain $K$ is
identified as a subset on a gauge fixing surface $\chi({\bf A})=0$.
Configurations from $K$ are used as local (affine) coordinates
on the orbit space \cite{singer,sob1,mitter}. 
 Clearly, the gauge fixing surface 
must have at least one point of intersection with every gauge
orbit. We take the Coulomb gauge
$\chi({\bf A}) = \pl_iA_i({\bf x}) =0$.  We adopt the method
and notations from the discussion of the two dimensional
case  in (\ref{dif}) and (\ref{metric}) where $a$ should be
replaced by a {\em transverse} potential ${\bf A}({\bf x})$,
$\pl_iA_i\equiv 0$, that is, the transverse potentials are 
chosen as local coordinates on the orbit space. 
We will use the same letter ${\bf A}$ for the transverse
connections (unless specified otherwise). In the new coordinates,
the functional differential of a generic connection can be written as
\be
\delta A_j \rightarrow \Omega\left(\delta A_j -\frac ig\nabla_j({\bf A})
\delta w\right)\Omega^{-1}\ ,
\ee
where $\pl_j\delta A_j \equiv 0$ in the right-hand side
and $\delta w=i\Omega^{-1}\delta\Omega$.
In contrast to (\ref{metric}), the metric is not block-diagonal
relative to the physical and nonphysical sectors. If $g_{AB}$ denotes
the metric tensor in the new coordinates where $A,B =1$ is
a collective index for the transverse connections $\delta {\bf A}$ and
$A,B=2$ is a collective index for pure gauge variables 
$\delta w({\bf x})$,
then the metric has a block form
\be
g_{AB}= \left(
\begin{array}{cc}
\delta_{jk}\ & - ig^{-1} P_{nm}\nabla_m({\bf A})\\
ig^{-1}\nabla_m({\bf A})P_{mn}\ & - g^{-2}\nabla^2({\bf A})
\end{array}
\right)
\label{rel2}
\ee
where $P_{jk}= \delta_{jk} - \pl_j\Delta^{-1}\pl_k$ is the projector
on transverse vector fields. It occurs through the simple
relation $\<\delta A_j, \nabla_j\delta w\>= 
\<\delta A_j, P_{jk}\nabla_k\delta w\>$ since 
by construction $\delta A_j$ is transverse, $P_{kj}\delta A_j=
\delta A_k$. 

The square root of the determinant of the metric (\ref{rel2}) is the
Jacobian of the change of variables. From the analysis of the simple
models one can naturally expect it to be proportional to the 
Faddeev-Popov determinant for the Coulomb gauge \cite{babelon}. Indeed,
making use of the formula for the determinant of the block matrix,
we find
\be
\mu[{\bf A}] = \left(\det g_{AB}\right)^{1/2}\! \sim\!
\left(\det\left\{\nabla_k^2 - \nabla_kP_{kn}\nabla_n\right\}
\right)^{1/2}\!\sim\! \det(-\pl_j\nabla_j({\bf A}))
\label{rel3}
\ee
which is the Faddeev-Popov determinant for the Coulomb gauge
as one might see by taking the determinant of the operator
whose kernel is determined by 
the Poisson bracket of the constraint $\sigma({\bf x})$ and the 
gauge fixing function  $\pl_iA_i({\bf x}')$. Thus, the Faddeev-Popov
determinant specifies a relative volume of a gauge orbit through
${\bf A }$. The singular points
of the change of variables are configurations where the determinant
vanishes (the Jacobian vanishes). For ${\bf A}=0$ the 
Faddeev-Popov operator
$\hat{M}_{FP}\equiv -\pl_j\nabla_j({\bf A})=-\Delta $ has no zero modes
in the space of functions decreasing to zero 
at spatial infinity. By perturbation
theory arguments one can also conclude that in the vicinity of the
zero configuration the operator $\hat{M}_{FP}$
has no zero modes. Given a configuration ${\bf A}$, consider
a ray $g{\bf A}$ in the functional space, where the ray parameter
$g$ may be frankly regarded as the gauge coupling constant in the  
operator $\hat{M}_{FP}$.  Gribov showed 
\cite{gribov} that
for sufficiently large $g$ the equation $\hat{M}_{FP}
\psi({\bf x})=0$ would always have a nontrivial solution, that is, the 
Faddeev-Popov operator would have a zero mode.
Therefore a ray from the zero configuration in any direction would 
reach the point where the Jacobian or the Faddeev-Popov determinant
vanishes. The singular points form a space of codimension one
in the space of transverse connections, which is called the Gribov
horizon (where the lowest eigenvalue of the Faddeev-Popov operator
vanishes (see below)). 

The plane waves associated with two transverse polarization
of gluons are solutions of the equations of motion in the
limit of the zero coupling constant. Therefore for dynamics
described by the perturbation theory of transverse gluons,
the coordinate singularities in the Coulomb gauge have no
effect. With the fact that the effective coupling constant
decreases in the high energy limit (see, e.g., \cite{zuber}), 
one can understand why
the perturbation theory based on the Faddeev-Popov path
integral in the Coulomb gauge was so successful. 
The relevant configurations are simply
far away from the coordinate singularities. In the strong coupling
limit, it is rather hard to determine the relative ``strength'' of the 
contributions to the dynamics which come from the coordinate
singularities (i.e. from the physical kinetic energy, or color electric
field energy)  and from
the strong self-interaction (i.e. from the color magnetic energy).
There is no technique to solve the Yang-Mills theory nonperturbatively
and compare the effects of the singular points in the Coulomb gauge
with those due to the self-interaction. This resembles the situation
discussed in section 3.5 (see also the remark
at the end of section 9.3) 
where the conic singularity of the physical
phase space  does not appear relevant for dynamics in the double
well potential in a certain regime: 
The classical ground state of the system is far from
the conic singularity so that small fluctuations around the ground state
are insensitive to it.  One should emphasize it again that, though 
the coordinate singularities are fully gauge dependent, they are
unavoidable. Therefore the singular
points should always be taken care of in any formalism which
relies on an explicit parameterization of the gauge orbit space.
However, they may or may not be relevant for 
a particular physical situation in question.

Returning to calculating the metric
on the gauge orbit space, we assume that ${\bf A}$ in 
(\ref{rel2}) is a generic configuration inside of the Gribov horizon,
so we can take the inverse of (\ref{rel2})
\be
g^{AB}= \left(
\begin{array}{cc}
\delta_{jk} + P_{jn}\nabla_n D^{-1}\nabla_mP_{mk}
\ \ & \  - ig P_{nm}\nabla_m D^{-1}\\
igD^{-1}\nabla_m P_{mn}\ ,& -g^2 D^{-1}
\end{array}
\right)
\label{rel4}
\ee
where $D = (\mbf{\pl},\mbf{\nabla})\Delta^{-1}(\mbf{\pl},\mbf{\nabla})$.
The metric $g^{ph}_{jk}$ 
on the gauge orbit space according to a general analysis
given in section 7.7 (see (\ref{b.6} and (\ref{b11})) is the inverse
of the upper left block $g^{11}$ in (\ref{rel4}). 
That is,
\be
g_{ph}^{jk} = \delta_{jk} + P_{jn}\nabla_n D^{-1}\nabla_mP_{mk}
\equiv \delta_{jk} + \Lambda_{jk}\ .
\label{rel5}
\ee
This metric specifies the
physical kinetic energy in our parameterization of the gauge orbit space.
The same result can be obtained by a solving the Gauss law for
the longitudinal components of the momenta $E_i$. Imposing the
gauge $\pl_iA_i=0$, one makes the decomposition
$E_i=E_i^\perp + \Delta^{-1}\pl_i (\pl_jE_j)$, where $\pl_iE_i^\perp
\equiv 0$. Substituting the latter into the Gauss law and solving
it for $\pl_iE_i$, one finds the expression of $E_i$ via 
the physical canonical variables $A_i^\perp(\equiv A_i)$ and $E_i^\perp$.
The physical metric is then extracted from the 
quadratic form $E_j^2$ (cf. (\ref{genmet})) and coincides
with (\ref{rel5}). The determinant of
the physical metric is {\em not} equal to the squared Faddeev-Popov
determinant, but rather we get \cite{babelon}
\be
\det g_{jk}^{ph}= \left[\det g^{jk}_{ph}\right]^{-1}\sim
\Delta_{FP}^2 \left[\det \Delta \det (\mbf{\nabla}, 
\mbf{\nabla})\right]^{-1}\ .
\label{rel6}
\ee
Formula (\ref{rel6}) can be derived by means of the exponential representation
of the determinant
$$
\det g_{ph}=\exp\left( {\rm tr}\,\ln g_{ph}\right)=
\exp\sum_{n=0}^\infty\frac{(-1)^n}{n}{\rm tr}\,\Lambda^n\ ,
$$
where  ${\rm tr}\,\Lambda^n=
{\rm tr}\,[(\mbf{\nabla}^2-D)D^{-1}]^n={\rm tr}\,(\mbf{\nabla}^2D^{-1}-1)^n$.
Therefore $\det g_{ph}= \det(\mbf{\nabla}^2D^{-1})$ which leads to
(\ref{rel6}). In the two-dimensional
case studied in section 5, 
the physical metric is proportional to the unit matrix, 
so its determinant is constant. If the space is one-dimensional,
then $\Delta_{FP}^2\sim \det(\pl\nabla)^2 \sim \det\pl^2\det(\nabla)^2$ 
and the determinant
(\ref{rel6}) equals one, indeed. The curvature of the gauge orbit
space is positive as has been shown by Singer \cite{singer2,jak2}.

\subsection{Gauge fixing and the Morse theory}

A representative of the gauge orbit in {\em classical} Yang-Mills
theory can be specified by means of the Morse theory as has been 
proposed by Semenov-Tyan-Shanskii and
Franke \cite{franke}. 
The idea is to minimize the $L^2$ norm of the vector potential
along the gauge orbit \cite{franke,zw1,zw2} 
\be
M_A(\Omega) =\left\< {\bf A}^\Omega\right\>^2=
\left\<\Omega {\bf A}\Omega^{-1} -ig^{-1}\Omega
\mbf{\pl}\Omega^{-1}\right\>^2 \ .
\ee
Here $\< F\>^2$ denotes $\int d^3x(F,F)$. The minima
of the Morse functional carry information about the topology
of the gauge orbit through ${\bf A}$. 
Taking $\Omega = e^{igw}$ and expanding the Morse functional 
around the critical point $w=0$, we find
\be
M_A(w) = \<{\bf A}\>^2 + 2\<w, \pl_jA_j\> -
\<w, (\pl_i\nabla_i)w\>
+ O(w^3)\ .
\label{rel7}
\ee
From (\ref{rel7}) it follows that the Morse function attains its relative
minima if the potentials satisfy the Coulomb gauge $\pl_jA_j=0$ and
the Faddeev-Popov operator 
$\hat{M}_{FP}=-\pl_j\nabla_j$ is a symmetric, positive
operator. The positivity of the Faddeev-Popov operator ensures
that the connection ${\bf A}$ has the property that 
$\Omega=e$ is a {\em minimum} of $M_A$.
The Gribov horizon is determined by the condition that
the {\em lowest} eigenvalue of 
the Faddeev-Popov operator vanishes.
The configurations on the Gribov horizon
are {\em degenerate} critical points of the Morse function.
A Gribov region $K_G$ is defined as the set of all minima
of the Morse functional. It has the property that
each gauge orbit intersects it at least once, and it is 
convex and bounded \cite{zw82}.

It may happen that two {\em relative} minima inside the Gribov
domain $K_G$  are related by a
gauge transformation, i.e., they are on the same gauge orbit. 
To obtain the modular domain $K$ which
contains only one representative of each gauge orbit, one has
to take only the absolute minima of 
the Morse functional. Let
${\bf A}$ and the gauge transform of it ${\bf A}^\Omega$ both be
from the  Gribov domain $K_G$. Then it is straightforward
to show \cite{baal92,baal97} that
\be
\<{\bf A}^\Omega\>^2 -\<{\bf A}\>^2 = \<\Omega^{-1}, \pl_i
\nabla_i^f({\bf A})\Omega\>\ ,
\label{rel8}
\ee
where $\mbf{\nabla}^f\Omega = \mbf{\pl}\Omega -ig{\bf A}\Omega$ 
is the covariant derivative in the fundamental representation. 
Since the Faddeev-Popov operator is positive in $K_G$, the absolute
minima of the Morse function can be defined in terms of the absolute
minima over the gauge group of the right-hand side of Eq. (\ref{rel8}).
A configuration ${\bf A}$ from the Gribov domain $K_G$ belongs
to the modular domain $K$ if the minimum of the functional 
(\ref{rel8}) over the gauge group vanishes. This condition simply
selects the absolute minima of the Morse function out of its
relative minima. Since the Faddeev-Popov operator 
for the Coulomb gauge is linear in
${\bf A}$, all configurations of the line segment $s{\bf A}_{(1)} + (1-s) 
{\bf A}_{(2)}$, where $s\in [0,1]$ and ${\bf A}_{(1,2)}\in K$,
also belong to $K$. That is, the modular domain is convex.
 
In a similar way the existence of the horizon and
the description of the modular domain have been
established in the background gauge $\nabla_i(\bar{\bf A})A_i=0$
($\bar{\bf A}$ is a background (fixed) connection). This
result is due to Zwanziger \cite{zw82}. In this case
the Morse functional is 
\be
M_A(\Omega)=\<{\bf A}^\Omega -\bar{\bf A}\>^2\ ,
\label{morse1}
\ee
and the Faddeev-Popov operator has the form $\hat{M}_{FP}=-
\nabla_k(\bar{\bf A})\nabla_k({\bf A})$.

The main properties of the modular domain are as follows \cite{baal97}.
First, its boundary has common points with the Gribov horizon,
i.e., it contains the coordinate singularities in the chosen gauge.
Second, the modular domain has a trivial topology as any 
convex subset in an affine space, but its
boundary contains gauge equivalent configurations.
Through their identification one obtains a nontrivial topology
of the gauge orbit space. 
In fact, the orbit space contains
non-contractable spheres of any dimension \cite{singer}.
Third, the gauge  transformations 
that relate configurations inside the Gribov domain $K_G$
may be homotopically {\em nontrivial}. 
Any point on the Gribov horizon has a {\em finite} distance
from the origin of the field space and one can derive a uniform
bound, as has been in done in the original work of Gribov \cite{gribov}
and later improved by Zwanziger and Dell' Antonio \cite{zw3}.

Although the above procedure to determine the modular domain
applies to general background connections, some properties
of $K_G$ and $K$ may depend on the choice of the background
connection. In particular, reducible and irreducible
background configurations have to be distinguished \cite{fuchs1,fuchs2}.
A connection ${\bf A}$ is said to be reducible if it has a 
nontrivial stationary group ${\cal G}_A$ (the stabilizer) such
that ${\bf A}^{\Omega}={\bf A}$ for all $\Omega\in {\cal G}_A$.
If ${\cal G}_A$ coincides with the center $Z_G$ of the structure group
$G$, then the connection is irreducible. From the identity
${\bf A}^\Omega = {\bf A} + ig^{-1}\Omega\mbf{\nabla}({\bf A})\Omega^{-1}$
it follows that $\Omega\in {\cal G}_A$ if $\mbf{\nabla}({\bf A})\Omega=0$.
Any stabilizer ${\cal G}_A$ is isomorphic to a closed subgroup
of $G$ \cite{4d4}. This can be understood as follows. We recall
that for any ${\bf A}$, ${\cal G}_A$ is isomorphic to the 
centralizer of the holonomy group of ${\bf A}$ relative to the 
structure group $G$ \cite{fiber}. By definition, the centralizer
$G'_c$ of a subgroup $G'$ of $G$ consists of all elements of $G$
which commute with all elements of $G'$. Clearly, $G'_c$ is
a subgroup of $G$. On the other hand, the holonomy group
is a Lie subgroup of $G$ (see, e.g., \cite{fiber}), i.e., its
centralizer is a subgroup of $G$.

The orbit space has the structure of a so called stratified
variety which can be regarded as the disjoint sum of strata
that are smooth manifolds \cite{4d6a,4d6b,4d6c}. Each stratum of the variety
consists of orbits of connections whose stabilizers are
conjugate subgroups of ${\cal G}$. In other words, the stabilizers
of the connections of a fixed stratum are isomorphic to one another.
A stratum that consists of orbits of all {\em irreducible}
connections is called a main stratum. The set of orbits
of {\em reducible} connections is a closed subset in the orbit
space which is {\em nowhere dense}. Accordingly, the main
stratum is dense in the orbit space, and any singular strata can be
approximated arbitrarily well by irreducible  connections \cite{fuchs1}.
If all reducible connections are excluded from the total configuration
space, then the orbit space is a manifold.

The Morse functional (\ref{morse1}) can also be regarded as
the distance between ${\bf A}^\Omega$ and $\bar{\bf A}$.
Let ${\bar{\bf A}}$ be an irreducible connection. Any two
connections ${\bf A}_{1,2}$, ${\bf A}_1\neq {\bf A}_2$, on the
gauge fixing surface $\nabla_i(\bar{\bf A}) A_i=0$ that are
sufficiently close to $\bar{\bf A}$ belong to distinct gauge
orbits. For reducible backgrounds, the gauge fixing surface
does not posses such a property. The Morse functional (\ref{morse1})
has a degeneracy for reducible backgrounds. Indeed, if 
$\bar{\Omega}\in {\cal G}_{\bar A}$, then we have
\be
M_A(\Omega\bar{\Omega}) =\left\<{\bf A}^{\Omega\bar{\Omega}}
-\bar{\bf A}\right\>^2=\left\<{\bf A}^{\Omega\bar{\Omega}}
-\bar{\bf A}^{\bar{\Omega}}\right\>^2= M_A(\Omega)\ ,
\label{morse2}
\ee
because the Morse functional is invariant under simultaneous
gauge transformations of ${\bf A}^{\Omega}$ and $\bar{\bf A}$. 
It is also not hard to see that, if 
$M_A(\Omega\bar{\Omega})= M_A(\Omega)$ holds true
for any ${\bf A}$, then $\bar{\Omega}$ should be an element
of the stabilizer ${\cal G}_{\bar{A}}$.

In the case of a reducible background, the Faddeev-Popov operator
always has zero modes. Let $\bar{\bf A}^{\Omega}=\bar{\bf A}$,
i.e., $\Omega\in {\cal G}_{\bar{A}}$. As ${\cal G}_{\bar{A}}$ is
isomorphic to a Lie group (a subgroup of $G$), there exists
a family $\Omega_\lambda\in {\cal G}_{\bar{A}}$ such that
all elements are connected to the group unity, $\Omega_{\lambda=0}=e$.
The Lie algebra valued function $\psi({\bf x})=\pl_\lambda
\Omega_\lambda({\bf x})\vert_{\lambda =0}$ is covariantly
constant, $\nabla_j(\bar{\bf A})\psi =0$, and therefore
it is a zero mode of the Faddeev-Popov operator,
$\hat{M}_{FP}\psi =-\nabla_j(\bar{\bf A})\nabla_j({\bf A})\psi
=-\nabla_j({\bf A})\nabla_j(\bar{\bf A})\psi =0$, thanks to
the symmetry of $\hat{M}_{FP}$. In particular, taking 
$\bar{\bf A}=0$ we get ${\cal G}_{\bar{A}=0}\sim G$, i.e.,
${\cal G}_{\bar{A}=0}$ is a group of constant gauge transformations.
By removing constant gauge transformations from the gauge
group, we remove a systematic degeneracy of the Faddeev-Popov
operator for the Coulomb gauge.

Next, we observe that the collection of all absolute minima
of the Morse functional cannot serve as the fundamental
modular domain $K$ because of the degeneracy (\ref{morse2}).
There are gauge equivalent configurations inside
the set of the absolute minima. The identification
in the interior precisely amounts to dividing
out the stabilizer ${\cal G}_{\bar{A}}$ \cite{fuchs1}.

We shall not pursue a further elaboration of the classical
physical configuration space because in quantum theory the
fields are distributions, and the relevance of the above
analysis to the quantum case is not yet clear because 
smooth (classical) configurations form a subset of zero
measure in the space of distibutions (see Section 10.4 
for details).
We refer to an excellent review by Daniel and Viallet
(see \cite{babelon}) where differential geometry
of the orbit space in classical Yang-Mills theory is
discussed. A stratification
of the orbit space of the classical SU(2) Yang-Mills theory
is studied in detail in the work of Fuchs, Schmidt and 
Schweigert \cite{fuchs1}. It is noteworthy that classical
trajectories in the Hamiltonian formalism are always contained
in one fixed stratum (in one smooth manifold) \cite{4d7}.
A final remark is that a principal bundle has isomorphism
classes characterized by the instanton number which can
be any integer. Connections with different instanton numbers
satisfy different asymptotic conditions at infinity. If we
allow asymptotic conditions associated with all instanton
numbers, then the fundamental modular domain will be 
the disjoint sum of modular domains for every instanton number
\cite{4d8}.

\subsection{The orbit space as a manifold. Removing
the reducible connections}

In the previous section we have seen that the main stratum
of the orbit space is a smooth manifold. On the other hand,
in 2D Yang-Mills theory  the orbit space appears to be an
orbifold (with trivial topology). The reason is that there
are reducible connections on the gauge fixing surface
whose stabilizers are subgroups of the group of constant
gauge transformations. If we restrict the gauge group by
excluding constant gauge transformations, then, as we shall
show, the orbit space is a manifold which is  
a {\em group manifold} \cite{raj}.
The group manifold is compact and has a nontrivial topology.
The latter occurs through the identification of gauge equivalent
points on the Gribov horizon (the example of the SU(2)
group has recently been studied in this regard by Heinzl and Pause
\cite{heinzl}). 
One should keep in mind, however, that such a truncation of the gauge
group cannot be done in the Lagrangian, and is added to 
the theory by hand as a supplementary condition on the gauge group.

If constant gauge transformations are excluded from the 
gauge group in the 2D Yang-Mills theory, then {\em all} zero modes
of the Faddeev-Popov operator $-\pl\nabla(A_0)$ are determined
by Eq. (\ref{zerm1}), where $a$ is replaced by $A_0$ being  
a generic element on the gauge fixing surface $\pl A=0, A=A_0\in X$, 
i.e., a constant connection in the whole Lie algebra (not in its
Cartan subalgebra). Since $A_0=\Omega a\Omega^{-1}$, for some
constant group element $\Omega$, the zero modes have the 
same form (\ref{zerm3}), where $\bar{\xi}\rightarrow \Omega^{-1}\bar{\xi}
\Omega$. Therefore Eq. (\ref{zerm5}) specifies zeros of the Faddeev-Popov
determinant because $\det[-\pl\nabla(a)]=\det[-\pl\nabla(A_0)]$, i.e.,
it does not depend on $\Omega$. The Cartan algebra element
$a$ related to $A_0$ by the adjoint action of the group
 has $r={\rm rank}\, X$ independent components
which can be expressed via the independent Casimir polynomials
$P_{\nu_i}(A_0)=\tr{A_0^{\nu_i}}=P_{\nu_i}(a)$. For example
in the SU(2) case, the only component of $a$ is proportional
to $[\tr{A_0^2}]^{1/2}$. Hence, the Faddeev-Popov determinant vanishes
at the concentric two-spheres $\tr A_0^2 = 2a_0^2n^2, n\neq 0$. The vacuum
configuration $A_0=0$ is {\em inside} the region bounded by
the {\em first} Gribov horizon $n=1$, which is the Gribov region.
The vacuum configuration $A_0=0$ 
is {\em no longer} a singular point because constant
gauge transformations are excluded. The Faddeev-Popov determinant
is proportional to the Haar measure \cite{hel} on the group manifold
\be
\Delta_{FP}(A_0)=\Delta_{FP}(a)=\prod\limits_{\alpha >0}
\frac{\sin^2(a,\alpha)}{(a,\alpha)^2}\ ,
\label{fphom}
\ee
which is {\em regular} at any hyperplane orthogonal to a root
and passing through the origin (vacuum $a=0$). Returning
to the SU(2) case, we remark that 
all the configurations $A_0$ such that $2a_0^2(n-1)^2\leq
\tr A_0^2\leq 2a_0^2n^2, n>1$ are gauge equivalent to those
in the Gribov region. In general, given a constant
connection $A_0$ one can find a group
element $\Omega$ and the Cartan subalgebra element $a$
such that $A_0=\Omega a\Omega^{-1}$.
To obtain a Gribov copy of $A_0$, we translate $a$ by 
an integral linear combination of the elements $\eta_\alpha$
defined by Eq. (\ref{eta}), and then bring the resulting
element back to the whole algebra by the inverse adjoint 
action generated by the group element $\Omega$.
 
In the SU(3) case, the first Gribov  
horizon is obtained by generic adjoint transformations 
of all the configurations that lie in the polyhedron
$B_1B_2\cdots B_6$ in Figure 5. It is a seven dimensional 
hypersurface manifold. The same holds in general. We take the 
polyhedron around the vacuum configurations $a=0$ whose
faces are portions of the hyperplanes $(a,\alpha)=a_0$
for all positive roots $\alpha$. Then each point of the
polyhedron is transformed by the adjoint action of
generic group elements. As the result, we obtain
the first Gribov horizon which is a compact hypersurface of
dimension $\dim X -1$. As should be, it has the
codimension one on the gauge fixing surface $\pl A=0$.
On the horizon the lowest eigenvalue of the Faddeev-Popov
operator  $-\pl\nabla(A)$ vanishes. The images of points
of intersection of the hyperplanes $(a,\alpha)=a_0$,
which are sets of codimensions $k$, $k\geq 2$, with
$k$ being the number of the distinct hyperplanes at
the point of intersection, form subsets of the Gribov
horizon where the zero eigenvalues of the Faddeev-Popov
operator are degenerate (with the multiplicity $k$).

As in the matrix model considered in section 4.8,
there are gauge equivalent configurations within
the Gribov horizon. To find them we observe that
the vacuum configuration $a=0$ can always be 
shifted to the first Gribov horizon by a homotopically
{\em nontrivial} gauge transformation (\ref{hm4})
with $n=1$. If we shift the vacuum configuration
by $\alpha a_0/(\alpha,\alpha)$ (see (\ref{hm4}))
and then rotate it as $\Omega \alpha\Omega^{-1}a_0/(\alpha,\alpha)
\equiv A_0^{(\alpha)}$,
where $\Omega\in G/G_H$, we obtain a portion of the Gribov
horizon that contains all possible images of the vacuum configuration
generated by the homotopically nontrivial transformations
associated with the root $\alpha$ (cf. (\ref{zg})).
All the points $A_0^{(\alpha)}$ of this portion of the horizon 
are related to one another by homotopically {\em trivial}
gauge transformations. Indeed, the homotopically nontrivial
gauge group element that transforms the vacuum configuration
to a generic configuration on the $\alpha$-portion of the horizon
is $\Omega(x, A_0^{(\alpha)})= \exp(igA_0^{(\alpha)}x)$. 
The gauge transformation that relates two configurations
$A_0^{(\alpha)}$ and $\tilde{A}_0^{(\alpha)}$ on the
$\alpha$-portion of the horizon is obtained by the composition
of the gauge transformation that shifts, say, $A_0^{(\alpha)}$
to the vacuum, and the gauge transformation that shifts
the vacuum to $ \tilde{A}_0^{(\alpha)}$. It is generated by
the group element $\Omega(x, \tilde{A}_0^{(\alpha)})
\Omega^{-1}(x,{A}_0^{(\alpha)})$ which is homotopically trivial:
\ba
\Omega(x+2\pi l, \tilde{A}_0^{(\alpha)})
\Omega^{-1}(x+2\pi l,{A}_0^{(\alpha)})
&=&z_\alpha\Omega(x, \tilde{A}_0^{(\alpha)})
\Omega^{-1}(x,{A}_0^{(\alpha)})z_\alpha^{-1}\nonumber\\
&=&
\Omega(x, \tilde{A}_0^{(\alpha)})\Omega^{-1}(x,{A}_0^{(\alpha)})\ ,
\ea
where $z_\alpha$ is the center element associated with the root 
$\alpha$ (cf. (\ref{zg})). In the case of the SU(2) group,
we have only one root. So all the points of the horizon, being
the two-sphere, are gauge equivalent. Identifying them we
get the gauge orbit space as the three-sphere, which is the group
manifold of SU(2). In the general case, we observe that
the Lie algebra elements $A_0=\Omega a\Omega^{-1}$ from the 
region bounded by the portions of the hyperplanes $(\alpha,a)=a_0$
serve as local affine coordinates on the group manifold.
Any element from the connected component of the group
has the form $\exp(2\pi iA_0/a_0)$ \cite{hel}. This coordinate
chart does not cover the center of the group. 
Singular points of the affine coordinate system are
zeros of the Haar measure (\ref{fphom}) \cite{hel}
and, therefore, form the first Gribov horizon.
The group manifold
is obtained by identifying all points in each of
the $\alpha$-portions of the horizon
so that the latter is shrunk to a finite number of points which
are distinct elements of the center of the group \cite{hel}, like 
in the SU(2) case the entire two-sphere $\tr A_0^2=2a_0$
is shrunk to a single point being the only nontrivial element
of the center $-e$.

Thus, if we exclude constant gauge transformations,
then the 2D Yang-Mills theory becomes an irreducible gauge
system. The corresponding gauge orbit space is a topologically
nontrivial (group) {\em manifold}. The above discussion
may serve as an illustration
for the classical Yang-Mills theory in four dimensions, where
the gauge orbit space exhibits the same features 
\cite{singer,babelon,baal92}. In particular, one needs more than one
coordinate chart to make a coordinate system on the gauge orbit
space. If one takes two geodesics outgoing from one point
on the orbit space, then they may have another point of intersection
which belongs to the Gribov horizon in the local 
affine coordinate system centered at the initial point of
the geodesics \cite{babelon}, thus indicating the singularity
 of the coordinate system. We emphasize that the existence of
conjugated points on the geodesics is an intrinsic feature of the theory.
We also point out that the use of several coordinate
charts allows one to avoid the Gribov singularities \cite{nahm,220}
(see also \cite{kvb2}) in principle, but does not lead to any
convenient method to calculate the path integral. 
A recent development of this approach in the framework of
stochastic quantization can be found in \cite{218}.

\subsection{Coordinate singularities in quantum
Yang-Mills theory}

The description of the parameterization of the gauge
orbit given in section 10.2 
applies to classical theory only. The configuration
space of in quantum field theory is much larger than the 
space of square integrable functions. It consists of distributions
\cite{kl99}.
Smooth classical functions form a subset of zero measure in
the space of distribution (a Sobolev functional space).
The Sobolev space ${\cal S}_k^p$, where 
$1\leq p<\infty, k=0,1,2,...$,
consists of fields all of whose derivatives 
up to and including order $k$ 
have integrable $p$-th power. 
The smaller the indices $p$ and $k$ the larger the 
space of the fields.
The result of Singer on the absence of a global
continuous gauge fixing for smooth classical field configurations
can be extended to the Sobolev space \cite{sob1,soloviev}, {\em provided}
\be
p(k+1)>n\ ,
\label{sobol}
\ee
where $n$ is the dimension of the base manifold.
Since the gauge transformation law of the connection 
involves the derivatives of the group elements,
the latter must have one derivative more than the
connections, i.e., they must be from the Sobolev
space (of the group valued functions) ${\cal S}_{k+1}^p$.
Only under the condition (\ref{sobol}) the gauge group
possesses the structure of  a finite-dimensional Lie
group and acts smoothly on the space of connections
${\cal S}_k^p$ \cite{sob2}. 
The condition (\ref{sobol}) is discussed in more details
in \cite{soloviev3}. Here we point out the following.
The condition (\ref{sobol}) is crucial for continuity of
gauge transformations as functions of a point of the base 
manifold. For instance \cite{soloviev}, the function
$|x|^{-\epsilon}$ is singular at the origin, but the p-th
power of its k-th derivative is integrable if $p(k+1)<n$
and $\epsilon <1$. If $p(k+1)=n$ ($p\neq 1$), then there
may exist a singularity $(-\ln |x|)^{1-1/p-\epsilon}$.
Thus, the necessity of the condition (\ref{sobol})
for continuity is clear. The condition (\ref{sobol}) ensures also the
existence of a local gauge fixing and the structure of the 
principal fiber bundle in the configuration space \cite{soloviev4}. 

If $p(k+1) <n$, the very notion
of the gauge fixing becomes meaningless \cite{soloviev}. Ordinary
conditions like the Coulomb gauge will not be any gauge fixing
even locally. Consider the transformation $A_i(x)\rightarrow
A_i^\lambda(x)=\lambda A_i(\lambda x)$ of connection in $\Rs^n$. Then
\ba
\left\vert\!\left\vert\pl_{j_1}\cdots\pl_{j_k}A^\lambda\right\vert\!
\right\vert_p &\equiv&
\left\{\int d^nx\left(\pl_{j_1}\cdots\pl_{j_k}A^\lambda,
\pl_{j_1}\cdots\pl_{j_k}A^\lambda\right)^{p/2}\right\}^{1/p}\nonumber\\
&=&\lambda^{k+1-n/p}
\left\vert\!\left\vert\pl_{j_1}\cdots\pl_{j_k}A\right\vert\!\right\vert_p
\label{sobol2}
\ea
The right-hand side of Eq. (\ref{sobol2}) tends to zero as 
$\lambda\rightarrow \infty$ if $p(k+1)<n$. If we take
a transverse connection and its Gribov copy and perform
the $\lambda$-transformation of them, then for sufficiently
large $\lambda$ both configurations will be arbitrary close
to zero field in the sense of the topology of ${\cal S}_k^p$,
and they will remain transverse. The noncompactness of the base
is not important here because both $A$ and its copy can be taken
near the vacuum configuration \cite{soloviev}. 

In the Sobolev space of connections satisfying (\ref{sobol})
there is an improved version of the theorem
of Singer which is due to Soloviev
\cite{soloviev}. It asserts that 
the gauge orbit fiber bundle in non-Abelian field theory
does not admit reduction to a finite-dimensional Lie group.
In other words, there is no gauge condition that would fix the 
gauge arbitrariness globally modulo some finite {\em subgroup}
of the gauge group. Observe that in all models we have discussed,
one can always find a gauge condition that removes the gauge
arbitrariness completely up to a discrete {\em subgroup} of the gauge
group. In contrast, in the Yang-Mills theory the residual gauge
symmetry in any gauge would not form a finite subgroup of the gauge
group. Soloviev's result gives the most precise characterization
of the Gribov problem in the Yang-Mills theory. 

The formal generalization of the path integral over the covering
space of the orbit space, though possible
\cite{plb91,corfu}, would be hard to use
since there is no way we could ever find all Gribov copies for
a generic configuration (being a distribution) satisfying a chosen gauge.
Moreover, a class of fields on which the functional integral
measure has support depends on the field model in question
(cf. the Minlos-Backner theorem \cite{gelfand}).
The property of continuity discussed above is decisive for 
Singer's analysis, while quantum field distributions in four dimensions would
typically have the singularity ``$|x|^{-1}$'' almost everywhere.
With such a poor state of affairs, we need some approximate
methods that would allow us to circumvent (or resolve) this
significant problem associated with the distributional
character of quantum fields.
We stress that the effects
in questions are essentially nonperturbative, so one of the conventional
ways of defining the path integral as a (renormalized) 
perturbative expansion with respect to the Gaussian 
measure does not apply here.

Since in any actual calculation on the gauge orbit space
the introduction of a (local) set of coordinates is unavoidable,
one should raise a natural question of how the coordinate singularities
can be interpreted in terms of quantum fields. 
Following the basic ideas of
(perturbative) quantum field theory, one may attempt
to interpret quantum Yang-Mills theory as the theory
of interacting gluons. This picture naturally results
from perturbation theory in the Coulomb gauge.
Consequently, the ``physical'' picture of the effects
caused by the coordinate singularities would strongly
depend on the choice of variables that are to describe
``physical'' (elementary) excitations in the theory.
There is, in fact, a great deal of the choice of physical variables,
especially in the nonperturbative region. For instance,
in the picture of self-interacting gluons, one may
expect some effects on the gluon propagator (cf. section 9.3)
caused by coordinate singularities in the Coulomb gauge 
as has been conjectured by Gribov. But with
another set of variables describing elementary excitations 
the physical picture would look differently in terms of the quanta
of the new fields because the singularities would also be different.
An example of this kind is provided by 't Hooft's Abelian
projection \cite{tmap}. The gauge is imposed on the field
components $F_{\mu\nu}$ ($\mu,\nu$ are fixed)
rather than on the potential, or on any local quantity $B({\bf A})$ that
transforms in the adjoint representation. It is required that
all non-Cartan components of $B$ vanish. Such a gauge 
restricts the gauge symmetry to a maximal Abelian subgroup
of the gauge group, which may be fixed further by the Coulomb
gauge without any singularities. A potential ${\bf A}$
gauge transformed to satisfy the Abelian projection
gauge would, in general, have singularities or topological
defects. They would have quantum numbers of magnetic monopoles with
respect to the residual Abelian gauge group. So, the effective
theory would look like QED with magnetic monopoles.
This is completely different interpretation of gluodynamics,
which leads to a different interpretation of the coordinate
singularities.
 Lattice simulations show that the monopole defects
of gauge fixed Yang-Mills fields are important in the 
nonperturbative regime and cannot be ignored \cite{lat1,lat2}.
Singularities in the path integral approach with a gauge
imposed on the field variables have also been observed in \cite{fd79}.
Thus, the bottom line is  that the singularities have a 
different ``physical'' appearance (or interpretation), 
depending on the choice
of the ``elementary'' excitations in the Yang-Mills theory.
However, whatever choice is made, they must be taken into
account in a complete (nonperturbative) quantum theory. 
Yet, it seems desirable to develop a formalism which is sort of
universal and  does not
rely on a particular choice of the physical variables, that is,
independent of the parameterization of the physical phase space.
A proposal based on the projection formalism is discussed in 
next section.

The existing approaches to analyze singularities of
local coordinates on the orbit space in quantum gauge
theories can be divided 
into two groups based respectively on
the functional Schr\"odinger equation and the path integral. 
In the first approach there are 
great complications as compared with the soluble two-dimensional
case we have discussed. First, there is a potential 
(color magnetic) energy in the Hamiltonian
which has terms cubic and quartic in the gauge potential. 
This would create nonperturbative  dynamical 
effects in the strong coupling limit, thus making it hard to distinguish
between the contributions of the kinetic and potential energies
to, say, the mass gap (the difference between the vacuum
and first excited state energies) in the quantum Yang-Mills theory
\cite{feyn21,jak2}. Second, the metric on the orbit space is  
not flat, so one should expect quantum corrections to the classical
potential stemming from the kinetic energy as predicted by Eqs (\ref{b11})
and (\ref{b12}). The quantum potential will be singular at the points where
the Jacobian (or the Faddeev-Popov determinant) vanishes as one
might see from its explicit form (\ref{b12}). Due to locality of the 
kinetic energy, the quantum corrections would contain
a nonphysical infinite factor $\hbar^2[\delta^3(0)]^2$ which typically
  results from the operator ordering in  
the kinetic energy operator of any local field theory 
that contains a non-Euclidean metric in the field space. 
Thus, the Schr\"odinger equation in field theory
requires a regularization of the local product of operators
involved. Needless to say about defining a proper Hilbert space
in this approach. Even in the case of a free field, which
is an infinite set of harmonic oscillators, solutions of the functional
Schr\"odinger equation are not without difficulty \cite{kl99}. 
Yet, in dimensional regularization one usually sets
$\delta^3(0)=0$. This
however would not justify throwing away the singular terms from
the Hamiltonian. The applicability of dimensional regularization
is proved within perturbation theory only.
Christ and Lee studied the effects of the operator ordering terms
resulting from solving the Gauss law in the Coulomb gauge \cite{christ}
in the (Hamiltonian) perturbation theory.
They did not find any effect for the physical  perturbative
S-matrix, though the operator ordering terms appeared to
be important for a renormalization of the two-loop {\em vacuum}
diagrams. Their work has been further extended by Prokhorov
and Malyshev \cite{prmal}.

It seems that a mathematically reasonable approach based on the 
Schr\"odinger equation can only be formulated if one truncates
the number of degrees of freedom. Cutkosky initiated one
such program attempting to investigate the effects of the 
coordinate singularities on the ground state \cite{cut1,cut2,cut3,cut4}.  
Another approach is due to L\"usher \cite{lus} which has been developed
further by Koller and van Baal \cite{kvb1,kvb2} (for recent
developments see \cite{hvb} and \cite{baal97} and references
therein). It is based on compactifying the space into a three-torus
and studying the limit of small torus size. The latter allows
one to use a perturbation theory for all excitations with higher
momenta, while the nonperturbative effects would be essential
for the low (or zero) momentum excitations. In all these approaches
the geometry of the modular domain appears to be
important for the spectrum of the truncated Yang-Mills
Hamiltonian just as for soluble Yang-Mills models we have
discussed. There is no reliable conclusion about what happens
when the torus size becomes large.

In the path integral approach,  Gribov proposed to modify the
original Fad\-de\-ev-Popov measure by inserting into it a
characteristic function of the domain where the Faddeev-Popov
operator  is positive \cite{gribov}. This would modify the path integral
substantially in the infrared region (the Green's functions,
e.g., the gluon propagator, 
derived from the path integral are modified) because
the horizon in the Coulomb gauge approaches the vacuum
confuguration from the infrared directions in the momentum space.
 Since later it was
understood that the modular domain is smaller than the Gribov
domain, the idea was appropriately elaborated by Zwanziger 
\cite{zw4a,zw4,zw5}
with a similar conclusion about the infrared behavior of
the gluon propagator. Instead of the conventional $G(k^2)\sim
[k^2]^{-1}$ it turned out to be  
\be
G(k^2)\sim \frac{k^2}{k^4+ m^4}\ ,
\label{gzp}
\ee
where $m^2$ is a dynamically generated mass scale.
In this approach the self-interaction
of gluon fields has been taken into account by perturbation
theory, so the entire effect on the gluon propagator came from
``horizon effects''. A propagator of the form (\ref{gzp})
has been also observed in the lattice simulations \cite{par}.
However the other group reported  a 
different result \cite{stella}: 
\be
G(k^2)\sim
Z^{-1}\left[m^2 + k^2(k^2/\Lambda^2)^\alpha\right]^{-1}\ , 
\ee
where $\alpha\sim 0.5$
and $m^2$ is compatible with zero. The constant $m^2$ has
been reported to be a
finite volume artifact. Thus, in the continuum limit, one has
$G(k^2) \sim (k^2)^{-1.5}$. So, it cannot be fitted as
a sum of single particles poles with positive residues, which
certainly unacceptable feature of the propagator of a physical
particle because it violates the K\"allen-Lehmann representation.
However it has been argued that it could be acceptable for a confined
particle \cite{mandula}. In this controversy it is also 
unclear which effect is most relevant 
for such a behavior of the nonperturbative gluon propagator: 
that of the Gribov horizon, or 
the effects of a strong self-interaction. 
For instance, the influence of the Gribov copies in the Coulomb
gauge on the correlation
functions in lattice QCD has been studied in \cite{vlad}.
It has been observed that the residual gauge symmetry
does not appear to be relevant. However, the authors
of \cite{vlad} have also noted that the effect may become
important on bigger lattices. 
Yet, invoking a special non-perturbative
technique of solving Schwinger-Dayson equations, Stingl \cite{stin} found
the expression (\ref{gzp}) for the non-perturbative gluon propagator
{\em without} taking into account the existence of the horizon. In his
approach the whole effect was due to the strong self-interaction
of gluon fields. In the aforementioned Abelian projection 
of QCD, the effects of Gribov copying has also been studied
on a lattice \cite{hart}. No significant effect has been found.
It is curios, however, that the singularities themselves
(``monopoles'') play the key role in the confinement scenario
in the maximal Abelian projection. It seems like the singularities
in that gauge serve as labels for configurations (or degrees of
freedom) that are most relevant for the confinement. As singularities
are gauge dependent it seems 
 very likely that in the nonperturbative
region the effects of coordinate singularities associated with
a generic gauge and 
those of the strong self-interaction would be hard to distinguish. 
The maximal Abelian gauge look more like an exception rather than a rule.

Returning to the Coulomb gauge, one may anticipate a 
potential problem in the path integral approach 
based on the formal restriction of the integration domain,
say, to configurations for which the Morse functional
attains absolute minima.
The point is that the modular domain found in {\em classical} 
theory cannot be applicable in the path integral 
whose measure has a support on the space of distributions. 
The formal extension of the classical results to the quantum
theory is questionable because classical configurations
have zero measure in the quantum configuration space. 
The way out is to go to lattice gauge theory. 
The above ideas of defining the modular domain
via the Morse theory and the restriction of the integration 
domain in the path integral has been implemented in the lattice
gauge theory by Zwanziger \cite{zwlat}. 
He also investigated the thermodynamic
limit of the modified path integral, i.e., when the number
of lattice sites becomes infinite (the limit of an infinite number 
of degrees of freedom).  The conclusion was that the existence
of the horizon alone (without a strong self-interaction) 
is sufficient to explain the area law of the Wilson loop, i.e.,
to fulfill the confinement criteria \cite{zwc}. This conclusion,
though being attractive, still remains a conjecture
since effects of a strong self-interaction have not been
estimated. 

Even if the lattice regularization is assumed
in the approach based on the restriction
of the functional integral measure 
to the modular domain, there is no obvious correspondence
to the operator formalism, which  should, as is believed,
be present since the operator and path integral formalism
are just two different representations of the same physical
model.  In section 8.2 it is argued that the topology and the 
boundaries in the configuration space cannot, in fact, be
taken into account simply by a formal restriction of the integration
domain, i.e., by inserting a characteristic
function of the modular domain into the path integral
measure. This would be in conflict with the operator formalism.
For soluble gauge models with a non-Euclidean geometry
of the physical phase space, the formal restriction of the integration
domain in the path integral turns them into {\em insoluble}
models because the integral is no longer Gaussian and leads
to results which are {\em not} consistent with the explicitly
gauge invariant approach. Recall that the partition function
on the lattice can be computed {\em without} gauge fixing.
The corresponding {\em gauge-fixed} path integral is given
by (\ref{gi}) and (\ref{pf}). It involves no formal restriction
of the integration domain.

In the work of Scholtz and Tupper a dynamical gauge fixing
has been proposed to circumvent the Gribov obstruction
to the path integral quantization \cite{st}. The idea
was to introduce a supersymmetric (auxiliary) multiplet 
coupled to the Yang-Mills fields in a special way that
the physical S-matrix is not modified. Then the gauge
is imposed on the bosonic components of the auxiliary
supermultiplet, while the Yang-Mills potentials
are left untouched. The operator version of such supersymmetric
quantization has been developed in \cite{fs,fs2}.

As one can see, it is rather hard to arrive at any definite
conclusion about the role of the orbit space geometry in quantum
gauge field theories. The reason is twofold. First, there is no
good understanding of the very notion of the orbit space in the 
quantum case. Distributional character of quantum fields imposes
severe restrictions on the use of conventional topological and
geometrical means based on continuity. Second, we do not know
how to solve strongly interacting quantum field theories, which
makes it impossible to distinguish between effects caused by the
geometrical structure of the orbit space and those due to the
strong interaction. The theory still needs more developmets from
both mathematical and physical sides. However, in various model
approximations, where the above difficulties can be resolved, we 
do see the importance of the orbit space geometry in quantum theory.

\subsection{The projection method in the Kogut-Susskind lattice
gauge theory}

A possible way
to extend the idea of combining the projection method
and the Kato-Trotter product formula to gauge field theories
is to make some regularization of a quantum field Hamiltonian.
A finite lattice 
regularization turns the quantum  field theory into quantum
mechanics. Since we still want to have the Schr\"odinger 
equation and the Hamiltonian formalism, which is essential
to control coordinate singularities in quantum theory on the orbit space, 
the only choice we have is the 
Kogut-Susskind lattice gauge theory \cite{kogut},
where the space is discretized, while the time remains continuous.

Let points of the three-dimensional periodic cubic lattice
be designated by three-vectors with integer components,
which we denote by $x, y$, etc. The total configuration space
is formed by the link variables $u_{xy} = u_{yx}^{-1}\in G$,
where $y=x + k$, and $k$ is the unit vector in the direction of 
the $k$th coordinate axis.
We also assume $G$ to be SU(N). If $A_k(x)$ is the 
(Lie algebra-valued) vector potential at the site $x$, then 
\be
u_{xy}\equiv u_{x,k} = e^{iga A_k(x)}\ .
\label{ks1}
\ee 
Here $a$ is the lattice spacing.
The gauge transformations of the link variables are
\be
u_{x,k}\rightarrow \Omega_x u_{x,k}\Omega^{-1}_{x+ka}\ ,
\label{ks2}
\ee
where $\Omega_x$ is the group element at the site $x$.
The variables conjugate to the link variables are electric
field operators associated with each link, which we denote
$E_{x,k}^b$, where the index $b$
 is a color index (the adjoint representation index in an
orthogonal basis of the Lie algebra). 
If the group element $u_{x,k}$ is parameterized
 by a set of variables $\varphi_{x,k}^b$ then the electric
field operator is the Lie algebra generator for each
link  
\be
E_{x,k}^b = -iJ^{bc}(\varphi_{x,k})\,\frac{\pl}{\pl \varphi^c_{x,k}}\ ,
\label{ks3}
\ee
where the functions $J^{bc}(\varphi_{x,k})$ are chosen so that
\be
\left[E_{x,k}^c, E_{y,j}^b\right] 
= i\delta_{xy}\delta_{kj}f^{bc}{}_e E^e_{x,k}\ .
\label{ks4}
\ee
The Kogut-Susskind Hamiltonian reads
\ba
H&=& H_0 +V\ ,
\label{ks5}\\
H_0  &=& \frac{g_0^2}{2a}\sum_{(x,k)} E_{x,k}^2\ ,
\label{ks6}\\
V &=& \frac{2N}{ag^2_0}\sum_p \left(1-N^{-1}{\rm Re}\,\tr u_p\right)\ ,
\label{ks7}
\ea
where $g_0$ is the bare coupling constant, $u_p$ is the product
of the link variables around the plaquette $p$. 

As it stands the kinetic energy $H_0$ is a sum of the quadratic
Casimir operators of the group at each link. 
It is a self-adjoint operator with respect to the natural 
measure on the configuration space being a product
of the Haar measures $d\mu_G(u_{x,k})$ over all links.
In the case of SU(2),
$H_0$ is nothing but the kinetic energy of
{\em free} quantum three-dimensional rotators. In general,
the kinetic energy describes a set of {\em non-interacting}
particles moving on the group manifold as follows from
the commutation relation (\ref{ks4}). We shall also call these
particles generalized rotators.
The magnetic potential
energy (\ref{ks7}) describes the coupling of the generalized rotators. 
Let $\{u'\}$ and $\{u\}$,  respectively, be collections
of initial and final configurations of the generalized rotators.
To construct the path integral representation of the transition
amplitude $U_t(\{u\},\{u'\})$, we make use of the modified Kato-Trotter
formula (\ref{pr4}) for gauge systems. The projector operator is
just the group average at each lattice site with the Haar measure
$d\mu_G(\Omega_x)$ normalized to unity.

As before, the crucial step is to establish the projected form
of the free transition amplitude. The entire information
about the geometry of the orbit space is encoded into it.
An important observation is that the free transition amplitude
is factorized into a product of the transition amplitude for each
generalized rotator. But the amplitude for a single free particle
on the group manifold is well known due to some nice work of
Marinov and Terentiev \cite{mar}. Let the amplitude for a single rotator
be $U_t^0 (u,u')$, then the gauge invariant transition amplitude
associated with the Dirac operator approach for a system
of rotators reads
\be
U_t^{0D}(\{u\},\{u'\}) = \int_G\prod_{x}d\mu_G(\Omega_x)
\, \prod_{x,k} U_t^0(u_{x,k}, \Omega_x u_{x,k}^\prime \Omega^{-1}_{x+k})
\label{ks8} 
\ee
Due to the invariance of the Casimir operator $H_0$ at each
site with respect to shifts on the group manifold, it is sufficient
to average only one of the arguments of the free transition 
amplitude. Simultaneous right (or left) group shifts of both 
arguments of the free transition amplitude leave the amplitude
unchanged. We now can see how a {\em kinematic} coupling 
of the generalized rotators occurs through the gauge group
average. The uncoupled rotators become coupled and 
factorization of the free transition amplitude over the degrees
of freedom disappears. This phenomenon we have already seen
in soluble gauge models. Observe that each group element 
$\Omega_{x}$ enters into {\em six} transition amplitudes
$U_t^0(u,u')$ associated with six links attached to the site $x$.
This is what makes the gauge average nontrivial even for
the ``free'' Kogut-Susskind quantum lattice gauge theory
(i.e., when the potential is set to zero).  
The projection of the transition amplitude on the gauge
orbit space (regardless of any explicit parameterization of
the latter) induces a nontrivial interaction between physical
degrees of freedom of the Yang-Mills theory. 
The difference
between the Abelian and non-Abelian cases is also clearly
seen in this approach. The projection implicitly enforces
the Gauss law in the path integral, i.e., {\em without} any
gauge fixing. In the Abelian case this is a trivial procedure
because the Gauss law merely requires vanishing of some
canonical momenta ($\pl_iE_i=0$), so the corresponding
part of the kinetic energy simply vanishes without any
effect of the redundant degrees of freedom. From the geometrical
point of view, the orbit space in QED is Euclidean and therefore
no coupling between physical degrees of freedom occurs through
the kinetic energy.

Once the averaging procedure has been defined,
one can proceed with introducing an explicit parameterization
of the physical configuration space. 
For instance, we can introduce the lattice analog
of the Morse functional \cite{zwlat}
\be
M_u(\Omega) = \sum_{x,k}\left[ 1 - N^{-1}{\rm Re}\, \tr\left(
\Omega_xu_{x,k}\Omega_{x+i}^{-1}\right)\right]\ .
\label{ks10}
\ee
The configurations $u_{x,k}$ at which the functional (\ref{ks10}) has
a critical point $\Omega_x=e$, $e$ is the group unity, 
relative minima form the gauge fixing surface, the lattice version of
the Coulomb  gauge.
The modular domain, being  a collection of unique representatives
of each gauge orbit, is 
\be
K = \left\{u_{x,k}: \, M_u(e) \leq M_u(\Omega), 
{\rm for\ all}\ \Omega\in G
\right\}\ .
\label{ks9}
\ee
Clearly, $K$ consists of configurations
at which the Morse function attains absolute minima. 
Let $u_{x,k}$ be from $K$. Then a generic link variable
can always be represented in the form $W_x u_{x,k}W^{-1}_{x+k}$
where $W_x$ is a group element. From the gauge invariance
of the amplitude  (\ref{ks8}) it follows that the initial and
final configurations can be taken from $K$, i.e., the 
amplitude does not depend on the set of  group elements
$W_x$. Having reduced the transition amplitude on the gauge
orbit space parameterized by the configurations (\ref{ks9}),
one may calculate the group average using the stationary
phase approximation in the limit $t=\epsilon\rightarrow 0$
and obtain the modified infinitesimal free transition amplitude
which would contain the information about the geometry and topology
of the orbit space and also an explicit form of the operator
ordering corrections resulting from the reduction of the kinetic
energy operator $H_0$ on the modular domain (\ref{ks9}). 
As in the general case, the amplitude has a unique gauge
invariant continuation outside the modular domain to
the entire gauge fixing surface. Consequently, the group
 averaging integral would have not only one stationary 
point. The sum over Gribov copies would emerge as the
sum over the stationary points of the gauge group average
integral, indicating a possible {\em compactification}
of the physical configuration space similar to what we have learned
with the two dimensional example. 
The structure of the path integral would  
be the same as that found in section 8.7 in the  general case.

Having proved the equivalence of the  path integral 
obtained by the projection method to the Dirac
operator approach and, thereby, ensured gauge invariance 
(despite  using a particular parameterization of the 
orbit space),  one could try to investigate the role
of the orbit space geometry in quantum theory, which partially
reveals itself through the coordinate singularities 
of the chosen parameterization.  This would
require studying the thermodynamic and continuum limits,
e.g., by the methods developed by Zwanziger \cite{zwlat}. 
It is also important
to note that the Coulomb gauge has recently been proved
to be renormalizable \cite{zwb}. This provides a tool to 
control ultraviolet
behavior of the theory in the continuum limit.
To separate the effects of the kinetic and potential energy
would be a hard problem in any approach. But in the strong coupling
limit, the kinetic energy dominates as one sees from the
Kogut-Susskind Hamiltonian \cite{kogut}. This leaves some hope that 
in this limit the effects of the kinetic energy reduced on the 
modular domain $K$ could be accurately studied in the path
integral approach. An investigation of the mass
gap \cite{feyn21} would be especially interesting.

The program can be completely realized in the two-dimensional case
(cf. section 8.8).
The gauge group average can be calculated explicitly 
by means of the decomposition of the transition
amplitude of a particle on a group manifold over
the characters of the irreducible representations
proposed by Marinov and Terentiev \cite{mar}. If the 
orbit space is parameterized by constant link
variables (the Coulomb gauge) $u_{x}=u_{x'}$ for any
$x,x'$, then in the continuum limit one obviously recovers
the transition amplitude (\ref{go3}).
Taking the resolvent of the evolution operator one can
find the mass gap, which would be 
impossible to see, had we neglected the true structure
of the physical configuration space. 
The compactness of the orbit space and the mass gap
follow from the very structure of the path integral
containing the sum over Gribov copies, which
appears as the result of the projection of the transition
amplitude onto the gauge orbit space. 
The projection formalism guarantees that the true geometry of
the gauge orbit space is always appropriately taken
into account in the path integral, whatever gauge is used, 
and thereby provides a right technical tool to study   
nonperturbative phenomena.

\section{Conclusions}

We have investigated the physical phase space structure
in gauge theories and found that its geometrical
structure has a significant effect on the corresponding quantum
theory. The conventional path integral requires a modification
to take into account the genuine geometry of the physical
phase space. Based on the projection method, the necessary
modification has been established, and its equivalence to
the explicitly gauge invariant operator formalism due to Dirac
has also been shown. 
Upon a quantum description of gauge systems, one usually
uses some explicit parameterization of the physical phase space
by local canonical coordinates. Because of a non-Euclidean
geometry of the physical phase space any coordinate description
would in general suffer from coordinate singularities. We
have developed a general procedure for how to cope with such
singularities in the operator and path integral formalisms
for gauge models of the Yang-Mills type. It appeared that
the singularities cannot generally be ignored and have to
be carefully taken into  account in quantum or classical theory
in order to provide the gauge invariance of the theory.

Though all the exact results have been
obtained for soluble gauge models, it is believed that
some essential features of quantum gauge dynamics
on the non-Euclidean physical phase space would also be present  
in the realistic theories. There are several important
problems yet to be solved in nonperturbative quantum
field theory to make some reliable conclusions about
the role of the physical phase space geometry in
quantum Yang-Mills theory.    
The way based on the projection formalism 
in the Kogut-Susskind lattice seems a rather natural
approach to this problem, which ensures agreement
with the operator formalism and leads to the functional
integral that does not depend on any explicit parameterization
of the gauge orbit space. The path integral formalism
based on the projection method gives evidence that
the compactness of the gauge orbit space might be important
for the existence of the mass gap in the theory and,
hence, for the gluon confinement, as has
been conjectured by Feynman.

When constructing the path integral over a non-Euclidean
physical phase space, we have always used a parameterization
where no restriction on the momentum variables has been
imposed. The reason for that is quite clear. The explicit
implementation of the projection on the gauge invariant states
is easier in the configuration space for gauge theories of
the Yang-Mills type. This latter restriction can be dropped, 
and a {\em phase-space}
path integral measure {\em covariant} under general canonical
transformations on the physical phase space can be found 
\cite{cfq1,cfq2,cfq3} for systems with a finite number  
of degrees of freedom. The corresponding path integral
does not depend on the parameterization of the physical
phase space and, in this sense, is coordinate-free.  
The problem remains open in quantum
gauge field theory. 

Despite many unsolved problems, it is believed
that the soluble examples studied above in detail
and the concepts introduced
would provide a good starting point for 
this exciting area of research.

\vskip 0.3cm
\noindent
{\bf Acknowledgments}

\vskip 0.3cm
I am deeply indebted to John Klauder whose encouragement,
support, interest and numerous comments have helped me
to accomplish this work. I also wish to thank him for
a careful reading of the manuscript, many suggestions
to improve it, and for stimulating discussions on the topics
of this review. I express my gratitude to Lev Prokhorov
from whom I learned a great deal about gauge theories and path
integrals. On this occasion I would like to thank
I.A. Batalin, L. Baulieu, A.A. Broyles,
T. Heinzl, M. Henneaux, H. H\"uffel,
J.C. Mourao, F.G. Scholtz, 
M. Shaden, T. Strobl, P. van Baal, C.-M. Viallet and D. Zwanziger
for fruitful discussions, references and comments that were
useful for me in this work.

It is a pleasure for me to thank 
the Departments of Physics and Mathematics of the University
of Florida for the warm hospitality
extended to me during my stay in Gainesville. 

\end{document}